\RequirePackage[T1]{fontenc}
\documentclass[12pt]{article}

\usepackage[height=8.85in,width=6.45in]{geometry}

\usepackage[utf8]{inputenc}
\usepackage{amsmath}
\usepackage{amssymb}
\usepackage{mathtools}
\numberwithin{equation}{section}
\usepackage{slashed}
\usepackage{braket}
\usepackage[svgnames]{xcolor}
\usepackage[colorlinks,citecolor=DarkGreen,linkcolor=FireBrick]{hyperref}
\usepackage{cite}
\usepackage{graphicx}
\usepackage{tikz}
\usepackage{tikz-cd}
\usepackage{times}
\usepackage{courier}
\usepackage{bm}
\usepackage{subfig}

\usepackage[normalem]{ulem}

\usepackage{soul}

\usetikzlibrary{decorations.pathreplacing,decorations.markings}

\tikzset{
    partial ellipse/.style args={#1:#2:#3}{
        insert path={+ (#1:#3) arc (#1:#2:#3)}
    }
}

\tikzset{
  on each segment/.style={
    decorate,
    decoration={
      show path construction,
      moveto code={},
      lineto code={
        \path [#1]
        (\tikzinputsegmentfirst) -- (\tikzinputsegmentlast);
      },
      curveto code={
        \path [#1] (\tikzinputsegmentfirst)
        .. controls
        (\tikzinputsegmentsupporta) and (\tikzinputsegmentsupportb)
        ..
        (\tikzinputsegmentlast);
      },
      closepath code={
        \path [#1]
        (\tikzinputsegmentfirst) -- (\tikzinputsegmentlast);
      },
    },
  },
  mid arrow/.style={postaction={decorate,decoration={
        markings,
        mark=at position .5 with {\arrow[#1]{stealth}}
      }}},
}

\usetikzlibrary{shapes.multipart}
\usetikzlibrary{decorations.pathmorphing}
\tikzset{snake it/.style={decorate, decoration=snake}}

\usepackage{xcolor}
\usepackage{mdframed}

\renewenvironment{figure}[1][]{
  \begin{originalfigure}[#1]
    \begin{mdframed}[linecolor=black!0,backgroundcolor=black!1]
}{
    \end{mdframed}
  \end{originalfigure}
}

\def\cC{\mathcal{C}}

\newcommand{\link}{\text{link}}
\definecolor{dgreen}{rgb}{0, 0.55, 0}

\def\CA{{\mathcal A}}

\def\CC{{\mathcal C}}

\def\CN{{\mathcal N}}
\def\CO{{\mathcal O}}
\def\CP{{\mathcal P}}

\def\CX{{\mathcal X}}

\def\CZ{{\mathcal Z}}
\newcommand{\Z}{\mathbb{Z}}
\newcommand{\ZZ}{\mathbb{Z}}

\usepackage{datetime}
\usepackage{dsfont}

\usetikzlibrary{shapes.geometric}
\newcommand{\bea}{\begin{eqnarray}}
\newcommand{\eea}{\end{eqnarray}}

\def\cX{{\mathcal X}}
\def\cN{{\mathcal N}}
\def\cC{{\mathcal C}}
\def\cP{{\mathcal P}}
\def\cD{{\mathcal D}}
\def\cZ{{\mathcal Z}}

\def\a{\alpha}
\def\cA{{\mathcal A}}
\def\eps{\epsilon}
\def\no{\nonumber}

\begin{document}
\begin{titlepage}

\begin{flushright}
\end{flushright}

\vskip 3cm

\begin{center}

{\Large \bfseries Symmetry TFTs for  Non-Invertible Defects}

\vskip 1cm
Justin Kaidi$^{1,2,3}$, Kantaro Ohmori$^4$, and Yunqin Zheng$^{3,5}$
\vskip 1cm

\begin{centering}
\begin{tabular}{ll}
$^1$&Department of Physics, \\& University of Washington, Seattle, WA, 98195, USA
\\
  $^2$&Simons Center for Geometry and Physics, \\& Stony Brook University, Stony Brook, NY 11794-3636, USA
  \\
  $^3$&Kavli Institute for the Physics and Mathematics of the Universe, \\
& University of Tokyo,  Kashiwa, Chiba 277-8583, Japan
  \\
  $^4$&Department of Physics, The University of Tokyo, \\&Bunkyo-ku, Tokyo 113-0033, Japan\\
$^5$&Institute for Solid State Physics, \\
&University of Tokyo,  Kashiwa, Chiba 277-8581, Japan\\
\end{tabular}
\end{centering}

\vskip 1cm

\end{center}

\noindent
Given any symmetry acting on a $d$-dimensional quantum field theory, there is an associated $(d+1)$-dimensional topological field theory known as the Symmetry TFT (SymTFT). The SymTFT is useful for decoupling the universal quantities of quantum field theories, such as their generalized global symmetries and 't Hooft anomalies, from their dynamics. In this work, we explore the SymTFT for theories with Kramers-Wannier-like duality symmetry in both $(1+1)$d and $(3+1)$d quantum field theories. 
After constructing the SymTFT, we use it to reproduce the non-invertible fusion rules of duality defects, and along the way we generalize the concept of duality defects to \textit{higher} duality defects.  We also apply the SymTFT to the problem of distinguishing intrinsically versus non-intrinsically non-invertible duality defects in $(1+1)$d.

\end{titlepage}

\setcounter{tocdepth}{2}
\tableofcontents

\section{Introduction and summary}

Kramers-Wannier duality, originally identified in the $(1+1)$d Ising model, is the simplest example of a so-called ``non-invertible symmetry." Non-invertible symmetries have been the subject of an extensive literature in $(1+1)$d (see e.g.\  \cite{verlinde1988fusion,Petkova:2000ip,Fuchs:2002cm,Davydov:2011kb,Bhardwaj:2017xup,Chang:2018iay,Lin:2022dhv,Komargodski:2020mxz,Tachikawa:2017gyf, Frohlich:2004ef, Frohlich:2006ch, Frohlich:2009gb, Carqueville:2012dk, Brunner:2013xna, Huang:2021zvu, Thorngren:2019iar, Thorngren:2021yso, Lootens:2021tet, Huang:2021nvb, Inamura:2022lun}), but have only recently been generalized to spacetime dimensions greater than two  \cite{Kaidi:2021xfk,Choi:2021kmx, Koide:2021zxj,Choi:2022zal,Hayashi:2022fkw,Arias-Tamargo:2022nlf,Roumpedakis:2022aik,Bhardwaj:2022yxj,Kaidi:2022uux,Choi:2022jqy,Cordova:2022ieu,Antinucci:2022eat,Bashmakov:2022jtl,Damia:2022rxw,Damia:2022bcd,Choi:2022rfe,Lu:2022ver,Bhardwaj:2022lsg,Lin:2022xod,Bartsch:2022mpm,Apruzzi:2022rei,GarciaEtxebarria:2022vzq, Benini:2022hzx, Wang:2021vki, Chen:2021xuc, DelZotto:2022ras, Heckman:2022muc}.
The constructions of non-invertible symmetries in higher dimensions that have appeared in the literature so far involve the following techniques, 
\begin{itemize}
\item Gauging a discrete symmetry in a theory with particular `t Hooft anomaly \cite{Kaidi:2021xfk};
\item Gauging a non-anomalous symmetry in half of the spacetime \cite{Choi:2021kmx, Choi:2022zal};
\item Gauging a non-normal finite subgroup of the global symmetry \cite{Bhardwaj:2022yxj,Nguyen:2021yld, Arias-Tamargo:2022nlf, Antinucci:2022eat,Arias-Tamargo:2022nlf};
\item Gauging a higher form symmetry along a higher co-dimension submanifold \cite{Roumpedakis:2022aik};
\item Gauging a diagonal symmetry between the quantum field theory and a lower dimensional topological theory on a defect \cite{Bhardwaj:2022lsg}. 
\end{itemize}
Applications of non-invertible symmetries have ranged from constraints on the IR phases of supersymmetric and non-supersymmetric gauge theories \cite{Choi:2022zal, Choi:2021kmx, Choi:2022rfe } to the generation of new strongly-coupled theories via twisted compactification \cite{Kaidi:2022uux}, and also include the obtaining of selection rules in real-world models such as QED and QCD \cite{Choi:2022jqy, Cordova:2022ieu}. 

Despite their newfound utility, sometimes statements made using non-invertible symmetries in a theory $\cX$ can be recast as statements involving only invertible symmetries in a theory $\phi(\cX)$, where $\phi$ is some appropriate topological manipulation. The set of topological manipulations $\phi$ includes gauging of finite (non-anomalous) symmetries, as well as stacking with invertible phases. Non-invertible symmetries which \textit{cannot} be recast as (potentially anomalous) invertible symmetries upon appropriate application of $\phi$ were dubbed ``intrinsically" non-invertible in \cite{Kaidi:2022uux}, whereas  non-invertible symmetries which \textit{can} be recast as invertible symmetries were dubbed ``non-intrinsically" non-invertible. The upshot is that the non-invertibility of a given symmetry, and indeed the full fusion (higher-)category, is not an invariant under topological manipulations. 

With this in mind, it is interesting to search for an object which \textit{is} invariant under topological operations $\phi$. Indeed, such an object is known to exist: given a $d$-dimensional theory with symmetry captured by a fusion $(d-1)$-category $\cC$, we may define a $(d+1)$-dimensional topological field theory SymTFT$(\cC)$ with the property that SymTFT$(\cC)=$SymTFT$(\phi(\cC))$ for any topological manipulation $\phi$. The topological theory SymTFT$(\cC)$ is known as the \textit{symmetry topological field theory} \cite{Freed:2012bs,Freed:2018cec,Gaiotto:2020iye,Apruzzi:2021nmk,Apruzzi:2022dlm, Burbano:2021loy, Freed:2022qnc}  (see also related developments in condensed matter physics \cite{Ji:2019jhk,Kong:2020cie,Ji:2021esj,Chatterjee:2022kxb, Chatterjee:2022tyg, Moradi:2022lqp}), and will be the main focus of this paper.

\subsection{Symmetry TFT and boundary conditions}

The SymTFT of a theory $\cX$ in $d$-dimensions with global symmetry $\CC$ is a topological theory SymTFT($\cC)$ in $(d+1)$-dimensions which, when compactified on an interval with appropriate boundary conditions, gives back the theory $\cX$ of interest. One of the many nice properties of the SymTFT is that it decouples the dynamics of $\cX$ from the symmetries. This makes the action of the various topological manipulations $\phi$ more transparent: indeed, by choosing appropriate topological boundary conditions for the SymTFT before compactification to $d$-dimensions, we can obtain not only the original $\cX$, but any theory of the form $\phi(\CX)$. 

\begin{figure}[!tbp]
	\centering
	\begin{tikzpicture}[scale=0.8]

	\shade[line width=2pt, top color=blue!30, bottom color=blue!5] 
	(0,0) to [out=90, in=-90]  (0,3)
	to [out=0,in=180] (6,3)
	to [out = -90, in =90] (6,0)
	to [out=180, in =0]  (0,0);
	
		\draw[very thick] (-7,0) -- (-7,3);
	\node[below] at (-7,0) {$Z_{\CX}[A]$};

	\draw[thick, snake it, <->] (-1.7,1.5) -- (-5, 1.5);
	
	\draw[thick] (0,0) -- (0,3);
	\draw[thick] (6,0) -- (6,3);
	\node at (3,1.5) {SymTFT($\cC$)};
	\node[below] at (0,0) {$\langle D(A)|$};
	\node[below] at (6,0) {$|\cX\rangle $}; 
	
	\end{tikzpicture}
	\\
	\vspace{0.5 in}
	\begin{tikzpicture}[scale=0.8]

	\shade[line width=2pt, top color=blue!30, bottom color=blue!5] 
	(0,0) to [out=90, in=-90]  (0,3)
	to [out=0,in=180] (6,3)
	to [out = -90, in =90] (6,0)
	to [out=180, in =0]  (0,0);
	
		\draw[very thick] (-7,0) -- (-7,3);
	\node[below] at (-7,0) {$Z_{\CX/G}[A]$};

	\draw[thick, snake it, <->] (-1.7,1.5) -- (-5, 1.5);
	
	\draw[thick] (0,0) -- (0,3);
	\draw[thick] (6,0) -- (6,3);
	\node at (3,1.5) {SymTFT($\cC$)};
	\node[below] at (0,0) {$\langle N(A)|$};
	\node[below] at (6,0) {$|\cX\rangle $}; 
	\end{tikzpicture} 
	
	\caption{Schematic picture of the SymTFT. By imposing  Dirichlet (resp. Neumann) boundary conditions on the left and the non-topological boundary condition (\ref{eq:dynamboundary}) on the right, we may compactify to obtain $\cX$ (resp. $\cX/G$). In general, there exist other topological boundary conditions as well.}
	\label{fig:symmTFTidea}
\end{figure}
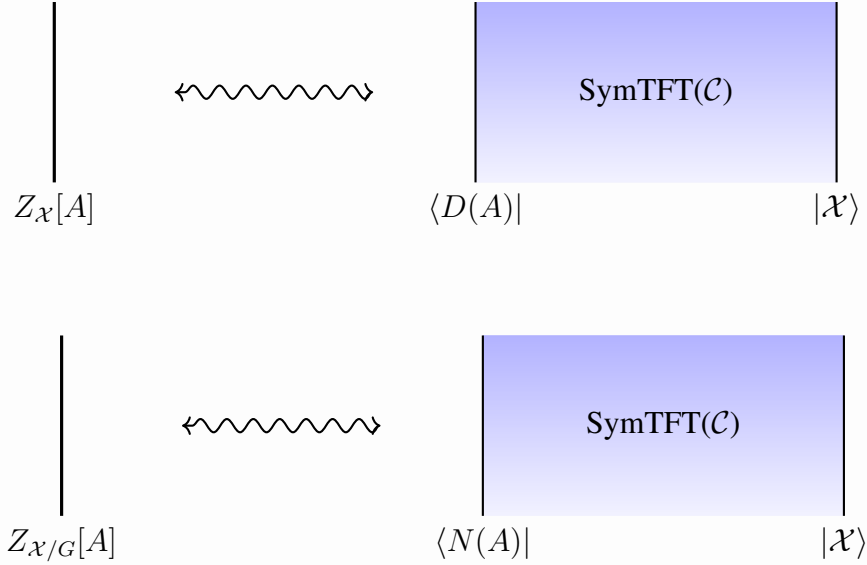

The basic idea is illustrated in Figure \ref{fig:symmTFTidea}. The $(d+1)$-dimensional SymTFT is placed on an interval, where the right boundary is endowed with non-topological ``enriched Neumann" boundary conditions capturing the dynamics of $\CX$, whereas the left boundary condition is topological. Both boundaries can be labeled by appropriate elements of the state space of SymTFT$(\cC)$. 
The dynamical boundary of SymTFT$(\cC)$ is taken to be 
\bea
\label{eq:dynamboundary}
| \cX\rangle = \sum_a Z_\cX[a] | a\rangle~. 
\eea
When $\CC$ is a finite group $G$, the $a$ above represents the collective set of flat connections of $G$, and $Z_{\cX}[a]$ denotes the partition function of $\cX$ coupled to gauge fields $a$ (on some fiducial $d$-manifold). When $\CC$ is a fusion (higher-)category rather than a group, the label $a$ is an appropriate collective label for the topological defects in the theory. 

The topological boundary of SymTFT$(\cC)$ can take a number of forms, with common options being Dirichlet or Neumann boundary conditions. Dirichlet boundary conditions fix the fields $a$ to certain values $A$, whereas Neumann boundary conditions allow $a$ to fluctuate freely. In the state notation, these can be written as 
\bea\label{eq:bc}
\begin{split}
    \mathrm{Dirichlet:}& \hspace{0.5 in} |D(A)\rangle = \sum_a \delta(a-A) |a\rangle ~,
\\
\mathrm{Neumann:}& \hspace{0.5 in} |N(A)\rangle = \sum_a \exp\left(-i \int a \cup A\right) |a\rangle ~. 
\end{split}
\eea
The normalization factor in $\int a\cup A$ depends on the symmetry and is suppressed  here.  We will restore it in the main text. 

Because the SymTFT is topological, the length of the interval is unimportant and can be shrank to zero size, upon which one obtains a $d$-dimensional theory whose partition function can be computed by taking the inner product between the states on the left and right boundaries. In the case of Dirichlet boundary conditions, one obtains 
\bea
\langle D(A)| \cX \rangle = \sum_{a, a'} Z_{\CX}(a) \,\delta(a'-A) \langle a' | a \rangle = \sum_{a, a'} Z_{\CX}(a) \,\delta(a'-A) \delta(a-a') = Z_{\cX}(A)
\eea
and hence one reproduces the original $d$-dimensional theory $\cX$, coupled to background fields $A$. 
On the other hand, by putting Neumann boundary conditions on the right, we obtain 
\begin{equation}
    \langle N(A)| \cX \rangle =\sum_{a, a'} Z_{\CX}(a)\,\exp\left(i \int a' \cup A\right) \langle a' | a \rangle = \sum_{a} Z_{\CX}(a) \,\exp\left(i \int a \cup A\right)  = Z_{\cX/G}(A)~.
\end{equation}
In other words, in this case we obtain a $d$-dimensional theory $\phi(\cX)$, where $\phi$ in this case represents gauging the discrete symmetry $G$ (which is indeed a topological operation).  Choosing mixed Dirichlet-Neumann boundary conditions will allow us to obtain $\cX$ with a subgroup of $G$ gauged. 

More generally, one can dress either boundary condition in \eqref{eq:bc} by a $d$-cocycle $\nu_d$.  The corresponding theories obtained by shrinking the slab are $\cX$ stacked with a counterterm, and $\cX/G$ with the gauging done with discrete torsion associated with the $d$-cocycle,
\begin{equation}
\begin{split}
    \ket{D(A)_{\nu_d}}= \sum_{a} \delta(a-A)\exp\left(-i \int \nu_d(a)\right)\ket{a}  &\longleftrightarrow  \langle D(A)_{\nu_d}| \cX \rangle= Z_{\cX}(A) \exp\left(i \int \nu_d(A)\right)\\
    \ket{N(A)_{\nu_d}}= \sum_{a} \exp\left(-i \int a\cup A -i \int \nu_d(a)\right)\ket{a}  &\longleftrightarrow  \langle N(A)_{\nu_d}| \cX \rangle= Z_{(\cX\times \nu_d)/G}(A) \\
\end{split}
\end{equation}
One therefore expects that any two theories $\CX$ and $\CX'$ that are related by a topological manipulation $\phi$, i.e. $\CX'=\phi(\CX)$, can be obtained by the slab construction with the \emph{same} SymTFT but with \emph{different} topological boundary conditions. In particular, this implies that the SymTFT in the bulk is  invariant under topological manipulations.

\subsection{Invertible symmetry and Dijkgraaf-Witten theories}
\label{sec:anomalyDW}

The SymTFT takes a particularly simple form when the symmetry in question is group-like. To see this, begin by considering a theory $\cX$ with background gauge fields $A$ and anomaly $\alpha$. The anomaly inflow paradigm \cite{Freed:2016rqq}, in modern language, says that such a theory may be realized on the boundary of an appropriate invertible phase Inv$^\alpha(A)$; see Figure \ref{fig:DWandanom}(a).

\hspace{-0.5 in}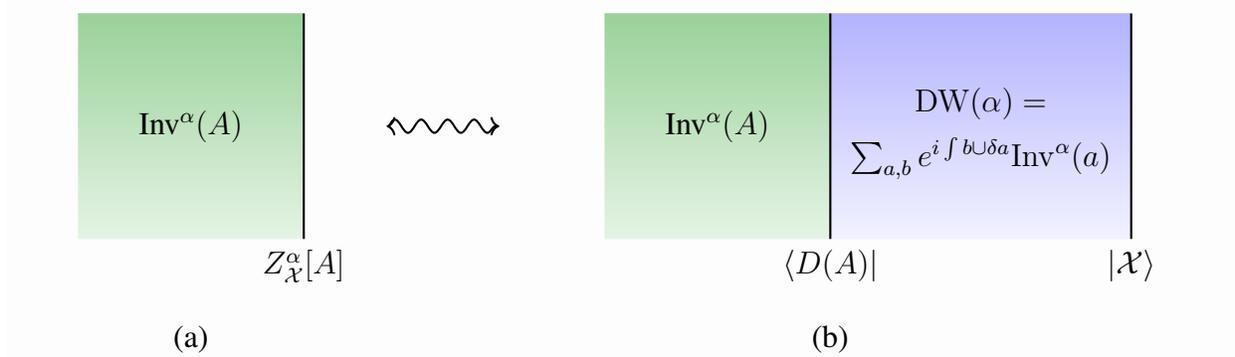
\begin{figure}[t]
	\centering
	\begin{tikzpicture}

	\shade[line width=2pt, top color=dgreen!40, bottom color=dgreen!10] 
	(0,0) to [out=90, in=-90]  (0,3)
	to [out=0,in=180] (3,3)
	to [out = -90, in =90] (3,0)
	to [out=180, in =0]  (0,0);

	\shade[line width=2pt, top color=blue!30, bottom color=blue!5] 
	(7+3,0) to [out=90, in=-90]  (7+3,3)
	to [out=0,in=180] (11+3,3)
	to [out = -90, in =90] (11+3,0)
	to [out=180, in =0]  (0,0);
	
	\shade[line width=2pt, top color=dgreen!40, bottom color=dgreen!10] 
	(11-4,0) to [out=90, in=-90]  (11-4,3)
	to [out=0,in=180] (14-4,3)
	to [out = -90, in =90] (14-4,0)
	to [out=180, in =0]  (0,0);
	
	\draw[thick] (3,0) -- (3,3);

	\node[below] at (1.5, -1) {(a)};
	
	\draw[thick, snake it, <->] (4.1,1.5) -- (5.6, 1.5);

	\draw[thick] (7+3,0) -- (7+3,3);
	\draw[thick] (11+3,0) -- (11+3,3);

	\node[below] at (10, -1) {(b)};
	
	\node at (1.5,1.5) {Inv$^\alpha(A)$};
	\node[below] at (3,0) {$Z^\alpha_{\CX}[A]$};
	\node at (9+3,1.8) {$\mathrm{DW}(\alpha) =$};
	\node[below] at (9+3,1.5) {$ \sum_{a,b}  e^{i \int b \cup\delta a}\mathrm{Inv}^\alpha(a) $};
	\node[below] at (7+3,0) {$\langle D(A)| $};
	\node[below] at (11+3,0) {$\ket{\cX}$}; 
	\node at (12.5-4,1.5) {Inv$^\alpha(A)$};
	
	\end{tikzpicture}
	
	\caption{A theory with anomaly $\alpha$ can be realized on the boundary of an invertible phase Inv$^\alpha(A)$. This is possible if and only if the Symmetry TFT is a (generalized) DW theory. }
	\label{fig:DWandanom}
\end{figure}

Given Figure \ref{fig:DWandanom}(a), one can promote the background gauge field $A$ to a dynamical gauge field $a$ both on the right boundary and in a slab in the bulk. The bulk theory within this slab is then a nontrivial TQFT, namely a
(generalized) Dijkgraaf-Witten (DW)   theory\footnote{The qualifier ``generalized" here refers to the fact that the relevant invertible phases are not restricted to group cohomology elements.}
\bea\label{eq:DW}
\mathrm{DW}(\alpha) = \sum_{a,b}  \exp\left(i \int b \cup\delta a\right) \mathrm{Inv}^\alpha(a)~,
\eea
where both $a$ and $b$ are invariant $G$-valued cochains, and as commented above we suppress the normalization in the BF coupling for now.   
On the right of the slab, one further imposes Dirichlet boundary conditions to pin the dynamical field $a$ to the background $A$.   
The theory $\cX$ is recovered by taking the thin slab limit where the non-topological boundary condition $\ket{\cX}$ and the Dirichlet boundary condition $\ket{D(A)}$ collide. 
We conclude that the SymTFT for an invertible symmetry with an anomaly is given by a (generalized) DW theory \eqref{eq:DW}, i.e. a discrete gauge theory, potentially with a non-trivial twist.

\subsection{Symmetry TFT for duality defects}

Having understood the SymTFT for theories with only group-like (i.e. invertible) symmetries, we may next proceed to consider theories with non-invertible symmetries. In $(1+1)$d, the simplest categories with non-invertible symmetries are the so-called \textit{Tambara-Yamagami categories} $TY(G)$ with $G$ an Abelian group. The objects of this category consist of invertible lines $L_g$ for each $g \in G$, together with a single non-invertible line $\cN$ with fusion rules\footnote{To fully specify the category $TY(G)$, one must specify more than just the fusion rules; the additional data required is the Frobenius-Schur indicator $\varepsilon\in \ZZ_2$ together with a bicharacter $\chi \in H^2(G,U(1))$. In this paper we will mostly work with the case of trivial $\varepsilon$ and $\chi$. 
}
\bea
L_{g_1}\times L_{g_2} = L_{g_1 \cdot g_2}~, \hspace{0.5 in} L_g \times \cN = \cN \times L_g = \cN~, \hspace{0.5 in} \cN \times \cN = \sum_{g \in G} L_g~. 
\eea
A higher-categorical analog of $TY(G)$ fusion rules exists in higher dimensions, as will be discussed for $G=\ZZ_N^{(1)}$ in $(3+1)$d in the main text. The goal of this paper is to construct the SymTFTs for $(1+1)$d theories with $TY(\ZZ_N)$ symmetry, as well as the analogs in $(3+1)$d. 

Before beginning this analysis, we should note that the SymTFT is, at least implicitly, already known for \textit{any} fusion (higher-)category. Indeed, given a theory with fusion (higher-)category $\cC$, the SymTFT is always given by Turaev-Viro theory on $\cC$, or equivalently as Reshetikhin-Turaev theory on the Drinfeld center $\cZ(\cC)$ (subject to important  caveats\footnote{It is known that in $(2+1)$d, given a spherical fusion category $\CC$, one can define the Turaev-Viro theory of $\CC$ as an extended TQFT, which is equivalent to Reshetikhin-Turaev theory on the Drinfeld center $\cZ(\cC)$ \cite{https://doi.org/10.48550/arxiv.1004.1533}. For fusion higher-categories, the authors expect that a similar statement holds---namely that the defects of the SymTFT are captured by the Drinfeld center of the corresponding fusion higher-category---though they are unaware of a definition of Reshetikhin-Turaev theory in higher dimensions. 
}). The latter definition makes it clear that the SymTFT of $\CC$ is the quantum double $\cZ(\cC)$ of $\cC$. This matches with the results obtained above for group-like symmetry $\cC = \mathrm{Vec}^\alpha(G)$ discussed above, since Turaev-Viro theory on $\mathrm{Vec}^\alpha(G)$ is known to be equivalent to DW theory with DW action $\alpha$. 

For the case of $\cC = TY(\ZZ_N)$ in $(1+1)$d, various properties of the Drinfeld center $\cZ(TY(\ZZ_N))$ are known in detail in the mathematics literature \cite{izumi2001structure,gelaki2009centers}, as well as in the physics literature \cite{Barkeshli:2014cna,2015arXiv150306812T}. As such, the results we present in $(1+1)$d are not new. However, the way in which we obtain these results will be new, and will carry the virtue of being more easily generalizable to the case of higher dimensions.

As will be described in the rest of this work, the SymTFTs for theories with duality symmetries can be obtained by starting with appropriate DW theories and gauging an electro-magnetic (EM) exchange symmetry. As part of our analysis, we will identify an explicit Lagrangian description of the EM-gauged theory, and will use it to obtain data about the spectrum of objects and (higher-)morphisms in the relevant categories $\mathcal{Z}(\cC)$.

\subsection{Organization and Conventions}

This paper is organized as follows. In the first half of the paper, we focus on the $(2+1)$d SymTFT for $(1+1)$d theories with $TY(\ZZ_N)$ symmetry. To prepare for this, we begin in Section \ref{sec:1+1ddualityinterface} by reviewing the construction of duality defects in $(1+1)$d via half-space gauging, as well as the derivation of their fusion rules. We pay special attention  to the normalizations appearing in the fusion rules, which we feel have not been carefully treated in the literature yet. After reviewing the fusion rules of $TY(\ZZ_N)$, we then construct the SymTFT in two steps. We begin in Section \ref{sec:symTFT3dZN} by constructing the SymTFT for a theory with non-anomalous $\ZZ_N^{(0)}$ symmetry. As discussed above, for theories with only group-like symmetries the SymTFT should be a DW theory, and in the current case we expect it to be $\ZZ_N$ gauge theory with trivial DW action. The spectrum of topological operators of this $\ZZ_N$ gauge theory, including the so-called ``twist defects," are studied in detail. Interestingly, the twist defects admit an interpretation as higher duality interfaces, generalizing the constructions of non-invertible defects/interfaces listed in the beginning of this introduction.   To obtain the SymTFT for $TY(\ZZ_N)$, we next gauge the $\ZZ_2^{\mathrm{EM}}$ symmetry of the bulk $\ZZ_N$ gauge theory. This is done in Section \ref{sec:3dDTYZN} in two different ways. In Section \ref{sec:ZTYZN}, the gauging is done at the level of the category by tracing the behavior of all of the simple objects of $\ZZ_N$ gauge theory under gauging of EM and carefully adding in the twisted sector objects. In Section \ref{Sec:3dcocycle}, the gauging is done at the level of the Lagrangian by promoting cocycles to twisted cocycles.

The second half of the paper gives the analogous construction of the $(4+1)$d SymTFT for $(3+1)$d theories with duality defects. As before, we begin in Section \ref{sec:3+1ddualityinterface} by reviewing the construction of duality defects in $(3+1)$d via half-space gauging, as well as the derivation of their fusion rules. Once again, we pay special attention to the normalization appearing in these fusion rules. Having done so, we then construct the SymTFT for the duality defects in two steps. First in Section \ref{sec:symTFT5d} we construct the SymTFT for $\ZZ_N^{(1)}$ one-form symmetry in $(3+1)$d, which is again simply a $\ZZ_N^{(1)}$ gauge theory, and analyze its spectrum of topological operators as well as their fusions. Then in Section \ref{sec:5dSymTFTTYZN} we gauge the $\ZZ_4^{\mathrm{EM}}$ EM duality of the $\ZZ_N^{(1)}$ gauge theory. This gauging is again done in two ways: in Section \ref{sec:5dgaugingmethod1} it is done by tracing the behavior of the various objects and morphisms of the ungauged theory under gauging, while in Section \ref{sec:5dcocycle} it is done by writing down an explicit Lagrangian in terms of twisted cocycles. 

Finally, we close in Section \ref{sec:SymTFTspinnonspin} by describing one application of the SymTFT, namely to the question of intrinsic vs non-intrinsic non-invertibility discussed in the beginning of this introduction. For example, we will see that when $N$ is not a perfect square, all duality defects in bosonic $(1+1)$d theories are intrinsically non-invertible. 

Various computations needed in the main text are relegated to appendices. In Appendix \ref{app:linecorr} we derive the commutation relations between $k$-dimensional operators in $(2k+1)$d $\Z_N^{(k-1)}$ gauge theory, which are crucial for understanding the fusion rules of objects in the SymTFT. In Appendix \ref{app:junctioncharge}, we explain how to measure the EM-charge of junctions in $\ZZ_N$ gauge theory, which is crucial for obtaining the fusion rules of the EM-gauged theory. In Appendix \ref{app:2dactioninvariance} we provide details for the proof of gauge invariance of the $(2+1)$d Lagrangian describing the EM-gauged theory, which is written in terms of twisted cocycles. In Appendix \ref{app:fusionDCTY}, we review results from the math literature regarding the modular $S$ and $T$ matrices for the Drinfeld center of $TY(\ZZ_N)$. These results provide a useful check of the results obtained in the first half of this paper. In Appendix \ref{app:DEM} we give details on the form of the EM duality defects in the $(4+1)$d SymTFT before gauging, which depend crucially on whether $N$ is odd, $N=2$, or $N$ is even with $N>2$. Finally, in Appendix \ref{app.absorbK} we provide a physical argument for the assignment of quantum defect lines after gauging $\Z_4^{\text{EM}}$ to operators whose constituents are invariant under the $\Z_2^{\text{EM}}$ normal subgroup of $\Z_4^{\text{EM}}$.

We close this introduction by listing some conventions that will be used throughout the paper:
\begin{enumerate}
    \item 1-form gauge fields are denoted as $a$ or $A$ (potentially with a hat or subscript) depending on whether they are dynamical or background fields. Likewise 2-form gauge fields are denoted by $b$ or $B$ depending on whether they are dynamical or background. 
    \item Cup products are mostly suppressed, e.g. $aA := a\cup A$. We will explicitly write down the cup product only when it is necessary to distinguish it from the twisted cup product. 
    \item We use $X_{d}$ to denote the $d$-dimensional spacetime, and $M_n$ to denote worldvolumes of $n$-dimensional topological defects/interfaces in spacetime. 
    \item We denote the global fusion by $\times$ and the global direct sum as $+$. On the other hand, we denote the local fusion by $\otimes$ and the local direct sum as $\oplus$.  We will also use the notation $\mathcal{O}_c \subset\mathcal{O}_a\otimes \mathcal{O}_b$ to indicate that there exists a junction between incoming topological defects $\mathcal{O}_a$ and $\mathcal{O}_b$ and an outgoing topological defect $\mathcal{O}_c$.  
\end{enumerate}

\section{Duality interfaces in $(1+1)$d}
\label{sec:1+1ddualityinterface}
 Before analyzing the SymTFT for theories with duality defects, we begin with a detailed discussion of the duality defects themselves, together with their fusion rules. Special attention will be paid to normalizations, which are surprisingly subtle. We begin in $(1+1)d$ where things are somewhat simpler. 

\subsection{Duality interfaces from half-space gauging}
\label{sec:1+1ddualityinterfacedef}

Consider a non-spin QFT $\CX$ in $(1+1)$d with an anomaly free $\Z_N^{(0)}$ zero-form  global symmetry, defined on a closed two-dimensional spacetime $X_2$.  We denote the $\Z_N^{(0)}$  background gauge field as $A$, and the partition function as $Z_{\CX}[X_2,A]$. Gauging $\Z_N^{(0)}$ gives a new theory $\CX/\Z_N$, 
\begin{eqnarray}\label{eq:2dgauging}
Z_{\CX/\Z_N}[X_2,A] = \frac{1}{|H^0(X_2, \Z_N)|} \sum_{a\in H^1(X_2, \Z_N)} Z_{\CX}[X_2,a]\, e^{ \frac{2\pi i}{N} \int_{X_2} a A}~,
\end{eqnarray}
where now $A\in H^1(X_2, \widehat \Z_N)$ is the background field of the quantum symmetry $\widehat\Z_N^{(0)}$ after gauging. The defect that generates this quantum symmetry is  the Wilson line of $a$, 
\begin{eqnarray}
\eta(\gamma) = \exp\left(\frac{2\pi i}{N}\oint_{\gamma} a\right)~.
\end{eqnarray}

Let us comment on the normalization factor in \eqref{eq:2dgauging}, which is introduced to subtract out gauge redundancies. It is straightforward to check that gauging $\Z^{(0)}_N$ twice maps the theory $\CX$ back to itself, up to an additional 
Euler counterterm $\chi[X_2,\Z_N]^{-1}$, with 
\begin{eqnarray}\label{eq:EulerCC2d}
\chi[X_2,\Z_N]=\frac{|H^0(X_2,\Z_N)||H^2(X_2,\Z_N)|}{|H^1(X_2,\Z_N)|}~.
\end{eqnarray}
The normalization of the partition function may be modified by multiplying by an arbitrary power of the Euler counterterm $\chi[X_2,\Z_N]^{\kappa}$. The case of $\kappa=\frac{1}{2}$ is of particular interest, since in this case the normalization in \eqref{eq:2dgauging} becomes $1/\sqrt{|H^1(X_2,\Z_N)|}$, and then gauging twice maps $\CX$ back to itself exactly, without any counterterm.\footnote{In obtaining this result, one uses the fact that $|H^n(X_2,\Z_N)|= |H^{2-n}(X_2,\Z_N)|$ for closed $X_2$. For the case of open manifolds $X_2^{\geq 0}$ with boundary $M_1|_0$, to be discussed below, absolute cohomology should be switched to relative cohomology, i.e. $|H^n(X_2^{\geq 0},\Z_N)|= |H^{2-n}(X_2^{\geq 0}, M_1|_{0}, \Z_N)|$.} We however will not include this factor, and will instead adopt the normalization in \eqref{eq:2dgauging}.

\begin{figure}[tbp]
\begin{center}

\begin{tikzpicture}[scale=0.4]

\shade[top color=gray!40, bottom color=gray!10] (0.03,-1) 
to [out=0, in=180] (-6.43,-1) 
to [out=270, in=90] (-6.43,1)
to [out=180, in=0] (0.03,1)
to [out=90, in=270] (0.03,-1) ;

\shade[top color=gray!40, bottom color=gray!10,rotate = 90]  (-1,0) coordinate (-left) 
to [out=260, in=60] (-3,-2) 
to [out=240, in=110] (-3,-4) 
to [out=290,in=180] (0,-7) 
to [out=0,in=250] (3,-4) 
to [out=70,in=300] (3,-2) 
to [out=120,in=280] (1,0)  coordinate (-right);

\draw[rotate = 90]  (-1,0) coordinate (-left) 
to [out=260, in=60] (-3,-2) 
to [out=240, in=110] (-3,-4) 
to [out=290,in=180] (0,-7) 
to [out=0,in=250] (3,-4) 
to [out=70,in=300] (3,-2) 
to [out=120,in=280] (1,0)  coordinate (-right);

\shade[top color=gray!40, bottom color=gray!10,rotate = 270,yshift=-2.5in]  (-1,0) coordinate (-left) 
to [out=260, in=60] (-3,-2) 
to [out=240, in=110] (-3,-4) 
to [out=290,in=180] (0,-7) 
to [out=0,in=250] (3,-4) 
to [out=70,in=300] (3,-2) 
to [out=120,in=280] (1,0)  coordinate (-right);

\draw[rotate = 270,yshift=-2.5in]  (-1,0) coordinate (-left) 
to [out=260, in=60] (-3,-2) 
to [out=240, in=110] (-3,-4) 
to [out=290,in=180] (0,-7) 
to [out=0,in=250] (3,-4) 
to [out=70,in=300] (3,-2) 
to [out=120,in=280] (1,0)  coordinate (-right);

\draw[] (0,-1)--(-6.4,-1);
\draw[] (0,1)--(-6.4,1);

\pgfgettransformentries{\tmpa}{\tmpb}{\tmp}{\tmp}{\tmp}{\tmp}
\pgfmathsetmacro{\myrot}{-atan2(\tmpb,\tmpa)}
\draw[rotate around={\myrot:(0,-2.5)},yshift=1.1in,xshift=1.4in] (-1.2,-2.4) to[bend right]  (1.2,-2.4);
\draw[fill=white,rotate around={\myrot:(0,-2.5)},yshift=1.1in,xshift=1.4in] (-1,-2.5) to[bend right] (1,-2.5) 
to[bend right] (-1,-2.5);

\draw[rotate around={\myrot:(0,-2.5)},yshift=0.5in,xshift=-4in] (-1.2,-2.4) to[bend right]  (1.2,-2.4);
\draw[fill=white,rotate around={\myrot:(0,-2.5)},yshift=0.5in,xshift=-4in] (-1,-2.5) to[bend right] (1,-2.5) 
to[bend right] (-1,-2.5);

\draw[rotate around={\myrot:(0,-2.5)},yshift=1.4in,xshift=-3.8in] (-1.2,-2.4) to[bend right]  (1.2,-2.4);
\draw[fill=white,rotate around={\myrot:(0,-2.5)},yshift=1.4in,xshift=-3.8in] (-1,-2.5) to[bend right] (1,-2.5) 
to[bend right] (-1,-2.5);

\draw[red, thick] (-3.2,-1) arc(-90:90:0.5cm and 1cm);
\draw[red, dotted, thick] (-3.2,-1) arc(270:90:0.5cm and 1cm);
\node[below] at (-3.2,-1) {$M_1|_0$};

\node[below] at (-12.5,4) {$X_2^{<0}$};
\node[below] at (6.5,4) {$X_2^{\geq 0}$};
\end{tikzpicture}

\caption{Decomposition of $X_2$ along a neck. The interface is located at $x=0$.}
\label{fig:X2}
\end{center}
\end{figure}
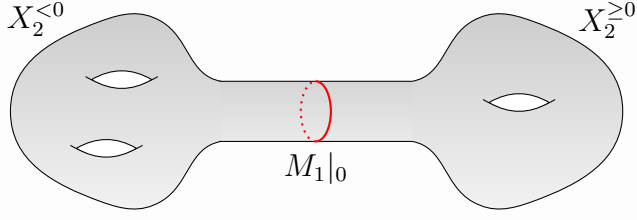

Instead of gauging $\Z_N^{(0)}$ over the entire $X_2$, one can gauge in half of the spacetime with Dirichlet boundary conditions. This defines a topological duality interface $\CN$ between $\CX$ and $\CX/\Z_N$ \cite{Choi:2021kmx,Kaidi:2021xfk}.\footnote{The duality interface defined here by half-gauging should not be confused with the topological interface (also termed ``duality wall") between two dual theories (e.g. mirror symmetry pairs) discussed in e.g. \cite{Dimofte:2013lba,Gaiotto:2015una,dimofte2018dual}. } Let us decompose the spacetime manifold $X_2$ into left and right parts, 
\begin{eqnarray}
X_2= X_2^{< 0} \cup X_2^{\geq 0}~, 
\end{eqnarray}
where $\partial X_{2}^{\geq 0}= M_1$ is the interface. Locally around the interface, the topology of the manifold is $M_1\times \mathbb{R}$, and we can use a local coordinate $x$ to parameterize $\mathbb{R}$. The interface is located at $x=0$, as shown in Figure \ref{fig:X2}. The theory $\CX$ lives on $X_2^{<0}$, while the theory $\CX/\Z_N$ lives on $X_2^{\geq 0}$.

\begin{figure}
    \centering
    \begin{tikzpicture}
      \shade[top color=red!20, bottom color=red!5]  (6,0) 
to [out=0, in=180] (12,0) 
to [out=90, in=270] (12,3) 
to [out=180,in=0] (6,3) 
to [out=270,in=90] (6,0);

    \draw[thick, ->] (0,0) -- (12,0);
    \draw[thick, dashed] (6,3) -- (6,-0.5); 
    \node[above] at (6,3) {$\CN$};
    \node[] at (3,2) {$Z_{\CX}[X_2^{<0},A]$};
    \node[] at (9,1.5) {\begin{tabular}{c}
          $Z_{\CX/\Z_N}[X_2^{\geq 0}, A]$ \bigskip\\ 
         \eqref{eq:halfgauging}
    \end{tabular}};
    \node[below] at (6,-0.5) {$x=0$};
    \node[right] at (12,0) {$x$};

    \end{tikzpicture}
    \caption{The duality defect from gauging $\Z_N$ over half of the spacetime $X_2^{\geq 0}$ with Dirichlet boundary conditions. }
    \label{fig:2ddualitydefect}
\end{figure}
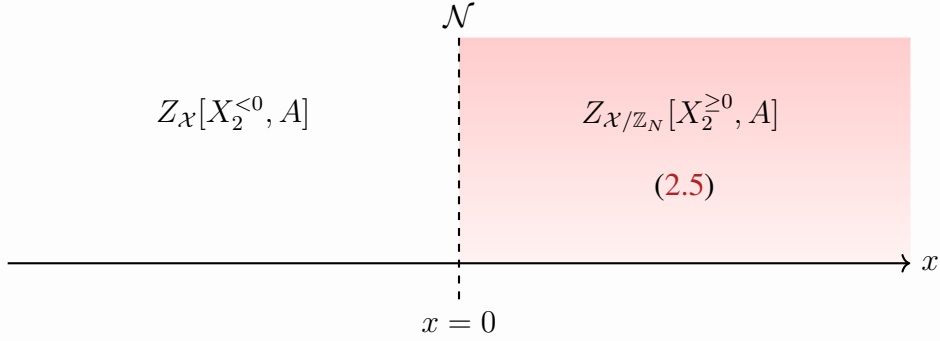

This is also shown in Figure \ref{fig:2ddualitydefect}, where the theory on the right side $X_2^{\geq 0}$ is defined to be
\begin{eqnarray}\label{eq:halfgauging}
Z_{\CX/\Z_N}[X_2^{\geq 0}, A]=\frac{1}{|H^0(X_2^{\geq 0},M_1|_0, \Z_N)|} \sum_{a\in H^1(X_2^{\geq 0}, M_1|_0, \Z_N)} Z_{\CX}[X_2^{\geq 0}, a]\, e^{ \frac{2\pi i}{N} \int_{X_2^{\geq 0} }a A }~.
\end{eqnarray}
The Dirichlet boundary condition implies that the dynamical gauge field $a$ is an element in relative cohomology $H^1(X_2^{\geq 0}, M_1|_0, \Z_N)$. Here, we use $M_1|_0$ to emphasize that $M_1$ is located at $x=0$.

The orientation reversal of the duality defect $\overline{\CN}$ is defined by exchanging the theories on the two sides of Figure \ref{fig:2ddualitydefect}. Concretely, $\CX/\Z_N$ lives on the left and $\CX$ lives on the right. By redefining the theory $\widetilde{\CX}= \CX/\Z_N$, the defect  $\overline{\CN}$ can be rewritten as having $\widetilde{\CX}$ on the left and $\CX= \chi \cdot \widetilde{\CX}/\Z_N$ on the right.  In other words, 
\begin{eqnarray}\label{eq:Ninvertible}
\overline{\CN} = \chi [X_2^{\geq 0}, \Z_N] \cdot \CN~.
\end{eqnarray}
Hence the duality interfaces in $(1+1)$d are orientation-reversal invariant, up to an Euler counterterm. This is also a consequence of the fact that gauging twice maps the theory to itself (up to a counterterm).

In the special case that the theory $\CX$ is self-dual under gauging, i.e. $\CX=\CX/\Z_N$, the duality \textit{interface} discussed above which connects two different theories becomes a duality \textit{defect} within a single theory. In this section we will not assume the self-duality condition; it will be the main focus of Section \ref{sec:3dDTYZN}.

\subsection{Fusion rules of duality interfaces}
\label{sec:2dfusionrule}

We now proceed to a discussion of the fusion rules of the duality interface $\CN$ defined in Section \ref{sec:1+1ddualityinterfacedef}. In particular, we will find that $\CN$ is non-invertible. We warn the reader that the following analysis is somewhat technical, and those interested only in the final answer may skip to (\ref{eq:NNfusionfinal}).

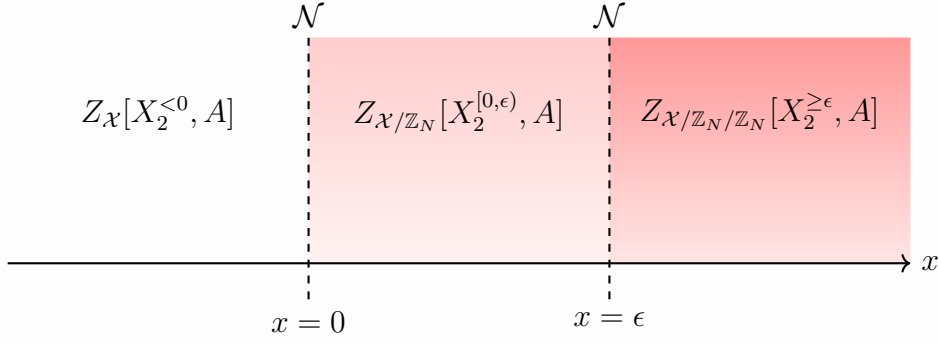
\begin{figure}
    \centering
    \begin{tikzpicture}
    
       \shade[top color=red!20, bottom color=red!5]  (4,0) 
to [out=0, in=180] (8,0) 
to [out=90, in=270] (8,3) 
to [out=180,in=0] (4,3) 
to [out=270,in=90] (4,0);

  \shade[top color=red!40, bottom color=red!10]  (8,0) 
to [out=0, in=180] (12,0) 
to [out=90, in=270] (12,3) 
to [out=180,in=0] (8,3) 
to [out=270,in=90] (8,0);

    \draw[thick, ->] (0,0) -- (12,0);
    \draw[thick, dashed] (4,3) -- (4,-0.5); 
    \draw[thick, dashed] (8,3) -- (8,-0.5); 
    \node[above] at (4,3) {$\CN$};
    \node[above] at (8,3) {$\CN$};
    \node[] at (2,2) {$Z_{\CX}[X_2^{<0},A]$};
    \node[] at (6,2) {\begin{tabular}{c}
          $Z_{\CX/\Z_N}[X_2^{[0,\epsilon)}, A]$ 
    \end{tabular}};
    \node[] at (10,2) {\begin{tabular}{c}
          $Z_{\CX/\Z_N/\Z_N}[X_2^{\geq \epsilon}, A]$ 
    \end{tabular}};
    \node[below] at (4,-0.5) {$x=0$};
    \node[below] at (8,-0.5) {$x=\epsilon$};
    \node[right] at (12,0) {$x$};
    \end{tikzpicture}
    \caption{Fusion of two duality interfaces. The partition function on $X^{\geq 0}$ is given by \eqref{eq:NN}.  }
    \label{fig:2ddualitydefectfusion}
\end{figure}

We first discuss the fusion rule $\eta \times \CN$. Since $a$ has Dirichlet boundary conditions on $M_1$, $a|_{M_1|_0}=0$, the $\Z^{(0)}_N$ symmetry defect $\eta$ is trivial on $M_1|_0$. This justifies the fusion rule
\begin{eqnarray}\label{2deq:etaNN}
\eta \times \CN =\CN~.
\end{eqnarray}
It is more interesting to study the fusion rule between two duality interfaces $\CN\times \CN$. Let us place the two duality defects at $x=0$ and $x= \epsilon$, and let $\epsilon \to 0^+$.   The two duality defects divide the spacetime $X_2$ into three regions, $X_2=X_2^{<0} \cup X_2^{[0,\epsilon)} \cup X_2^{\geq \epsilon}$. Since $\epsilon$ is small, we may always take $X_2^{[0,\epsilon)}= M_1\times I_{[0,\epsilon)}$. The theories living on the three regions are as shown in Figure \ref{fig:2ddualitydefectfusion}. Instead of defining $Z_{\CX/\Z_N}[X_2^{[0,\epsilon)}, A]$  and  $Z_{\CX/\Z_N/\Z_N}[X_2^{\geq \epsilon}, A]$  separately and discussing how to glue them together along $M_1|_{\epsilon}$, it suffices to discuss the theory on $X_2^{\geq 0}$ all together. The theory living on $X_2^{\geq 0}$ is given by 
\begin{equation}\label{eq:NN}
{1 \over |H^0(X_2^{\geq 0}, M_1|_0, \Z_N)||H^0(X_2^{\geq \eps}, M_1|_\eps,\Z_N)|} \sum_{\substack{a \in H^1(X_2^{\geq 0}, M_1|_0, \Z_N) \\ \widetilde a \in H^1(X_2^{\geq \eps}, M_1|_\eps, \Z_N)}} Z_{\CX}[X_2^{\geq 0}, a]\, e^{{2 \pi i \over N}\int_{X_2^{[0,\eps)}} a A + {2 \pi i \over N}\int_{X_2^{\geq \eps}} (a-A)\widetilde a}
\end{equation}
and we will now evaluate \eqref{eq:NN} in detail. 

Since $Z_{\CX}$ in the summand does not depend on $\widetilde{a}$, we would first like to sum over $\widetilde{a}$, which intuitively should enforce $a=A$ on $X_{2}^{\geq \eps}$. To perform this summation rigorously, one must convert the sum over cohomologies into a sum over cochains (for reasons that will be explained below). Note that $H^1= Z^1/B^1$, $|B^1|=|C^0|/|Z^0|$, and $|B^0|=1$;\footnote{We remind the reader that $C^n$ is the set of $n$-cochains, $Z^n\subset C^n$ is the set of $n$-cocycles, and $B^n\subset Z^n$ is the set of exact $n$-cocycles.} these equations hold for both absolute and relative cohomologies. The sum $\frac{1}{|H^0|}\sum_{a\in H^1}$ can thus be rewritten as $\frac{1}{|H^0||B^0|} \sum_{a\in Z^1}= \frac{1}{|C^0|} \sum_{a\in Z^1}$. Equation \eqref{eq:NN} then becomes 
\begin{equation}\label{eq:NN2d1}
{1 \over |C^0(X_2^{\geq 0}, M_1|_0, \Z_N)||C^0(X_2^{\geq \eps}, M_1|_\eps,\Z_N)|} \sum_{\substack{a \in Z^1(M_2^{\geq 0}, M_1|_0, \Z_N) \\ \widetilde a \in Z^1(X_2^{\geq \eps}, M_1|_\eps, \Z_N)}} Z_{\CX}[X_2^{\geq 0}, a]\, e^{{2 \pi i \over N}\int_{X_2^{[0,\eps)}} a A + {2 \pi i \over N}\int_{X_2^{\geq \eps}} (a-A)\widetilde a}~.
\end{equation}
The cocycle condition in the sum can further be relaxed by introducing additional dynamical scalars $\phi\in C^0(X_2^{\geq 0}, \Z_N)$ and $\widetilde{\phi}\in C^0(X_2^{\geq \eps}, \Z_N)$ with BF couplings acting as Lagrange multiplier fields. The relative condition can likewise be eliminated by including the couplings $\phi a$ and $\widetilde{\phi} \widetilde{a}$ on the subloci $M_1|_0$ and $M_1|_\eps$. Then all the variables in the sum are cochains,
\begin{equation}
    \begin{split}
        &{1 \over |C^0(X_2^{\geq 0}, M_1|_0, \Z_N)||C^0(X_2^{\geq \eps}, M_1|_\eps, \Z_N)|}{1 \over |C^0(X_2^{\geq 0}, \Z_N)| |C^0(X_2^{\geq \eps}, \Z_N)|}\times\\
        &\sum_{\substack{a \in C^1(X_2^{\geq 0}, \Z_N), \widetilde a \in C^1(X_2^{\geq \eps}, \Z_N) \\ \phi \in C^0(X_2^{\geq 0}, \Z_N), \widetilde \phi \in C^0(X_2^{\geq \eps}, \Z_N) }} Z_{\CX}[X_2^{\geq 0},a] e^{{2 \pi i \over N}\int_{X_2^{[0,\eps)}} a A + {2 \pi i \over N}\int_{X_2^{\geq \eps}} (a-A)\widetilde a} e^{{2\pi i \over N}\int_{X_2^{>0}}\phi \delta a - {2\pi i \over N}\int_{M_1|_0} \phi a}e^{{2\pi i \over N}\int_{X_2^{>\eps}}\widetilde\phi \delta\widetilde a - {2\pi i \over N}\int_{M_1|_\eps} \widetilde\phi \widetilde a}
    \end{split}
\end{equation}
Indeed, summing over $\phi$ in the bulk $X_2^{\geq 0}$ enforces $a$ to be a cocycle due to the coupling $\int_{X_2^{> 0}}\phi \delta a$, and summing over $\phi$ on the boundary $X_1|_0$ enforces $a=0$, which makes the cocycle relative to $M_1|_0$. The same comments apply to $\widetilde{a}$. 

We are now ready to perform the sum over $\widetilde{a}$. By rewriting the final factor of the summand as
\bea
 e^{{2\pi i \over N}\int_{X_2^{>\eps}}\widetilde\phi \delta\widetilde a - {2\pi i \over N}\int_{M_1|_\eps} \widetilde\phi \widetilde a} = e^{- {2 \pi i \over N} \int_{X_2^{\geq \eps}} \delta \widetilde \phi \widetilde a}~,
\eea
we see that summing over $\widetilde{a}$ produces a factor of $|C^1(X_2^{\geq \eps}, \Z_N)|$ and enforces that $a - A - \delta \widetilde \phi = 0$ on $X_2^{\geq \eps}$,
\bea
&\vphantom{.}& {1 \over |C^0(X_2^{\geq 0}, M_1|_0, \Z_N)||C^0(X_2^{\geq \eps}, M_1|_\eps, \Z_N)|}{|C^1(X_2^{\geq \eps}, \Z_N)| \over |C^0(X_2^{\geq 0}, \Z_N)| |C^0(X_2^{\geq \eps}, \Z_N)|}\times
\\
&\vphantom{.}&
 \sum_{\substack{a \in C^1(X_2^{\geq 0}, \Z_N) \\ \phi \in C^0(X_2^{\geq 0}, \Z_N),\,\,\, \widetilde \phi \in C^0(X_2^{\geq \eps}, \Z_N) }} Z_{\CX}[X_2^{\geq 0}, a] e^{{2 \pi i \over N}\int_{X_2^{[0,\eps)}} a A}e^{{2\pi i \over N}\int_{X_2^{>0}}\phi \delta a - {2\pi i \over N}\int_{M_1|_0} \phi a} \delta(a-A-\delta\widetilde \phi)|_{X_2^{\geq \eps}}~.
 \no
 \eea
Next we integrate out $\phi$, which produces a factor of $| C^0(X_2^{\geq 0},\Z_N)|$ and enforces $a \in Z^1(X_2^{\geq 0}, M_1|_0, \ZZ_N)$, 
\begin{equation}
    \begin{split}
        &{1 \over |C^0(X_2^{\geq 0}, M_1|_0,\Z_N)||C^0(X_2^{\geq \eps}, M_1|_\eps,\Z_N)|}{|C^1(X_2^{\geq \eps},\Z_N)| \over |C^0(X_2^{\geq \eps},\Z_N)|} \\&\hspace{3cm}\sum_{\substack{a \in Z^1(X_2^{\geq 0}, M_1|_0,\Z_N)\\ \widetilde \phi \in C^0(X_2^{\geq \eps},\Z_N) }} Z_{\CX}[X_2^{\geq 0}, a] e^{{2 \pi i \over N}\int_{X_2^{[0,\eps)}} a A} \delta(a-A-\delta\widetilde \phi)|_{X_2^{\geq \eps}}~.
    \end{split}
\end{equation}
The summand is now manifestly independent of $\widetilde \phi$, so we may set $\widetilde \phi$ to zero in the delta function and replace the sum with a factor of $|C^0(X_2^{\geq \eps},\Z_N)|$.  The remaining delta function then fixes $a = A$ in $X_2^{\geq \eps}$, which makes $a$ an element of $a \in Z^1(X_2^{[0,\eps]}, M_1|_0 \cup M_1|_\eps, \Z_N)$, 
\bea\label{eq:NN2d2}
 {|C^1(X_2^{\geq \eps},\Z_N)| \over |C^0(X_2^{\geq 0}, M_1|_0,\Z_N)||C^0(X_2^{\geq \eps}, M_1|_\eps,\Z_N)|} \sum_{\substack{a \in Z^1(X_2^{[0,\eps]}, M_1|_0 \cup M_1|_\eps,\Z_N) }} Z_{\CX}[X_2^{\geq 0}, a+A|_{M_2^{\geq \eps}}] e^{{2 \pi i \over N}\int_{X_2^{[0,\eps]}} a A}~.
\eea
where $A|_{M_2^{\eps}}$ is equal to $A$ if we're on $M_2^{\geq \eps}$, and vanishes elsewhere. It is useful reorganize the normalization by factorizing out the Euler counter term, 
\begin{eqnarray}\label{eq:NN2d3}
\begin{split}
    {|C^0(X_2^{\geq \eps},\Z_N)| |C^2(X_2^{\geq \eps},\Z_N)| \over |C^0(X_2^{\geq 0}, M_1|_0,\Z_N)||C^0(X_2^{\geq \eps}, M_1|_\eps,\Z_N)|} \cdot {|C^1(X_2^{\geq \eps},\Z_N)| \over |C^0(X_2^{\geq \eps},\Z_N)| |C^2(X_2^{\geq \eps},\Z_N)|}~.
\end{split}
\end{eqnarray}
The second factor above is the inverse of the Euler counterterm $\chi[M_2^{\geq \eps}, \Z_N]^{-1}$ on $X_2^{\geq \eps}$, which can be seen by direct evaluation
\begin{eqnarray}
\chi[X_2^{\geq \eps}, \Z_N]= {|H^0(X_2^{\geq \eps},\Z_N)| |H^2(X_2^{\geq \eps},\Z_N)| \over |H^1(X_2^{\geq \eps},\Z_N)|} = {|C^0(X_2^{\geq \eps},\Z_N)| |Z^2(X_2^{\geq \eps},\Z_N)| \over |C^1(X_2^{\geq \eps},\Z_N)|}
\end{eqnarray}
and by further noting that $Z^2(X_2^{\geq \eps},\Z_N)= C^2(X_2^{\geq \eps},\Z_N)$ since all top forms are closed. Moreover, the first factor in \eqref{eq:NN2d3} can be simplified by using $|C^2(X_2^{\geq \eps},\Z_N)|= |C^0(X_2^{\geq \eps}, M_1|_\eps,\Z_N)|$ since the elements in $C^0(X_2^{\geq \eps}, M_1|_\eps,\Z_N)$ and the elements in $C^2(X_2^{\geq \eps},\Z_N)$ are Fourier partners. We finally use the fact that 
\begin{eqnarray}\label{eq:NN2d4}
|C^n( X_2^{\geq 0}, M_1|_0, \Z_N)| = |C^n( X_2^{[0,\epsilon]}, M_1|_0 \cup  M_1|_\epsilon, \Z_N)| |C^n( X_2^{\geq \epsilon}, \Z_N)| ~,
\end{eqnarray}
which is simply a decomposition of cochains on $X_2^{\geq 0}$ into the sum of cochains on $X_2^{[0,\eps]}$ with fixed boundary condition at $M_1|_{\eps}$ and cochains on $X_2^{\geq \eps}$ with free boundary conditions. Note crucially that such decomposition can \textit{not} be achieved at the level of cocycles or cohomologies, because the flatness condition is violated at $M_1|_{\eps}$. The ability to use this decomposition is the main reason for our reformulation of the sum over cohomologies in \eqref{eq:NN} as a sum over cochains. 

Using (\ref{eq:NN2d4}), we may now simplify the first term in \eqref{eq:NN2d3} to $1/|C^0( X_2^{[0,\epsilon]}, M_1|_0 \cup  M_1|_\epsilon, \Z_N)|$. Substituting the simplified normalization in \eqref{eq:NN2d3} into \eqref{eq:NN2d2}, we get
\begin{equation}
\label{eq:N218}
\chi[X_2^{\geq \eps}, \Z_N]^{-1} {1 \over |C^0( X_2^{[0,\epsilon]}, M_1|_0 \cup  M_1|_\epsilon, \Z_N)|} \sum_{\substack{a \in Z^1(X_2^{[0,\eps]}, M_1|_0 \cup M_1|_\eps,\Z_N) }} Z_{\CX}[X_2^{\geq 0}, a+A|_{X_2^{\geq \eps}}] e^{{2 \pi i \over N}\int_{X_2^{[0,\eps]}} a A}~.
\end{equation}
This formula admits a simple physical interpretation. First, in the region $M_2^{\geq \eps}$ we have two overlapping gaugings, which ``annihilate" to produce a factor of $\chi[M_2^{\geq \eps}, \ZZ_N]^{-1}$. This is a consequence of the fact, mentioned before, that gauging twice  takes the theory $\CX$ back to itself up to an Euler counterterm.\footnote{As mentioned around \eqref{eq:EulerCC2d}, this piece could be removed by modifying the definition of gauging by an appropriate Euler counterterm.} As for the rest of (\ref{eq:N218}), we notice that after the two gaugings ``annihilate" to give the Euler counterterm, we are left with a gauging in the strip $X_2^{[0,\eps]}$, with Dirichlet boundary conditions on both boundaries $M_1|_{0} \cup M_1|_{\eps}$. This gives the remaining portions of the formula. 

By taking the limit $\eps\to 0$, it now follows that the $\CN\times \CN$ fusion rule is
\begin{equation}\label{eq:NNfusionfinal}
\begin{split}
    \CN\times \CN &= \chi[X_2^{\geq \eps}, \Z_N]^{-1} \frac{1}{|C^0( X_2^{[0,\epsilon]}, M_1|_0 \cup  M_1|_\epsilon, \Z_N)|}  \sum_{\substack{a \in Z^1(X_2^{[0,\eps]}, M_1|_0 \cup M_1|_\eps,\Z_N) }}  \eta(\mathrm{LD}(a))\\
    &= \chi[X_2^{\geq \eps}, \Z_N]^{-1} \frac{1}{|H^0( X_2^{[0,\epsilon]}, M_1|_0 \cup  M_1|_\epsilon, \Z_N)|}  \sum_{\substack{a \in H^1(X_2^{[0,\eps]}, M_1|_0 \cup M_1|_\eps,\Z_N) }}  \eta(\mathrm{LD}(a))\\
    &= \chi[X_2^{\geq \eps}, \Z_N]^{-1}  \sum_{\substack{\gamma \in H_1(M_1,\Z_N) }}  \eta(\gamma)
\end{split}
\end{equation}
where LD stands for the Lefschetz dual, and we have used $|H^0( X_2^{[0,\epsilon]}, M_1|_0 \cup  M_1|_\epsilon, \Z_N)|= 1$. Using \eqref{eq:Ninvertible}, we likewise find 
\begin{eqnarray}
\CN\times \overline{\CN} = \sum_{\substack{\gamma \in H_1(M_1,\Z_N) }}  \eta(\gamma)~.
\end{eqnarray}
This fusion rule more straightforwardly corresponds to gauging in a slab, and indeed this ``gauging in a slab" is precisely how the fusion rule of duality defects was originally derived in \cite{Choi:2021kmx,Kaidi:2021xfk}. The method used here was more roundabout, but gives important insight into the correct way to keep track of normalization factors.

In summary, the fusion rules for duality interfaces in $(1+1)$d are as follows,
\bea
\label{eq:FR2d}
&\vphantom{.}&\CN\times \overline{\CN}= \sum_{\gamma\in H_1(M_1, \Z_N)} \eta(\gamma)~,\hspace{0.5 in}\eta \times \CN =\CN,\no\\
&\vphantom{.}&
\overline{\CN} = \chi [X_2^{\geq0},\Z_N]\, \CN ~,\hspace{2.5cm}  \eta^N=1 ~.
\eea
These are, not coincidentally, the fusion rules of the $\Z_N$ Tambara-Yamagami fusion categories.

\section{$(2+1)$d Symmetry TFT for $\Z_N^{(0)}$ symmetry}
\label{sec:symTFT3dZN}

In Section \ref{sec:1+1ddualityinterface}, we defined the duality interface $\CN$ and studied its fusion rules. In this section, we study the properties of the duality interface from the SymTFT point of view. In particular, we find that the duality interface in a $(1+1)$d QFT can be realized as a twist defect in a $(2+1)$d $\Z_N$ TQFT, from which the fusion rules \eqref{eq:FR2d} can be reproduced. We also study the F-symbols of the duality interfaces using the SymTFT. Throughout this section, we will only assume that the $(1+1)$d QFT has a $\Z_N^{(0)}$ zero-form global symmetry; in particular, we do \textit{not} assume invariance under gauging $\Z_N^{(0)}$.

\subsection{$\Z_N$ gauge theory as the Symmetry TFT}
\label{sec:3dZNgaugetheoryasSymTFT}

\begin{figure}[!tbp]
	\centering
	\begin{tikzpicture}[scale=0.8]

	\draw[very thick, blue] (-7,0) -- (-7,1.5);
	\draw[very thick, blue] (-7,1.5) -- (-7,3);
	\node[below] at (-7,0) {$Z_{\CX}[A]$};
    \node[above] at (-7,3) {$x=0$};
	
	\draw[thick, snake it, <->] (-1.7,1.5) -- (-5, 1.5);
	\node[above] at (-3.3,1.6) {Expand\,/\,Shrink};
	
		\shade[line width=2pt, top color=blue!30, bottom color=blue!5] 
	(0,0) to [out=90, in=-90]  (0,3)
	to [out=0,in=180] (6,3)
	to [out = -90, in =90] (6,0)
	to [out=180, in =0]  (0,0);
	
	\draw[thick] (0,0) -- (0,3);
	\draw[thick] (6,0) -- (6,3);
	\node[] at (3,1.5) {$\frac{2\pi}{N} \widehat{a} \delta a$};
	\node[below] at (0,0) {$\langle D(A)| $};
	\node[below] at (6,0) {$|\cX\rangle $}; 
	\node[above] at (0,3) {$x=0$};
	\node[above] at (6,3) {$x=\varepsilon$};

	\end{tikzpicture}
	
	\caption{A $(1+1)$d QFT $\cX$ with $\Z_N$ symmetry can be expanded into a $(2+1)$d slab. The bulk is the $(2+1)$d $\Z_N$ SymTFT, the right boundary encodes the dynamical information of the $(1+1)$d QFT, and the left boundary is a topological Dirichlet boundary condition for the bulk field $a$.}
	\label{fig:2d3dSymTFT}
\end{figure}
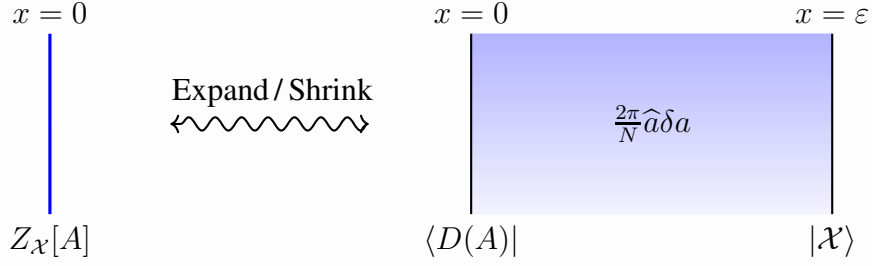

Suppose $\CX$ is a $(1+1)$d QFT with an anomaly-free $\Z_N^{(0)}$ global symmetry, whose partition function is $Z_{\CX}[X_2,A]$. As discussed in the introduction, we can expand the theory into a $(2+1)$d slab, as shown in Figure \ref{fig:2d3dSymTFT}. The SymTFT living in the bulk of the slab is a $(2+1)$d $\Z_N$ gauge theory, whose action is 
\begin{eqnarray}\label{eq:3dBFtheory}
S= \frac{2\pi}{N}\int_{X_3} \widehat{a} \delta a~.
\end{eqnarray}
Both $a$ and $\widehat{a}$ are dynamical $\Z_N^{(0)}$-valued 1-cochains. We will take the bulk to be a product $X_3= X_2\times I_{[0,\varepsilon]}$, and will use the coordinate $x$ to parameterize the interval $I_{[0,\varepsilon]}$. The two boundaries are $X_2|_{\varepsilon}$ and $X_2|_0$.

The boundary conditions on the left ($x=0$) and right ($x=\varepsilon$) boundaries of the slab are specified by appropriate boundary states. On the right boundary, the state is  
\begin{eqnarray}
\ket{\CX} = \sum_{a\in H^1(X_2|_{\varepsilon}, \Z_N)} Z_{\CX}[X_2|_{\varepsilon}, a] \ket{a}~.
\end{eqnarray}
It corresponds to a non-topological enriched Neumann boundary condition that encodes all of the dynamical information of the $(1+1)$d QFT $\CX$. 
On the left boundary, the state is 
\begin{eqnarray}
\bra{D(A)} = \sum_{a\in H^1(X_2|_0, \Z_N)} \bra{a} \delta(a-A) ~,
\end{eqnarray}
which corresponds to a topological Dirichlet boundary condition. 
Note that the background field dependence only enters in the left, topological boundary. 
By shrinking the slab, i.e. taking $\varepsilon\to 0$, the partition function of the $(1+1)$d theory is reproduced by the inner product of the two boundary states, 
\begin{eqnarray}
\braket{D(A)| \CX} = \sum_{a,a'\in H^1(X_2|_0, \Z_N)} Z_{\CX}[X_{2}|_{0}, a] \delta(a'-A)  \braket{a'|a}= Z_{\CX}[X_2|_0, A] ~,
\end{eqnarray}
where we have used $\braket{a'|a}=\delta(a'-a)$.

Gauging the $\Z_N^{(0)}$ symmetry of $\CX$ in $(1+1)$d corresponds to changing the topological boundary condition on the left from Dirichlet to Neumann. To see this, we define the Neumann boundary condition as 
\begin{eqnarray}
\bra{N(A)} = \frac{1}{|H^0(X_2|_0, \Z_N)|}\sum_{a\in H^1(X_2|_0, \Z_N)} \bra{a} e^{\frac{2\pi i}{N}\int_{X_2|_0} aA }
\end{eqnarray}
and check explicitly that
\begin{equation}
\begin{split}
    \braket{N(A)|\CX} =\frac{1}{|H^0(X_2|_0, \Z_N)|} \sum_{a,a'\in H^1(X_2|_0, \Z_N)}  Z_{\CX}[X_2|_0, a] e^{\frac{2\pi i}{N}\int_{X_2|_0} a'A } \braket{a'|a} = Z_{\CX/\Z_N}[X_2, A]~.
\end{split}
\end{equation}
More generally, when the $\Z_N^{(0)}$ symmetry has a nontrivial 't Hooft anomaly, the SymTFT will be a $\Z_N$ gauge theory with a non-trivial twist, i.e. a Dijkgraaf Witten theory. In this case, one cannot gauge $\Z_N^{(0)}$ in $(1+1)$d because of the anomaly. Correspondingly, there is no Neumann topological boundary condition.

\subsection{Lines and surfaces in $\Z_N$ gauge theory}

We now consider the line and surface operators in $\Z_N$ gauge theory, focusing for the moment on the ones without topological boundaries. The operators with boundaries will be discussed in subsequent subsections.

\subsubsection{Line operators}

The $\Z_N$ gauge theory \eqref{eq:3dBFtheory} has $N^2$ genuine topological line operators,
\bea
\label{eq:beforegaugingLs}
L_{(e,m)}(\gamma) =  \exp\left(\frac{2\pi i}{N} \oint_\gamma e a\right) \exp\left( \frac{2\pi i}{N} \oint_\gamma m \widehat{a}\right)~, \hspace{0.5 in} (e, m) \in \ZZ_N \times \ZZ_N~,
\eea
where $L_{(1,0)}$ and $L_{(0,1)}$ together generate a $\Z_N^{(1)}\times \Z_N^{(1)}$ 1-form symmetry, or in other words $\CZ(\text{Vec}_{\Z_N})$.  They satisfy the following fusion and braiding rules:
\begin{enumerate}
    \item \textbf{Fusion rule:}  The lines $L_{(e,m)}(\gamma)$ are invertible, and have straightforward fusion rules
\bea
\label{eq:fusLL}
L_{(e,m)}(\gamma) \times L_{(e',m')}(\gamma) = L_{(e+e',m+m')}(\gamma)~.
\eea
    \item \textbf{Commutation relation:} 
    As derived in Appendix \ref{app:linecorr}, the correlation function between $L_{(e,m)}$ on a closed, contractible loop $\gamma$ and $L_{(e',m')}$ on another closed, contractible loop $\gamma'$ is 
\begin{eqnarray}\label{eq:linking}
\langle L_{(e,m)}(\gamma) L_{(e',m')}(\gamma')\dots\rangle = \exp\left( -\frac{2\pi i}{N}(em'+ me')\link(\gamma,\gamma')\right)\braket{\dots}~,
\end{eqnarray}
where $\link(\gamma,\gamma')$ is the Hopf linking number between $\gamma$ and $\gamma'$. The phase on the right-hand side gives the braiding phase between the two anyons whose worldlines are $L_{(e,m)}$ and $L_{(e',m')}$. Here, $\gamma$ and $\gamma'$ are contractible loops that do not intersect each other and we also assume that the operators represented by ``$\dots$" in \eqref{eq:linking} do not link with $\gamma$ and $\gamma'$. Equation
\eqref{eq:linking} can be equivalently rewritten as an equal time commutation relation  by pushing $\gamma$ and $\gamma'$ to a 2d plane \cite{Gaiotto:2014kfa},
\bea
\label{eq:corrLL}
 L_{(e,m)}(\gamma) L_{(e',m')}(\gamma') = \exp\left(-{2 \pi i \over N} (e m' +  m e') \langle \gamma, \gamma'\rangle \right) L_{(e',m')}(\gamma')  L_{(e,m)}(\gamma) ~,
\eea
where  $\langle \gamma, \gamma'\rangle$ represents the intersection number between $\gamma$ and $\gamma'$. Equation \eqref{eq:corrLL} makes sense even when $\gamma$ and $\gamma'$ are line intervals.  Note that the order of lines in the correlation function matters: the operators are ``time-ordered." The correlation function $L_{(e,m)}(\gamma) L_{(e',m')}(\gamma')$ on the left hand side of \eqref{eq:corrLL}  means that the line segment $\gamma'$ is at time $t$, while the line segment $\gamma$ is at time $t+\epsilon$ with $\epsilon$  an infinitesimal  positive real number. Hence $L_{(e,m)}$ crosses $L_{(e',m')}$ from above. Likewise, the correlation function $L_{(e',m')}(\gamma')  L_{(e,m)}(\gamma) $ on the right-hand side of \eqref{eq:corrLL}  means that the line  $L_{(e,m)}$ crosses $L_{(e',m')}$ from below. See Figure  \ref{fig:braidingcrossing} for a pictorial explanation. 

\begin{figure}
    \centering
    {\begin{tikzpicture}[baseline=-0.4]
 \draw[dashed] (0,0) circle (1);
 \draw [thick,blue, ->] (-0.7,-0.7)  to (0.7,0.7);
  \draw [thick,dgreen, <-] (-0.7,0.7)  to (-0.1,0.1);
    \draw [thick,dgreen] (0.1,-0.1)  to (0.7,-0.7);
    \node[below] at (-1,-0.9) {$L_{(e,m)}(\gamma)$};
    \node[below] at (1,-0.9) {$L_{(e',m')}(\gamma')$};
\end{tikzpicture}}
$=$ 
{\begin{tikzpicture}[baseline=-0.4]
 \draw[dashed] (0,0) circle (1);
 \draw[thick,rotate=45,blue, <-] (0,0.3) [partial ellipse=-80:260:0.4cm and 0.15cm];

  \draw [thick,dgreen,<-] (-0.7,0.7)  to (-0.4,0.4);
    \draw [thick,dgreen] (-0.2,0.2)  to (0.7,-0.7);
    \draw [thick,blue ] (-0.7,-0.7)  to (-0.1,-0.1);
  \draw [thick,blue, ->] (0.1,0.1)  to (0.7,0.7);
  \node[below] at (-1,-0.9) {$L_{(e,m)}(\gamma)$};
    \node[below] at (1,-0.9) {$L_{(e',m')}(\gamma')$};
\end{tikzpicture}}
$=$ $e^{-2\pi i (em'+me') /N}$
{\hspace{-0.3cm}\begin{tikzpicture}[baseline=-0.4]
 \draw[dashed] (0,0) circle (1);
 \draw [thick,blue ] (-0.7,-0.7)  to (-0.1,-0.1);
  \draw [thick,blue,->] (0.1,0.1)  to (0.7,0.7);
  \draw [thick,dgreen,<-] (-0.7,0.7)  to (0.7,-0.7);
  \node[below] at (-1,-0.9) {$L_{(e,m)}(\gamma)$};
    \node[below] at (1,-0.9) {$L_{(e',m')}(\gamma')$};
\end{tikzpicture}}
\quad.
    \caption{Braiding between lines in the 3d bulk gives crossing between lines on a 2d plane.  }
    \label{fig:braidingcrossing}
\end{figure}
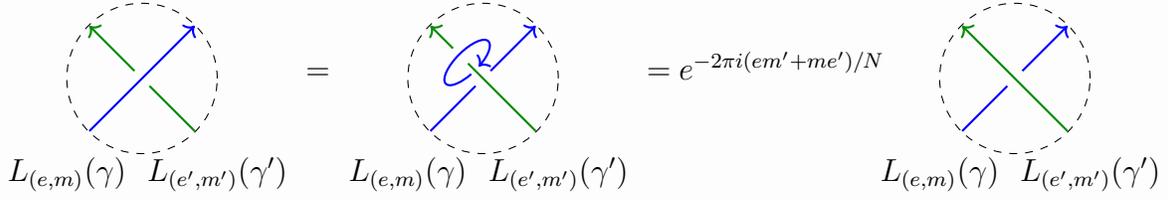

\item \textbf{Quantum torus algebra:} The correlation function \eqref{eq:corrLL}  implies a quantum torus algebra on the plane. To see this, consider two lines of the same type but supported on two seperate segments, $L_{(e,m)}(\gamma)$ and $L_{(e,m)}(\gamma')$. From the definition in \eqref{eq:beforegaugingLs}, it is obvious that $L_{(1,0)}(\gamma+\gamma')= L_{(1,0)}(\gamma)L_{(1,0)}(\gamma')$ and similarly for $L_{(0,1)}$. However, this is not true for general $L_{(e,m)}$ due to the non-commutativity \eqref{eq:corrLL}. 
By using the definition \eqref{eq:beforegaugingLs} to expand $L_{(e,m)}= \left(L_{(1,0)}\right)^e \left(L_{(0,1)}\right)^m$, and applying the commutation relations mentioned above,  the quantum torus algebra is
\begin{eqnarray}\label{eq:quanttoral}
L_{(e,m)}(\gamma) L_{(e,m)}(\gamma') = \exp\left( -\frac{2\pi i}{N}  em \braket{\gamma,\gamma'}\right)  L_{(e,m)}(\gamma+\gamma')~.
\end{eqnarray}
This will be useful below when discussing the symmetry defect of the $\Z_2^{\text{EM}}$ electro-magnetic exchange symmetry.

\end{enumerate}

\subsubsection{$\Z_2^{\text{EM}}$ symmetry and condensation defects}
\label{sec:EMsurfaceop}

The $\Z_N$ gauge theory \eqref{eq:3dBFtheory} has an electromagnetic exchange symmetry $\Z_2^{\text{EM}}$ which interchanges the two gauge fields
\bea
\label{eq:EMexchangedef}
a \rightarrow \widehat{a} ~, \hspace{0.8 in} \widehat{a} \rightarrow a~.
\eea
Indeed, this manifestly leaves the action (\ref{eq:3dBFtheory}) (on a closed manifold)  invariant. This symmetry is a zero-form symmetry, and there is a corresponding codimension-one surface defect $D_{\mathrm{EM}}$ in the bulk implementing the symmetry transformation. It was proven in \cite{Fuchs:2012dt} and rediscovered in \cite{Roumpedakis:2022aik} that the symmetry defect of any zero-form symmetry in $(2+1)$d TQFT with a single vacuum can be realized as a condensation defect, including the defect generating $\Z_2^{\text{EM}}$. In particular, it was pointed out in \cite{Roumpedakis:2022aik} that the defect for $\Z_2^{\text{EM}}$ in a $\Z_N$ gauge theory can be realized by condensing $L_{(1,N-1)}$ on a surface. Below, we provide further motivation for why $D_{\mathrm{EM}}$ should be a condensation defect, and then provide a rigorous definition. We then perform some consistency checks. 

\begin{figure}[t]
    \centering
    {\begin{tikzpicture}[baseline=35]
 \shade[top color=red!40, bottom color=red!10,rotate=90]  (0,-0.7) -- (2,-0.7) -- (2.6,-0.2) -- (0.6,-0.2)-- (0,-0.7);
 \draw[thick,rotate=90] (0,-0.7) -- (2,-0.7);
\draw[thick,rotate=90] (0,-0.7) -- (0.6,-0.2);
\draw[thick,rotate=90]  (0.6,-0.2)--(2.6,-0.2);
\draw[thick,rotate=90]  (2.6,-0.2)-- (2,-0.7);
\draw[thick,blue,postaction={on each segment={mid arrow=blue}}] (-0.9,1.4)--++(1.3,0) coordinate (blue_r);
\draw[thick,dgreen,dotted] (blue_r)--++(0.28,0) coordinate(green_dot);
\draw[thick,dgreen,postaction={on each segment={mid arrow=dgreen}}] (green_dot) ++ (0.04,0)--++(1.3,0) coordinate(green_r);
\node[right] at (-2.1,1.4) {$L_{(e,m)}$};
\node[right] at (2,1.4) {$L_{(m,e)}$};
\node[below] at (0.4, -0) {$D_{\text{EM}}$};
\end{tikzpicture}}
    \caption{The piercing action of the $\Z_2^{\text{EM}}$ symmetry defect $D_{\text{EM}}$ on the line $L_{(e,m)}$. }
    \label{fig:pierce}
\end{figure}
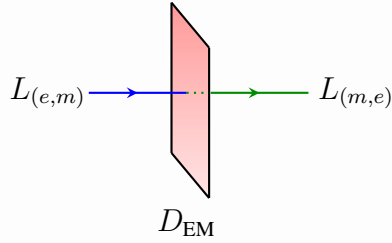

\begin{figure}
    \centering
        \begin{tikzpicture}[baseline=-30]
     \shade[top color=red!40, bottom color=red!10,draw,thick]  (0,0) -- (2,0) -- (2,-2) -- (0,-2)--cycle;
     \node[anchor=north] at (1,-2) {$D_{\text{EM}}$};
        \end{tikzpicture}
        \quad = \quad
        \begin{tikzpicture}[baseline=-30]
     \shade[top color=red!40, bottom color=red!10,draw,thick]  (0,0) -- (2,0) -- (2,-2) -- (0,-2)--cycle;
     \draw[thick,red,fill=white] (1,-1) circle(.3); 
     \node[anchor=north] at (1,-2) {$D_{\text{EM}}$};
        \end{tikzpicture}
    \caption{One can punch a hole in the $\mathbb{Z}_2^{\text{EM}}$ symmetry defect.} 
    \label{fig:hole}
\end{figure}
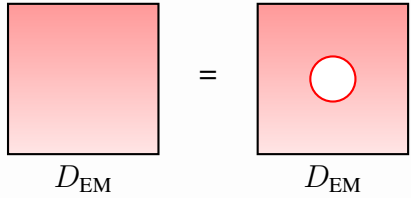

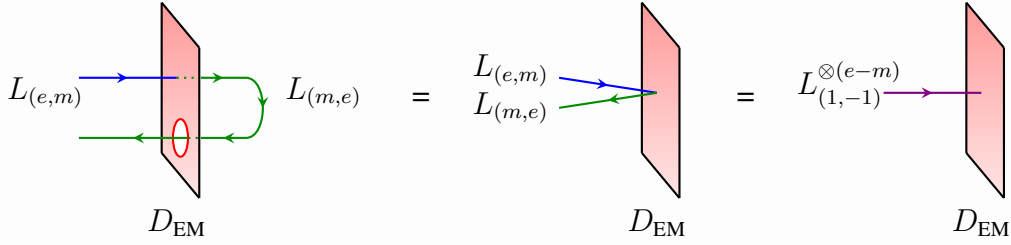
\begin{figure}[t]
    \centering
    {\begin{tikzpicture}[baseline=35]
 \shade[top color=red!40, bottom color=red!10,rotate=90]  (0,-0.7) -- (2,-0.7) -- (2.6,-0.2) -- (0.6,-0.2)-- (0,-0.7);
 \draw[thick,rotate=90] (0,-0.7) -- (2,-0.7);
\draw[thick,rotate=90] (0,-0.7) -- (0.6,-0.2);
\draw[thick,rotate=90]  (0.6,-0.2)--(2.6,-0.2);
\draw[thick,rotate=90]  (2.6,-0.2)-- (2,-0.7);
\draw[thick,blue,postaction={on each segment={mid arrow=blue}}] (-0.9,1.6)--++(1.3,0) coordinate (blue_r);
\draw[thick,dgreen,dotted] (blue_r)--++(0.28,0) coordinate(green_dot);
\draw[thick,dgreen,postaction={on each segment={mid arrow=dgreen}}] (green_dot) ++ (0.04,0)--++(0.6,0) coordinate(green_r) ..  controls ++(0.3,0) and ++ (0.3,0) .. ++(0,-.8) coordinate(green_d) ;
\draw[thick,red,fill=white,thick] (green_d) ++(-.87,.25) arc(90:450:.1 and .25);
\draw[thick,dgreen,postaction={on each segment={mid arrow=dgreen}}] (green_d) -- ++(-.6,0) coordinate(green_d2);
\draw[dotted,dgreen,thick] (green_d2) ++(-.04,0) -- ++(-.14,0) coordinate(green_d3);
\draw[dgreen,thick,postaction={on each segment={mid arrow=dgreen}}] (green_d3) -- ++(-1.44,0);
\node[right] at (-2,1.4) {$L_{(e,m)}$};
\node[right] at (1.7,1.4) {$L_{(m,e)}$};
\node[below] at (0.4, -0) {$D_{\text{EM}}$};
\end{tikzpicture}}
\quad=\quad
{\begin{tikzpicture}[baseline=35]
 \shade[top color=red!40, bottom color=red!10,rotate=90]  (0,-0.7) -- (2,-0.7) -- (2.6,-0.2) -- (0.6,-0.2)-- (0,-0.7);
 \draw[thick,rotate=90] (0,-0.7) -- (2,-0.7);
\draw[thick,rotate=90] (0,-0.7) -- (0.6,-0.2);
\draw[thick,rotate=90]  (0.6,-0.2)--(2.6,-0.2);
\draw[thick,rotate=90]  (2.6,-0.2)-- (2,-0.7);
\draw[thick,blue,postaction={on each segment={mid arrow=blue}}] (-0.9,1.6)--(0.4,1.4) coordinate(r);
\draw[thick,dgreen,postaction={on each segment={mid arrow=dgreen}}] (r)--(-0.9,1.2);
\node[right] at (-2.2,1.7) {$L_{(e,m)}$};
\node[right] at (-2.2,1.2) {$L_{(m,e)}$};
\node[below] at (0.4, -0) {$D_{\text{EM}}$};
\end{tikzpicture}}
\quad=\quad
{\begin{tikzpicture}[baseline=35]
 \shade[top color=red!40, bottom color=red!10,rotate=90]  (0,-0.7) -- (2,-0.7) -- (2.6,-0.2) -- (0.6,-0.2)-- (0,-0.7);
 \draw[thick,rotate=90] (0,-0.7) -- (2,-0.7);
\draw[thick,rotate=90] (0,-0.7) -- (0.6,-0.2);
\draw[thick,rotate=90]  (0.6,-0.2)--(2.6,-0.2);
\draw[thick,rotate=90]  (2.6,-0.2)-- (2,-0.7);
\draw[thick,violet,postaction={on each segment={mid arrow=violet}}] (-0.9,1.4)--(0.4,1.4);
\node[right] at (-2.2,1.5) {$L_{(1,-1)}^{\otimes(e-m)}$};
\node[below] at (0.4, -0) {$D_{\text{EM}}$};
\end{tikzpicture}}\quad
    \caption{The $\Z_2^{\text{EM}}$ symmetry defect $D_{\text{EM}}$ can absorb the line $L_{(1,-1)}$. }
    \label{fig:Z2emdefect}
\end{figure}

We first motivate the condensation construction of the $\Z_2^{\text{EM}}$ symmetry defect $D_{\text{EM}}$.
Note that in a $(2+1)$d TQFT with a single vacuum, there are no non-trivial local operators. This means that $D_\text{EM}$ acts only on line operators via the piercing action depicted in Figure \ref{fig:pierce}, resulting from \eqref{eq:EMexchangedef}. 
Furthermore, it means that $D_\text{EM}$ admits topological boundary conditions that can be used to punch a hole in the defect (see Figure \ref{fig:hole}), since there are no local operators which could detect such a hole in the first place (see \cite[Theorem 6.7]{Fuchs:2012dt} for the case of $2+1$ dimensions, and \cite[Theorem 4]{johnson2020classification} for TQFTs in more general dimensions).
When a line $L_{(e,m)}$ enters $D_{\text{EM}}$ perpendicularly from the left, the line $L_{(m,e)}$ should leave $D_{\text{EM}}$ perpendicularly from the right. We may then consider a configuration in which the line $L_{(m,e)}$ is folded back through a hole to the left side of $D_{\text{EM}}$, as shown in Figure \ref{fig:Z2emdefect}.
By reversing the orientation of the folded line, which flips the sign of both electric and magnetic charges, we arrive at a configuration where a line $L_{(e-m,m-e)}=L_{(1,-1)}^{\otimes(e-m)}$ enters from the left and is absorbed into the surface defect. In other words, $D_{\text{EM}}$ can absorb $L_{(1,-1)}$. This implies that $D_{\text{EM}}$ should be a condensate of $L_{(1,-1)}$.

With the above motivation, we now give a precise definition of the $\Z_2^{\text{EM}}$ condensation defect supported on a surface $M_2$, following \cite{Roumpedakis:2022aik}, 
\bea\label{eq:DEMdef}
D_{\mathrm{EM}}(M_2) = {1\over {|H^0(M_2, \ZZ_N)|}} \sum_{\gamma \in H_1(M_2, \ZZ_N)} L_{(1,-1)}(\gamma)~.
\eea
Note that condensing $L_{(1,-1)}$ on $M_2$ amounts to gauging the $\Z_N^{(1)}$ one-form symmetry only on the surface $M_2$, which amounts to gauging a $\Z_N^{(0)}$ zero-form symmetry from the point of view of the surface. The normalization in \eqref{eq:DEMdef} comes precisely from the gauge redundancy of gauging the $\Z_N^{(0)}$ zero-form symmetry on $M_2$. However, as noted in Section \ref{sec:1+1ddualityinterfacedef}, such a gauging is always subjected to an Euler counterterm ambiguity. For example, the convention adopted in \cite{Roumpedakis:2022aik} is to further multiply \eqref{eq:DEMdef} by $\chi[M_2, \Z_N]^{1/2}$ such that $D_{\text{EM}}^2=1$. It turns out that the standard convention as in \eqref{eq:DEMdef} is more convenient when discussing operators with boundaries, so we work with that convention.

With this definition, we may verify that the fusion of $L_{(e,m)}$ with $D_{\mathrm{EM}}$ from the left is equivalent to fusion of $L_{(m,e)}$ with $D_{\mathrm{EM}}$ from the right, as desired. To see this, we consider $L_{(e,m)}(M_1)$ and $D_{\mathrm{EM}}(M_2)$, where $M_1$ is located to the left of $M_2$ and parallel to it.  We begin by noting that
\begin{equation}
\begin{split}
L_{(e,m)}(M_1)\times D_{\mathrm{EM}}(M_2)&= \frac{1}{|H^0(M_2, \ZZ_N)|}\sum_{\gamma \in H_1(M_2, \Z_N)} L_{(e,m)}(M_1) L_{(1,-1)}(\gamma)\\
&=\frac{1}{|H^0(M_2, \ZZ_N)|} \sum_{\gamma \in H_1(M_2, \Z_2)} e^{{2 \pi i \over N}(e-m) \langle M_1, \gamma \rangle} L_{(1,-1)}(\gamma) L_{(e,m)}(M_1) \\
\end{split}
\end{equation}
where we have used (\ref{eq:corrLL}). We may then use the fusion rules (\ref{eq:fusLL}) to split $L_{(e,m)}(M_1) = L_{(e-m,m-e)}(M_1) \times L_{(m,e)}(M_1) = L_{(1,N-1)}(M_1)^{ (e-m)} \times L_{(m,e)}(M_1)$.

Using the quantum torus algebra (\ref{eq:quanttoral}), we obtain 
\begin{equation}
\begin{split}
    L_{(e,m)}(M_1)\times D_{\mathrm{EM}}(M_2)&= \frac{1}{|H^0(M_2, \ZZ_N)|}\sum_{\gamma \in H_1(M_2, \Z_N)} L_{(1,-1)}(\gamma+(e-m)M_1) \times  L_{(m,e)}(M_1)
\\
&=D_{\mathrm{EM}}(M_2) \times L_{(m,e)}(M_1)~,
\end{split}
\end{equation}
giving the desired result.

We may also verify the invertibility of $D_{\text{EM}}$ by computing $D_{\mathrm{EM}}(M_2)\times D_{\mathrm{EM}}(M_2)$, 
\begin{eqnarray}
\begin{split}
    D_{\text{EM}}(M_2)\times D_{\text{EM}}(M_2)&= \frac{1}{|H^0(M_2, \ZZ_N)|^2}\sum_{\gamma, \gamma'\in H_1(M_2, \Z_N)}L_{(1,N-1)}(\gamma) L_{(1,N-1)}(\gamma')\\
    &=\frac{1}{|H^0(M_2, \ZZ_N)|^2}\sum_{\gamma, \gamma'\in H_1(M_2, \Z_N)} e^{\frac{2\pi i}{N}\braket{\gamma,\gamma'}}L_{(1,N-1)}(\gamma+\gamma')\\
    &= \frac{|H^1(M_2, \ZZ_N)|}{|H^0(M_2, \ZZ_N)|^2} = \chi[M_2, \Z_N]^{-1}
\end{split}
\end{eqnarray}
where we used the quantum torus algebra \eqref{eq:quanttoral} in the second equality. The result is an Euler counterterm, which confirms that $D_{\text{EM}}$  is an invertible condensation defect.

As discussed in \cite{Roumpedakis:2022aik}, there are actually more condensation defects in $\Z_N$ gauge theory than just $D_{\mathrm{EM}}$. These are obtained by condensing different line operators with or without discrete torsion. Most of these condensation defects are non-invertible. In the present work, we will only study the particular invertible condensation defect $D_{\text{EM}}$ implementing the $\Z_2^{\text{EM}}$ symmetry.

\subsection{Twist defects as higher duality interfaces}
\label{sec:Defectending}

So far we have only discussed topological defects without boundaries. We now consider those with boundaries. By gauge invariance, the simple line operator $L_{(e,m)}$ cannot be given a boundary (unless it ends at a non-trivial junction involving other lines, as discussed below). On the other hand, there is a topological boundary condition for the condensation operator $D_{\text{EM}}$, which can be used construct a new topological defect while maintaining gauge invariance. This is known as a \emph{twist defect} \cite{2015arXiv150306812T, Barkeshli:2014cna}. As we will see below, such twist defects can be interpreted as higher duality interfaces.

\subsubsection{Twist defects}

A twist defect can be defined by condensing $L_{(1,-1)}$ on a surface $M_2$ with a nontrivial boundary $M_1=\partial M_2$ and imposing Dirichlet boundary conditions on $M_1$. 
We denote this ``minimal" twist defect by $\Sigma_{(0)}$. Concretely,
\begin{eqnarray}\label{eq:twistdefect}
\Sigma_{(0)}(M_1, M_2)= \frac{1}{{|H^0(M_2, M_1, \Z_N)|}} \sum_{\gamma\in H_1(M_2, \Z_N)} L_{(1,-1)}(\gamma)~,
\end{eqnarray}
where $M_2$ has a boundary. Equation \eqref{eq:twistdefect} is almost identical to the definition of the condensation defect \eqref{eq:DEMdef}, with the only difference being that absolute cohomology $H^n(M_2,\Z_N)$ is replaced by relative cohomology $H^n(M_2,M_1,\Z_N)$ due to the Dirichlet boundary conditions.\footnote{Note that via Lefschetz duality $H^n(M_2,M_1,\Z_N) \simeq H_{2-n}(M_2,\Z_N)$, so the homology in the summand is still absolute. } Moreover, thanks to the Dirichlet boundary conditions, the twist defect is topological. The twist defect $\Sigma_{(0)}(M_1,M_2)$ can be understood as a non-genuine topological line operator on $M_1$ with a condensation surface on $M_2$ attached it.

Whereas fusing $L_{(1,-1)}$ with $D_{\mathrm{EM}}$ gives a new quantum operator (i.e. a line supported on a surface), because of the Dirichlet boundary conditions  the non-genuine line operator $\Sigma^{(0)}$ can absorb $L_{(1,-1)}$, and more generally any line of the form $L_{(n,-n)}$. On the other hand, it cannot absorb lines $L_{(e,m)}$ for $e + m \neq 0 \,\,\, \mathrm{mod}\,\, N$. We may thus obtain new twist operators by beginning with $\Sigma_{(0)}$ and fusing with such $L_{(e,m)}$. Due to the Dirichlet boundary conditions, the resulting operator does not depend on $e$ and $m$ individually, but only on the combination $e+m\mod N$, and hence we have $N$ distinct twist defects, labeled by $e+m=0,1,...,N-1$, each being writable in $N$ equivalent ways. Picking one representative, we can define
\begin{eqnarray}\label{eq:defoftwdefects}
\Sigma_{(e)}(M_1,M_2):= L_{(n+e,-n)}(M_1)\times \Sigma_{(0)}(M_1,M_2), \hspace{1cm} \forall \,\,n=0, 1, \dots, N-1~.
\end{eqnarray}
It is useful to note that $\Sigma_{(0)}$ is $\Z_2^{\text{EM}}$ invariant. As a consequence, $\Sigma_{(e)}$ is also $\Z_2^{\text{EM}}$ invariant, since
\begin{eqnarray}
\begin{split}
    \Z_2^{\text{EM}}:\,\, \Sigma_{(e)}(M_1, M_2) &\to L_{(-n,n+e)}(M_1)\times \Sigma_{(0)}(M_1,M_2)\\
&= L_{(n+e,-n)}(M_1)\times L_{(-2n-e, 2n+e)}(M_1)\times \Sigma_{(0)}(M_1,M_2)\\
&= L_{(n+e,-n)}(M_1)\times \Sigma_{(0)}(M_1,M_2) = \Sigma_{(e)}(M_1, M_2)~.
\end{split}
\end{eqnarray}

\subsubsection{Higher duality interfaces}

Before discussing the fusion of twist defects, let us compare their construction with that of duality interfaces \cite{Choi:2021kmx,Kaidi:2021xfk}, which was reviewed in Section \ref{sec:1+1ddualityinterface}. Duality interfaces are constructed by gauging a global symmetry in half of the entire spacetime and imposing  Dirichlet boundary conditions for the gauge fields. Likewise, the twist defect is defined by  gauging a global symmetry (a one-form symmetry in the present context) along half of a codimension-$q$ spacetime submanifold ($q=1$ in the present context), with Dirichlet boundary conditions on the codimension-$(q+1)$ boundary. Hence the twist defects can naturally be interpreted as \emph{higher duality interfaces}---duality interfaces associated with higher gauging.

In Section \ref{sec:3dDTYZN}, we will gauge the $\Z_2^{\text{EM}}$ symmetry of the $(2+1)$d gauge theory. After gauging, the condensation defect \eqref{eq:DEMdef} becomes transparent, and the twist defects become genuine line operators on $M_1$. Accordingly, the higher duality interface becomes a higher duality \textit{defect} since both sides of $M_1$ in $M_2$ support trivial operators after gauging.

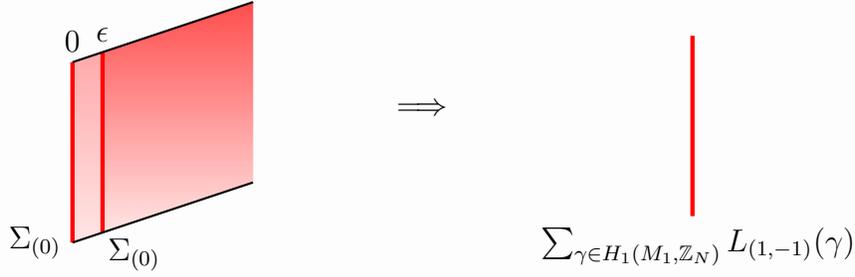
\begin{figure}[!tbp]
	\centering
	\[{\begin{tikzpicture}[scale=0.8,baseline=-20]
	 \shade[top color=red!40, bottom color=red!10]  (0,0) -- (0,-3) -- (3,-2) -- (3,1)-- (0,0);
 \draw[ultra thick,red] (0,0) -- (0,-3);

 \shade[top color=red!70, bottom color=red!10]   (0.5,0.5/3) --(0.5,0.5/3-3) -- (3,-2)-- (3,1)-- (0.5,0.5/3);
 
 \draw[ultra thick,red] (0.5,0.5/3) -- (0.5,0.5/3-3);
 
 \draw[thick] (0,-3) -- (3,-2);
 \draw[thick] (0,0) --(3,1);
 
 \node[above] at (0,0) {$0$};
 
 \node[above] at (0.5, 0.5/3) {$\eps$};

      \node[left] at (0,-3) {$\Sigma_{(0)}$};
       \node[left] at (1.65,-3.2) {$\Sigma_{(0)}$};
	\end{tikzpicture}}
	\hspace{0.7in}\Longrightarrow\qquad
	{\begin{tikzpicture}[scale=0.8,baseline=-30]
	 \draw[ultra thick,red] (0,0) -- (0,-3);
	 \node[below] at (0,-3) {\,\,$\sum_{\gamma\in H_1(M_1,\Z_N)} L_{(1,-1)}(\gamma)$};
	\end{tikzpicture}}
	\]
	\caption{Fusion of two twist defects $\Sigma_{(0)}$. }
	\label{fig:fusiontwdefect}
\end{figure}

\subsubsection{Fusion rules of the twist defects}

We now proceed to study the fusion rules of the twist defects. Interestingly, although the condensation defect $D_{\text{EM}}$ on a closed surface is invertible, the twist defect obeys non-invertible fusion rules.

The fusion between $\Sigma_{(e)}$ and the invertible defects $L_{(e',m')}$ follows directly from the definition \eqref{eq:defoftwdefects}, 
\begin{eqnarray}\label{sigma2fusion2}
\Sigma_{(e)}(M_1,M_2)\times L_{(e',m')}(M_1) = \Sigma_{(e+e'+m')}(M_1, M_2)~.
\end{eqnarray}
More nontrivial are the fusion rules between the twist defects themselves $\Sigma_{(e)}\times \Sigma_{(e')}$. It is simplest to first understand the fusion rule for $\Sigma_{(0)}\times \Sigma_{(0)}$, and then to fuse the outcome with an appropriate invertible line  $L_{(e+e',0)}$. 
The fusion rule $\Sigma_{(0)}\times \Sigma_{(0)}$ can be obtained by similar calculations as in Section \ref{sec:1+1ddualityinterface}. Near the boundary of $M_2$, the manifold is locally $M_1\times \mathbb{R}_{+}$. As before, we parameterize $\mathbb{R}_+$ by $x$ and label the manifolds as $M_2^{\geq 0}$ and $M_1|_0$ in the bulk and on the boundary, respectively. We take the two twist defects to be $\Sigma_{(0)}(M_1|_0, M_2^{\geq 0})$ and $\Sigma_{(0)}(M_1|_\eps, M_2^{\geq \eps})$. They overlap on the same two-dimensional manifold $M_2^{\geq \eps}$. Fusing two operators amounts to taking $\eps\to 0$, as seen in the left panel of Figure \ref{fig:fusiontwdefect}.

To proceed, we directly compute the fusion rule
\begin{eqnarray}\label{eq:TTfusion2d}
\begin{split}
    &\Sigma_{(0)}(M_1|_0,M_2^{\geq 0})\times \Sigma_{(0)}(M_1|_{\eps},M_2^{\geq \eps})\\&= \frac{1}{|H^0(M_2^{\geq 0}, M_1|_{0},\Z_N)| |H^0(M_2^{\geq \eps}, M_1|_{\eps},\Z_N)|} \sum_{\substack{\gamma\in H_1(M_2^{\geq 0}, \Z_N)\\
    \gamma'\in H_1(M_2^{\geq \eps}, \Z_N)}} L_{(1,-1)}(\gamma) L_{(1,-1)}(\gamma')~.
\end{split}
\end{eqnarray}
Evaluating this sum is analogous to evaluating \eqref{eq:NN}. The idea is to first rewrite the sum over lines $\gamma, \gamma'$ in absolute homology as a sum over the Lefschetz duals $a_{\gamma}, a_{\gamma'}$ in relative cohomology. Then we further convert the sums over relative cohomology into  sums over cochains by introducing additional fields $\phi, \phi'$ and appropriate BF terms. Integrating  out $a_{\gamma'}$ on $M_2^{\geq \eps}$, the right-hand side of \eqref{eq:TTfusion2d} becomes
\begin{eqnarray}
\begin{split}
     \chi[M_2^{\geq \eps}, \Z_N]^{-1} \frac{1}{|H^0(M_2^{[0,\eps]}, M_1|_0 \cup M_1|_\eps, \Z_N)|} \sum_{a_{\gamma}\in H^1(M_2^{[0,\eps]}, M_1|_0\cup M_1|_{\eps}, \Z_N)} L_{(1,-1)}(\gamma)
\end{split}
\end{eqnarray}
which is of precisely the same form as \eqref{eq:NNfusionfinal}. The first factor is the Euler counterterm and can be removed by appropriate field redefinition. The remaining factors have the physical interpretation of condensing $L_{(1,-1)}$ only on $M_2^{[0,\eps]}$, with Dirichlet boundary conditions on both $M_1|_0\cup M_1|_{\eps}$. Interpreting this in terms of higher gauging, it is the 1-gauging of the one-form symmetry generated by $L_{(1,-1)}$ on the slab $M_2^{[0,\eps]}$.  Taking the limit $\eps\to 0$, we find the fusion rule
\begin{eqnarray}\label{sigma2fusion0}
\Sigma_{(0)}(M_1,M_2)\times \Sigma_{(0)}(M_1,M_2) =\chi[M_2, \Z_N]^{-1} \sum_{\gamma\in H_1(M_1,\Z_N)} L_{(1,-1)}(\gamma)~,
\end{eqnarray}
where we have used $|H^0(M_2^{[0,\eps]}, M_1|_0 \cup M_1|_\eps, \Z_N)|=1$. One can likewise define the orientation reversal of $\Sigma_{(e)}$ by switching the two sides of  $M_1|_0$. As in \eqref{eq:Ninvertible}, we have 
\begin{eqnarray}
\overline{\Sigma}_{(e)}(M_1, M_2) = \chi[M_2, \Z_N]\cdot  \Sigma_{(-e)}(M_1, M_2)~.
\end{eqnarray}
Thus \eqref{sigma2fusion0} becomes 
\begin{eqnarray}
\Sigma_{(0)}(M_1,M_2)\times \overline{\Sigma}_{(0)}(M_1,M_2) =\sum_{\gamma\in H_1(M_1,\Z_N)} L_{(1,-1)}(\gamma)~.
\end{eqnarray}
The right hand side can be interpreted as 2-gauging of the one-form symmetry generated by $L_{(1,-1)}$ on a codimension two manifold $M_1$. When $M_1= S^1$, the right-hand side is simply the sum over $L_{(e,-e)}$ for all $e\in \Z_N$.

Given \eqref{sigma2fusion0}, we obtain the $\Sigma_{(e)}\times \Sigma_{(e')}$ fusion rule by further fusing with $L_{(e,0)}(M_1)$ and $L_{(e',0)}(M_1)$,
\begin{eqnarray}\label{sigma2fusion1}
\begin{split}
    \Sigma_{(e)}(M_1,M_2)\times \Sigma_{(e')}(M_1,M_2) &= \chi[M_2, \Z_N]^{-1} \sum_{\gamma\in H_1(M_1,\Z_N)} L_{(e+e',0)}(M_1) L_{(1,-1)}(\gamma)\\
    & 
    \mathrel{\stackrel{\makebox[0pt]{\mbox{\normalfont \footnotesize $M_1\to S^1$ }}}{=}} \hspace{0.8cm} \chi[M_2, \Z_N]^{-1} \sum_{n=0}^{N-1} L_{(n+e+e', -n)}(S^1)~.
\end{split}
\end{eqnarray}
In the special case of $N=2$, these fusion rules were discussed in \cite{2015arXiv150306812T} from a lattice Hamiltonian perspective. It immediately follows from \eqref{sigma2fusion1} that the twist defect $\Sigma_{(e)}$ is non-invertible, and has quantum dimension $\sqrt{N}$.

\subsection{Symmetry defects/interfaces in $(1+1)$d from topological defects in $(2+1)$d }
\label{sec:symdef1+12+1}

Having discussed the closed and open topological defects in $\Z_N$ gauge theory, we now insert these operators into the $(2+1)$d slab in Figure \ref{fig:2d3dSymTFT} and examine their behavior upon shrinking the slab. Schematically, we find the following correspondences
\begin{eqnarray}
\begin{split}
    \text{Twist defects } \Sigma_{(e)} &\longleftrightarrow \text{Duality interface } \CN\\
    \text{Magnetic line } L_{(0,1)} & \longleftrightarrow \text{$\Z_N^{(0)}$ symmetry defect } \eta\\
    \text{Electric line } L_{(1,0)} & \longleftrightarrow \text{$\Z_N^{(0)}$ order parameter } \CO 
\end{split}
\end{eqnarray}
These correspondences have also been discussed in e.g.  \cite{Gaiotto:2020iye, Burbano:2021loy,Moradi:2022lqp, Freed:2022qnc}.

\subsubsection{$\Z_N$ symmetry defects and order parameters from the bulk line operators}

We first insert a line operator into the $(2+1)$d slab. One can either place the line operator parallel to the topological boundary $X_2|_{0}$, or orthogonal to the boundary. Because of the Dirichlet boundary condition of $a$, i.e. $a|_{X_2|_{0}}=0$, the electric line $L_{(1,0)}$ can either end on the boundary perpendicularly, or be completely absorbed into the boundary (in the sense that it becomes a trivial operator) if it is parallel to it. Thus the only way that it can survive upon shrinking the slab is to place it orthogonal to the boundary, with one end anchored on the Dirichlet boundary on the left, and the other end anchored on the non-topological boundary on the right. As a consequence of ending on the non-topological boundary, one generically obtains a non-topological point-like operator $\CO$ upon shrinking the slab. This point operator will be charged under the $\Z^{(0)}_N$ symmetry of the $(1+1)$d theory, and hence serves as the non-topological $\Z^{(0)}_N$ order parameter.

On the other hand, the magnetic line $L_{(0,1)}$ can neither be absorbed into the Dirichlet boundary nor terminate on it. Hence it survives as a line defect upon shrinking the slab. This is the topological line $\eta$ generating the $\Z_N^{(0)}$ symmetry of the $(1+1)$d QFT $\CX$. Indeed, because $L_{(0,1)}$ and $L_{(1,0)}$ have a nontrivial correlation function when linked, it follows that the correlation function between $\eta$ and $\CO$ is also nontrivial when $\eta$ links  with the $\CO$. This link measures the $\Z^{(0)}_N$ charge of the order parameter $\CO$.

\begin{figure}[t]
	\centering
	\[{\begin{tikzpicture}[scale=0.8,baseline=-20]
	 \shade[top color=blue!40, bottom color=blue!10]  (0,0) -- (0,-3) -- (3,-2) -- (3,1)-- (0,0);
 \draw[ thick] (0,0) -- (0,-3);
\draw[thick] (0,-3) -- (3,-2);
 \draw[thick] (0,0) --(3,1);
 \draw[thick] (3,-2) -- (3,1);

 \shade[top color=red!40, bottom color=red!10]   (0.5,-0.1) --(0.5,-3.1) -- (3.5,-2.1)-- (3.5,0.9)-- (0.5,-0.1);
   \draw[ thick] (0.5,-0.1) -- (0.5,-3.1);
   \draw[thick] (0.5,-3.1) -- (3.5,-2.1);
      \draw[thick] (0.5,-0.1)-- (3.5,0.9);
      \draw[thick] (3.5,-2.1)-- (3.5,0.9);
      
      \node[left] at (0,-3) {$\bra{D(A)}$};
       \node[left] at (0.95,-3.6) {$D_{\text{EM}}$};
	\end{tikzpicture}}
	\hspace{0.7in}\Longrightarrow\qquad
	{\begin{tikzpicture}[scale=0.8,baseline=-30]
	 \shade[top color=violet!40, bottom color=violet!10]  (0,0) -- (0,-3) -- (3,-2) -- (3,1)-- (0,0);
 \draw[ thick] (0,0) -- (0,-3);
\draw[thick] (0,-3) -- (3,-2);
 \draw[thick] (0,0) --(3,1);
 \draw[thick] (3,-2) -- (3,1);
 \node[left] at (0,-3) {$\bra{N(A)}$};
	\end{tikzpicture}}
	\]
	\caption{Colliding $D_{\text{EM}}$ with Dirichlet boundary condition produces Neumann boundary condition.  }
	\label{fig:DEMcollide}
\end{figure}
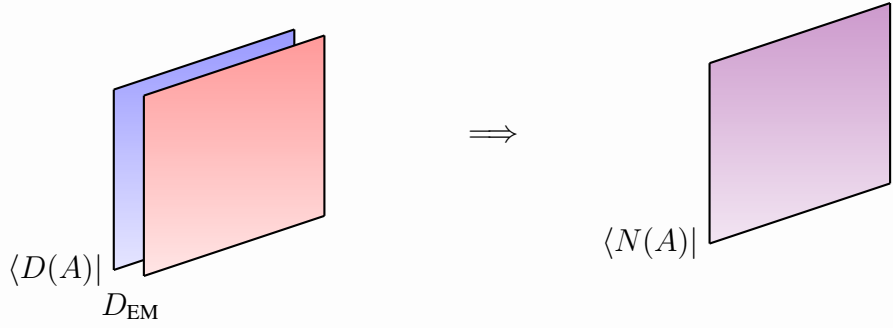

\subsubsection{Duality interfaces from the twist defects}

We next insert the condensation defect $D_{\text{EM}}(M_2)$ into the slab. When the condensation defect $D_{\text{EM}}(M_2)$ collides with the Dirichlet boundary condition $\bra{D(A)}$ on $X_2$, it produces the Neumann boundary condition $\bra{N(A)}$; see Figure \ref{fig:DEMcollide}.  One can see this explicitly as follows,
\begin{equation}\label{eq:emexchangebc}
\begin{split}
    \bra{D(A)}  &= \sum_{a\in H^1(X_2, \Z_N)} \bra{a} \delta(a-A)\,\,\,\, 
\xrightarrow{D_{\text{EM}}} \,\,\,\, \sum_{\widehat{a}\in H^1(X_2, \Z_N)} \bra{\widehat{a}} \delta(\widehat{a}-A)\\& = \frac{1}{|H^0(X_2,\Z_N)|} \sum_{a\in H^1(X_2, \Z_N)} \bra{a} e^{ \frac{2\pi i}{N} \int a A}  = \bra{N(A)}~,
\end{split}
\end{equation}
where we have used the fact that the EM dual bases are related by a discrete Fourier transform, i.e.
\bea
\bra{\widehat{a}} = \frac{1}{|H^0(X_2,\Z_N)|} \sum_{a\in H^1(X_2, \Z_N)} \bra{a} e^{i \frac{2\pi}{N}  \int a\widehat{a}} ~.
\eea

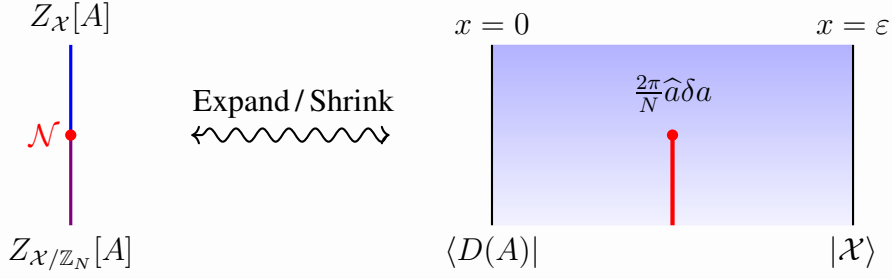
\begin{figure}[!tbp]
	\centering
	\begin{tikzpicture}[scale=0.8]
	
		\shade[line width=2pt, top color=blue!30, bottom color=blue!5] 
	(0,0) to [out=90, in=-90]  (0,3)
	to [out=0,in=180] (6,3)
	to [out = -90, in =90] (6,0)
	to [out=180, in =0]  (0,0);
	
	\draw[very thick, violet] (-7,0) -- (-7,1.5);
	\draw[very thick, blue] (-7,1.5) -- (-7,3);
	\node[above] at (-7,3) {$Z_{\CX}[A]$};
	\node[below] at (-7,0) {$Z_{\CX/\ZZ_N}[A]$};
	\node at (-7,1.5) [circle,fill,red, inner sep=1.5pt]{};
	\node[left,red] at (-7,1.5) {$\cN$};

	\draw[thick, snake it, <->] (-1.7,1.5) -- (-5, 1.5);
	\node[above] at (-3.3,1.6) {Expand\,/\,Shrink};
	
	\draw[thick] (0,0) -- (0,3);
	\draw[thick] (6,0) -- (6,3);
	\draw[ultra thick,red] (3,0) -- (3,1.5);
	\node[above] at (3,1.7) {$\frac{2\pi}{N} \widehat{a} \delta a$};
	\node at (3,1.5) [circle,fill,red, inner sep=1.5pt]{};
	\node[below] at (0,0) {$\langle D(A)| $};
	\node[below] at (6,0) {$|\cX\rangle $}; 
	
	\node[above] at (0,3) {$x=0$};
	\node[above] at (6,3) {$x=\varepsilon$};

	\end{tikzpicture}
	
	\caption{A $(1+1)$d QFT $\cX$ with $\Z_N$ symmetry and another $(1+1)$d QFT $\cX/\Z_N$ with quantum $\widehat{\Z}_N$ symmetry are separated by a topological interface $\cN$. This setup can be expanded into a $(2+1)$d slab, where the $(2+1)$d $\Z_N$ SymTFT has an insertion of a twist defect parallel to the Dirichlet boundary.  }
	\label{fig:twistdefshrink}
\end{figure}

We may now consider inserting a twist defect $\Sigma_{(e)}(M_1, M_2)$ into the slab. We place it parallel to the Dirichlet boundary, as shown in Figure \ref{fig:twistdefshrink}.\footnote{In \cite{Burbano:2021loy}, a similar configuration is considered. Their configuration can be obtained from our Figure \ref{fig:twistdefshrink} by bringing the endpoint of $D_{\text{EM}}$ to the left boundary and letting $D_{\text{EM}}$ stretch across the bulk towards the right boundary.} 
For convenience, we use $y$ to parameterize the horizontal direction in the slab, and $x$ to parameterize the vertical direction near the boundary of the twist defect, with the boundary $M_1$ located at $x=0$.  By colliding the twist defect with the left boundary of the slab, the lower half of the Dirichlet boundary conditions (i.e. in $X_2^{\geq 0}$) is transformed to Neumann boundary conditions. Thus upon shrinking the slab, from Section \ref{sec:3dZNgaugetheoryasSymTFT} the $(1+1)$d QFT on  $X_2^{\geq 0}$ becomes $\CX/\Z_N$, as shown in the left panel of Figure \ref{fig:twistdefshrink}. This implies that the twist defect $\Sigma_{(e)}$, when collided with the Dirichlet boundary, becomes a duality interface $\CN$.

Because there are $N$ twist defects in $(2+1)$d,  one may initially expect $N$ different duality interfaces in  $(1+1)$d. However, all $N$ types of twist defects actually reduce to a single type of duality interface upon colliding with the Dirichlet boundary condition. To see this, note that
the generic twist defect $\Sigma_{(e)}$ is related to the minimal twist defect $\Sigma_{(0)}$ by fusing an electric line $L_{(e,0)}$ on its boundary. Such electric line can be absorbed by the Dirichlet boundary condition of the SymTFT. As a consequence, when $\Sigma_{(e)}$ is brought to the Dirichlet boundary, it reduces to $\Sigma_{(0)}$. We thus have
\begin{eqnarray}
\CN(M_1) = \Sigma_{(e)}(M_1, M_2)|_{x\to 0}~, \hspace{1.2cm} \forall~e\in \Z_N~.
\end{eqnarray}

Furthermore, the fusion rules of the twist defects given in \eqref{sigma2fusion2} and \eqref{sigma2fusion1} descend to the fusion rules of the duality interfaces. Upon colliding with the Dirichlet boundary, $\Sigma_{(e)}$ and $L_{(e',m')}$  reduce to $\CN$ and  $\eta^{m'}$ respectively, and  \eqref{sigma2fusion2} thus simplifies to 
\begin{eqnarray}
\CN \times \eta^{m'} = \CN~,
\end{eqnarray}
which reproduces \eqref{2deq:etaNN}. Moreover, \eqref{sigma2fusion1} simplifies to 
\begin{eqnarray}
\CN \times \CN = \chi[M_2, \Z_N]^{-1}\sum_{\gamma\in H_1(M_1, \Z_N)} \eta(\gamma)~,
\end{eqnarray}
which reproduces \eqref{eq:NNfusionfinal}.

\subsection{F-symbols of twist defects and duality interfaces}
\label{sec:3dFsymbols}

We close this section by discussing the F-symbols for lines, both genuine and non-genuine, of the $\Z_N$ gauge theory, as well as the F-symbols of the duality interfaces. Additional details on the techniques used here are given in Appendix \ref{app:junctioncharge}.

 \begin{figure}[tbp]
\begin{center}

{\begin{tikzpicture}[baseline=0,square/.style={regular polygon,regular polygon sides=4},scale=0.6]
\draw [blue, thick, decoration={markings, mark=at position 0.5 with {\arrow{>}}}, postaction={decorate}] (1,1) -- (2,2);
\draw [blue, thick, decoration={markings, mark=at position 0.5 with {\arrow{>}}}, postaction={decorate}] (0,0) -- (1,1);
\draw [blue, thick, decoration={markings, mark=at position 0.5 with {\arrow{>}}}, postaction={decorate}] (-1,-1) -- (0,0);
\draw [blue, thick, decoration={markings, mark=at position 0.5 with {\arrow{>}}}, postaction={decorate}] (1,-1) -- (0,0);
\draw [blue, thick, decoration={markings, mark=at position 0.5 with {\arrow{>}}}, postaction={decorate}] (3,-1) -- (1,1);

\filldraw[blue] (0,0) circle (2pt);
\filldraw[blue] (1,1) circle (2pt);

\node[below] at (-1.2,-1) {\footnotesize \color{blue}$L_{(e_1,m_1)}$};
\node[below] at (1.1,-1) {\footnotesize \color{blue}$L_{(e_2,m_2)}$};
\node[below] at (3.6,-1) {\footnotesize \color{blue}$L_{(e_3,m_3)}$};
\node[left] at (0.4,0.8) {\footnotesize \color{blue}$L_{(e_1+e_2,m_1+m_2)}$};
\end{tikzpicture}}
= \hspace{0.1 in}
{\begin{tikzpicture}[baseline=0,square/.style={regular polygon,regular polygon sides=4},scale=0.6]
\draw [blue, thick, decoration={markings, mark=at position 0.5 with {\arrow{>}}}, postaction={decorate}] (1,1) -- (2,2);
\draw [blue, thick] (0,0) -- (1,1);
\draw [blue, thick, ->] (-1,-1) -- (0,0);
\draw [blue, thick, decoration={markings, mark=at position 0.5 with {\arrow{>}}}, postaction={decorate}] (1,-1) -- (2,0);
\draw [blue, thick, decoration={markings, mark=at position 0.5 with {\arrow{>}}}, postaction={decorate}] (3,-1) -- (2,0);
\draw [blue, thick, decoration={markings, mark=at position 0.5 with {\arrow{>}}}, postaction={decorate}] (2,0) -- (1,1);

\filldraw[blue] (2,0) circle (2pt);
\filldraw[blue] (1,1) circle (2pt);

\node[below] at (-1.2,-1) {\footnotesize \color{blue}$L_{(e_1,m_1)}$};
\node[below] at (1.1,-1) {\footnotesize \color{blue}$L_{(e_2,m_2)}$};
\node[below] at (3.6,-1) {\footnotesize \color{blue}$L_{(e_3,m_3)}$};
\node[right] at (1.3,0.8) {\footnotesize \color{blue}$L_{(e_2+e_3,m_2+m_3)}$};
\end{tikzpicture}}
\\\vspace{0.2 in}
\hspace{0.0 in}{\begin{tikzpicture}[baseline=-5,rotate=180]
\draw [blue,thick,decoration={markings, mark=at position 0.5 with {\arrow{>}}}, postaction={decorate}] (0,-.5)  to (0,0);
\draw [blue,thick] (0,0) arc [radius=.3, start angle=240, end angle=120];
\draw [blue,thick, decoration={markings, mark=at position 0.7 with {\arrow{>}}}, postaction={decorate}] (0,0.52) to +(30:.5) ;
\draw [blue,thick] (0,0) arc [radius=.3, start angle=-60, end angle=42];
\draw [blue,thick, decoration={markings, mark=at position 0.6 with {\arrow{>}}}, postaction={decorate}] (0,0.52) ++(150:.1) to +(150:.4) ;

\filldraw[blue] (0,0) circle (1.2pt);

\node[below] at (0.8,0.7) {\footnotesize \color{blue}$L_{(e_1,m_1)}$};
\node[below] at (-0.9,0.7) {\footnotesize \color{blue}$L_{(e_2,m_2)}$};
\end{tikzpicture}}
\hspace{-0.2 in}= \,$e^{-{2 \pi i \over N} e_1 m_2}$\hspace{-0.2 in}
{\begin{tikzpicture}[baseline=5,rotate=180]
\draw [blue,thick,decoration={markings, mark=at position 0.5 with {\arrow{>}}}, postaction={decorate}] (0,-.8) to (0,0);
\draw [blue,thick, decoration={markings, mark=at position 0.7 with {\arrow{>}}}, postaction={decorate}] (0,0) to (-.6,.5);
\draw [blue,thick, decoration={markings, mark=at position 0.7 with {\arrow{>}}}, postaction={decorate}] (0,0)--(.6,.5) ;
\filldraw[blue] (0,0) circle (1.2pt);

\node[below] at (0.8,0.4) {\footnotesize \color{blue}$L_{(e_1,m_1)}$};
\node[below] at (-0.9,0.4) {\footnotesize \color{blue}$L_{(e_2,m_2)}$};
\end{tikzpicture}}
\hspace{0.2 in}{\begin{tikzpicture}[baseline=-5,rotate=180]
\draw [blue,thick,decoration={markings, mark=at position 0.5 with {\arrow{>}}}, postaction={decorate}] (0,-.5)  to (0,0);
\draw [blue,thick] (0,0) arc [radius=.3, start angle=240, end angle=120];

\draw [blue,thick] (0,0) arc [radius=.3, start angle=-60, end angle=42];

\draw [blue,thick, decoration={markings, mark=at position 0.8 with {\arrow{>}}}, postaction={decorate}] (-0.1,0.42) -- (0.4,0.7) ;

\draw [white, line width = 0.15cm] (0.1,0.42) -- (-0.4,0.7) ;

\draw [blue,thick, decoration={markings, mark=at position 0.8 with {\arrow{>}}}, postaction={decorate}] (0.1,0.42) -- (-0.4,0.7) ;

\filldraw[blue] (0,0) circle (1.2pt);

\node[below] at (0.8,0.7) {\footnotesize \color{blue}$L_{(e_1,m_1)}$};
\node[below] at (-0.9,0.7) {\footnotesize \color{blue}$L_{(e_2,m_2)}$};
\end{tikzpicture}}
\hspace{-0.2 in}= \,$e^{{2 \pi i \over N} e_2 m_1}$\hspace{-0.2 in}
{\begin{tikzpicture}[baseline=5,rotate=180]
\draw [blue,thick,decoration={markings, mark=at position 0.5 with {\arrow{>}}}, postaction={decorate}] (0,-.8) to (0,0);
\draw [blue,thick, decoration={markings, mark=at position 0.7 with {\arrow{>}}}, postaction={decorate}] (0,0) to (-.6,.5);
\draw [blue,thick, decoration={markings, mark=at position 0.7 with {\arrow{>}}}, postaction={decorate}] (0,0)--(.6,.5) ;
\filldraw[blue] (0,0) circle (1.2pt);

\node[below] at (0.8,0.4) {\footnotesize \color{blue}$L_{(e_1,m_1)}$};
\node[below] at (-0.9,0.4) {\footnotesize \color{blue}$L_{(e_2,m_2)}$};
\end{tikzpicture}}

\caption{F-symbols and half-braidings for genuine lines $L_{(e,m)}$.}
\label{fig:Fsymbhalfbraid}
\end{center}
\end{figure}
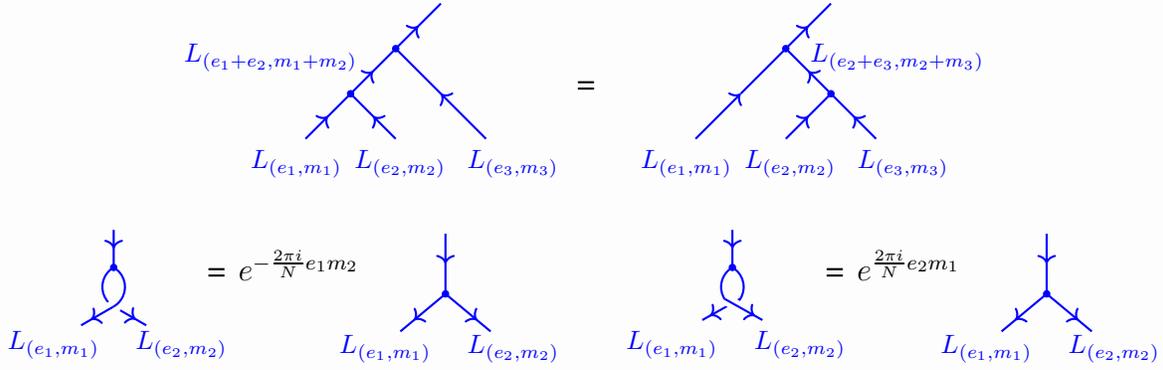

\subsubsection{F-symbols of twist defects}

In the current context, it is possible to work in a basis of the junction vector space such that the F-symbols for the genuine lines $L_{(e,m)}$ are trivial, and the half-braiding is as shown in Figure \ref{fig:Fsymbhalfbraid}. Note in particular that this half-braiding correctly reproduces the full braiding given in Figure \ref{fig:braidingcrossing}. Our goal is to now incorporate twist defects.

Before considering the F-symbols involving twist defects, we must first discuss the relevant trivalent junctions. 
 From the fusion rules in (\ref{sigma2fusion1}), it is clear that each junction should involve two twist defects $\Sigma_{(e)}$ and $\Sigma_{(e')}$, together with one genuine line $L_{(e+e'+n,-n)}$ for arbitrary $n$. A crucial fact is that these junctions actually depend on the \textit{angle} $\theta$ between $L_{(e+e'+n,-n)}$ and the surface $D_{\mathrm{EM}}$ on which $\Sigma_{(e)}$ and $\Sigma_{(e')}$ are anchored, the signature of framing dependence. We will consider two angles for our current analysis, namely $\theta = 0^+$ and $\pi$.  The junction with $\theta = \pi$ will be labelled with a triangle, while the junction with $\theta = 0^+$ will be denoted by a square. The notation $\theta=0^+$ is to emphasize that the $D_{\text{EM}}$ surface is \textit{behind} the line anchored on the square junction. Both of these configurations are illustrated in Figure \ref{fig:squarevstrianglejunc}. They will be studied in more detail in Appendix \ref{app:junctioncharge}.

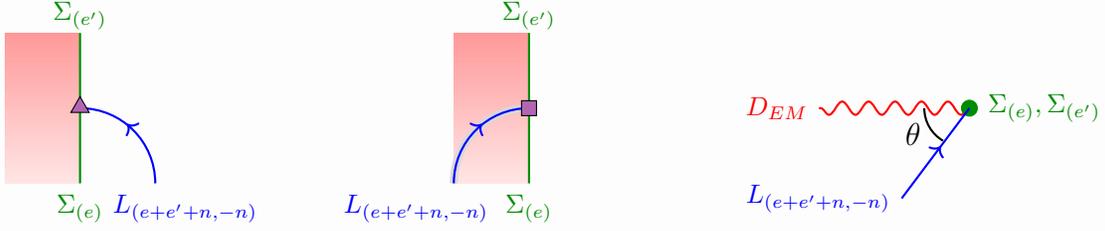
\begin{figure}[tbp]
\begin{center}
 {\begin{tikzpicture}[baseline=0]

 \shade[top color=red!40, bottom color=red!10]  (0,-1) -- (-1,-1) -- (-1,1) -- (0,1)-- (0,-1);
 
\draw[dgreen,thick](0,-1) -- (0,1); 
\draw [blue, thick,decoration={markings, mark=at position 0.5 with {\arrow{<}}}, postaction={decorate}] (0,0) to[out=0,in=90 ]  (1,-1);
\node[dgreen, above] at (0,0.9) {\footnotesize{$\Sigma_{(e')}$}};
\node[dgreen, below] at (0,-1) {\footnotesize{$\Sigma_{(e)}$}};
\node[blue, below] at (1.4,-1) {\footnotesize{$L_{(e+e'+n,-n)}$}};

\node[isosceles triangle,scale=0.4,
    isosceles triangle apex angle=60,
    draw,fill=violet!60,
    rotate=90,
    minimum size =0.01cm] at (0,0){};
    
\end{tikzpicture}}\hspace{0.3in}
{\begin{tikzpicture}[baseline=0,square/.style={regular polygon,regular polygon sides=4}]

 \shade[top color=red!40, bottom color=red!10]  (0,-1) -- (-1,-1) -- (-1,1) -- (0,1)-- (0,-1);
\draw[dgreen,thick](0,-1) -- (0,1); 
\draw [gray!30, line width=3pt] (0,0) to[out=180,in=90 ]  (-1,-1);
\draw [blue, thick,decoration={markings, mark=at position 0.5 with {\arrow{<}}}, postaction={decorate}] (0,0) to[out=180,in=90 ]  (-1,-1);
\node[dgreen, above] at (0,0.9) {\footnotesize{$\Sigma_{(e')}$}};
\node[dgreen, below] at (0,-1) {\footnotesize{$\Sigma_{(e)}$}};
\node[blue, below] at (-1.5,-1) {\footnotesize{$L_{(e+e'+n,-n)}$}};

\node at (0,0) [square,draw,fill=violet!60,scale=0.5] {}; 
   
\end{tikzpicture}}
\hspace{0.8in}
{\begin{tikzpicture}[baseline=0]
\draw[red, thick, snake it, -] (0,0) -- (-2,0);
\filldraw[dgreen] (0,0) circle (3pt);
\node[dgreen,right] at (0.1,0) {\footnotesize{$\Sigma_{(e)},\Sigma_{(e')}$}};
\node[red,left] at (-2,0) {\footnotesize{$D_{EM}$}};

\draw[blue, thick, decoration={markings, mark=at position 0.5 with {\arrow{<}}}, postaction={decorate}] (0,0) -- (-0.9,-1.2);

\draw[thick] (0,0)+(-0.6,0cm) arc[start angle=180, end angle=240, radius=0.5cm];
\node[blue,left] at (-0.9,-1.2) {\footnotesize{$L_{(e+e'+n,-n)}$}};
\node[left] at (-0.5,-0.35) {$\theta$};
\end{tikzpicture}}
\caption{The triangle junction with $\theta = \pi$, together with the square junction with $\theta = 0^+$. The $\ZZ_2^{\mathrm{EM}}$ duality surface $D_{\mathrm{EM}}$, which is a mesh of  algebra objects $\CA=\bigoplus_{n=0}^{N-1} L_{(n,-n)}$ that can end on $\Sigma_{(e)}$, is drawn in red.  When there is no ambiguity, we will suppress the $D_{\text{EM}}$ surface in our figures, taking it to go out to the left. }
\label{fig:squarevstrianglejunc}
\end{center}
\end{figure}

Another important fact is that the junctions should not depend on where along $\Sigma_{(e)}$ the line $L_{(e+e'+n,-n)}$ is anchored. To phrase this more concretely, we begin by defining the algebra object $\cA= \bigoplus_{n=0}^{N-1} L_{(n,-n)}$.
The condensation defect $D_{\mathrm{EM}}$ can be resolved into a mesh of such algebra objects \cite{Fuchs:2002cm, Kapustin:2010if, Carqueville:2017ono}, and in terms of this mesh the invariance under change in the anchor point is equivalent to Figure \ref{fig:squaretriangconsconds}. The $\mu_L$ and $\mu_L^\vee$ appearing in Figure \ref{fig:squaretriangconsconds} are certain junction vectors, defined in Appendix \ref{app:junctioncharge}, which will not be important to us here. What is important for us is that, given a solution to the consistency condition for the triangle junction, a solution for the square junction can be obtained by defining the junction as in Figure \ref{fig:tritosquare}. This is discussed in more detail in Appendix \ref{app:junctioncharge}. We will take this as the definition of the square junction from now on.

\begin{figure}[tbp]
\begin{center}
{\begin{tikzpicture}[baseline=0]

\draw[dgreen,thick](0,-1) -- (0,1); 
\draw [blue, thick, decoration={markings, mark=at position 0.5 with {\arrow{<}}}, postaction={decorate}] (0,0) to[out=0,in=90 ]  (1,-1);
\draw [red, thick] (0,-0.5) to[out=180,in=270 ]  (-0.5,0);
\draw [red, thick] (-0.5,0) to[out=90,in=180 ]  (0,0.5);

\node[red, left] at (-0.5,0) {\footnotesize{$\mathcal{A}$}};
\node[dgreen, below] at (0,-1) {\footnotesize{$\Sigma_{(e)}$}};
\node[blue, below] at (1.5,-1) {\footnotesize{$L_{(e+e'+n,-n)}$}};
\node[red, right] at (0,0.5) {\footnotesize{$\mu_L$}};
\node[red, right] at (0,-0.5) {\footnotesize{$\mu_L^\vee$}};

\filldraw[red] (0,0.5) circle (1.5pt);
\filldraw[red] (0,-0.5) circle (1.5pt);

\node[isosceles triangle,scale=0.4,
    isosceles triangle apex angle=60,
    draw,fill=violet!60,
    rotate=90,
    minimum size =0.01cm] at (0,0){};
    
\end{tikzpicture}}
\hspace{-0.2in}$=$\,\,\,\,
 {\begin{tikzpicture}[baseline=0]

\draw[dgreen,thick](0,-1) -- (0,1); 
\draw [blue, thick, decoration={markings, mark=at position 0.5 with {\arrow{<}}}, postaction={decorate}] (0,0) to[out=0,in=90 ]  (1,-1);

\node[dgreen, below] at (0,-1) {\footnotesize{$\Sigma_{(e)}$}};
\node[blue, below] at (1.5,-1) {\footnotesize{$L_{(e+e'+n,-n)}$}};


\node[isosceles triangle,scale=0.4,
    isosceles triangle apex angle=60,
    draw,fill=violet!60,
    rotate=90,
    minimum size =0.01cm] at (0,0){};
    
\end{tikzpicture}}\hspace{0.3in}
{\begin{tikzpicture}[baseline=0,square/.style={regular polygon,regular polygon sides=4}]

\draw[dgreen,thick](0,-1) -- (0,1); 
\draw [red, thick] (0,-0.5) to[out=180,in=270 ]  (-0.5,0);
\draw [red, thick] (-0.5,0) to[out=90,in=180 ]  (0,0.5);
\draw [white, line width=5pt] (0,0) to[out=180,in=90 ]  (-1,-1);
\draw [blue, thick, decoration={markings, mark=at position 0.5 with {\arrow{<}}}, postaction={decorate}] (0,0) to[out=180,in=90 ]  (-1,-1);

\node[red, left] at (-0.5,0) {\footnotesize{$\mathcal{A}$}};
\node[dgreen, below] at (0,-1) {\footnotesize{$\Sigma_{(e)}$}};
\node[blue, below] at (-1.5,-1) {\footnotesize{$L_{(e+e'+n,-n)}$}};
\node[red, right] at (0,0.5) {\footnotesize{$\mu_L$}};
\node[red, right] at (0,-0.5) {\footnotesize{$\mu_L^\vee$}};

\filldraw[red] (0,0.5) circle (1.5pt);
\filldraw[red] (0,-0.5) circle (1.5pt);

\node at (0,0) [square,draw,fill=violet!60,scale=0.5] {}; 
   
\end{tikzpicture}}
\,\,\,\,$=$\hspace{-0.2in}
{\begin{tikzpicture}[baseline=0,square/.style={regular polygon,regular polygon sides=4}]

\draw[dgreen,thick](0,-1) -- (0,1); 
\draw [white, line width=5pt] (0,0) to[out=180,in=90 ]  (-1,-1);
\draw [blue, thick, decoration={markings, mark=at position 0.5 with {\arrow{<}}}, postaction={decorate}] (0,0) to[out=180,in=90 ]  (-1,-1);

\node[dgreen, below] at (0,-1) {\footnotesize{$\Sigma_{(e)}$}};
\node[blue, below] at (-1.5,-1) {\footnotesize{$L_{(e+e'+n,-n)}$}};


\node at (0,0) [square,draw,fill=violet!60,scale=0.5] {}; 
   
\end{tikzpicture}}

\caption{Consistency conditions for the triangle and square junctions.}
\label{fig:squaretriangconsconds}
\end{center}
\end{figure}
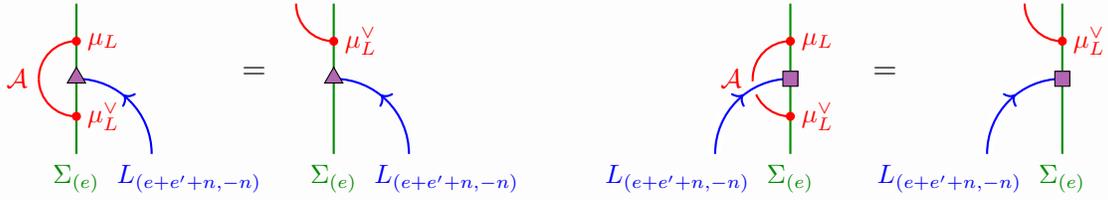

\begin{figure}[tbp]
\begin{center}
{\begin{tikzpicture}[baseline=0,square/.style={regular polygon,regular polygon sides=4}]

\draw[dgreen,thick](0,-1) -- (0,1); 
\draw [white, line width=5pt] (0,0) to[out=180,in=90 ]  (-1,-1);
\draw [blue, thick, decoration={markings, mark=at position 0.5 with {\arrow{<}}}, postaction={decorate}] (0,0) to[out=180,in=90 ]  (-1,-1);

\node at (0,0) [square,draw,fill=violet!60,scale=0.5] {}; 
   
\end{tikzpicture}}
\hspace{0.4in}$=$\hspace{0.2in}
{\begin{tikzpicture}[baseline=0,square/.style={regular polygon,regular polygon sides=4}]

\draw[dgreen,thick](0,-0.45) -- (0,1); 
\draw[dgreen,thick](0,-0.55) -- (0,-1); 
\draw [white, line width=5pt] (0,0) to[out=180,in=90 ]  (-1,-1);
\draw [blue, thick] (0,0) to[out=0,in=0,distance=0.2 in]  (0,-0.5);
\draw [blue, thick, decoration={markings, mark=at position 0.5 with {\arrow{<}}}, postaction={decorate}] (0,-0.5) to[out=180,in=90,distance=0.2 in]  (-1,-1);

\node[isosceles triangle,scale=0.4,
    isosceles triangle apex angle=60,
    draw,fill=violet!60,
    rotate=90,
    minimum size =0.01cm] at (0,0){};
   
\end{tikzpicture}}

\caption{Defining the square junction in terms of the triangle junction.}
\label{fig:tritosquare}
\end{center}
\end{figure}
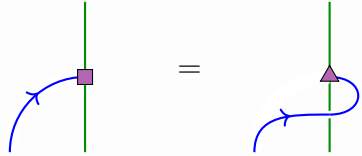

\begin{figure}[htbp]
\begin{center}
{\begin{tikzpicture}[baseline=0,square/.style={regular polygon,regular polygon sides=4},scale=0.6]

 \shade[top color=red!30, bottom color=red!5]  (1,-1) -- (0,0) -- (2,2) -- (-1,2)-- (-1,-1)-- (1,-1);

\draw [dgreen, thick] (0,0) -- (1,1);
\draw [dgreen, thick] (1,1) -- (2,2);
\draw [blue, thick, decoration={markings, mark=at position 0.5 with {\arrow{>}}}, postaction={decorate}] (-1,-1) -- (0,0);
\draw [dgreen, thick] (1,-1) -- (0,0);
\draw [blue, thick, decoration={markings, mark=at position 0.5 with {\arrow{>}}}, postaction={decorate}] (3,-1) -- (1,1);

\node[isosceles triangle,scale=0.5,
    isosceles triangle apex angle=60,
    draw,fill=violet!60,
    rotate=90,
    minimum size =0.01cm] at (1,1){};

\node at (0,0) [square,draw,fill=violet!60,scale=0.6] {}; 

\node[below] at (-1,-1) {\footnotesize \color{blue}$L_{(e_1,m_1)}$};
\node[below] at (1,-1) {\footnotesize \color{dgreen}$\Sigma_{(e)}$};
\node[below] at (3,-1) {\footnotesize \color{blue}$L_{(e_2,m_2)}$};
\end{tikzpicture}}\hspace{0 in} \footnotesize{$=$} \hspace{0 in}
{\begin{tikzpicture}[baseline=0,square/.style={regular polygon,regular polygon sides=4},scale=0.6]
\draw [dgreen, thick] (1,1) -- (2,2);
\draw [dgreen, thick] (0.5,0.5)-- (1,1);

\draw [blue, thick, decoration={markings, mark=at position 0.5 with {\arrow{>}}}, postaction={decorate}] (3,-1) -- (1,1);

\node[isosceles triangle,scale=0.5,
    isosceles triangle apex angle=60,
    draw,fill=violet!60,
    rotate=90,
    minimum size =0.01cm] at (1,1){};

\draw [dgreen, thick] (0.5,0.5) to[out=225,in=90 ]  (-0,-1);
\draw [white, line width = 0.05 in] (0.22,0) to[out=-1,in=0,distance = 0.5cm ] (0,-0.5);
\draw [white,  line width = 0.05 in] (0,-0.5) to[out=180,in=90,distance = 0.5cm ] (-1,-1);
\draw [blue, thick] (0.15,0) to[out=-1,in=0,distance = 0.5cm ] (0,-0.5);
\draw [blue, thick, decoration={markings, mark=at position 0.5 with {\arrow{<}}}, postaction={decorate}] (0,-0.5) to[out=180,in=90,distance = 0.5cm ] (-1,-1);

\node[isosceles triangle,scale=0.5,
    isosceles triangle apex angle=60,
    draw,fill=violet!60,
    rotate=90,
    minimum size =0.01cm] at (0.15,0){};
    
    \node[below] at (-1,-1) {\footnotesize \color{blue}$L_{(e_1,m_1)}\,\,\,$};
\node[below] at (0.4,-1) {\footnotesize \color{dgreen}$\Sigma_{(e)}$};
\node[below] at (3,-1) {\footnotesize \color{blue}$L_{(e_2,m_2)}$};
\end{tikzpicture}}
\hspace{0 in} \hspace{0.15 in}$=\,\, \sum_n$
{\begin{tikzpicture}[baseline=0,square/.style={regular polygon,regular polygon sides=4},scale=0.6]

\draw [blue, thick, decoration={markings, mark=at position 0.5 with {\arrow{>}}}, postaction={decorate}] (3,-1) -- (1,1);
\draw[blue, thick] (0.5,0.5)-- (1,1);
\draw [blue, thick, decoration={markings, mark=at position 0.5 with {\arrow{>}}}, postaction={decorate}] (1,1) -- (2,2);

\filldraw[blue] (1,1) circle (2pt);
\filldraw[blue] (0.15,0) circle (2pt);

\draw [blue, thick, <-, decoration={markings, mark=at position 0.85 with {\arrow{<}}}, postaction={decorate}] (0.5,0.5) to[out=225,in=90 ]  (-0,-1);
\draw [white, line width = 0.05 in] (0.22,0) to[out=-1,in=0,distance = 0.5cm ] (0,-0.5);
\draw [white,  line width = 0.05 in] (0,-0.5) to[out=180,in=90,distance = 0.5cm ] (-1,-1);

\draw [blue, thick] (0.15,0) to[out=-1,in=0,distance = 0.5cm ] (0,-0.5);
\draw [blue, thick, decoration={markings, mark=at position 0.5 with {\arrow{<}}}, postaction={decorate}] (0,-0.5) to[out=180,in=90,distance = 0.5cm ] (-1,-1);

        \node[below] at (-1.6,-1) {\footnotesize \color{blue}$L_{(e_1,m_1)}$};
\node[below] at (0.7,-1) {\footnotesize \color{blue}$L_{(e+n, -n)}$};
\node[below] at (3.4,-1) {\footnotesize \color{blue}$L_{(e_2,m_2)}$};
\end{tikzpicture}}
\\\vphantom{.}\hspace{-0.3 in} \footnotesize{ $=$} \,\,$ \footnotesize{\sum_n e^{ {2\pi i \over N}e_1 n}}$
{\begin{tikzpicture}[baseline=0,square/.style={regular polygon,regular polygon sides=4},scale=0.6]
 \draw[blue,thick] (0.5,0.5)-- (1,1);
 \draw [blue, thick, decoration={markings, mark=at position 0.5 with {\arrow{>}}}, postaction={decorate}] (1,1) -- (2,2);

 \draw [blue, thick, decoration={markings, mark=at position 0.5 with {\arrow{>}}}, postaction={decorate}] (3,-1) -- (1,1);

 \filldraw[blue] (1,1) circle (2pt);
 \filldraw[blue] (0.15,0) circle (2pt);

 \draw [blue, thick, <-, decoration={markings, mark=at position 0.75 with {\arrow{<}}}, postaction={decorate}] (0.5,0.5) to[out=225,in=90 ]  (-0,-1);
\draw [blue, thick, decoration={markings, mark=at position 0.5 with {\arrow{<}}}, postaction={decorate}] (0.15,0) to[out=180,in=90,distance = 0.5cm ] (-1,-1);

 \node[below] at (-1.6,-1) {\footnotesize \color{blue}$L_{(e_1,m_1)}$};
 \node[below] at (0.7,-1) {\footnotesize \color{blue}$L_{(e+n, -n)}$};
\node[below] at (3.4,-1) {\footnotesize \color{blue}$L_{(e_2,m_2)}$};
\end{tikzpicture}}
\hspace{-0.2 in}  \footnotesize$=$  \,\,$\footnotesize{\sum_n e^{ {2\pi i \over N}e_1 n}}$\hspace{-0.1 in}
{\begin{tikzpicture}[baseline=0,square/.style={regular polygon,regular polygon sides=4},scale=0.6]
\draw [blue, thick, decoration={markings, mark=at position 0.25 with {\arrow{>}}, mark=at position 0.65 with {\arrow{>}},mark=at position 0.85 with {\arrow{>}}}, postaction={decorate}] (0.4,-1) -- (2,2);

\draw [blue, thick, decoration={markings, mark=at position 0.5 with {\arrow{>}}}, postaction={decorate}] (3,-1) -- (1.15,0.35);

\filldraw[blue] (1.15,0.35) circle (2pt);

\draw [blue, thick, decoration={markings, mark=at position 0.5 with {\arrow{<}}}, postaction={decorate}] (1.6,1.2) to[out=180,in=90,distance = 0.9cm ] (-1,-1);

\filldraw[blue] (1.6,1.2) circle (2pt);

        \node[below] at (-1.6,-1) {\footnotesize \color{blue}$L_{(e_1,m_1)}$};
\node[below] at (0.7,-1) {\footnotesize \color{blue}$L_{(e+n, -n)}$};
\node[below] at (3.4,-1) {\footnotesize \color{blue}$L_{(e_2,m_2)}$};
\node[right] at (1.15,0.4) {\footnotesize \color{blue}$L_{(e_2+e+n,m_2-n)}$};
\end{tikzpicture}}
\\
\hspace{-0.0 in}  \,\,\footnotesize$=$\,\, \hspace{0 in} $ \footnotesize{\sum_n e^{{2\pi i \over N}e_1 n} e^{{2\pi i \over N}e_1(m_2-n)}}$\hspace{-0.3 in} 
{\begin{tikzpicture}[baseline=0,square/.style={regular polygon,regular polygon sides=4},scale=0.6]
\draw [blue, thick, decoration={markings, mark=at position 0.25 with {\arrow{>}}, mark=at position 0.6 with {\arrow{>}}, mark=at position 0.9 with {\arrow{>}}}, postaction={decorate}] (0.4,-1) -- (2,2);

\draw [blue, thick, decoration={markings, mark=at position 0.5 with {\arrow{>}}}, postaction={decorate}] (3,-1) -- (1.15,0.35);

\filldraw[blue] (1.15,0.35) circle (2pt);

\draw [white, thick,line width = 0.05 in] (1.8,1.2) to[out=0,in=0,distance = 0.6cm ] (1.2,0.8);
\draw [white, thick,line width = 0.05 in] (1.2,0.8) to[out=180,in=90,distance = 0.6cm ] (-1,-1);

\draw [blue, thick] (1.6,1.2) to[out=0,in=0,distance = 0.7cm ] (1.2,0.8);
\draw [blue, thick, decoration={markings, mark=at position 0.5 with {\arrow{<}}}, postaction={decorate}] (1.2,0.8) to[out=180,in=90,distance = 0.6cm ] (-1,-1);

\filldraw[blue] (1.6,1.2) circle (2pt);

\node[below] at (-1.6,-1) {\footnotesize \color{blue}$L_{(e_1,m_1)}$};
\node[below] at (0.7,-1) {\footnotesize \color{blue}$L_{(e+n, -n)}$};
\node[below] at (3.4,-1) {\footnotesize \color{blue}$L_{(e_2,m_2)}$};
\end{tikzpicture}}\hspace{-0.1 in} 
$=$ \hspace{0 in} $\footnotesize e^{{2\pi i\over N } e_1 m_2}$
{\begin{tikzpicture}[baseline=0,square/.style={regular polygon,regular polygon sides=4},scale=0.6]
\shade[top color=red!30, bottom color=red!5]  (1,-1) -- (2,0) -- (1,1) -- (2,2)--(-1,2)-- (-1,-1)-- (1,-1);
\draw [dgreen, thick] (1,1) -- (2,2);
\draw [blue, thick, decoration={markings, mark=at position 0.5 with {\arrow{>}}}, postaction={decorate}] (-1,-1) -- (1,1);
\draw [dgreen, thick] (1,1) -- (2,0);
\draw [dgreen, thick] (1,-1) -- (2,0);
\draw [blue, thick, decoration={markings, mark=at position 0.5 with {\arrow{>}}}, postaction={decorate}] (3,-1) -- (2,0);

\node[isosceles triangle,scale=0.5,
    isosceles triangle apex angle=60,
    draw,fill=violet!60,
    rotate=90,
    minimum size =0.01cm] at (2,0){};

\node at (1,1) [square,draw,fill=violet!60,scale=0.6] {}; 

\node[below] at (-1,-1) {\footnotesize \color{blue}$L_{(e_1,m_1)}$};
\node[below] at (1,-1) {\footnotesize \color{dgreen}$\Sigma_{(e)}$};
\node[below] at (3,-1) {\footnotesize \color{blue}$L_{(e_2,m_2)}$};
\end{tikzpicture}}

\caption{Computation of the F-symbol for two genuine lines and one twist defect. The surface $D_{\mathrm{EM}}$ is suppressed at intermediate steps.}
\label{fig:Fsymbol1}
\end{center}
\end{figure}
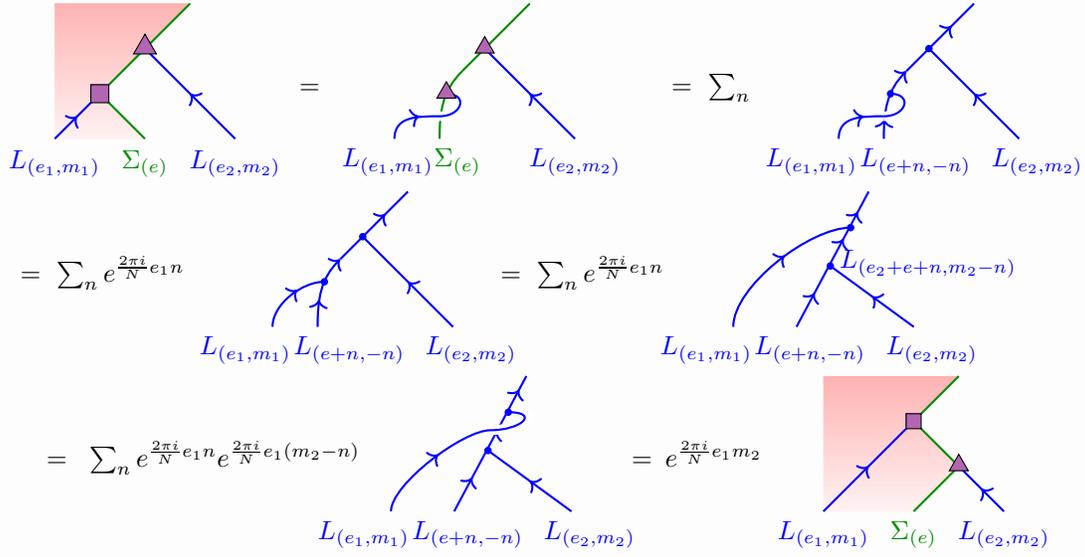

We now return to the F-symbols. We begin with the case of two external genuine lines and two external twist defect. From the form of the fusion rules in (\ref{sigma2fusion1}), it is clear that the internal line is necessarily non-genuine, i.e. it is a twist defect. We may now calculate the F-symbols using the tools developed above. The computation is shown in Figure \ref{fig:Fsymbol1}. Summarizing it in words, one begins by using the definition of the square junction in terms of the triangle junction, and then decomposes the boundary of $\Sigma_{(e)}$ into $\cA\otimes L_{(e,0)} = \bigoplus_{n=0}^{N-1} L_{(e+n,-n)}$. The factor of $\cA$ is included here so that the mesh of $\cA$ making up $D_{\mathrm{EM}}$ can end on $\Sigma_{(e)}$  from the left. One then uses the half-braiding and fusion rules of the genuine lines, given in Figure \ref{fig:Fsymbhalfbraid}, to obtain the  right-hand side of the second line of  Figure \ref{fig:Fsymbol1}. One finally does a half-braiding and reassembles $\Sigma_{(e)}$ to get the end result.

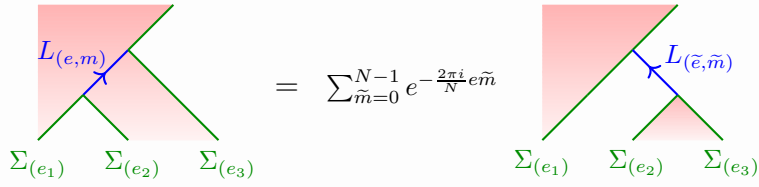
\begin{figure}[htbp]
\begin{center}

{\begin{tikzpicture}[baseline=0,square/.style={regular polygon,regular polygon sides=4},scale=0.6]

\shade[top color=red!30, bottom color=red!5]  (1,-1) -- (3,-1) -- (1,1) -- (2,2)--(-1,2)-- (-1,-1)-- (0,0)--(1,-1);

\draw [dgreen, thick] (1,1) -- (2,2);
\draw [blue, thick, decoration={markings, mark=at position 0.5 with {\arrow{>}}}, postaction={decorate}] (0,0) -- (1,1);
\draw [dgreen, thick] (-1,-1) -- (0,0);
\draw [dgreen, thick] (1,-1) -- (0,0);
\draw [dgreen, thick] (3,-1) -- (1,1);

\node[below] at (-1,-1) {\footnotesize \color{dgreen}$\Sigma_{(e_1)}$};
\node[below] at (1.1,-1) {\footnotesize \color{dgreen}$\Sigma_{(e_2)}$};
\node[below] at (3.2,-1) {\footnotesize \color{dgreen}$\Sigma_{(e_3)}$};
\node[left] at (0.8,0.9) {\footnotesize \color{blue}$L_{(e,m)}$};
\end{tikzpicture}}
$= $ \hspace{0.1 in}\footnotesize{$ \sum_{\widetilde{m}=0}^{N-1}e^{-{2 \pi i \over N}e \widetilde m}$}
{\begin{tikzpicture}[baseline=0,square/.style={regular polygon,regular polygon sides=4},scale=0.6]

\shade[top color=red!30, bottom color=red!5]  (1,-1) -- (2,0) -- (3,-1) --(1,-1);

\shade[top color=red!30, bottom color=red!5]  (-1,-1) -- (2,2) -- (-1,2) --(-1,-1);

\draw [dgreen, thick] (1,1) -- (2,2);
\draw [dgreen, thick] (0,0) -- (1,1);
\draw [dgreen, thick] (-1,-1) -- (0,0);
\draw [dgreen, thick] (1,-1) -- (2,0);
\draw [dgreen, thick] (3,-1) -- (2,0);
\draw [blue, thick, decoration={markings, mark=at position 0.5 with {\arrow{<}}}, postaction={decorate}] (1,1) -- (2,0);

\node[below] at (-1,-1) {\footnotesize \color{dgreen}$\Sigma_{(e_1)}$};
\node[below] at (1.1,-1) {\footnotesize \color{dgreen}$\Sigma_{(e_2)}$};
\node[below] at (3.2,-1) {\footnotesize \color{dgreen}$\Sigma_{(e_3)}$};
\node[right] at (1.5,0.8) {\footnotesize \color{blue}$L_{(\widetilde{e},\widetilde{m})}$};
\end{tikzpicture}}

\caption{F-symbol for three incoming twist defects. The diagrams are non-vanishing only when the intermediate legs $L_{(e,m)}$ and $L_{(\widetilde{e},\widetilde{m})}$ satisfy $e_1+e_2=e+m$, $e_2+e_3=\widetilde{e}+\widetilde{m}$. The $D_{\text{EM}}$ surface attached to the twist defects are assumed to be to the left of $\Sigma_{(e_1)}$ as well as between $\Sigma_{(e_2)}$ and $\Sigma_{(e_3)}$. }
\label{fig:Fsymbol2}
\end{center}
\end{figure}

We next consider the F-symbols involving four external twist defects. Note that because of the fusion rules (\ref{sigma2fusion1}), the internal line is guaranteed to be genuine. Let us take the three incoming twist defects to be $\Sigma_{(e_1)},\Sigma_{(e_2)},$ and $\Sigma_{(e_3)}$, and place the mesh of $\CA$ as shown in Figure \ref{fig:Fsymbol2}.  The coefficient relating the configuration with internal line $L_{(e,m)}$ to that with internal line $L_{(\widetilde e, \widetilde m)}$ is then given in Figure \ref{fig:Fsymbol2}, and derived in Figure \ref{fig:Fsymbol3}. The various charges appearing here satisfy $e_1+e_2=e+m$, $e_2+e_3=\widetilde{e}+\widetilde{m}$. For simplicity we neglect real number normalization factors, and focus only on the phases. In the special case of $N=2$, the F-symbols in Figure \ref{fig:Fsymbol1} and \ref{fig:Fsymbol2} are consistent with those found in \cite{2015arXiv150306812T}.

\begin{figure}[htbp]
\begin{center}

{\begin{tikzpicture}[baseline=0,square/.style={regular polygon,regular polygon sides=4},scale=0.6]
\shade[top color=red!30, bottom color=red!5]  (1,-1) -- (3,-1) -- (1,1) -- (2,2)--(-1,2)-- (-1,-1)-- (0,0)--(1,-1);
\draw [dgreen, thick] (1,1) -- (2,2);
\draw [blue, thick, decoration={markings, mark=at position 0.5 with {\arrow{>}}}, postaction={decorate}] (0,0) -- (1,1);
\draw [dgreen, thick] (-1,-1) -- (0,0);
\draw [dgreen, thick] (1,-1) -- (0,0);
\draw [dgreen, thick] (3,-1) -- (1,1);

\node[below] at (-1,-1) {\footnotesize \color{dgreen}$\Sigma_{(e_1)}$};
\node[below] at (1.1,-1) {\footnotesize \color{dgreen}$\Sigma_{(e_2)}$};
\node[below] at (3.2,-1) {\footnotesize \color{dgreen}$\Sigma_{(e_3)}$};
\node[left] at (0.8,0.9) {\footnotesize \color{blue}$L_{(e,m)}$};
\end{tikzpicture}}
= \hspace{0.1 in}
{\begin{tikzpicture}[baseline=0,square/.style={regular polygon,regular polygon sides=4},scale=0.6]
\draw [dgreen, thick] (1.2,1.2) -- (2,2);
\draw [dgreen, thick] (3,-1) -- (1.2,0.8);

\draw [dgreen, thick] (1.2,1.2) to[out=225,in=135,distance = 0.2cm ] (1.2,0.8);
\draw [blue, thick, decoration={markings, mark=at position 0.5 with {\arrow{<}}}, postaction={decorate}] (1.1,1) to[out=180,in=180,distance = 1cm ] (-0.45,-0.5) ;

\draw [dgreen, thick]  (-1,-1) to[out=45,in=135,distance = 1.5cm ] (1,-1);
\node at (1.1,1)[square,draw,fill=violet!60,scale=0.5] {}; 
\node at (-0.45,-0.5)[square,draw,fill=violet!60,scale=0.5] {}; 

\node[below] at (-1,-1) {\footnotesize \color{dgreen}$\Sigma_{(e_1)}$};
\node[below] at (1.1,-1) {\footnotesize \color{dgreen}$\Sigma_{(e_2)}$};
\node[below] at (3.2,-1) {\footnotesize \color{dgreen}$\Sigma_{(e_3)}$};
\node[left] at (0.25,0.9) {\footnotesize \color{blue}$L_{(e,m)}$};
\end{tikzpicture}}
= \hspace{0.1 in}
{\begin{tikzpicture}[baseline=0,square/.style={regular polygon,regular polygon sides=4},scale=0.6]
\draw [dgreen, thick] (1.2,1.2) -- (2,2);
\draw [dgreen, thick] (3,-1) -- (1.2,0.8);

\draw [dgreen, thick] (1.2,1.2) to[out=225,in=135,distance = 0.2cm ] (1.2,0.8);

\draw [dgreen, thick]  (-1,-1) to[out=45,in=135,distance = 1.5cm ] (1,-1);

\draw [white, line width = 0.05in] (1.1,1) to[out=0,in=45,distance = 0.5cm ] (1.5,0.5) ;
\draw [white, line width = 0.05in] (1.5,0.5) to[out=225,in=45,distance = 0.3cm ] (0.6,-0.5) ;
\draw [white,line width = 0.05in] (-0.45,-0.5) to[out=-45,in=225,distance = 0.3cm ] (0.6,-0.5) ;

\draw [blue, thick] (1.1,1) to[out=0,in=45,distance = 0.5cm ] (1.5,0.5) ;
\draw [blue, thick, decoration={markings, mark=at position 0.5 with {\arrow{<}}}, postaction={decorate}] (1.5,0.5) to[out=225,in=45,distance = 0.3cm ] (0.6,-0.5) ;
\draw [blue, thick] (-0.45,-0.5) to[out=-45,in=225,distance = 0.3cm ] (0.6,-0.5) ;

\node[isosceles triangle,scale=0.4,
    isosceles triangle apex angle=60,
    draw,fill=violet!60,
    rotate=90,
    minimum size =0.01cm] at (1.1,1){};
\node[isosceles triangle,scale=0.4,
    isosceles triangle apex angle=60,
    draw,fill=violet!60,
    rotate=90,
    minimum size =0.01cm] at (-0.45,-0.5){};

\node[below] at (-1,-1) {\footnotesize \color{dgreen}$\Sigma_{(e_1)}$};
\node[below] at (1.1,-1) {\footnotesize \color{dgreen}$\Sigma_{(e_2)}$};
\node[below] at (3.2,-1) {\footnotesize \color{dgreen}$\Sigma_{(e_3)}$};
\node[left] at (0.25,0.9) {\footnotesize \color{blue}$L_{(e,m)}$};
\end{tikzpicture}}
\\
= \hspace{0.1 in}\footnotesize{$\sum_{n_1,n_2,n_3} e^{{2\pi i\over N} e (n_2 + n_3)}$}
{\begin{tikzpicture}[baseline=0,square/.style={regular polygon,regular polygon sides=4},scale=0.6]
\draw [blue, thick, decoration={markings, mark=at position 0.5 with {\arrow{>}}}, postaction={decorate}] (1,1) -- (2,2);
\draw [blue, thick, decoration={markings, mark=at position 0.5 with {\arrow{>}}}, postaction={decorate}] (0,0) -- (1,1);
\draw [blue, thick, decoration={markings, mark=at position 0.5 with {\arrow{>}}}, postaction={decorate}] (-1,-1) -- (0,0);
\draw [blue, thick, decoration={markings, mark=at position 0.5 with {\arrow{>}}}, postaction={decorate}] (1,-1) -- (0,0);
\draw [blue, thick, decoration={markings, mark=at position 0.5 with {\arrow{>}}}, postaction={decorate}] (3,-1) -- (1,1);

\filldraw[blue] (0,0) circle (2pt);
\filldraw[blue] (1,1) circle (2pt);

\node[below] at (-2,-1) {\footnotesize \color{blue}$L_{(e_1+n_1,-n_1)}$};
\node[below] at (1.1,-1) {\footnotesize \color{blue}$L_{(e_2+n_2,-n_2)}$};
\node[below] at (4.2,-1) {\footnotesize \color{blue}$L_{(e_3+n_3,-n_3)}$};
\node[left] at (0.8,0.9) {\footnotesize \color{blue}$L_{(e,m)}$};
\end{tikzpicture}}\\
$=$ \hspace{0.1 in}\footnotesize{$\sum_{n_1,n_2,n_3} e^{{2\pi i\over N}e (n_2 + n_3)}$}
{\begin{tikzpicture}[baseline=0,square/.style={regular polygon,regular polygon sides=4},scale=0.6]
\draw [blue, thick, decoration={markings, mark=at position 0.5 with {\arrow{>}}}, postaction={decorate}] (1,1) -- (2,2);
\draw [blue, thick] (0,0) -- (1,1);
\draw [blue, thick, ->] (-1,-1) -- (0,0);
\draw [blue, thick, decoration={markings, mark=at position 0.5 with {\arrow{>}}}, postaction={decorate}] (1,-1) -- (2,0);
\draw [blue, thick, decoration={markings, mark=at position 0.25 with {\arrow{>}}, mark=at position 0.75 with {\arrow{>}}}, postaction={decorate}] (3,-1) -- (1,1);

\filldraw[blue] (2,0) circle (2pt);
\filldraw[blue] (1,1) circle (2pt);

\node[below] at (-2,-1) {\footnotesize \color{blue}$L_{(e_1+n_1,-n_1)}$};
\node[below] at (1.1,-1) {\footnotesize \color{blue}$L_{(e_2+n_2,-n_2)}$};
\node[below] at (4.2,-1) {\footnotesize \color{blue}$L_{(e_3+n_3,-n_3)}$};
\node[right] at (1.5,0.8) {\footnotesize \color{blue}$L_{(e_2+e_3+n_2+n_3,-n_2-n_3)}$};
\end{tikzpicture}}
\\
$=$ \hspace{0.1 in}\footnotesize{ $\sum_{\widetilde{m}=0}^{N-1}e^{-{2\pi i\over N}e \widetilde{m}}$}
{\begin{tikzpicture}[baseline=0,square/.style={regular polygon,regular polygon sides=4},scale=0.6]
\draw [dgreen, thick] (1,1) -- (2,2);
\draw [dgreen, thick] (0,0) -- (1,1);
\draw [dgreen, thick] (-1,-1) -- (0,0);
\draw [dgreen, thick] (1,-1) -- (2,0);
\draw [dgreen, thick] (3,-1) -- (2,0);
\draw [blue, thick, decoration={markings, mark=at position 0.5 with {\arrow{<}}}, postaction={decorate}] (1,1) -- (2,0);

\node[below] at (-1,-1) {\footnotesize \color{dgreen}$\Sigma_{(e_1)}$};
\node[below] at (1.1,-1) {\footnotesize \color{dgreen}$\Sigma_{(e_2)}$};
\node[below] at (3.2,-1) {\footnotesize \color{dgreen}$\Sigma_{(e_3)}$};
\node[right] at (1.5,0.8) {\footnotesize \color{blue}$L_{(\widetilde{e},\widetilde{m})}$};

\node[isosceles triangle,scale=0.4,
    isosceles triangle apex angle=60,
    draw,fill=violet!60,
    rotate=90,
    minimum size =0.01cm] at (1,1){};

\node[isosceles triangle,scale=0.4,
    isosceles triangle apex angle=60,
    draw,fill=violet!60,
    rotate=90,
    minimum size =0.01cm] at (2,0){};

\end{tikzpicture}} $=$\hspace{0.1 in}\footnotesize{$\sum_{\widetilde{m}=0}^{N-1}e^{-{2\pi i\over N}e \widetilde{m}}$}
{\begin{tikzpicture}[baseline=0,square/.style={regular polygon,regular polygon sides=4},scale=0.6]
\shade[top color=red!30, bottom color=red!5]  (1,-1) -- (2,0) -- (3,-1) --(1,-1);

\shade[top color=red!30, bottom color=red!5]  (-1,-1) -- (2,2) -- (-1,2) --(-1,-1);
\draw [dgreen, thick] (1,1) -- (2,2);
\draw [dgreen, thick] (0,0) -- (1,1);
\draw [dgreen, thick] (-1,-1) -- (0,0);
\draw [dgreen, thick] (1,-1) -- (2,0);
\draw [dgreen, thick] (3,-1) -- (2,0);
\draw [blue, thick, decoration={markings, mark=at position 0.5 with {\arrow{<}}}, postaction={decorate}] (1,1) -- (2,0);

\node[below] at (-1,-1) {\footnotesize \color{dgreen}$\Sigma_{(e_1)}$};
\node[below] at (1.1,-1) {\footnotesize \color{dgreen}$\Sigma_{(e_2)}$};
\node[below] at (3.2,-1) {\footnotesize \color{dgreen}$\Sigma_{(e_3)}$};
\node[right] at (1.5,0.8) {\footnotesize \color{blue}$L_{(\widetilde{e},\widetilde{m})}$};
\end{tikzpicture}}

\caption{Computation of the F-symbol for three incoming twist defects. For the diagrams in the second and third line to be non-vanishing, we require $e_1+e_2=m+e, m=-n_1-n_2, e_2+e_3=\widetilde{e}+\widetilde{m}$ and $\widetilde{m}=-n_2-n_3$. 
The surface $D_{\mathrm{EM}}$ is suppressed at intermediate steps.}
\label{fig:Fsymbol3}
\end{center}
\end{figure}

\subsubsection{F-symbols of duality interfaces}

We finally obtain the F-symbols of duality interfaces in $(1+1)$d. As explained in Section \ref{sec:symdef1+12+1}, upon shrinking the slab the twist defects of $\Z_N$ gauge theory become the duality interface of a $(1+1)$d QFT. As a consequence, the F-symbols of the duality interface naturally follow from those of the twist defects.

We first consider the F-symbols of duality interfaces descending from Figure \ref{fig:Fsymbol1}. Upon shrinking, $\Sigma_{(e)}$ reduces to $\CN$. Moreover, note that $L_{(e_2,m_2)}$ collides with the Dirichlet boundary condition upon shrinking, and hence only its magnetic components survive the $\Z_N$ symmetry generators, i.e. $L_{(e_2,m_2)}$ reduces to $\eta^{m_2}$. On the other hand, $L_{(e_1, m_1)}$ is on top of $D_{\text{EM}}$, and hence upon shrinking it collides with the Neumann boundary condition and its electric component survives, i.e. $L_{(e_1, m_1)}$ reduces to $\widehat\eta^{e_1}$, where $\widehat{\eta}$ is the generator of the quantum $\Z_N$ symmetry.  Hence the F-symbol involving one duality interface and two $\Z_N$ symmetry generators distributed on its two sides is as shown in Figure \ref{fig:Fsymbol4}. Note that the duality interface is unoriented, i.e. $\CN=\overline{\CN}$ (up to an Euler counterterm, c.f. \eqref{eq:Ninvertible}), and hence we do not place an arrow on it. In the next section we will gauge $\ZZ_2^{\mathrm{EM}}$ to obtain the SymTFT for the $TY(\Z_N)$ category, for which the phase appearing in Figure \ref{fig:Fsymbol4} implies a nontrivial bi-character.

\begin{figure}[tbp]
\begin{center}
{\begin{tikzpicture}[baseline=0,square/.style={regular polygon,regular polygon sides=4},scale=0.6]

 \shade[top color=red!30, bottom color=red!5]  (1,-1) -- (0,0) -- (2,2) -- (-1,2)-- (-1,-1)-- (1,-1);
 
\draw [red, very thick] (0,0) -- (1,1);
\draw [red, very thick] (1,1) -- (2,2);
\draw [blue, thick, decoration={markings, mark=at position 0.5 with {\arrow{>}}}, postaction={decorate}] (-1,-1) -- (0,0);
\draw [red, very thick] (1,-1) -- (0,0);
\draw [blue, thick, decoration={markings, mark=at position 0.5 with {\arrow{>}}}, postaction={decorate}] (3,-1) -- (1,1);

\node[below] at (-1,-1) {\footnotesize
\color{blue}$\widehat\eta^{e_1}$};
\node[below] at (1,-1) {\footnotesize \color{red}$\CN$};
\node[below] at (3,-1) {\footnotesize \color{blue}$\eta^{m_2}$};
\end{tikzpicture}}\hspace{0 in} \hspace{-0.1 in} 
$=$ \hspace{0 in} $\footnotesize e^{{2\pi i\over N } e_1 m_2}$
{\begin{tikzpicture}[baseline=0,square/.style={regular polygon,regular polygon sides=4},scale=0.6]

\shade[top color=red!30, bottom color=red!5]  (1,-1) -- (2,0) -- (1,1) -- (2,2)--(-1,2)-- (-1,-1)-- (1,-1);

\draw [red, very thick] (1,1) -- (2,2);
\draw [blue, thick, decoration={markings, mark=at position 0.5 with {\arrow{>}}}, postaction={decorate}] (-1,-1) -- (1,1);
\draw [red, very thick] (1,1) -- (2,0);
\draw [red, very thick] (1,-1) -- (2,0);
\draw [blue, thick, decoration={markings, mark=at position 0.5 with {\arrow{>}}}, postaction={decorate}] (3,-1) -- (2,0);

\node[below] at (-1,-1) {\footnotesize \color{blue}$\widehat\eta^{e_1}$};
\node[below] at (1,-1) {\footnotesize \color{red}$\CN$};
\node[below] at (3,-1) {\footnotesize \color{blue}$\eta^{m_2}$};
\end{tikzpicture}}

\caption{F-symbol involving two external duality interfaces and two external $\Z_N$ symmetry generators. }
\label{fig:Fsymbol4}
\end{center}
\end{figure}

\begin{figure}[tbp]
\begin{center}

{\begin{tikzpicture}[baseline=0,square/.style={regular polygon,regular polygon sides=4},scale=0.6]

\shade[top color=red!30, bottom color=red!5]  (1,-1) -- (3,-1) -- (1,1) -- (2,2)--(-1,2)-- (-1,-1)-- (0,0)--(1,-1);
\draw [dgreen, thick] (1,1) -- (2,2);

\draw [red, very thick] (1,1) -- (2,2);
\draw [blue, thick, decoration={markings, mark=at position 0.5 with {\arrow{>}}}, postaction={decorate}] (0,0) -- (1,1);
\draw [red, very thick] (-1,-1) -- (0,0);
\draw [red, very thick] (1,-1) -- (0,0);
\draw [red, very thick] (3,-1) -- (1,1);

\node[below] at (-1,-1) {\footnotesize \color{red}$\CN$};
\node[below] at (1.1,-1) {\footnotesize \color{red}$\CN$};
\node[below] at (3.2,-1) {\footnotesize \color{red}$\CN$};
\node[left] at (0.8,0.9) {\footnotesize \color{blue}$\widehat\eta^{e}$};
\end{tikzpicture}}
$=$ \hspace{0.1 in}\footnotesize{$\sum_{\widetilde{m}=0}^{N-1}e^{-{2 \pi i \over N}e \widetilde m}$}
{\begin{tikzpicture}[baseline=0,square/.style={regular polygon,regular polygon sides=4},scale=0.6]

\shade[top color=red!30, bottom color=red!5]  (1,-1) -- (2,0) -- (3,-1) --(1,-1);

\shade[top color=red!30, bottom color=red!5]  (-1,-1) -- (2,2) -- (-1,2) --(-1,-1);

\draw [red, very thick] (1,1) -- (2,2);
\draw [red, very thick] (0,0) -- (1,1);
\draw [red, very thick] (-1,-1) -- (0,0);
\draw [red, very thick] (1,-1) -- (2,0);
\draw [red, very thick] (3,-1) -- (2,0);
\draw [blue, thick, decoration={markings, mark=at position 0.5 with {\arrow{<}}}, postaction={decorate}] (1,1) -- (2,0);

\node[below] at (-1,-1) {\footnotesize \color{red}$\CN$};
\node[below] at (1.1,-1) {\footnotesize \color{red}$\CN$};
\node[below] at (3.2,-1) {\footnotesize \color{red}$\CN$};
\node[right] at (1.5,0.8) {\footnotesize \color{blue}$\eta^{\widetilde{m}}$};
\end{tikzpicture}}

\caption{F-symbol for four external duality interfaces. 
}
\label{fig:Fsymbol5}
\end{center}
\end{figure}

We next consider the F-symbols involving three duality interfaces, which follows from Figure \ref{fig:Fsymbol2}. The result is shown in Figure \ref{fig:Fsymbol5}. Note that $e$ and $\widetilde{m}$ can be arbitrary integers. The nontrivial phase on the right hand side of Figure \ref{fig:Fsymbol5} implies that, once we gauge $\ZZ_2^{\mathrm{EM}}$ to make $\CN$ into a duality defect, the defect $\CN$ is ``self-anomalous," in the sense that it is an obstruction to the existence of a trivially gapped phase. This is consistent with the results in \cite{Choi:2021kmx}.

\section{$(2+1)$d Symmetry TFT for duality defects}
\label{sec:3dDTYZN}

In Section \ref{sec:symTFT3dZN} we discussed the SymTFT for a theory with $\Z_N^{(0)}$ symmetry, and derived the fusion rules and F-symbols for the duality interface implementing $\Z_N^{(0)}$ gauging. In the current section, we will demand more symmetries of the $(1+1)$d theory $\cX$: namely, we require that it not only be $\Z_N$ symmetric, but also invariant under gauging $\Z_N^{(0)}$, i.e. $\cX=\cX/\Z_N^{(0)}$, or equivalently $D_\text{EM}\ket{\cX}=\ket{\cX}$. This means that the duality interface $\cN$ defined in Section \ref{sec:1+1ddualityinterface} is not just a topological interface between two distinct theories, but a topological \textit{defect} within the theory $\cX$ itself. In this case the full symmetry is the Tambara-Yamagami fusion category $TY(\Z_N)$ (for particular choice of the Frobenius-Schur index and bicharacter). Our goal now is to find the SymTFT for  $TY(\Z_N)$.

\subsection{Symmetry TFT for $TY(\ZZ_N)$}
\label{sec:ZTYZN}

From a mathematical point of view, the SymTFT for $TY(\Z_N)$ 
is the Turaev-Viro theory specified by the Tambara-Yamagami $TY(\ZZ_N)$ category, or alternatively the Reshetikhin-Turaev theory for the Drinfeld center 
$\cZ(TY(\ZZ_N))$. As we have mentioned already in the introduction, the simple objects and fusion rules of $\cZ(TY(\ZZ_N))$ have appeared already in the math and physics literature, so the results to be obtained below are not, strictly speaking, new.

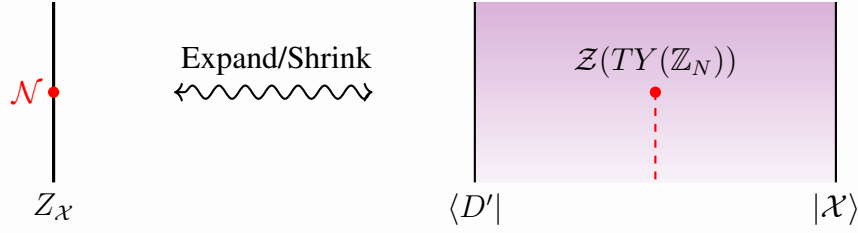
\begin{figure}[!tbp]
	\centering
	\begin{tikzpicture}[scale=0.8]

	\draw[thick, snake it, <->] (-1.7,-3.5) -- (-5, -3.5);
	\node[above] at (-3.3,-3.4) {Expand/Shrink};
	
	\shade[line width=2pt,	top color=violet!30, bottom color=violet!5] 
	(0,-5) to [out=90, in=-90]  (0,-2)
	to [out=0,in=180] (6,-2)
	to [out = -90, in =90] (6,-5)
	to [out=180, in =0]  (0,-5);

	\draw[thick] (0,-5) -- (0,-2);
	\draw[thick] (6,-5) -- (6,-2);
	\node at (3,-3) {$\cZ(TY(\Z_N))$};
	\node at (3,-3.5) [circle,fill,red, inner sep=1.5pt]{};
	\draw[ thick,red,dashed] (3,-3.5) -- (3,-5);
	\node[below] at (0,-5) {$\langle{D'}|$};

	\node[below] at (6,-5) {$|\cX\rangle $};

	\draw[very thick] (-7,-5) -- (-7,-2);
	\node[below] at (-7,-5) {$Z_{\CX}$};
	\node at (-7,-3.5) [circle,fill,red, inner sep=1.5pt]{};
	\node[left,red] at (-7,-3.5) {$\cN$};

	\end{tikzpicture}
	
	\caption{The SymTFT of $TY(\Z_N)$ is obtained by gauging the $\Z_2^{\text{EM}}$ symmetry of the $\Z_N$ gauge theory in Figure \ref{fig:twistdefshrink}. Mathematically, it is given by RT theory on the Drinfeld center of $TY(\Z_N)$.  The twist defect represented by the red dot attached to a red line (which eas actually a line living on the boundary of a 2d surface) in Figure \ref{fig:twistdefshrink} now becomes a genuine line operator (represented by a red dot) in the current figure. After shrinking, this gives a genuine line defect $\cN$ within the theory $\cX$. 
}
	\label{fig:dualitySTFT}
\end{figure}

The presentation given here emphasizes the fact that the SymTFT of $TY(\Z_N)$ can be obtained from that of $\mathrm{Vec}(\ZZ_N)$ by gauging the $\Z_2^{\text{EM}}$   electro-magnetic symmetry of the latter.\footnote{Let us mention that in $(2+1)$d, one can twist the gauging by stacking with an SPT for the $\Z_N^{(0)}$ zero-form symmetry, which is classified by $H^3(B\Z_N, U(1))$. This could give rise to different variants of the Tambara-Yamagami category. In this work, we consider only the untwisted gauging. }
This fact is simple to see physically: indeed, recall that in the case of $\mathrm{Vec}(\ZZ_N)$, the duality interface $\cN$ descended from the twist defects $ \Sigma_{(e)}$ of the SymTFT, which were the boundaries of the $\Z_2^{\text{EM}}$ duality surface $D_{\mathrm{EM}}$. This was illustrated in Figure \ref{fig:twistdefshrink}. Upon gauging  the $\Z_2^{\text{EM}}$ symmetry of the SymTFT this duality surface becomes transparent, and hence the duality interface $\cN$ becomes a duality defect, as shown in Figure \ref{fig:dualitySTFT}. 

Note that the procedure of gauging a zero-form symmetry in a $(2+1)$d TFT has been fully studied in the condensed matter literature \cite{Barkeshli:2014cna,2015arXiv150306812T}, where the power of modular tensor categories was utilized. Since the ultimate goal of this paper will  be to generalize to higher dimensions, and since a full-fledged theory of higher modular tensor categories has not been fully developed, we will not adopt this approach here. Instead, we will give a more physical rederivation of the known results from both the math \cite{izumi2001structure,gelaki2009centers} and physics \cite{Barkeshli:2014cna,2015arXiv150306812T} literature, being content with the operator contents and their fusion rules (without worrying about braiding properties). Our rederivation will have the virtue of being generalizable to higher dimensions.\footnote{Our derivation is similar to the approach in \cite{Bhardwaj:2022yxj}, where only the fusion rules of  topological lines with integer quantum dimensions were discussed. Here we will extensively discuss the fusion rules involving  topological lines with non-integer quantum dimensions, i.e. those descending from twist defects. } We will also present a second derivation of the SymTFT of $TY(\Z_N)$ based on discrete twisted-cocycles. This will be (at least partially) generalizable to higher dimensions as well.

\subsubsection{Lines defects in the SymTFT of $TY(\ZZ_N)$ }
\label{sec:ZTYZNlines}

As we have just claimed, the SymTFT for $TY(\ZZ_N)$ can be obtained from $\ZZ_N$ gauge theory by gauging the $\ZZ_2^{\mathrm{EM }}$ zero-form symmetry. This gauging may be split into two steps. First, we begin by keeping only the lines invariant under $\ZZ_2^{\mathrm{EM }}$. This leaves us with lines $L_{(e,e)}$ with equal electric and magnetic charge. Next, we note that upon gauging the $\ZZ_2^{\mathrm{EM }}$ zero-form symmetry, we obtain a quantum one-form symmetry $\widehat \ZZ_2^{(1)}$ generated by an operator $K$. Note that $K$ is labelled by a representation of $\ZZ_2^{\text{EM}}$.  If in the pre-gauged theory a simple line is invariant under $\ZZ_2^{\mathrm{EM }}$, then in the gauged theory it can be assigned a $\ZZ_2$ representation distinguished by whether or not it is stacked with a $K$-line. We therefore denote the invertible lines of the gauged theory by $\widehat L_{(e)}^q$ where the $q=\pm$ labels whether $K$ is stacked with it or not, i.e. $\widehat L_{(e)}^+ := L_{(e,e)}$ and $\widehat L_{(e)}^- :=K\, L_{(e,e)}$. There are a total of $2N$ invertible lines. 

In additional to the invertible lines, we should also allow for the combinations $L_{(e,m)}\oplus L_{(m,e)}$, which are invariant under $\ZZ_2^{\mathrm{EM }}$ when viewed as a single object.\footnote{We will be mainly discussing things at a local level, with each line operator $L_{(e,m)}$ is supported on a small interval. The direct sum/product between operators in such a small patch will be denoted by $\oplus, \otimes$. Note however that since all the operators in the SymTFT of $TY(\Z_N)$ are lines, the local fusion coincides with the global fusion, and thus it is equally fine to change $\oplus,\otimes$ to $+, \times$. The distinction will only be important in higher dimensions.}  Such combinations $L_{(e,m)}\oplus L_{(m,e)}$ are not simple before gauging, but become simple after gauging, and will be denoted by $\widehat L_{[e,m]}$.\footnote{We use the word ``simple" here in the sense of category theory, where an object $a$ is simple if $\mathrm{Hom}(a,a)$ is one-dimensional.} Noting that when $e=m$ the two factors in the direct sum are individually $\Z_2^{\text{EM}}$ invariant, and because by definition $\widehat L_{[e,m]} \equiv \widehat{L}_{[m,e]}$, we may assume without loss of generality that $0\leq e<m\leq N-1$. 
Unlike for $\widehat L_{(e)}^\pm$, in the current case we cannot stack with a $K$ defect since the constituent lines $L_{(e,m)}$ and $L_{(m,e)}$ in the pre-gauged theory are not invariant under $\ZZ_2^{\mathrm{EM}}$. 
Another way to say this is that $\widehat L_{[e,m]}$ can absorb $K$.

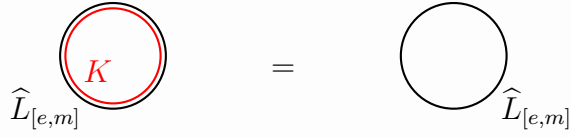
\begin{figure}[!tbp]
	\centering
\[
 \raisebox{-2em}{\begin{tikzpicture}
 \draw[thick] (0,0) circle (20pt);
  \draw[thick, red] (0,0) circle (18pt);
\node at (-0.9,-0.7) {$\widehat L_{[e,m]}$};
\node[red] at (-0.2,-0.2) {$K$};
\end{tikzpicture}}
\hspace{0.5 in} = \hspace{0.5 in}
 \raisebox{-2em}{\begin{tikzpicture}
\draw[thick] (0,0) circle (20pt);
\node at (1.1,-0.7) {$\widehat L_{[e,m]}$};
\end{tikzpicture}}
\]
	\caption{The line $\widehat L_{[e,m]}$ can absorb the line $K$ since whenever the $K$ loop (red) is non-trivial, the $\widehat L_{[e,m]}$ loop (black) vanishes.}
	\label{fig:LemabsorbK}
\end{figure}

A physical picture for why this absorption can occur is given in Figure \ref{fig:LemabsorbK}. We consider a configuration of coincident loops of $\widehat L_{[e,m]}$ and $K$ and ask if this configuration can be distinguished from a loop of $\widehat L_{[e,m]}$ in isolation. To answer this, we fix an arbitrary EM gauge configuration and compute the vev of the loops. In order for the $K$ loop to give a non-trivial result, it must link with an odd number of units of EM flux. However, in the presence of an odd number of units of EM flux, the loop of $\widehat L_{[e,m]}$ gives vanishing contribution, since each consitutent of $\widehat L_{[e,m]}$ is not EM invariant and hence cannot form a closed loop around such flux. The configuration of $\widehat L_{[e,m]}$ stacked with $K$ is thus indistinguishable from the configuration with only $\widehat L_{[e,m]}$. We conclude that there are only ${1 \over 2} N(N-1)$ distinct non-invertible defects of quantum dimension 2, labeled by the symmetric pair $[e,m]\equiv [m,e]$ with $e\neq m$.

Finally, there are lines descending from the twist defects in the pre-gauged theory. As shown in Figure \ref{fig:dualitySTFT}, after gauging $\Z_2^{\text{EM}}$ the surface attached to the twist defect becomes transparent, and hence the twist defect become a genuine line defect. 
Since, as discussed in Section \ref{sec:Defectending}, before gauging $\Z_2^{\text{EM}}$ the twist defect $\Sigma_{(e)}$ is $\Z_2^{\text{EM}}$ invariant, the genuine lines after gauging can be stacked with a $K$ line. 
We will denote the resulting genuine lines by $\widehat{\Sigma}_{(e)}^q$, where $q=\pm$ again indicates whether or not we have stacked with $K$. There are $2N$ lines of this type. Below, we will find that the fusion rules of these genuine lines are almost identical to the fusion rules of the twist defects in \eqref{sigma2fusion2}, \eqref{sigma2fusion0}, and \eqref{sigma2fusion1}, from which we can tell that the quantum dimensions are $\sqrt{N}$. The only new ingredient in determining the fusion rules is determining how to assign factors of $q$.

To summarize the discussion so far, the theory with $\ZZ_2^{\mathrm{EM }}$ gauged has the following simple objects: 
\begin{itemize}
\item $2N$ invertible lines $\widehat L_{(e)}^\pm$~,
\item  $ {1\over 2} N(N-1)$ lines $\widehat L_{[e,m]}$ of quantum dimension 2~,
\item $2N$ lines $\widehat{\Sigma}_{(e)}^\pm$ of quantum dimension $\sqrt{N}$ ~.
\end{itemize}
A first-order consistency check is that the total quantum dimension for these lines is $ 2N\times (1)^2 + {1\over 2}N(N-1)\times (2)^2 + 2N \times  (\sqrt{N})^2  = (2N)^2$, which is the square of the total quantum dimension of $\mathrm{TY}(\ZZ_N)$. We now obtain the fusion rules for these objects.

\subsubsection{Fusion rules involving only $\widehat{L}_{(e)}^{\pm}$ and $\widehat{L}_{[e,m]}$}
\label{sec:fusionLL}

The fusion rule involving only the lines $\widehat{L}_{(e)}^{\pm}$ and $\widehat{L}_{[e,m]}$ largely follow from those in the pre-gauged theory. The only new ingredient is to determine the  distribution of $K$ lines (i.e. the value of $q$'s).  To determine this, we need to determine the $\Z_2^{\text{EM}}$ charge localized at junctions in the pre-gauged theory; if the junction is $\Z_2^{\text{EM}}$ even, then there should be an even number of $K$ lines anchored on the junction after gauging, whereas if the junction is $\Z_2^{\text{EM}}$ odd there should be an odd number of $K$ lines anchored on the junction. The junction charges of the pre-gauged theory can be measured by wrapping $D_{\text{EM}}$ surfaces around them, as discussed in Appendix \ref{app:junctioncharge}.  Using the tools developed there, one finds that e.g. the junction between three operators ${L}_{(e,e)}, {L}_{(e',e')}$ and ${L}_{(e+e', e+e')}$ is $\ZZ_2^\mathrm{EM}$ even, and hence the number of $K$ lines is conserved under fusion, i.e. $\widehat{L}_{(e)}^{q}\otimes \widehat{L}_{(e')}^{q'}= \widehat{L}_{(e+e')}^{qq'}$.

Another interesting case is the fusion between $\widehat{L}_{[e,m]}\otimes \widehat{L}_{[e',m']}$ when $e+e'=m+m', m+e'=e+m'$. Before gauging $\Z_2^{\text{EM}}$, the right hand side is a direct sum of two lines $L_{(e+e',e+e')}$, $L_{(m+e', m+e')}$. Each of them are $\Z_2^{\text{EM}}$ invariant separately, and hence we must determine the assignment of $q$ after gauging. One could again determine this via computation of the junction charge, but a nice trick to circumvent this computation is to observe that since $\widehat{L}_{[e,m]}\otimes \widehat{L}_{[e',m']}$ can absorb a $K$ line, the right-hand side of the fusion rules must also be able to absorb a $K$ line. The only possibility is then to sum over all possible $q$ values, i.e. $\widehat{L}_{[e,m]}\otimes \widehat{L}_{[e',m']}= \widehat{L}_{(e+e')}^{+}\oplus \widehat{L}_{(e+e')}^{-}\oplus \widehat{L}_{(m+e')}^{+}\oplus \widehat{L}_{(m+e')}^{-}$. See \cite{Bhardwaj:2022yxj} for more systematic discussions. In summary, we are able to determine the fusion rules as follows,
\begin{equation}
\label{eq:3daftergaugingLfusion}
\begin{split}
    \widehat{L}_{(e)}^{q}\otimes \widehat{L}_{(e')}^{q'}&= \widehat{L}_{(e+e')}^{qq'}~,\\
    \widehat{L}_{(e)}^{q}\otimes \widehat{L}_{[e',m']}&= \widehat{L}_{[e+e',e+m']}~,\\
    \widehat{L}_{[e,m]}\otimes \widehat{L}_{[e',m']}&=
    \begin{cases}
    \widehat{L}_{(e+e')}^{+}\oplus \widehat{L}_{(e+e')}^{-}\oplus \widehat{L}_{(m+e')}^{+}\oplus \widehat{L}_{(m+e')}^{-}& 
    e+e'=m+m',m+e'=e+m'\\
    \widehat{L}_{(e+e')}^{+}\oplus \widehat{L}_{(e+e')}^{-}\oplus \widehat{L}_{[m+e',e+m']}&
        e+e'=m+m',m+e'\neq e+m'\\
    \widehat{L}_{[e+e',m+m']} \oplus 
    \widehat{L}_{(m+e')}^{+}\oplus \widehat{L}_{(m+e')}^{-}&
    e+e'\neq m+m',m+e'= e+m'\\
    \widehat{L}_{[e+e',m+m']} \oplus
    \widehat{L}_{[m+e',e+m']}&
    e+e'\neq m+m',m+e'\neq  e+m'\\
    \end{cases}
\end{split}
\end{equation}
Note that since all of the operators above are lines, the local fusion is identical to the global fusion.

\subsubsection{Fusion rules involving $\widehat{\Sigma}_{(e)}^{q}$}
\label{sec:3dfusionruleswithsigma}

We now describe the fusion rules involving the $\widehat{\Sigma}_{(e)}^{q}$ lines of quantum dimension $\sqrt{N}$.

\paragraph{Fusion rule $\widehat{L}_{(e)}^{q}\otimes \widehat{\Sigma}_{(e')}^{q'}$:}
We start by considering the fusion between the invertible line $\widehat{L}_{(e)}^{q}$ and the non-invertible line $\widehat{\Sigma}_{(e')}^{q'}$. The cases of $N$ odd and even are qualitatively different, and we begin by analyzing the former.

For odd $N$, it is useful to observe that before gauging, none of the twist defects map to themselves under fusing with a non-trivial $\Z_2^{\text{EM}}$ invariant invertible line,
\bea
\label{eq:intermediatefusionrule}
L_{(e,e)}\otimes \Sigma_{(e')} =  \Sigma_{(2e+e')}~,
\eea
where $e' \neq e'+2e$ for $e\neq 0\mod N$.  This implies that after gauging $\Z_2^{\text{EM}}$ one can \emph{define} the twist defect $\widehat{\Sigma}_{(e)}^{q}$ by fusing $\widehat{L}_{(\widetilde{e})}^{q}$ with the minimal twist defect $\widehat{\Sigma}_{(0)}^{+}$, 
\begin{eqnarray}\label{eq:LSigmaaftergauging}
\widehat{\Sigma}_{(e)}^{q}= \widehat{L}_{(\widetilde{e})}^{q}\otimes \widehat{\Sigma}_{(0)}^{+}~, \hspace{1cm} e=0, ..., N-1, \hspace{1cm} q=\pm
\end{eqnarray}
where $\widetilde{e}= e/2$ for even $e$, and $\widetilde{e}= (e+N)/2$ for odd $e$. Note that since the operators are always subject to relabeling, such a definition is always allowed. The fusion rule then follows from \eqref{eq:LSigmaaftergauging}, 
\begin{eqnarray}\label{eq:LSfusionodd}
\widehat{L}^q_{(e)} \otimes \widehat{\Sigma}^{q'}_{(e')}= \widehat{L}^{qq'}_{(e+\widetilde{e}')} \otimes \widehat{\Sigma}^{+}_{(0)}= \widehat{\Sigma}^{qq'}_{(2e+e')}~, \hspace{1cm} N\,\, \text{ odd}~.
\end{eqnarray}

We now proceed to the case of $N$ even. The main conceptual difference is that we can no longer obtain all  defects $\widehat \Sigma^q_{(e)}$  from $\widehat \Sigma^0_{(0)}$ by fusing with an appropriate choice of $\widehat L_{(e')}^q$. Before gauging $\Z_2^{\text{EM}}$, every twist defect maps to itself upon fusing with $L_{(N/2, N/2)}$. This means that after gauging one can define the twist defects as 
\begin{eqnarray}
\widehat \Sigma_{(2e)}^{q}=\widehat L_{(e)}^q \otimes \widehat \Sigma_{(0)}^{+} ~, \hspace{0.5 in} \widehat\Sigma_{(2e+1)}^{q}= \widehat L_{(e)}^q \otimes \widehat \Sigma_{(1)}^{+} ~, \hspace{1cm} e=0, 1, ..., N/2-1~.
\end{eqnarray}
Once this definition is fixed though, we are no longer free to choose the sign that appears in the fusion of $\widehat L_{(N/2)}^q$ with $\widehat \Sigma_{(0)}^{+}$ and $\widehat \Sigma_{(1)}^{+}$. Indeed, because before gauging the twist defects are stabilized under this fusion, the $q', q''$ in the following fusion rules
\begin{eqnarray}\label{eq:q'q''}
\widehat \Sigma_{(0)}^{q'}=\widehat L_{(N/2)}^q \otimes \widehat \Sigma_{(0)}^{+} ~, \hspace{0.5 in} \widehat\Sigma_{(1)}^{q''}= \widehat L_{(N/2)}^q \otimes \widehat \Sigma_{(1)}^{+}
\end{eqnarray}
are unambiguously defined. In other words, it is not possible to relabel the twist defects to change $q' $ and $q''$ once $q$ is given. To determine $q'$ and $q''$, 
one must measure the $\ZZ_2^{\mathrm{EM}}$ charge of the relevant junction in the theory before gauging EM. Relegating the details to Appendix \ref{app:junctioncharge}, we find that the junction associated with the first fusion rule in \eqref{eq:q'q''} is $\Z_2^{\text{EM}}$ even, and hence the number of $K$ lines is conserved. On the other hand, the junction associated with the second fusion rule in \eqref{eq:q'q''} is $\Z_2^{\text{EM}}$ odd, and hence the number of $K$ lines jumps by one. In terms of the relation between $q,q'$ and $q''$, we have
\begin{eqnarray}
q'=q~, \hspace{1cm} q''=-q~.
\end{eqnarray}
As a consequence, fusing $\widehat{L}^q_{(e)}$ and $\widehat\Sigma^+_{(0,1)}$ for the remaining choices of $e$, i.e. $e=N/2+1, ..., N-1$, are given by 
\begin{eqnarray}
\widehat \Sigma_{(2e)}^{q}=\widehat L_{(e)}^q \otimes \widehat \Sigma_{(0)}^{+} ~, \hspace{0.5 in} \widehat\Sigma_{(2e+1)}^{-q}= \widehat L_{(e)}^q \otimes \widehat \Sigma_{(1)}^{+} ~, \hspace{1cm} e=N/2+1, ..., N-1~.
\end{eqnarray}
The fusion between generic $\widehat{L}^q_{(e)}$ and $\widehat\Sigma_{(e')}^{q'}$ then follows straightforwardly. 

To summarize the results for even $N$, 
we have
\begin{equation}\label{eq:LSfusion}
\begin{split}
    \widehat{L}_{(e)}^{q}\otimes \widehat{\Sigma}_{(e')}^{q'}= 
    \begin{cases}
    \widehat{\Sigma}_{(2e+e')}^{qq'}~, &  \{e'\in 2\Z\} \cup \{ e'\in 2\Z+1,\,\, 0\leq [e+\frac{e'-1}{2}]_N\leq N/2-1\}\\
    \widehat{\Sigma}_{(2e+e')}^{-qq'}~, &  \{ e'\in 2\Z+1,\,\, N/2\leq [e+\frac{e'-1}{2}]_N\leq N-1\}
    \end{cases}
\end{split}
\end{equation}
where $[x]_N$ is the mod $N$ reduction of $x$.

\paragraph{Fusion rule $\widehat{L}_{[e,m]}\otimes \widehat{\Sigma}_{(e)}^{q}$:} 
Since $\widehat{L}_{[e,m]}$ before gauging $\Z_2^{\text{EM}}$ consists of two lines exchanged under $\Z_2^{\text{EM}}$, after gauging $\Z_2^{\text{EM}}$ the fusion rules $\widehat{L}_{[e,m]}\otimes \widehat{\Sigma}_{(e)}^{q}$ contains two fusion channels which differ by a $\Z_2^{\text{EM}}$ representation. This leads to the fusion rule
\begin{eqnarray}\label{eq:hLSfusion}
\begin{split}
    \widehat{L}_{[e,m]}\otimes \widehat{\Sigma}_{(e')}^{q'} = \widehat{\Sigma}_{(e+m+e')}^{+} \oplus \widehat{\Sigma}_{(e+m+e')}^{-}~, \hspace{1cm} q'=\pm~.
\end{split}
\end{eqnarray}
Another way to argue for the sum of $\widehat{\Sigma}_{(e+m+e')}^{\pm}$ on the right-hand side is to notice that since $\widehat L_{[e,m]}$ can absorb a factor of $K$, the right-hand side must also be able to absorb a factor of $K$. This can only be achieved by summing over the defects with and without $K$ stacked.

\paragraph{Fusion rule $\widehat{\Sigma}_{(e)}^{q}\otimes \widehat{\Sigma}_{(e')}^{q'}$:}
The fusion rule $\widehat{\Sigma}_{(e)}^{q}\otimes \widehat{\Sigma}_{(e')}^{q'}$ can be obtained from the fusion rules derived in \eqref{eq:LSfusionodd}, \eqref{eq:LSfusion}, and \eqref{eq:hLSfusion}. When $N$ is odd, the fusion rules are 
\begin{eqnarray}\label{eq:SSfusionNodd}
\begin{split}
    \widehat{\Sigma}_{(e)}^{q} \otimes \widehat{\Sigma}_{(e')}^{q'} &= \widehat{L}_{(\tilde{e})}^{q} \otimes \widehat{\Sigma}_{(0)}^{+} \otimes \widehat{L}_{(\tilde{e}')}^{q'} \otimes \widehat{\Sigma}_{(0)}^{+}\\
    &= \widehat{L}_{(\tilde{e}+ \tilde{e}')}^{qq'} \otimes \left( \widehat{L}_{(0)}^{+} \oplus  \bigoplus_{n=1}^{(N-1)/2} \widehat{L}_{[n,N-n]}\right)\\
    &= \widehat{L}_{(\tilde{e}+\tilde{e}')}^{qq'} \oplus \bigoplus_{n=1}^{(N-1)/2} \widehat{L}_{[\tilde{e}+\tilde{e}'+n,\tilde{e}+\tilde{e}'-n]}~,
\end{split}
\end{eqnarray}
where in the second line, we used the commutativity of lines (since they are objects with codimension higher than one). The $\tilde{e}$ appearing here is defined below \eqref{eq:LSigmaaftergauging} 
and similarly for $\tilde{e}'$. When $N$ is even, one obtains the fusion rules 
\begin{equation}\label{eq:SSfusionNeven}
\begin{split}
    &\widehat{\Sigma}_{(e)}^{q}\otimes \widehat{\Sigma}_{(e')}^{q'}\\&\,\,\,= 
    \begin{cases}
    \widehat{L}_{((e+e')/2)}^{qq'}\oplus \widehat{L}_{((e+e'+N)/2)}^{qq'} \oplus\bigoplus_{n=1}^{N/2-1} \widehat{L}_{[(e+e')/2+n, (e+e')/2-n]}~, & (e,e')=(\text{even}, \text{even})\\
    \bigoplus_{n=0}^{N/2-1} \widehat{L}_{[(e+e'-1)/2+n+1, (e+e'-1)/2-n]}~, & (e,e')= (\text{even}, \text{odd})\\
    \bigoplus_{n=0}^{N/2-1} \widehat{L}_{[(e+e'-1)/2+n+1, (e+e'-1)/2-n]}~, & (e,e')= (\text{odd}, \text{even})\\
    \widehat{L}_{((e+e')/2)}^{qq'}\oplus \widehat{L}_{((e+e'+N)/2)}^{-qq'} \oplus\bigoplus_{n=0}^{N/2-2} \widehat{L}_{[(e+e')/2+n+1, (e+e')/2-n-1]}~, & (e,e')=(\text{odd}, \text{odd})\\
    \end{cases}
\end{split}
\end{equation}
The lines in the SymTFT of $TY(\Z_N)$ form objects in the Drinfeld center of the fusion category $TY(\Z_N)$, i.e. $\cZ(TY(\Z_N))$. This center $\cZ(TY(\Z_N))$ has been studied in the mathematics literature, see e.g. \cite{izumi2001structure,gelaki2009centers}. In particular, the explicit expressions of the modular $S$ matrices have been derived. 
In Appendix \ref{app:fusionDCTY}, we collect the modular $S$ matrices, and use the Verlinde formula to verify the fusion rules above.

\subsubsection{Example: SymTFT for $TY(\Z_2)$}
\label{sec:TYZ2}

To illustrate the results we have obtained, let us apply them to the special case of $N=2$. We first discuss the SymTFT for a theory with only $\Z_2$ symmetry, which as described in Section \ref{sec:symTFT3dZN} is simply $\Z_2$ gauge theory. There are four invertible lines 
\begin{eqnarray}\label{eq:N2invertibleline}
L_{(0,0)} \leftrightarrow 1~, \hspace{1cm} L_{(1,0)} \leftrightarrow e~, \hspace{1cm} L_{(0,1)} \leftrightarrow m~, \hspace{1cm} L_{(1,1)} \leftrightarrow \psi~.
\end{eqnarray}
There is also a $\Z_2^{\text{EM}}$ zero-form global symmetry, generated by the 2d surface defect $D_{\text{EM}}$. In the current case $D_{\text{EM}}$ is obtained by condensing the $\psi$ line on a 2d surface. There are also twist defects obtained by condensing $\psi$ on half of a 2d surface with Dirichlet boundary conditions, and fusing with an appropriate invertible line in \eqref{eq:N2invertibleline},
\begin{eqnarray}
\Sigma_{(0)} \leftrightarrow \sigma_+~, \hspace{1cm} \Sigma_{(1)} \leftrightarrow \sigma_-~.
\end{eqnarray}
The notation given on the right-hand side of $\leftrightarrow$ is that used in  \cite{2015arXiv150306812T, Barkeshli:2014cna}. 
By the results in Section \ref{sec:symTFT3dZN}, the fusion rules among the lines and the twist defects are 
\begin{eqnarray}
\begin{split}
     &e \otimes e = m \otimes m = \psi \otimes \psi = 1~,\\ 
     &e\otimes m = \psi~,\\
&e \otimes \psi =  m~,\\ 
&m \otimes \psi = e~,\\
&\psi \otimes \sigma_+ = \sigma_{+}~,\\
&\psi \otimes \sigma_{-} = \sigma_{-}~,\\
&e \otimes \sigma_{+} = m \otimes \sigma_{+} = \sigma_{-}~,\\ 
&e \otimes \sigma_{-} = m \otimes \sigma_{-} = \Sigma_{+}~,\\
&\sigma_{+}\otimes \sigma_{+}= \sigma_{-}\otimes \sigma_{-}=1\oplus \psi~,\\
&\sigma_{+}\otimes \sigma_{-}= \sigma_{-}\otimes \sigma_{+}=e\oplus m~.
\end{split}
\end{eqnarray}

We now obtain the SymTFT for $TY(\ZZ_2)$ by gauging  $\ZZ_2^{\mathrm{EM}}$. Upon gauging, we obtain a quantum $\widehat \ZZ_2^{(1)}$ symmetry generated by $K$. There are four line operators of quantum dimension 1,
\begin{eqnarray}
\begin{split}
    \widehat{L}_{(0)}^{+} \leftrightarrow (I, +)~, \hspace{1cm} \widehat{L}_{(0)}^{-} \leftrightarrow (I, -)= K~,\hspace{1cm} \widehat{L}_{(1)}^{+} \leftrightarrow (\psi, +)~, \hspace{1cm} \widehat{L}_{(1)}^{-} \leftrightarrow(\psi, -)~,
\end{split}
\end{eqnarray}
 one line operator of quantum dimension 2, 
\begin{eqnarray}
\widehat{L}_{[0,1]}\leftrightarrow [em]~,
\end{eqnarray}
and four line operators of quantum dimension $\sqrt{2}$,
\begin{eqnarray}
\widehat{\Sigma}_{(0)}^{+}\leftrightarrow (\sigma_+, +)~, \hspace{1cm} \widehat{\Sigma}_{(0)}^{-}\leftrightarrow (\sigma_+, -)~, \hspace{1cm}
\widehat{\Sigma}_{(1)}^{+}\leftrightarrow (\sigma_-, +)~, \hspace{1cm} \widehat{\Sigma}_{(1)}^{-}\leftrightarrow (\sigma_-, -)~.
\end{eqnarray}
By the formulas given in this section, the fusion rules are found to be,
\begin{eqnarray}
\begin{split}
    &(I, q) \otimes (X, q')= (X, qq'), \hspace{1cm} X=I~, \psi, \sigma_+, \sigma_-\\
    &(I, q)\otimes [em] = [em]~,\\
    &(\psi, q)\otimes (\psi, q')= (I, qq')~,\\
    &(\psi, q)\otimes (\sigma_\pm, q')= (\sigma_{\pm}, \pm qq')~,\\
    &(\psi, q)\otimes [em] = [em]~,\\
    &(\sigma_{\pm}, q)\otimes (\sigma_{\pm}, q')= (I, qq')\oplus (\psi, \pm qq')~,\\
    &(\sigma_{\pm}, q)\otimes (\sigma_{\mp}, q')= [em]~,\\
    &(\sigma_{\pm}, q)\otimes [em] = (\sigma_{\mp}, +)\oplus (\sigma_{\mp}, -)~,\\
    &[em]\otimes [em]= (I, +)\oplus (I, -)\oplus (\psi, +)\oplus (\psi, -)~.
\end{split}
\end{eqnarray}
These fusion rules coincide with the ones given in \cite{2015arXiv150306812T, Barkeshli:2014cna}, where it was further noted that 
the fusion rules exactly match those of $\text{Ising}\times \overline{\text{Ising}}$, if we identify
\begin{eqnarray}
\begin{split}
    &(I,+)\leftrightarrow 1~, \qquad (I,-) \leftrightarrow \bar\eta\eta~, \qquad (\psi,+) \leftrightarrow \eta~, \qquad (\psi,-) \leftrightarrow \bar\eta~, \qquad [em] \leftrightarrow \bar\CN \CN~,\\
    &(\sigma_+,+) \leftrightarrow \CN, \qquad (\sigma_+,-) \leftrightarrow \bar\eta \CN~, \qquad (\sigma_-,+) \leftrightarrow \bar\CN~, \qquad (\sigma_-,-) \leftrightarrow \eta\bar\CN~.
\end{split}
\end{eqnarray}
This identification also matches with the expectation that the SymTFT should have symmetry given by the Drinfeld center of $TY(\ZZ_N)$. Indeed, for a modular tensor category (of which $TY(\ZZ_2)=\mathrm{Ising}$ is an example) the Drinfeld center is known to be the tensor product of the original category with its orientation reversal.

\subsection{Twisted cocycle description}
\label{Sec:3dcocycle}

In Section \ref{sec:ZTYZN} we obtained the simple objects and fusion rules for the SymTFT of $TY(\ZZ_N)$. We now reobtain these results in a different way, using an explicit Lagrangian description of the SymTFT.  This in particular allows us to give a concrete twisted cocycle description of all simple objects.

\subsubsection{The action and restricted gauge transformations}
The starting point is once again the $\ZZ_N$ gauge theory in (\ref{eq:3dBFtheory}).  We will find it convenient to 
symmetrize the two gauge fields, and rewrite the BF theory \eqref{eq:3dBFtheory} in the K-matrix form:
\begin{eqnarray}\label{3dZN}
S= \frac{2\pi}{2N}\int \mathbf{a}^T \cup K\, \delta \mathbf{a}~,
\end{eqnarray}
where $\mathbf{a}= (a, \widehat{a})$ is a two-component cochain valued in $\Z$, and $K=\sigma^x$.\footnote{We hope that the reader will not confuse the $K$-matrix here with the quantum line $K$ defined above.} In terms of $\mathbf{a}$,  the $\Z_2^{\mathrm{EM}}$ symmetry acts as  $\mathbf{a} \rightarrow K \mathbf{a}$. Beginning in the pre-gauged theory, the action is as given in (\ref{3dZN}), and the line operators are as given in (\ref{eq:beforegaugingLs}). We now attempt to gauge $\Z_2^{\mathrm{EM}}$. To do so, we must first understand how to couple the action \eqref{3dZN} to a background field $C$ for $\Z_2^{\mathrm{EM}}$.

The coupling to $C$ can be achieved by promoting the cochain $\mathbf{a}$ to a $\Z_2^{\mathrm{EM}}$-twisted cochain. Let us begin by recalling this notion. A 1-cochain is most easily understood as a link between sites on a lattice. Such a 1-cochain can be labelled by two indices $\alpha_{ij}$, which label the lattice sites between which the link is stretched. In this index notation, the usual cup product and coboundary operator are defined as 
\bea
(\alpha \cup \beta)_{ijk} = \alpha_{ij}  \beta_{jk} ~, \hspace{0.5 in}(\delta \alpha)_{ijk} = \alpha_{jk} - \alpha_{ik} + \alpha_{ij}~.
\eea
It is straightforward to check that $\delta^2 = 0$. The gauge transformation $\a \rightarrow \a  + \delta g$  in index notation becomes
\bea
\a_{ij} \rightarrow \a_{ij}  +  g_j - g_i~.
\eea

For a $\ZZ_2$-\textit{twisted} cocycle  $\boldsymbol{\alpha}_{ij}$ with background $C$ and matrix $K$, the twisted cup product and twisted coboundary operations are defined as 
\bea
(\boldsymbol{\alpha} \cup_C \boldsymbol{\beta})_{ijk} :=  \boldsymbol{\alpha}_{ij} K^{C_{ij}} \boldsymbol{\beta}_{jk} ~, \hspace{0.5 in}(\delta_C \boldsymbol{\alpha})_{ijk} := K^{C_{ij}}\boldsymbol{\alpha}_{jk} - \boldsymbol{\alpha}_{ik} + \boldsymbol{\alpha}_{ij}~, 
\eea
and the twisted gauge transformation is given by 
\begin{eqnarray}\label{3dGT}
\boldsymbol{\alpha} _{ij}\to\boldsymbol{\alpha} _{ij} + K^{C_{ij}} \mathbf{g}_j - \mathbf{g}_i~; 
\end{eqnarray}
see Appendix B of \cite{Benini:2018reh} for more details. 
One can again check that $\delta_C^2 = 0$ as long as $C$ is flat. Intuitively, in the current context promoting $\alpha$ to a $\Z_2$-twisted cocycle $\boldsymbol{\alpha}$ means that we take $\boldsymbol{\alpha}$ to be a 2-vector, and allow the transition functions between patches to include a matrix swapping the entries of $\boldsymbol{\alpha}$.

We may now write the action of (\ref{3dZN}) coupled to a background for $\Z_2^{\mathrm{EM}}$ as
\begin{eqnarray}\label{3dZNC}
S[C] = \frac{2\pi}{2N}\int \mathbf{a}^T \cup_C K \delta_C \mathbf{a}~,
\end{eqnarray}
where the integrand, in components, is
\begin{eqnarray}\label{3dgaugeinvariant}
\left(\mathbf{a}^T \cup_C K \delta_C \mathbf{a}\right)_{ijkl} = \mathbf{a}_{ij}^T K^{C_{ij}+1} \left(K^{C_{jk}}\mathbf{a}_{kl}- \mathbf{a}_{jl} + \mathbf{a}_{jk}\right)~.
\end{eqnarray}
It is a good exercise to verify that \eqref{3dZNC} is invariant under dynamical gauge transformations \eqref{3dGT}. Details of this exercise are provided in Appendix \ref{app:2dactioninvariance}. The action is furthermore invariant under background gauge transformations of $C$, which are given by
\begin{eqnarray}\label{3dbgdTransformation}
C_{ij}\to C_{ij}+ \omega_j - \omega_i, \hspace{0.5 in }\mathbf{a}_{ij} \to K^{-\omega_i} \mathbf{a}_{ij}, \hspace{0.5 in } \mathbf{g}_i\to K^{-\omega_i}\mathbf{g}_i~.
\end{eqnarray}
The form of these transformations can be understood pictorially as in Figure \ref{fig.cocycles}.

\begin{figure}[t]
	\centering
 	\begin{tikzpicture}[scale=1.5,baseline=40]
	
	\draw[very thick] (-0.2,0) -- (3.2,0);
	\draw[very thick] (-0.2,1.732) -- (3.2,1.732);
	\draw[very thick] (-0.2,2.598) -- (3.2,2.598);
	
	\draw[very thick] (0,0) -- (1.5,2.598);
	\draw[very thick] (1,0) -- (2.5,2.598);
	\draw[very thick] (2,0) -- (0.5,2.598);
	\draw[very thick] (3,0) -- (1.5,2.598);
	
	\draw[very thick] (2,0) -- (3,1.732);
	\draw[very thick] (1,0) -- (0,1.732);
	\draw[very thick] (2.5,2.598) -- (3,1.732);
	\draw[very thick] (0.5,2.598) -- (0,1.732);
	
	\draw[very thick, blue,dotted] (1,2.598)--(2.5,0);
		
	\node[below,red] at (1.5,0.84) {$j$};
	\node[below,red] at (0.5,0.84) {$i$};
	\node[below,dgreen] at (2.5,0.84) {$k$};
	
	\draw[very thick,red] (-0.2,0.866) -- (2.5,0.866);
	\draw[very thick,dgreen] (2.5,0.866) -- (3.2,0.866);
	\node[above,red] at (1,0.866) {$a_{ij}$};
	\node[above,red] at (2,0.866) {$a_{jk}$};
	\node[above,dgreen] at (3,0.866) {$\widehat{a}_{k\ell}$};
	
	\node at (2.5,0.866) [circle,fill,yellow, inner sep=1.5pt]{};

	\end{tikzpicture} 
	\hspace{0.4 in}
	$\xrightarrow[C_{jk} \rightarrow C_{jk} + {\color{orange}\omega_k} - {\color{orange} \omega_j}]{\mathbf{a}_{jk} \rightarrow K^{- \color{orange}{\omega_j}} \mathbf{a}_{jk}}$
	\hspace{0.4 in}
	\begin{tikzpicture}[scale=1.5,baseline=40]
	
	\draw[very thick] (-0.2,0) -- (3.2,0);
	
	\draw[very thick] (-0.2,1.732) -- (3.2,1.732);
	\draw[very thick] (-0.2,2.598) -- (3.2,2.598);
	
	\draw[very thick] (0,0) -- (1.5,2.598);
	\draw[very thick] (1,0) -- (2.5,2.598);
	\draw[very thick] (2,0) -- (0.5,2.598);
	\draw[very thick] (3,0) -- (1.5,2.598);
	
	\draw[very thick] (2,0) -- (3,1.732);
	\draw[very thick] (1,0) -- (0,1.732);
	\draw[very thick] (2.5,2.598) -- (3,1.732);
	\draw[very thick] (0.5,2.598) -- (0,1.732);
	
	\draw[very thick, blue,dotted] (1,2.598)--(1.5,1.5);
	\draw[very thick, blue,dotted] (1,1)--(1.5,1.5);
	\draw[very thick, blue,dotted] (1,1)--(1.5,0.2);
	\draw[very thick, blue,dotted] (1.5,0.2)--(2,0.7);
	\draw[very thick, blue,dotted] (2,0.7)--(2.5,0);

	\node[below,dgreen] at (1.5,0.84) {$j$};
	\node[below,red] at (0.5,0.84) {$i$};
	\node[below,dgreen] at (2.5,0.84) {$k$};
	
	\draw[very thick,red] (-0.2,0.866) -- (1.5,0.866);
	\draw[very thick,dgreen] (1.5,0.866) -- (3.2,0.866);
	\node[above,red] at (1,0.866) {$a_{ij}$};
	\node[above,dgreen] at (2,0.866) {$\widehat{a}_{jk}$};
	\node[above,dgreen] at (3,0.866) {$\widehat{a}_{k\ell}$};
	
	\node at (1.5,0.866) [circle,fill,yellow, inner sep=1.5pt]{};

	\end{tikzpicture} 
	\caption{The schematic form of $\ZZ_2^{\mathrm{EM}}$ gauge transformations. The background gauge field is $C$. The red and green lines are Wilson lines of $\mathbf{a}$, while the blue dotted lines represent the surface defects implementing the $\Z_2^{\text{EM}}$ transformation. }
	\label{fig.cocycles}
\end{figure}

Since the action (\ref{3dZNC}) is completely invariant under background gauge transformations, we may now gauge $\Z_2^{\mathrm{EM}}$. To do so, we promote the background gauge field $C$ to a dynamical gauge field $c$, upon which the action \eqref{3dZNC} becomes
\begin{eqnarray}
\label{eq:2dfinalanswer}
S_{\mathrm{SymTFT}}=\frac{2\pi}{2N}\int \mathbf{a}^T \cup_c K \delta_c \mathbf{a} + \pi \int x\cup \delta c~. 
\end{eqnarray} 
The last term is a BF term, such that integrating out $x$ enforces that $c$ is a $\Z_2$ cocycle. Because of the complicated dependence on $c$ in the kinetic term, we see that $c$ is now flat only on-shell. This means that the action is no longer invariant under \eqref{3dGT}, since without $c$ being flat the first term in \eqref{eq:2dfinalanswer} transforms by
\begin{eqnarray}
\begin{split}
\label{eq:nongaugeinv}
\delta S \,\,=\,\, &\frac{2\pi}{2N}\left(\mathbf{g}_i^T \left(K^{c_{ij}+c_{jk}+c_{kl}} - K^{c_{ik}+ c_{kl}} - K^{c_{ij}+ c_{jk}} + K^{c_{ik}}\right) K \mathbf{a}_{kl}\right)\\
&+ \frac{2\pi}{2N}\left((\mathbf{g}_j^T K^{-c_{ij}} - \mathbf{g}_i^T) K^{c_{ij} +1} (K^{c_{jk}+ c_{kl}}- K^{c_{jl}}) \mathbf{g}_l \right)~,
\end{split}
\end{eqnarray}
as follows from the computation in Appendix \ref{app:2dactioninvariance}. For generic $\mathbf{g}$, this is non-vanishing, and our action would seemingly not be gauge invariant. However, for $\mathbf{g}$ satisfying 
\bea
\label{eq:constraintong}
(K^{\delta c_{ijk} }- \mathds{1}) \mathbf{g}_k = 0~,
\eea
the problematic term (\ref{eq:nongaugeinv}) vanishes. Thus as long as we restrict to gauge transformations satisfying this constraint, the action is indeed gauge invariant. Note that this is a non-trivial constraint on $\mathbf{g}$ only at points where $c$ is not flat. We will see below that this constraint has a simple physical interpretation.

\subsubsection{Cocycle description of line operators}
\label{sec:cocyclelines}
We now give a cocycle description for all of the lines identified in Section \ref{sec:ZTYZNlines}. The combined gauge transformations  \eqref{3dGT} and \eqref{3dbgdTransformation} take the form
\begin{eqnarray}\label{3dGT2}
\mathbf{a} _{ij}\to K^{-\omega_i }\left(\mathbf{a} _{ij} + K^{c_{ij}} \mathbf{g}_j - \mathbf{g}_i \right)~, \hspace{0.2in} c_{ij}\to c_{ij}+ \omega_j - \omega_i~, \hspace{0.2 in} x_{ij}\to x_{ij}+ \eta_j-\eta_i~, \,\,
\end{eqnarray}
with $\mathbf{g}$ constrained as in (\ref{eq:constraintong}), and our goal is to construct the full spectrum of gauge-invariant line operators. 

\paragraph{Invertible Operators}
We first study the invertible operators. Begin by rewriting the line operators (\ref{eq:beforegaugingLs}) of the pre-gauged theory as\footnote{This definition of $L_{\mathbf{n}}$ is the same as the definition \eqref{eq:beforegaugingLs}, as can be seen by using (\ref{eq:corrLL}) together with the fact that $\langle \gamma, \gamma \rangle$ is vanishing in $(2+1)$d.} 
\bea
L_{\mathbf{n}}(\gamma)= e^{i\frac{2\pi}{N} \oint_\gamma \mathbf{n}^T\cdot\, \mathbf{a}}~, \hspace{0.5 in} \mathbf{n}\in \Z_N\times \Z_N~.
\eea
Under the gauge transformations given in \eqref{3dGT2}, these lines transform as
\begin{eqnarray}
L_{\mathbf{n}}(\gamma) \hspace{0.2 in}\longrightarrow \hspace{0.2 in}e^{i \frac{2\pi}{N} \sum_{i \in \gamma} \mathbf{n}^T  K^{-\omega_i} (\mathbf{a}_{i, i+1} + K^{c_{i, i+1}} \mathbf{g}_{i+1} - \mathbf{g}_{i})} ~.
\end{eqnarray}
For $L_{\mathbf{n}}$ to be gauge-invariant, we must first remove the $\omega_i$ dependence from the right-hand side, which means enforcing
\begin{eqnarray}
\mathbf{n}^T K = \mathbf{n}^T \hspace{0.5 in} \Rightarrow  \hspace{0.5 in} \mathbf{n} = (e,e)^T~,  \hspace{0.5 in}  e\in \Z_N~. 
\end{eqnarray}
This immediately removes the $\mathbf{g}_i$ dependence as well, since subject to this condition we have
\bea
\sum_i \mathbf{n}^T K^{-\omega_i}(K^{c_{i,i+1}} \mathbf{g}_{i+1} - \mathbf{g}_i) = \mathbf{n}^T \sum_i (\mathbf{g}_{i+1} - \mathbf{g}_i) = 0~.
\eea
 We thus conclude that the following line operators
 \bea
L_{(e,e)}(\gamma) = e^{i {2 \pi  \over N} \oint_\gamma e    (a + \widehat a)} ~, \hspace{0.5 in} e=0,\dots, N-1
  \eea
  are the only gauge invariant operators of this form. 
  
 There exist two other obvious invertible gauge-invariant line operators, namely
\begin{eqnarray}
\label{eq:chiK}
\chi(\gamma) = e^{i \pi \oint_\gamma x}~, \hspace{0.5 in} K(\gamma) =  e^{i \pi \oint_\gamma c }~.
\end{eqnarray} 
The line $\chi(\gamma)$ is naively not topological, since $\delta x$ is not generically zero---this is because $c$ cannot act as a Lagrange multiplier due to its appearance in the kinetic term. The line $\chi(\gamma)$ will reappear below in our analysis of non-invertible lines, but for the moment we will ignore it. On the other hand, the line $K(\gamma)$ is topological, and is the generator of the dual $\widehat \ZZ_2^{(1)}$ symmetry which we have already met before. Hence we conclude that there are $2N$ invertible topological line operators $\widehat L_{(e)}^+ := L_{(e,e)}$ and $\widehat L_{(e)}^- := K \, L_{(e,e)}$.

\paragraph{Non-invertible Operators}

\begin{figure}[!tbp]
 	\[\widehat L_{[e,m]} = \sum_{\widetilde{c}_{j_i,j_{i+1}}}\mathrm{exp}\left[{2\pi i \over N}\sum_{\{j_i\}}
	{\begin{tikzpicture}[scale=2.2,baseline=0]
	\draw[very thick,blue] (-0.7,0) -- (0,0);
	\draw[very thick,dgreen] (0,0) -- (0.7,0);
	\draw[very thick,dgreen] (0.7,0) -- (1.4,0);
	\draw[very thick,dgreen] (1.4,0) -- (2.1,0);
	\draw[very thick,blue] (2.1,0) -- (2.8,0);
	\node at (0,0) [circle,fill,orange, inner sep=1.5pt]{};
	\node at (0.7,0) [circle,fill,black, inner sep=1.5pt]{};
	\node at (1.4,0) [circle,fill,black, inner sep=1.5pt]{};
	\node at (2.1,0) [circle,fill,orange, inner sep=1.5pt]{};
	\node[below,orange] at (0,0) {$K^{\widetilde{c}_{j_1,j_{2}}}$};
	\node[below,orange] at (2.1,0) {$K^{\widetilde{c}_{j_n,j_{n+1}}}$};
	
	\node[above,blue] at (-0.25,0) {$(ea+m\widehat a)_{j_0,j_1}$};
	\node[above,dgreen] at (1.05,0) {$(ma+e\widehat a)_{\dots}$};
	\node[above,blue] at (2.8,0) {$(ea+m\widehat a)_{j_{n+1},j_{n+2}}$};
	
	\end{tikzpicture}}\right] \]
	\caption{The schematic form of $\widehat L_{[e,m]}$. This definition effectively corresponds to insertion of $L_{(e,m)}\oplus L_{(m,e)}$ on each link.}
	\label{fig.QD2guys}
\end{figure}

We now proceed to the non-invertible operators. The non-invertible operators with quantum dimension 2 are the simplest to construct, and are obtained by noting that alternating chains of $e a + m \widehat a$ and $m a + e \widehat a$ on links can be made gauge-invariant by inserting appropriate factors of $K$. 
Concretely, the line
\bea
\widehat L_{[e,m]} = \prod_j \sum_{\widetilde{c}_{j,j+1}\in \{0,1\}} e^{{2 \pi i \over N} (e,m) K^{\widetilde{c}_{j,j+1}} \mathbf{a}_{j,j+1}}~,
\eea
is gauge invariant. As shown in Figure \ref{fig.QD2guys}, this intuitively corresponds to the insertion of $L_{(e,m)} \oplus L_{(m,e)}$ on each link. There are ${1\over 2} N(N-1)$ distinct such operators. These operators can absorb the line $K$, as follows from precisely the same argument as was given in Section \ref{sec:ZTYZNlines}.

We next consider the non-invertible lines with quantum dimension $\sqrt{N}$. In fact, searching for gauge invariant operators involving only $\mathbf{a}$ and $c$ does not turn up anything new besides the operators identified above. We are thus led back to consideration of the non-topological operator $\chi(\gamma)$ defined in (\ref{eq:chiK}). 
To understand this better, we may begin by using the equations of motion of $c$ to obtain 
\bea
\label{eq:deltax}
\delta_{i \tilde{i}} \delta_{j \tilde{j}}(\delta x)_{jk\ell} = {1\over N} \left[ \delta_{i \tilde{i}} \delta_{j \tilde{j}}(\mathbf{a}^T \cup_c \,(K-\mathds{1}) \delta_c \mathbf{a})_{ i jk \ell} + \delta_{j \tilde{i}} \delta_{k \tilde{j}} \mathbf{a}_{ij}^T\, K^{c_{ij}+c_{jk}}  (K-\mathds{1})\, \mathbf{a}_{k\ell}\right]~.
\eea
Note that the repeated indices are not summed over.
In obtaining this, we used
\bea
{\partial \over \partial c_{\tilde{i}\tilde{j}}} K^{c_{ij}} := \delta_{i\tilde{i}}\delta_{j\tilde{j}} (K^{c_{ij}+1}-K^{c_{ij}})= \delta_{i\tilde{i}}\delta_{j\tilde{j}} K^{c_{ij}} (K-\mathds{1})
\eea
where the derivative with respect to the $\Z_2$ valued cochain $c_{ij}$ is defined to be the finite difference.
Thanks to the factor of $(K-\mathds{1})$, we see that $\delta x$ actually \textit{can} be made to vanish as long as $a = \widehat a$ at the location of the line. This condition is also physically sensible, since $ x$ acts as a source of $c$ flux, meaning that traversing any cycle linking it causes $(a, \widehat a) \rightarrow (\widehat a, a)$.  Subject to this condition, the line $\chi(\gamma)$ becomes topological and can be used to define the non-invertible lines of quantum dimension $\sqrt{N}$. We may also stack with $K(\gamma)$ or $L_{(e,0)}$ to obtain 
\bea
\widehat \Sigma_{(e)}^q(\gamma) := \sqrt{|H^0(\gamma, \ZZ_N)|}\, \chi(\gamma) L_{(e,0)}(\gamma) K(\gamma)^{\log q/ i\pi} \,\delta(a - \widehat a)|_\gamma~. \eea
Note that stacking with $L_{(e,0)}$ still gives something gauge invariant since $a$ and $\widehat a$ are identified on the locus $\gamma$. The factor of $\sqrt{H^0(\gamma, \ZZ_N)|}$ included here is such that we reproduce the correctly normalized fusion rules obtained in Section \ref{sec:ZTYZNlines}.

To reproduce the results for parallel fusion identified before, we may first note that the delta function above can be rewritten as\footnote{We assume $\gamma$ to be connected so that $H^0(\gamma,\ZZ_N)=\ZZ_N$.} 
\bea
\delta(a - \widehat a)\big|_{\gamma} &=& {1\over |H^0(\gamma,\ZZ_N)|} \sum_{\lambda \in  H^0(\gamma,\ZZ_N)} e^{ {2 \pi i \over N} \int_{\gamma} \lambda(a - \widehat a)} = {1\over |H^0(\gamma,\ZZ_N)|} \sum_{n \in \ZZ_N} L_{(n,-n)}(\gamma)~.\no\\
\eea
For e.g. $N$ odd, we then note that 
\bea
\sum_{n \in \ZZ_N} L_{(n,-n)}(\gamma) = \sum_{n = - {N-1 \over 2}}^{{N-1 \over 2}} L_{(n,-n)}(\gamma) =  \widehat L_{(0)}^+(\gamma) + \sum_{n=1}^{{N-1 \over 2}} \widehat L_{[n,N-n]}(\gamma)~,
\eea
 which allows us to reproduce the results for parallel fusion given in (\ref{eq:SSfusionNodd}). The other results for parallel fusion follow straightforwardly.

\subsubsection{Trivalent junctions}
\label{sec:3djunc}
In addition to studying parallel fusion, we may also study gauge-invariant trivalent junctions. Once one has all of the trivalent junctions, the parallel fusion may be reproduced. This analysis is conceptually straightforward but somewhat technical, and can be skipped on a first read.

To begin, consider $\widehat L_{(e)}^\pm$ on a chain $\gamma$ with boundary. Such a configuration is not gauge invariant under (\ref{3dGT}), but rather transforms as 
\bea
\widehat L_{(e)}^\pm (\gamma)\rightarrow \widehat L_{(e)}^\pm(\gamma)\, e^{i {2 \pi \over N} (e,e) \mathbf{g} |_{\partial \gamma}}~.
\eea
However, this configuration can be made consistent if 
\begin{itemize}
    \item It ends on a point with other lines $\widehat L_{(e')}^\pm (\gamma)$ such that the total charge cancels.
    \item It ends on a locus with non-zero $c$ flux.
    \item A mix of the above. 
\end{itemize}
The first of these allows for gauge-invariant junctions between three invertible defects $\widehat L_{(e)}^\pm (\gamma)$, $\widehat L_{(e')}^\pm (\gamma)$, and $\widehat L_{(e+e')}^\pm (\gamma)$; we will write this in the notation 
\bea
\label{eq:juncLL}
\widehat L_{(e+e')}^{qq'} \subset \widehat L_{(e)}^q \otimes \widehat L_{(e')}^{q'} ~,
\eea
where we take the convention that $\widehat L_{(e)}^q$ and $\widehat L_{(e')}^{q'}$ are incoming and  $\widehat L_{(e+e')}^{qq'}$ is outgoing. The factors of $q, q'$ shown follow from the fact that, under $c$ gauge transformations the lines $K(\gamma)$ transform as 
\bea 
\label{eq:Kendpointtransf}
K(\gamma)\rightarrow K(\gamma)\, e^{i \pi \omega|_{\partial \gamma}}~
\eea
and hence cannot end unless the junction point itself transforms under gauge transformations of $c$. When the junction point is invariant, as is the case here, there must be an even number of $K$-lines.

Note that this is the \textit{only} gauge-invariant junction between $\widehat L_{(e)}^q$ and $\widehat L_{(e')}^{q'}$. Indeed, no $\widehat L_{[e,m]}$  or $\widehat \Sigma^q_{(e)}$ can appear on the right-hand side: for $\widehat L_{[e,m]}$ this follows from charge conservation, while for $\widehat \Sigma^q_{(e)}$ this is because there does not exist a gauge-invariant endpoint for $\chi(\gamma)$, and hence any trivalent junction involving $\widehat \Sigma^q_{(e)}$  must contain both an incoming and outgoing $\widehat \Sigma$. Because (\ref{eq:juncLL}) is the only allowed gauge-invariant junction, we may conclude that the local fusion rules are 
\bea
\widehat L_{(e)}^q \otimes \widehat L_{(e')}^{q'}  = \widehat L_{(e+e')}^{qq'}~.
\eea
The overall normalization is fixed by the fact that $\widehat L_{(e)}^q$ is invertible, and hence has quantum dimension 1. This reproduces the result in (\ref{eq:3daftergaugingLfusion}). 

The second option above relies on the fact that, on a locus with non-zero $c$ flux, we have $(e,-e) \mathbf{g}=0$ by the constraint (\ref{eq:constraintong}). This gives rise to a gauge-invariant configuration of $\widehat L_{(e)}^\pm$ ending on $\widehat \Sigma_{(e')}^\pm$ as long as $ e = - e \,\,\,\mathrm{mod}\,\, N$. This happens whenever $N$ is even and $e = {N \over 2}$, giving rise to the junction  
\bea
\label{eq:juncLSig1}
\widehat\Sigma_{(e')}^{q''}\subset \widehat L_{(N/2)}^q \otimes \widehat\Sigma_{(e')}^{q'}~,
\eea
where $q''=\pm qq'$ depends on $e'$ being even/odd. This junction contains both an incoming and outgoing factor of $\chi(\gamma)$, consistent with the fact that $\chi(\gamma)$ cannot end. 

For more general $e \neq N/2$, the lines $\widehat L_{(e)}^\pm$ cannot end on $\widehat \Sigma_{(e')}^\pm$ alone, but by the third option there can be a trivalent junction between $\widehat L_{(e)}^\pm$, $\widehat \Sigma_{(e')}^\pm$, and $\widehat \Sigma_{(2e+e')}^\pm$,
\bea
\label{eq:juncLSig2}
\widehat \Sigma_{(2e+e')}^{q''}\subset \widehat L_{(e)}^q \otimes \widehat\Sigma_{(e')}^{q'}  ~.
\eea
 The sign $q''$ must be fixed by checking if the junction point transforms under $c$ gauge transformations. If not, then by (\ref{eq:Kendpointtransf}) the $K$ line cannot end and thus $q''=q q'$. On the other hand, if the junction point does transform non-trivially, then the $K$ line must end, and we have $q''=-q q'$. 
Unfortunately, it is not clear to us how to  understand this aspect of the junction point in the cocycle description, so here we simply quote the result from the analysis in Section \ref{sec:ZTYZN},
\begin{equation}
\begin{split}
    q''= 
    \begin{cases}
    qq'~, &  \{e'\in 2\Z\} \cup \{ e'\in 2\Z+1,\,\, 0\leq [e+\frac{e'-1}{2}]_N\leq N/2-1\}\\
    -qq'~, &  \{ e'\in 2\Z+1,\,\, N/2\leq [e+\frac{e'-1}{2}]_N\leq N-1\}
    \end{cases}
\end{split}
\end{equation}
These turn out to be the only junctions between $\widehat L_{(e)}^q$ and $\widehat\Sigma_{(e')}^{q'}$, and hence we may now reproduce the fusion rules in (\ref{eq:LSfusion}). 
As we will see momentarily though, these are \textit{not} the only gauge-invariant junction involving two $\widehat\Sigma_{(e)}^{q}$, so we cannot yet read of the fusion rules for $\widehat\Sigma_{(e)}^{q} \otimes \widehat\Sigma_{(e')}^{q'}$. For future purposes though, let us note that (\ref{eq:juncLSig2}) can be rewritten as 
\bea
\label{eq:juncSigSigL1}
\widehat L_{((e+e')/2)}^{q''}\subset \widehat\Sigma_{(e)}^{q} \otimes \widehat \Sigma_{(e')}^{q'}~, 
\eea
for $e+e'$ even, being careful about the conventions for orientation of lines at the junction.

We next consider junctions involving $\widehat L_{[e,m]}$. To do so, let us first notice that the given three curves $\gamma_1, \gamma_2, \gamma_3$ with the same boundary point $\partial \gamma_i$, we have the following shift under gauge transformations, 
\bea
&\vphantom{.}& L_{(e,m)}(\gamma_1) L_{(e',m')}(\gamma_2) \overline L_{(e'',m'')}(\gamma_3) 
\no\\
&\vphantom{.}&\hspace{0.8 in}\longrightarrow L_{(e,m)}(\gamma_1) L_{(e',m')}(\gamma_2) \overline L_{(e'',m'')}(\gamma_3)  e^{{2 \pi i \over N} (e + e' - e'', m+ m'-m'') \mathbf{g}|_{\partial \gamma_i}}~.
\eea 
If the boundary is not located on a locus of non-trivial $c$ flux, then we conclude that gauge invariant junctions must have 
\bea
e+e' - e'' = 0 \,\,\, \mathrm{mod}\,\,N~, \hspace{0.5 in} m+m' - m'' = 0 \,\,\, \mathrm{mod}\,\,N~,
\eea
which leads to junctions of the form 
\bea
\widehat L_{[e+e', m+m']} \subset \widehat L_{(e)}^q \otimes \widehat L_{[e', m']}  ~. 
\eea
These are the only junctions involving $\widehat L_{(e)}^q$ and $\widehat L_{[e', m']}$, and hence we determine the fusion rules 
\bea
\widehat L_{(e)}^q \otimes \widehat L_{[e', m']} =  \widehat L_{[e+e', m+m']}~,
\eea
where the coefficient is again fixed by knowing that the quantum dimension of $\widehat L_{(e)}^q$ is 1. One may similarly determine the fusion rules between $\widehat L_{[e, m]}$ and $\widehat L_{[e', m']}$.

On the other hand, assume that the boundary \textit{is} located on a locus of non-trivial $c$ flux, i.e. on an (ingoing) $\widehat \Sigma^q_{(e')}$. Since there is no gauge-invariant endpoint for $\chi(\gamma)$, the third (outgoing) leg in the trivalent junction must also be a $\widehat \Sigma^{q''}_{(e'')}$. In this case $m' = m'' = 0$ and $(1,-1) \mathbf{g} = 0$, so the condition for gauge-invariance is that 
\bea 
\label{eq:beforeorientchange}
e + m + e' - e'' = 0\,\,\, \mathrm{mod} \,\, N~.
\eea
We thus obtain junctions of the form 
\bea
\label{eq:juncLSigSig1}
\widehat \Sigma^{q''}_{(e+m+e')}\subset \widehat L_{[e,m]} \otimes \widehat \Sigma^{q'}_{(e')} ~.
\eea
Note that in this case the sign $q'$ is uncorrelated with $q''$, since as discussed in Section \ref{sec:cocyclelines} the line $\widehat L_{[e,m]}$ can absorb a $K$ line. Hence for every pair of $\widehat L_{[e,m]}$ and $ \widehat \Sigma^{q'}_{(e')}$, there are two possible gauge invariant junctions, with the outgoing leg being either $\widehat \Sigma^{+}_{(e+m+e')}$ or $\widehat \Sigma^{-}_{(e+m+e')}$. The local fusion rules involve a sum over these possibilities, with the result given by 
\bea
\widehat L_{[e,m]} \otimes \widehat \Sigma^{q'}_{(e')} = \widehat \Sigma^{+}_{(e+m+e')}\oplus \widehat \Sigma^{-}_{(e+m+e')}~.
\eea

We have now finally obtained the full set of gauge-invariant junctions, and can use this to read off the fusion rules of $\widehat \Sigma^q_{(e)}$ and $\widehat \Sigma^{q'}_{(e)}$. 
First, being careful about orientation of lines in the junction, we may rewrite  (\ref{eq:juncLSigSig1}) as 
\bea
\label{eq:juncSigSigL2}
\widehat L_{[ e''_n,  m''_n]}\subset \Sigma^q_{(e)}  \otimes \Sigma^{q'}_{(e')} 
\eea
where $( e''_n,  m''_n)$ are the solutions, parameterized by $n \in \ZZ_N$, to the equation 
\bea 
e''_n+ m''_n = e + e' \,\,\,\mathrm{mod}\,\,N~.
\eea
This equation descends from (\ref{eq:beforeorientchange}) by appropriate relabeling and changing of orientation. The solutions are given by
\bea
( e''_n,  m''_n) =\left\{ \begin{matrix}  ({e+e'\over 2} + n, {e+e'\over 2} - n) & & e+e' \,\,\,\mathrm{even} \\
 ({e+e'+N \over 2} + n, {e+e'+N \over 2} - n) & & e+e' \,\,\mathrm{odd},\,\,N \,\,\mathrm{odd} \\
({e+e'-1 \over 2} + n, {e+e'+1 \over 2} - n) & & e+e' \,\,\mathrm{odd},\,\,N \,\,\mathrm{even} \\
\end{matrix} \right.
\eea
We may restrict to $n = 1, \dots, \lfloor{N-1 \over 2}\rfloor$ to obtain the full set of non-redundant pairs. 

The junctions (\ref{eq:juncSigSigL1}) and (\ref{eq:juncSigSigL2}) give the full set of gauge-invariant junctions involving $\widehat \Sigma^q_{(e)}$ and $\widehat \Sigma^{q'}_{(e)}$. We may thus obtain the fusion rules by summing over these possibilities, successfully reproducing the results in (\ref{eq:SSfusionNodd}) and (\ref{eq:SSfusionNeven}).

\subsubsection{$S$-matrix elements}

\begin{figure}[!tbp]
	\centering
\[
T_{XX}= \frac1{d_X} \;\; \raisebox{-1.5em}{\begin{tikzpicture}
\draw [thick] (0,0) ++(45:.5) arc (45:315:.5);
\draw [thick] (0,0) ++(-45:.5) to +(45:1);
\draw [thick, decoration = {markings, mark=at position .46 with {\arrowreversed[scale=1.5,rotate=10]{stealth}}}, postaction=decorate] (0,0) ++(-45:1) ++(45:.5) arc (-135:135:.5);
\node at (2.20,0) {\small $X$};
\draw [thick] (0,0) ++(45:.5) to +(-45:.3); \draw [thick] (0,0) ++(45:.5) ++ (-45:.7) to +(-45:.3);
\end{tikzpicture}}\hspace{0.8 in}
S_{XY}\,\,=\,\, {1\over{\cal D}} \raisebox{-1.1em}{\begin{tikzpicture}
\draw [thick, decoration = {markings, mark=at position .33 with {\arrowreversed[scale=1.5,rotate=10]{stealth}}}, postaction=decorate] (0,0) ++(60:.5) arc (60:390:.5);
\node at (-.8,0) {\small $X$};
\draw [thick, decoration = {markings, mark=at position .88 with {\arrowreversed[scale=1.5,rotate=10]{stealth}}}, postaction=decorate] (.7,0) ++(-120:.5) arc (-120:210:.5);
\node at (-.05,0) {\small $Y$};
\end{tikzpicture}}
\]
	\caption{Pictorial representation of entries of the $S$ and $T$ matrices. Note that $T_{XX}$ is also denoted $\theta(X)$ and referred to as the spin.}
	\label{fig:SandT}
\end{figure}

We have now reproduced the fusion rules of Section \ref{sec:ZTYZN} via both parallel fusion and analysis of gauge-invariant junctions. Before moving on to higher dimensions, let us briefly mention a third strategy towards obtaining the fusion rules, which is to first compute the S-matrix elements and then utilize the Verlinde formula. This is relatively straightforward in $(2+1)$d since the $S$-matrix element $S_{XY}$ between two line operators $X$ and $Y$ is given by the amplitude of the Hopf link shown in Figure \ref{fig:SandT}, up to an overall factor of the total quantum dimension
\bea
\cD := \sqrt{\sum_X d_X^2}= 2N~.
\eea
The Hopf link amplitude can in turn be obtained directly from the path integral. Unfortunately, for the $(4+1)$d case this third strategy will no longer be viable, since there is not yet an established analog of the S-matrix (though see \cite{Johnson-Freyd:2021chu,reuttertalk} for work in this direction) or of the Verlinde formula. Because of this, we will not give an exhaustive treatment of this strategy, but only a flavor of it.

 Begin by considering the $S$-matrix element between $\widehat L_{(e)}^q$ and $\widehat L_{(e')}^{q'}$. Take $\widehat L_{(e)}^q$ to be inserted on $\gamma$ and $\widehat L_{(e')}^{q'}$ to be inserted on $\gamma'$, such that $\mathrm{link}(\gamma, \gamma') = 1$ as in Figure \ref{fig:SandT}. The action in the presence of these insertions becomes 
\bea
S &=&{2 \pi \over 2N} \int_{M_3} \mathbf{a}^T \cup_c K \delta_c \mathbf{a} +  \pi \int x \cup \delta c+ {2 \pi  \over N} (e, e) \oint_\gamma \mathbf{a}-i\log q \, \oint_\gamma c 
\no\\
&\vphantom{.}&\hspace{0.5 in} + {2 \pi  \over N} (e', e') \oint_{\gamma'} \mathbf{a}-i\log q' \, \oint_{\gamma'} c
\no\\
&=&{2 \pi  \over 2N} \int_{M_3} \mathbf{a}^T \cup_c K \delta_c \mathbf{a} +  \pi \int x \cup \delta c+{2 \pi  \over N} \int_{M_3} \left[ (e,e) \omega_\gamma +  (e',e') \omega_{\gamma'} \right]\cup \mathbf{a}
\no\\
&\vphantom{.}&\hspace{0.5 in}
-i\log q \, \oint_{\gamma} c-i\log q' \, \oint_{\gamma'} c
\eea
where $\omega_\gamma$ is the Poincar{\'e} dual to the one-cycle $\gamma$ and we take $\log q = 0 , i \pi$ for $q= 1, -1$.

We would now like to carry out manipulations similar to those used in Appendix \ref{app:linecorr} to compute the braidings. Because of the complicated appearance of $c$ in the kinetic term for $\mathbf{a}$, this would at first sight seem impossible, but under certain circumstances it turns out to not be a problem. Indeed,  the  $S$-matrix element $S_{XY}$ involves only the linking of the lines $X$ and $Y$, and hence as long as we are considering lines $X$ and $Y$ which do not act as sources for $c$, we are allowed to consider the link in a small patch of $M_3$ in which $c$ is turned off. This means that the computation proceeds exactly as in Appendix \ref{app:linecorr}; namely, we can focus on a patch $U_3 \subset M_3$ in which we have  
\bea
S ={2 \pi \over N} \int_{U_3} \widehat a \cup \delta a + {2 \pi  \over N}  \int_{U_3} ( e \omega_\gamma+ e' \omega_{\gamma'}) \cup a + {2 \pi  \over N}  \int_{U_3}  ( e \omega_\gamma+ e' \omega_{\gamma'}) \cup \widehat a~\no,
\eea
and integrating out $\widehat a$ then gives 
\bea
\delta a = - (e \omega_\gamma + e' \omega_{\gamma'})~.
\eea
Defining $M_2$ such that $\partial M_2 = e \gamma + e' \gamma'$ and plugging back into the action then gives a factor of
\bea
- {4 \pi \over N} e e' \, \mathrm{link}(\gamma, \gamma') = - {4 \pi \over N} e e'~.
\eea
We thus conclude that 
\bea
S_{\widehat{L}_{(e)}^q , \widehat{L}_{(e')}^{q'}} = {1\over 2N} e^{- {4 \pi i \over N} e e'}~.
\eea

We next consider the $S$-matrix element between $\widehat L_{(e)}^\pm$ and $\widehat L_{[e',m']}$. Since neither of these is a source of $c$ flux, we may again focus on a region in which $c$ is trivial. Using $\widehat L_{[e',m']} = L_{(e',m')} \oplus  L_{(m',e')}$, we see that the relevant $S$-matrix element is given by the sum of amplitudes for the Hopf link between $\widehat L_{(e)}^\pm$ and $L_{(e',m')}$ or $L_{(m',e')}$. Using the general formula in (\ref{eq:Hopflinkgen}), we conclude that
\bea
S_{\widehat L_{(e)}^\pm, \widehat L_{[e',m']}} = {1 \over 2 N} e^{-{2 \pi i\over N} e(e'+m')} +  {1 \over 2 N} e^{-{2 \pi i\over N} e(e'+m')} = {1\over N }e^{-{2 \pi i\over N} e(e'+m')}~.
\eea
Finally, the S-matrix element between $\widehat L_{[e,m]}$ and $\widehat L_{[e',m']}$ can be easily computed, giving rise to
\bea
S_{\widehat L_{[e,m]},\widehat L_{[e',m']}}&=&\frac{1}{N} \left( e^{-\frac{2\pi i}{N}(em'+me')}+ e^{-\frac{2\pi i}{N}(ee' + mm')}\right)~.
\eea
All of the results so far match precisely with those identified in the math literature for $\cZ(TY(\ZZ_N))$; see Appendix \ref{app:fusionDCTY} and in particular (\ref{eq:mathSmatrices}) for a summary of those results. Plugging them into the Verlinde formula (\ref{eq:Verlindeformula}), they also match with the fusion rules obtained before. 

We now turn towards the more difficult case involving $\widehat \Sigma_{(e)}^q$. These defects involve $\chi = e^{i \pi \oint x}$, and hence act as a source for $c$ flux. This means that our previous strategy of evaluating the link in a patch with zero $c$ is invalid. We will not try to give a complete treatment here, but instead just note that a surprising amount of information about the $S$-matrices can be obtained even without a deep understanding of this complication. 

Begin by considering the $S$-matrix element between $\widehat L_{(e)}^q$ and the basic twist line $\widehat \Sigma_{(0)}^{q'} \sim \chi \times K^{\log q'/i \pi} \delta(a-\widehat a)|_\gamma$. The action in the presence of these insertions is given by 
\bea
S &=&{2 \pi \over 2N} \int_{M_3} \mathbf{a}^T \cup_c K \delta_c \mathbf{a} +  \pi \int_{M_3} x \cup \delta c+  \pi \oint_{\gamma'} x-i\log q' \oint_{\gamma'} c
\no\\
&\vphantom{,}&\hspace{0.5 in}+ {2 \pi  \over N} (e, e) \oint_\gamma \mathbf{a} -i \log q \oint_\gamma c +{2 \pi  \over N} \oint_\gamma \lambda (a - \widehat a)
\eea
where as before $\log q = 0, i \pi$ for $ q = 1,-1$ and $\lambda$ is a Lagrange multiplier field implementing the delta function. To evaluate this, we first path integrate over $x$, which restricts 
\bea
\delta c = - \omega_{\gamma'}~.
\eea
As before, we may define $M_2$ such that $\partial M_2 = \gamma'$, in which case $c = - \mathrm{PD}(M_2)$. Plugging this back into the action gives a term $\log q \, \link(\gamma, \gamma')$; note that there is no term proportional to $\log q'$ since the unknot does not have self-linking. The amplitude thus has an overall factor of $e^{\log q \, \link(\gamma, \gamma')}= q$. When we consider the more general twist lines $\widetilde \Sigma_{(e)}^{q'}$, we must also consider the linking between $\widehat L_{(e)}^q$ and $L_{(e,0)}$, which by (\ref{eq:Hopflinkgen}) gives a factor of $e^{- {2 \pi i \over N} e e'}$. To summarize then, thus far we find 
\bea
S_{\widehat L_{(e)}^q, \widetilde \Sigma_{(e')}^{q'}} \stackrel{?}{=} {q\over 2 N} e^{- {2 \pi i \over N} e e'} ~.
\eea
This is almost the correct result, c.f. (\ref{eq:mathSmatrices}), up to an overall factor of $\sqrt{N}$, which comes from the quantum dimension of $\widehat \Sigma_{(e)}^{q'}$. Obtaining this factor would require a more careful treatment of the path integral over $\mathbf{a}$ in the presence of $c$ flux. 

Next, we consider the $S$-matrix between $\widehat \Sigma_{(e)}^\pm$ and $\widehat L_{[e'm']}$. In fact, in this case the result is rather easy to understand. Because $\widehat L_{[e'm']} = L_{(e,m)}\oplus L_{(m,e)}$, computing this $S$-matrix element would involve summation over the Hopf link of $\widehat \Sigma_{(e)}^\pm$ with $L_{(e,m)}$, together with the Hopf link of $\widehat \Sigma_{(e)}^\pm$ with $L_{(m,e)}$. However, both of these configurations are not gauge invariant, and hence both correlators must individually vanish. Indeed, because $\widehat \Sigma_{(e)}^\pm(\gamma)$ acts as a source for $c$ flux on $\gamma$, then traversing a cycle $\gamma'$ with linking $\mathrm{link}(\gamma, \gamma') =1$ leads to exchange $(a, \widehat a) \rightarrow (\widehat a, a)$. Thus only the lines $L_{(e,m)}$ with $e = m$ can consistenly wrap $\gamma'$, and we conclude that $S_{\widetilde \Sigma_{(e)}^\pm, L_{[e'm']}}=0$, matching with the known results given in (\ref{eq:mathSmatrices}).

Finally, let us briefly mention the $S$-matrix between $\widehat \Sigma_{(e)}^q$ and $\widehat \Sigma_{(e')}^{q'}$. In this case the naive considerations above are not enough to obtain the full form of the matrix elements, but we can at least correctly predict that we should have $S_{\widehat \Sigma_{(e)}^q,\widehat \Sigma_{(e)}^{q'}} \propto q q'$. This follows since the $\chi = e^{i \pi \oint x}$ of each line can link with the  $K^{\log q / i \pi} = e^{\log q \oint c}$ of the other. It would be interesting to carry out this analysis more thoroughly.

\section{Duality interfaces in $(3+1)$d}
\label{sec:3+1ddualityinterface}

Having introduced the relevant tools for studying duality defects in $(1+1)$d, we may now proceed to the analogous discussion in $(3+1)$d. As before, we will first review the construction of the duality interfaces/defects themselves, and then move on to the construction of the corresponding SymTFT. 
Note that the fusion rules of duality defects in $(3+1)$d have already been discussed in \cite{Choi:2021kmx, Choi:2022zal, Kaidi:2021xfk}, though the discussion here will be more explicit about the precise normalization and counterterms involved.

\subsection{Duality interfaces from half-space gauging}
\label{sec:3+1ddualityinterfacedef}

We begin by considering a non-spin QFT $\CX$ in $(3+1)$d with an anomaly free $\Z_N^{(1)}$ one-form global symmetry, defined on a closed four-dimensional spacetime $X_4$. We denote the $\Z_N^{(1)}$ background gauge field as $B$, and the partition function as $Z_{\CX}[X_4, B]$. Gauging $\Z_N^{(1)}$ gives a new theory $\CX/\Z_N$, 
\begin{eqnarray}\label{eq:4dgauging}
Z_{\CX/\Z_N}[X_4, B] = \frac{|H^0(X_4, \Z_N)|}{|H^1(X_4, \Z_N)|} \sum_{b\in H^2(X_4, \Z_N)} Z_{\CX}[X_4, b] e^{\frac{2\pi i}{N}\int_{X_4} bB}~,
\end{eqnarray}
where $B\in H^2(X_4, \Z_N)$ is the background field of the quantum symmetry $\Z_N^{(1)}$ after gauging. The defect that generates this quantum symmetry is the Wilson surface of $b$, 
\begin{eqnarray}
\eta(\sigma) = \exp\left(\frac{2\pi i}{N} \oint_{\sigma} b\right).
\end{eqnarray}

The denominator of the normalization in \eqref{eq:4dgauging} comes from the volume of the gauge redundancy, while the numerator comes from the volume of the gauge redundancies for the previous gauge redundancies. It is straightforward to check that gauging $\Z_N^{(1)}$ twice does not quite map the theory $\CX$ back to itself, but rather to the charge conjugate of $\CX$, up to an Euler counterterm $\chi[X_4, \Z_N]$, where 
\begin{eqnarray}\label{eq:Euler4d}
\chi[X_4, \Z_N] = \frac{|H^0(X_4, \Z_N)| |H^2(X_4, \Z_N)| |H^4(X_4, \Z_N)|}{|H^1(X_4, \Z_N)||H^3(X_4, \Z_N)|}~.
\end{eqnarray}
This means that gauging $\Z_N^{(1)}$ is, up to the Euler counterterm, an order four operation. 
As in $(1+1)$d, we are allowed to redefine our gauging such that it is exactly order four, which involves multiplying  \eqref{eq:4dgauging} by $\chi[X_4, \Z_N]^{-\frac{1}{2}}$. In this case the normalization becomes $1/\sqrt{|H^2(X_4, \Z_N)|}$, and gauging twice maps the theory $\CX$ exactly to the charge conjugate of  $\CX$. We will however continue to work with the normalization \eqref{eq:4dgauging}.

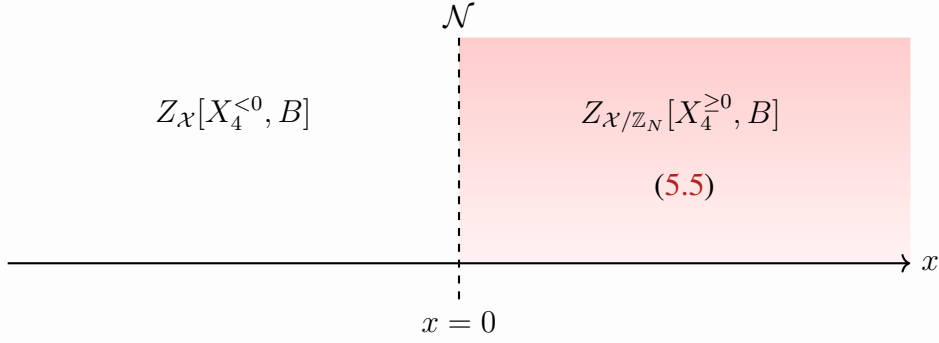
\begin{figure}
    \centering
    \begin{tikzpicture}
    
      \shade[top color=red!20, bottom color=red!5]  (6,0) 
to [out=0, in=180] (12,0) 
to [out=90, in=270] (12,3) 
to [out=180,in=0] (6,3) 
to [out=270,in=90] (6,0);

    \draw[thick, ->] (0,0) -- (12,0);
    \draw[thick, dashed] (6,3) -- (6,-0.5); 
    \node[above] at (6,3) {$\CN$};
    \node[] at (3,2) {$Z_{\CX}[X_4^{<0},B]$};
    \node[] at (9,1.5) {\begin{tabular}{c}
          $Z_{\CX/\Z_N}[X_4^{\geq 0}, B]$ \bigskip\\ 
         \eqref{eq:halfgauging4d}
    \end{tabular}};
    \node[below] at (6,-0.5) {$x=0$};
    \node[right] at (12,0) {$x$};
    \end{tikzpicture}
    \caption{The duality defect from gauging $\Z_N^{(1)}$ over half of the spacetime $X_4^{\geq 0}$ with Dirichlet boundary condition. }
    \label{fig:4ddualitydefect}
\end{figure}

Instead of gauging $\Z_N^{(1)}$ over the entire $X_4$, one can gauge in half of the spacetime with Dirichlet boundary conditions. This defines a topological duality interface $\CN$ between $\CX$ and $\CX/\Z_N$. We decompose the spacetime into two parts, 
\begin{eqnarray}
X_4= X_4^{<0} \cup X_4^{\geq 0}
\end{eqnarray}
where $\partial X_4^{\geq 0} = X_3$ is the interface. 
Depending on the situation, we sometimes also include $X_3$ in $X_4^{\leq 0}$. Around the interface, the geometry is $X_3\times \mathbb{R}$, and we use the coordinate $x$ to parameterize $\mathbb{R}$. The interface sits at $x=0$, and we denote this locus by $M_3|_0$. The duality interface is defined by specifying the theories on its two sides. 
The theory $\CX$ lives on $X_4^{<0}$, while the theory $\CX/\Z_N$ lives on $X_4^{\geq 0}$.
A definition of the duality defect is given in Figure \ref{fig:4ddualitydefect}, where the theory on the right side $X_4^{\geq 0}$ is
\begin{eqnarray}\label{eq:halfgauging4d}
Z_{\CX/\Z_N}[X_4^{\geq 0}, B]=\frac{|H^0(X_4^{\geq 0},M_3|_0, \Z_N)|}{|H^1(X_4^{\geq 0},M_3|_0, \Z_N)|} \sum_{b\in H^2(X_4^{\geq 0}, M_3|_0, \Z_N)} Z_{\CX}[X_4^{\geq 0}, b] e^{ \frac{2\pi i}{N} \int_{X_4^{\geq 0} }b B }.
\end{eqnarray}
The Dirichlet boundary condition implies that the dynamical gauge field $b$ is an element in relative cohomology $H^2(X_4^{\geq 0}, M_3|_0, \Z_N)$.

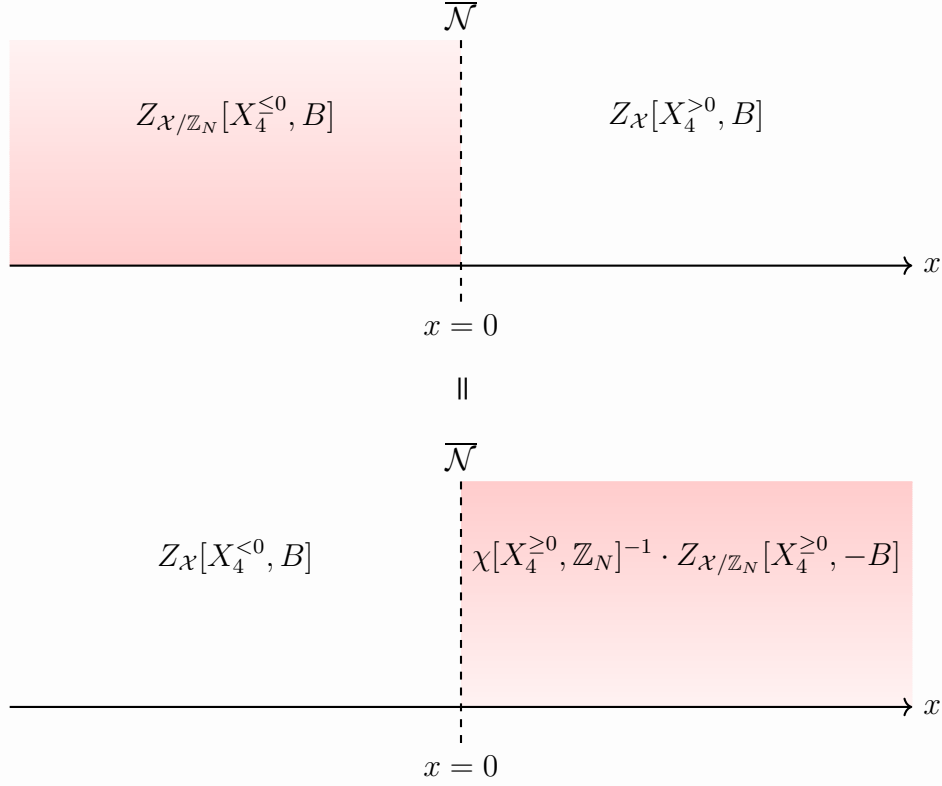
\begin{figure}
    \centering
    \begin{tikzpicture}
    
      \shade[top color=red!5, bottom color=red!20]  (0,0) 
to [out=0, in=180] (6,0) 
to [out=90, in=270] (6,3) 
to [out=180,in=0] (0,3) 
to [out=270,in=90] (0,0);

    \draw[thick, ->] (0,0) -- (12,0);
    \draw[thick, dashed] (6,3) -- (6,-0.5); 
    \node[above] at (6,3) {$\overline\CN$};
    \node[] at (3,2) {$Z_{\CX/\Z_N}[X_4^{\leq 0},B]$};
    \node[] at (9,2) {\begin{tabular}{c}
          $Z_{\CX}[X_4^{> 0}, B]$ 
    \end{tabular}};
    \node[below] at (6,-0.5) {$x=0$};
    \node[right] at (12,0) {$x$};
    \end{tikzpicture}
    \\\vspace{0.1 in}
    \hspace{-0.18in}\begin{tikzpicture}
    \node[] at (0,0) {||};
    \end{tikzpicture}
    \\\vspace{0.1 in}
    \begin{tikzpicture}
      \shade[top color=red!20, bottom color=red!5]  (6,0) 
to [out=0, in=180] (12,0) 
to [out=90, in=270] (12,3) 
to [out=180,in=0] (6,3) 
to [out=270,in=90] (6,0);

    \draw[thick, ->] (0,0) -- (12,0);
    \draw[thick, dashed] (6,3) -- (6,-0.5); 
    \node[above] at (6,3) {$\overline\CN$};
    \node[] at (3,2) {$Z_{\CX}[X_4^{< 0},B]$};
    \node[] at (9,2) {\begin{tabular}{c}
          $\chi[X_4^{\geq 0}, \Z_N]^{-1}\cdot {Z}_{\CX/\Z_N}[X_4^{\geq  0}, -B]$ 
    \end{tabular}};
    \node[below] at (6,-0.5) {$x=0$};
    \node[right] at (12,0) {$x$};
    \end{tikzpicture}
    \caption{Orientation reversal of the duality interface.}
    \label{fig:4ddualitydefectbar}
\end{figure}

The orientation reversal of the duality defect is defined by exchanging the theories on the two sides of Figure \ref{fig:4ddualitydefect}. This is illustrated in the upper panel of Figure \ref{fig:4ddualitydefectbar}. By reorganizing $\CX$ to the left of the defect, we obtain an equivalent expression of $\overline{\CN}$ as shown in the lower panel of Figure \ref{fig:4ddualitydefectbar}. 
Therefore we have
\begin{eqnarray}\label{eq:CNbarCN4d}
\overline{\CN} = \chi[X_4^{\geq 0}, \Z_N]^{-1} C\cdot \CN~,
\end{eqnarray}
where $C$ is the charge conjugation operator mapping $B\to -B$.

\subsection{Fusion rule of duality interfaces}
\label{sec:fusionruledualityinterfaces4d}

We proceed to discuss the fusion rules of the duality interface $\CN$ defined in Section \ref{sec:3+1ddualityinterfacedef}. In particular, we will find that $\CN$ is non-invertible. The discussion here is somewhat technical, and the reader interested in only the answer may skip to (\ref{eq:FR4d}).

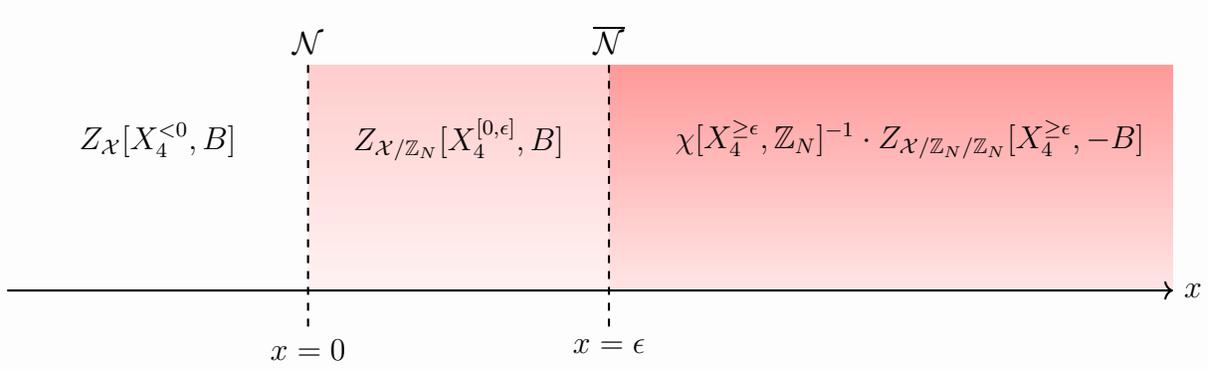
\begin{figure}
    \centering
    \begin{tikzpicture}
      \shade[top color=red!20, bottom color=red!5]  (4,0) 
to [out=0, in=180] (8,0) 
to [out=90, in=270] (8,3) 
to [out=180,in=0] (4,3) 
to [out=270,in=90] (4,0);

  \shade[top color=red!40, bottom color=red!10]  (8,0) 
to [out=0, in=180] (15.5,0) 
to [out=90, in=270] (15.5,3) 
to [out=180,in=0] (8,3) 
to [out=270,in=90] (8,0);

    \draw[thick, ->] (0,0) -- (15.5,0);
    \draw[thick, dashed] (4,3) -- (4,-0.5); 
    \draw[thick, dashed] (8,3) -- (8,-0.5); 
    \node[above] at (4,3) {$\CN$};
    \node[above] at (8,3) {$\overline{\CN}$};
    \node[] at (2,2) {$Z_{\CX}[X_4^{<0},B]$};
    \node[] at (6,2) {\begin{tabular}{c}
          $Z_{\CX/\Z_N}[X_4^{[0,\epsilon]}, B]$ 
    \end{tabular}};
    \node[] at (12,2) {\begin{tabular}{c}
          $\chi[X_4^{\geq \eps}, \Z_N]^{-1}\cdot Z_{\CX/\Z_N/\Z_N}[X_4^{\geq \epsilon}, -B]$ 
    \end{tabular}};
    \node[below] at (4,-0.5) {$x=0$};
    \node[below] at (8,-0.5) {$x=\epsilon$};
    \node[right] at (15.5,0) {$x$};
    \end{tikzpicture}
    \caption{Fusion of two duality interfaces. The partition function on $X^{\geq 0}$ is given by \eqref{eq:NN}.  }
    \label{fig:4ddualitydefectfusion}
\end{figure}

We first discuss the fusion rule $\eta \times \CN$. Since $b$ has Dirichlet boundary conditions on $M_3|_0$, i.e. $b|_{M_3|_0}=0$, the $\Z_N^{(1)}$ symmetry defect $\eta$ is trivial on $M_3|_0$. This justifies
\begin{eqnarray}\label{eq:etaNN}
\eta \times \CN =\CN~.
\end{eqnarray}
It is more interesting to study the fusion rules between two duality interfaces. We will first consider $\CN\times \overline{\CN}$, from which $\CN\times \CN$ can be derived using \eqref{eq:CNbarCN4d}. The derivation is similar to the computation in Section \ref{sec:2dfusionrule}.  We begin by placing $\CN$ and $\overline{\CN}$ at $x=0$ and $x= \epsilon$ respectively, and let $\epsilon \to 0^+$.   The two duality defects divide the spacetime $X_4$ into three regions, $X_4=X_4^{<0} \cup X_4^{[0,\epsilon)} \cup X_4^{\geq \epsilon}$. Since $\epsilon$ is small, we always take $X_4^{[0,\epsilon)}= M_3\times I_{[0,\epsilon)}$. The theories living on the three regions are as shown in Figure \ref{fig:4ddualitydefectfusion}. Instead of defining the theories in the two regions to the right of $\CN$ separately and  
discussing how to glue them together along $M_3|_{\epsilon}$, we will instead discuss the theory on $X_4^{\geq 0}$ all together. The theory living on $X_4^{\geq 0}$ is given by 
\begin{equation}\label{eq:NN4d}
\begin{split}
    &\chi[X_4^{\geq \eps}, \Z_N]^{-1}{|H^0(X_4^{\geq 0}, M_3|_0, \Z_N)| |H^0(X_4^{\geq \eps}, M_3|_\eps, \Z_N)| \over |H^1(X_4^{\geq 0}, M_3|_0, \Z_N)| |H^1(X_4^{\geq \eps}, M_3|_\eps, \Z_N)|} \times  \\& \hspace{3cm}\sum_{\substack{b \in H^2(X_4^{\geq 0}, M_3|_0, \Z_N) \\ \widetilde b \in H^2(X_4^{\geq \eps}, M_3|_\eps, \Z_N)}} Z_{\CX}[X_4^{\geq 0}, b]\, e^{{2 \pi i \over N}\int_{X_4^{[0,\eps)}} b B + {2 \pi i \over N}\int_{X_4^{\geq \eps}} (b-B)\widetilde b}~.
\end{split}
\end{equation}
The techniques used to evaluate \eqref{eq:NN4d} are similar to those used to evaluate \eqref{eq:NN}, and hence we will only mention the main steps, highlighting new features. 

To evaluate \eqref{eq:NN4d}, the first step is to convert the sum over relative cohomologies to absolute cochains by introducing additional 1-form cochains and BF couplings. This gives
\begin{equation}
\begin{split}
    &\chi[X_4^{\geq \eps}, \Z_N]^{-1}  {|C^0(X_4^{\geq 0}, M_3|_0, \Z_N)| \over |C^1(X_4^{\geq 0}, M_3|_0, \Z_N)|}{|C^0(X_4^{\geq \eps}, M_3|_\eps, \Z_N)| \over |C^1(X_4^{\geq \eps}, M_3|_\eps, \Z_N)|}{1 \over |C^1(X_4^{\geq 0}, \Z_N)||C^1(X_4^{\geq \eps}, \Z_N)| }\times\\
    &\sum_{\substack{b\in C^2(X_4^{\geq 0}, \Z_N), \widetilde{b}\in C^2(X_4^{\geq \eps}, \Z_N)\\
    u\in C^1(X_4^{\geq 0}, \Z_N), \widetilde{u}\in C^1(X_4^{\geq \eps}, \Z_N)}} Z_{\CX}[X_4^{\geq 0},b] e^{{2 \pi i \over N}\int_{X_4^{[0,\eps)}} b B + {2 \pi i \over N}\int_{X_4^{\geq \eps}} (b-B)\widetilde b} e^{{2\pi i \over N}\int_{X_4^{>0}}u \delta b + {2\pi i \over N}\int_{M_3|_0} u b}e^{{2\pi i \over N}\int_{X_4^{>\eps}}\widetilde u \delta\widetilde b + {2\pi i \over N}\int_{M_3|_\eps} \widetilde u \widetilde b}.
\end{split}
\end{equation}
We then integrate out $\widetilde{b}$, $u$, and $\widetilde{u}$ subsequently. The final result is 
\begin{eqnarray}\label{eq:4dNNbarfusion1}
\begin{split}
    &\chi[X_4^{\geq \eps}, \Z_N]^{-1}  {|C^0(X_4^{\geq 0}, M_3|_0, \Z_N)||C^0(X_4^{\geq \eps}, M_3|_\eps, \Z_N)|  |C^2(X_4^{\geq \eps}, \Z_N)| \over |C^1(X_4^{\geq 0}, M_3|_0, \Z_N)||C^1(X_4^{\geq \eps}, M_3|_\eps, \Z_N)|}\\
    & \hspace{4cm}\sum_{\substack{b \in Z^2(X_4^{[0,\eps]}, M_3|_0 \cup M_3|_\eps, \Z_N)}} Z_{\CX}[X_4^{\geq 0},b+B|_{M_4^{\geq \eps}}] e^{{2 \pi i \over N}\int_{X_4^{[0,\eps]}} b B }
\end{split}
\end{eqnarray}
where $b=B$ on $X_4^{\geq \eps}$. Using the identity for cochains $|C^n(X_4, M_3, \Z_N)| = |C^{4-n}(X_4, \Z_N)|$ as well as the analogue of \eqref{eq:NN2d4},
\bea|C^n( X_4^{\geq 0}, M_3|_0, \Z_N)| = |C^n( X_4^{[0,\epsilon]}, M_3|_0 \cup  M_3|_\epsilon, \Z_N)| |C^n( X_4^{\geq \epsilon}, \Z_N)|~,
\eea
the  normalization factors in the first line of \eqref{eq:4dNNbarfusion1} can be simplified to
\begin{eqnarray}
\begin{split}
    &{|C^0(X_4^{[0,\eps]}, M_3|_0\cup M_3|_{\eps}, \Z_N)|  \over |C^1(X_4^{[0,\eps]}, M_3|_0\cup M_3|_{\eps}, \Z_N)|}  \sum_{\substack{b \in Z^2(X_4^{[0,\eps]}, M_3|_0 \cup M_3|_\eps, \Z_N)}} Z_{\CX}[X_4^{\geq 0},b+M_4^{\geq \eps}] e^{{2 \pi i \over N}\int_{X_4^{[0,\eps]}} b B }.
\end{split}
\end{eqnarray}
Converting the sum back to cohomologies, we find the fusion rule 
\begin{eqnarray}
\begin{split}
    \CN\times \overline{\CN} &= {|H^0(X_4^{[0,\eps]}, M_3|_0\cup M_3|_{\eps}, \Z_N)|  \over 
    |H^1(X_4^{[0,\eps]}, M_3|_0\cup M_3|_{\eps}, \Z_N)|}  
    \sum_{\substack{b \in H^2(X_4^{[0,\eps]}, M_3|_0 \cup M_3|_\eps, \Z_N)}} 
    \eta(\mathrm{LD}(b)) \\
    &= {1  \over |H^0(M_3, \Z_N)|}  
    \sum_{\substack{\sigma \in H_2(M_3, \Z_N)}} \eta(\sigma) ~,
\end{split}
\end{eqnarray}
where LD stands for Lefschetz dual. 
The right-hand side is  the condensation defect \cite{Choi:2022zal} associated with 1-gauging of the $\Z_N^{(1)}$ one-form symmetry on a codimension-one submanifold $M_3$.

In summary, the fusion rules are
\begin{equation}\label{eq:FR4d}
\begin{split}
    &\CN\times \overline{\CN}= \frac{1}{|H^0(M_3,\Z_N)|}\sum_{\sigma\in H_2(M_3, \Z_N)} \eta(\sigma)~,\\&\eta \times \CN =\CN~, \hspace{1cm} \eta^N=1~,\hspace{1cm} \overline{\CN} = \chi[X_4^{\geq 0}, \Z_N]^{-1} C\cdot \CN~.
\end{split}
\end{equation}
Note that the right-hand side of the first equation is the condensation defect of the algebra made of $\eta$ \cite{Carqueville:2017aoe,Roumpedakis:2022aik,Choi:2022zal}.
This is reminiscent of the fusion rules for the Tambara-Yamagami category $TY(\ZZ_N)$ given in (\ref{eq:FR2d}), which in particular contained an object whose fusion with its conjugate was an algebra made of invertible symmetry defects.
For this reason, we consider the fusion rules in \eqref{eq:FR4d} to be the higher-dimensional analogues of the fusion rules for the Tambara-Yamagami category. The reason that we have a condensation defect on the right-hand side of the first equation,  as opposed to the algebra object itself, is to match the dimensionality of the defects on both sides. To better understand this higher-category, we now turn to the SymTFT.

\section{$(4+1)$d Symmetry TFT for $\Z_N^{(1)}$ symmetry}
\label{sec:symTFT5d}

In Section \ref{sec:3+1ddualityinterface}, we defined the duality interface $\CN$ in $(3+1)$d and studied its (global) fusion rules. In this section we study properties of the duality interface from the SymTFT point of view. As in the lower-dimensional case, we will aim to classify the twist defects, exhibit their relation with the duality interface upon shrinking, and derive their fusion rules.

\subsection{$\Z_N^{(1)}$ gauge theory as the Symmetry TFT}
\label{sec:ZNgaugetheoryasSymTFT}

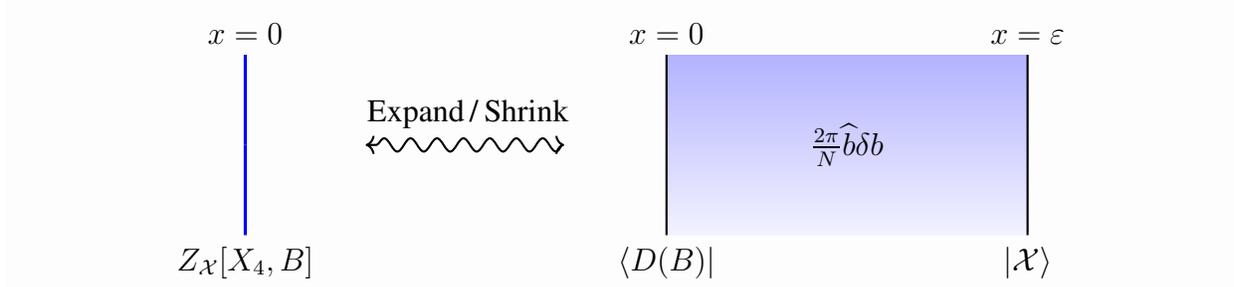
\begin{figure}[!tbp]
	\centering
	\begin{tikzpicture}[scale=0.8]
	
	\draw[very thick, blue] (-7,0) -- (-7,1.5);
	\draw[very thick, blue] (-7,1.5) -- (-7,3);
	\node[below] at (-7,0) {$Z_{\CX}[X_4, B]$};

    \node[above] at (-7,3) {$x=0$};
	
	\draw[thick, snake it, <->] (-1.7,1.5) -- (-5, 1.5);
	\node[above] at (-3.3,1.6) {Expand\,/\,Shrink};

		\shade[line width=2pt, top color=blue!30, bottom color=blue!5] 
	(0,0) to [out=90, in=-90]  (0,3)
	to [out=0,in=180] (6,3)
	to [out = -90, in =90] (6,0)
	to [out=180, in =0]  (0,0);

	\draw[thick] (0,0) -- (0,3);
	\draw[thick] (6,0) -- (6,3);

	\node[] at (3,1.5) {$\frac{2\pi}{N} \widehat{b} \delta b$};

	\node[below] at (0,0) {$\langle D(B)| $};
	\node[below] at (6,0) {$|\cX\rangle $}; 
	\node[above] at (0,3) {$x=0$};
	\node[above] at (6,3) {$x=\varepsilon$};

	\end{tikzpicture}
	
	\caption{A $(3+1)$d QFT $\cX$ with $\Z_N^{(1)}$ one-form symmetry can be expanded into a $(4+1)$d slab. The bulk is a $(4+1)$d $\Z_N^{(1)}$ gauge theory, with the right boundary encoding the dynamical information of the $(3+1)$d QFT and the left boundary being a topological Dirichlet boundary condition for the bulk field $b$.}
	\label{fig:4d5dSymTFT}
\end{figure}

Suppose $\CX$ is a $(3+1)$d QFT with an anomaly-free $\Z_N^{(1)}$ global symmetry, whose partition function is $Z_{\CX}[X_4,B]$. In the current section we will not assume that $\CX$ is self-dual under gauging $\Z_N^{(1)}$. It is always possible to expand this theory into a $(4+1)$d slab, as shown in Figure \ref{fig:4d5dSymTFT}. The theory in the bulk of the slab is a $(4+1)$d $\Z_N^{(1)}$ gauge theory with action
\begin{eqnarray}\label{eq:5dBFtheory}
S= \frac{2\pi}{N}\int_{X_5} \widehat{b} \delta b~.
\end{eqnarray}
Both $b$ and $\widehat{b}$ are dynamical $\Z_N$ valued 2-cochains. We will take the bulk to be the product $X_5= X_4\times I_{[0,\varepsilon]}$ and will use the coordinate $x$ to parameterize the interval $I_{[0,\varepsilon]}$. The two boundaries are $X_4|_{\varepsilon}$ and $X_4|_0$ respectively.

The boundary conditions on the left ($x=0$) and right ($x=\varepsilon$) boundaries of the slab are specified by appropriate boundary states. On the right boundary, the state is  
\begin{eqnarray}
\ket{\CX} = \sum_{b\in H^b(X_4|_{\varepsilon}, \Z_N)} Z_{\CX}[X_4|_{\varepsilon}, b] \ket{b}.
\end{eqnarray}
This state encodes all the dynamical information of the $(3+1)$d QFT $\CX$. 
On the left boundary, the state is 
\begin{eqnarray}
\bra{D(B)} = \sum_{b\in H^2(X_4|_0, \Z_N)} \bra{b} \delta(b-B)~.
\end{eqnarray}
Note that the background field dependence only enters at the topological boundary. 
By shrinking the slab $\varepsilon\to 0$, the partition function of the $(3+1)$d theory is reproduced by the inner product of the two boundary states, 
\begin{eqnarray}
\braket{D(B)| \CX} = \sum_{b,b'\in H^2(X_4|_0, \Z_N)} Z_{\CX}[X_{4}|_{0}, b] \delta(b'-B)  \braket{b'|b}= Z_{\CX}[X_4|_0, B] 
\end{eqnarray}
where we have used $\braket{b'|b}=\delta(b'-b)$.

Gauging $\Z_N^{(1)}$ of $\CX$ in $(3+1)$d amounts to changing the topological boundary condition on the left boundary from Dirichlet to Neumann. To see this, we define the Neumann boundary condition as 
\begin{eqnarray}
\bra{N(B)} = \frac{|H^0(X_4|_0, \Z_N)|}{|H^1(X_4|_0, \Z_N)|}\sum_{b\in H^2(X_4|_0, \Z_N)} \bra{b} e^{\frac{2\pi i}{N}\int_{X_4|_0} bB }
\end{eqnarray}
and check explicitly that
\begin{equation}
\begin{split}
    \braket{N(B)|\CX} =\frac{|H^0(X_4|_0, \Z_N)|}{|H^1(X_4|_0, \Z_N)|} \sum_{b,b'\in H^2(X_4|_0, \Z_N)}  Z_{\CX}[X_4|_0, b] e^{\frac{2\pi i}{N}\int_{X_4|_0} b'B } \braket{b'|b} = Z_{\CX/\Z_N}[X_4, B]~.
\end{split}
\end{equation}

\subsection{Extended defects in $\Z_N^{(1)}$ gauge theory}

In this section, we consider the extended defects in $\Z_N^{(1)}$ gauge theory that do not have topological boundaries. The operators with boundaries will be discussed in Sections \ref{sec:6.3}, \ref{sec:6.5} and \ref{sec:6.7}.

\subsubsection{Fusion, braiding and higher quantum torus algebra of surfaces}

The $\Z_N^{(1)}$ gauge theory has $N^2$ genuine topological surfaces, defined by
\begin{eqnarray}\label{eq:4dsigma}
\begin{split}
    &S_{(e,m)}(\sigma)\equiv \exp\left(\frac{2\pi i }{N} \oint_{\sigma} e a^{(2)}\right)\exp\left(\frac{2\pi i }{N} \oint_{\sigma} m b^{(2)}\right) \equiv  S_{(e,0)}(\sigma) S_{(0,m)}(\sigma)~,\\ &(e,m)\in \Z_N\times \Z_N~.
\end{split}
\end{eqnarray}
Above, $\sigma$ is a two-dimensional surface. The surfaces $S_{(1,0)}$ and $S_{(0,1)}$ together generate a $\Z_N^{(2)}\times \Z_N^{(2)}$ two-form symmetry in the bulk. From canonical quantization, we have the commutation relations
\begin{eqnarray}\label{eq:4dcomm}
S_{(1,0)}(\sigma) S_{(0,1)}(\sigma')= e^{-2\pi i/N \braket{\sigma,\sigma'}}S_{(0,1)}(\sigma')S_{(1,0)}(\sigma)~,
\end{eqnarray}
where $\braket{\sigma,\sigma'}$ is the intersection pairing between the two surfaces $\sigma,\sigma'$ within the four manifold; see Appendix \ref{app:linecorr} for a derivation. Since $\sigma$ is two-dimensional, the pairing is symmetric---this is to be contrasted with the case in $(2+1)$d, in which case the relevant pairing was anti-symmetric. The fact that the pairing is symmetric implies that we are allowed to consider the self-pairing $\braket{\sigma,\sigma}$, which will be important below. 
Based on the definition \eqref{eq:4dsigma} and the basic commutation relation \eqref{eq:4dcomm}, one can derive the fusion rule 
\begin{eqnarray}\label{eq:4dfusionrule}
S_{(e,m)}(\sigma) S_{(e',m')}(\sigma) = \exp\left( \frac{2\pi i}{N} m e' \braket{\sigma,\sigma}\right) S_{(e+e',m+m')}(\sigma)~.
\end{eqnarray}
We should view the phase $\exp\left( \frac{2\pi i}{N} m e' \braket{\sigma,\sigma}\right)$ as a counterterm living on $\sigma$. This is related to the ambiguity in the definition of $S_{(e,m)}$ in \eqref{eq:4dsigma}. Indeed, one can instead define $S_{(e,m)}(\sigma)$ in the opposite order as $S_{(0,m)}(\sigma)S_{(e,0)}(\sigma)$, which differs from \eqref{eq:4dsigma} by a phase $\exp\left(-2\pi i em/N \braket{\sigma,\sigma}\right)$. 

Combining \eqref{eq:4dsigma} and \eqref{eq:4dcomm}, we also have the commutation relation between arbitrary two surfaces $S_{(e,m)}(\sigma)$ and $S_{(e',m')}(\sigma')$, 
\begin{eqnarray}\label{eq:5dcomm}
S_{(e,m)}(\sigma) S_{(e',m')}(\sigma') = \exp\left(- \frac{2\pi i}{N} (em' - m e')\braket{\sigma, \sigma'}\right) S_{(e',m')}(\sigma')S_{(e,m)}(\sigma)
\end{eqnarray}
and the higher quantum torus algebra
\begin{eqnarray}\label{eq:4dtorusquantumalgebra}
\begin{split}
    S_{(e,m)}(\sigma) S_{(e,m)}(\sigma')&=\exp\left( \frac{2\pi i}{N}e m\braket{\sigma,\sigma'}\right) S_{(e,m)}(\sigma+\sigma')~.
\end{split}
\end{eqnarray}
The $\sigma$ and $\sigma'$ in these equations can be either local patches of surfaces or entire surfaces without boundary.

\subsubsection{$\Z_4^{\text{EM}}$ symmetry and co-dimension one condensation defect in $(4+1)$d}
\label{sec:Z4EM4d}

Just as the $\Z_N$ gauge theory in $(2+1)$d had a $\Z_2^{\text{EM}}$ symmetry, so too does the $\Z_N^{(1)}$ gauge theory \eqref{eq:5dBFtheory} have an  electro-magnetic exchange symmetry.  The symmetry is defined by acting on the gauge fields as 
\begin{eqnarray}\label{eq:Z4EM}
b\to \widehat{b}, \hspace{1cm} \widehat{b}\to - b~,
\end{eqnarray}
and thus acts on the invertible surface operators as
\begin{eqnarray}\label{eq:Z4EMoper}
S_{(e,m)}(\sigma)   \to S_{(0,e)}(\sigma) S_{(-m,0)}(\sigma)= \exp\left( -\frac{2\pi i}{N} em \braket{\sigma,\sigma}\right) S_{(-m,e)}(\sigma)~.
\end{eqnarray}
This operation leaves the action \eqref{eq:5dBFtheory} invariant on a closed 5d manifold. The group which the operation \eqref{eq:Z4EM} generates depends on the value of $N$. When $N>2$, it generates $\Z_4^{\text{EM}}$. Moreover, the square of \eqref{eq:Z4EM} generates the charge conjugation symmetry, transforming both $b$ and $\widehat{b}$ by a minus sign. When $N=2$, since the charge conjugation transformation is trivial, \eqref{eq:Z4EM} generates $\Z_2^{\text{EM}}$.

Like the $\Z_2^{\text{EM}}$ symmetry in $(2+1)$d for which the corresponding surface was a condensation defect, the $\Z_4^{\text{EM}}$ symmetry generator in $(4+1)$d can also be constructed as a condensation defect. The intuition is analogous to that in Figure \ref{fig:Z2emdefect}. As shown in \eqref{eq:Z4EMoper}, when a surface $S_{(e,m)}$ intersects the 4d defect $D_{\text{EM}}$ from the left, an operator $S_{(-m,e)}$ intersects the 4d defect $D_{\text{EM}}$ from the right (up to a phase to be specified later). Using the folding trick, we see that the surface $S_{(e+m, m-e)} = S^{\otimes e}_{(1,-1)} \otimes S^{\otimes m}_{(1,1)}$ can be absorbed into the 4d defect for arbitrary $e,m$. This implies that the defect is now a condensation of \textit{two} species of surfaces, with charges 
\begin{eqnarray}\label{eq:surfacesin4dM}
(1,-1)~, \hspace{1cm} (1,1)~.
\end{eqnarray}
This should be contrasted with the defect \eqref{eq:DEMdef},  for which there was only one species of defect condensed. 

The two types of surfaces generate different higher form symmetries depending on the value of $N$. 
When $N$ is odd, the surfaces $S_{(1,-1)}$ and $ S_{(1,1)}$ generate a $\Z_N^{(2)}\times \Z_N^{(2)}$ two-form symmetry in the bulk, or a $\Z_N^{(1)}\times \Z_N^{(1)}$ one-form symmetry on the worldvolume $M_4$ of the defect $D_{\text{EM}}$. When $N$ is even, because of the identification $S_{(N/2,-N/2)}=S_{(N/2,N/2)}$, the surfaces $S_{(1,-1)}$ and $S_{(1,1)}$ generate a $(\Z_N^{(2)}\times \Z_N^{(2)})/\Z_2$ two-form symmetry in the bulk, or a $(\Z_N^{(1)}\times \Z_N^{(1)})/\Z_2$ one-form symmetry on the worldvolume of the defect $M_4$. When $N=2$, the two surfaces coincide, and generate a $\Z_2^{(1)}$ symmetry on $M_4$.  In summary, the symmetries generated by the surfaces in \eqref{eq:surfacesin4dM} within the worldvolume of the $\Z_4^{\text{EM}}$ defect are 
\begin{eqnarray}\label{eq:3cases}
\text{odd } N:\,\, \Z_N^{(1)}\times \Z_N^{(1)}~, \hspace{1cm} N=2:\,\, \Z_2^{(1)}~, \hspace{1cm}\text{even } N\geq 4:\,\, \frac{\Z_N^{(1)}\times \Z_N^{(1)}}{\Z_2}~.
\end{eqnarray}

With the above intuition, we now give a precise definition of the $\Z_4^{\text{EM}}$ condensation defect on a 4d manifold $M_4$. When $N$ is odd, the condensation defect is
\begin{equation}\label{eq:Z4EMcond}
D_{\text{EM}}(M_4) =
\frac{|H^0(M_4, \Z_N)|^2}{|H^1(M_4, \Z_N)|^2} \sum_{\sigma, \sigma'\in H_2(M_4, \Z_N)} \exp\left(\frac{2\pi i}{N} (\braket{\sigma', \sigma'} + \braket{\sigma, \sigma'})\right) S_{(1,-1)}(\sigma) S_{(1,1)}(\sigma')~.
\end{equation}
When $N=2$, the condensation defect is 
\begin{eqnarray}\label{eq:DEMN=2}
D_{\text{EM}}(M_4) = \frac{|H^0(M_4,\Z_2)|}{|H^1(M_4,\Z_2)|} \sum_{\sigma\in H_2(M_4,\Z_2)} S_{(1,0)}(\sigma) S_{(0,1)}(\sigma+[w_2^{TM}])~.
\end{eqnarray}
When $N$ is even and $N\geq 4$, the condensation defect is
\begin{equation}\label{eq:DEMN>2}
\begin{split}
    D_{\text{EM}}(M_4) =&\frac{|H^0(M_4, (\Z_N\times \Z_N)/\Z_2)|}{|H^1(M_4, (\Z_N\times \Z_N)/\Z_2)|} \sum_{(\sigma, \sigma')\in H_2(M_4, (\Z_N\times \Z_N)/\Z_2)} \exp\left(\frac{2\pi i}{N} (\braket{\sigma', \sigma'} + \braket{\sigma, \sigma'})\right) \\& \,\,\,\times  S_{(1,-1)}(\sigma) S_{(0,1)}(\frac{N}{2}[w_2^{TM}]) S_{(1,1)}(\sigma')~.
\end{split}
\end{equation}
 We relegate the derivation of these condensation defects to Appendix \ref{app:DEM}, but will 
comment on a couple of interesting features which are new compared to the $\Z_2^{\text{EM}}$ defect in $(2+1)$d $\Z_N$ gauge theory: 
\begin{enumerate}
    \item 
    The defect depends on the parity of $N$. 
    Summing over surfaces on $M_4$ amounts to 1-gauging of  $\Z_N^{(2)}\times \Z_N^{(2)}$ for odd $N$, and 1-gauging of $(\Z_N^{(2)}\times \Z_N^{(2)})/\Z_2$ for even $N$. 
    \item In contrast to the condensation defect in $(2+1)$d $\Z_N$ gauge theory, for which there was no discrete torsion due to $H^2(B\Z_N, U(1))$ being trivial, in the $(4+1)$d $\Z_N^{(1)}$ gauge theory we are allowed to include discrete torsion when constructing the condensation defect. In fact, it turns out that the discrete torsion as shown in \eqref{eq:Z4EMcond} is required in order to produce the correct fusion rules between $S_{(e,m)}$ and $D_{\text{EM}}$. Furthermore, the discrete torsion depends on the ordering of the surfaces. According to \eqref{eq:5dcomm}, moving $S_{(1,1)}(\sigma')$ to the left of $S_{(1,-1)}(\sigma)$ produces an additional phase $\exp\left(-\frac{4\pi i}{N}\braket{\sigma,\sigma'}\right)$. 
\end{enumerate}
The fusion rules involving the defect $D_{\text{EM}}(M_4)$ can be worked out by using the definitions \eqref{eq:Z4EMcond}, \eqref{eq:DEMN=2}, and \eqref{eq:DEMN>2}, as well as the properties of the surface operators \eqref{eq:4dfusionrule}, \eqref{eq:4dcomm}, and \eqref{eq:4dtorusquantumalgebra}. The fusion rules are found to be 
\begin{eqnarray}\label{eq:4dDfusion}
\begin{split}
&S_{(e,m)}(\tau) D_{\text{EM}}(M_4)= \exp\left( - \frac{2\pi i}{N} em \braket{\tau,\tau}\right) D_{\text{EM}}(M_4) S_{(-m,e)}(\tau)~, \\
&N=2: \hspace{0.5cm} D_{\text{EM}}(M_4)^2 = \chi[M_4,\Z_2]~,\\
&N\geq 3: \hspace{0.5cm} D_{\text{EM}}(M_4)^{ 4} = \chi[M_4,G]^2 \CZ_{Y}[M_4,G]^2~.
\end{split}
\end{eqnarray}
For the third line, the group $G$ is given by \eqref{eq:3cases}. Here $\chi[M_4,G]$ is the 
Euler characteristic of $M_4$ defined in \eqref{eq:Euler4d}, while $\CZ_{Y}[M_4,G]$ is the partition function of an invertible TQFT defined in \eqref{eq:Euler4dNeven}.   Concretely, for odd $N$ one has $G=\Z_N\times \Z_N$ and
the invertible TQFT $\CZ_{Y}[M_4,\Z_N\times \Z_N]:= \CZ_{Y}[M_4,\Z_N]^2$ where $\CZ_{Y}[M_4,\Z_N]$ is given by \eqref{eq:Yfunc}.  
For $N=2$, one has $G=\Z_2$ and we don't include any invertible TQFT. 
Finally, for even $N$  and $N\geq 4$, one has $G= (\Z_N\times \Z_N)/\Z_2$, 
and the invertible TQFT 
$\CZ_{Y}[M_4, (\Z_N\times \Z_N)/\Z_2]$ is defined as in \eqref{eq:Euler4dNeven}.

The first fusion rule in \eqref{eq:4dDfusion} is exactly as expected from \eqref{eq:Z4EMoper}. 
The last two fusion rules in \eqref{eq:4dDfusion} confirm that the condensation defect $D_{\text{EM}}$ is an invertible defect, generating a $\Z_4^{\text{EM}}$ symmetry for $N>2$ and a $\Z_2^{\text{EM}}$ symmetry for $N=2$.

\subsection{Twist defects  and their fusion rules for odd $N$}
\label{sec:6.3}

We next discuss the twist defects in $(4+1)$d $\Z_N^{(1)}$ gauge theory. Since the cases of odd $N$, $N=2$, and even $N$ for $N\geq 4$  differ significantly, we will treat them separately, beginning with the case of $N$ odd.

\subsubsection{Twist defects as higher duality interfaces}

As in $(2+1)$d, twist defects are defined by placing the $\Z_4^{\text{EM}}$ symmetry operator $D_{\text{EM}}$ defined in \eqref{eq:Z4EMcond} on a manifold $M_4$ with boundary $M_3=\partial M_4$, with Dirichlet boundary conditions imposed on $M_3$. We first define the ``minimal" twist defect as
\begin{equation}\label{eq:tw4d}
\begin{split}
    &V_{(0)}(M_3,M_4)= \frac{|H^0(M_4, M_3, \Z_N)|^2}{|H^1(M_4, M_3, \Z_N)|^2}
    \sum_{\sigma,\sigma'\in H_2(M_4,\Z_N)} \exp\left(\frac{2\pi i}{N} (\braket{\sigma',\sigma'}+ \braket{\sigma,\sigma'})\right) S_{(1,-1)}(\sigma) S_{(1,1)}(\sigma')~.
\end{split}
\end{equation}
Let us point out that, despite the fact that we have a boundary with Dirichlet boundary conditions, the sum is still over elements of \textit{absolute} homology, not \textit{relative} homology, as would be the case if we were working with cohomology.\footnote{
\label{footnote:relcoh} Indeed, by rewriting the surface operator as $S_{(1,1)}(\sigma)= e^{2\pi i/N \int_{\sigma} b} e^{2\pi i/N \int_{\sigma} \widehat{b}} = e^{2\pi i/N \int_{M_4} b_{\sigma}\cup b}e^{2\pi i/N \int_{M_4} b_{\sigma}\cup \widehat{b}}$ where $b_{\sigma}$ is the Lefschetz dual of $\sigma$ on $M_4$, summing over the surfaces $\sigma$ amounts to summing over 2-cohomologies $b_{\sigma}$. The Dirichlet boundary condition means that $b_{\sigma}$ takes value in \textit{relative cohomology} $H^2(M_4,M_3,\Z_N)$. By Lefschetz duality, $\sigma$ takes value in standard, absolute homology $H_2(M_4, \Z_N)$. Similar statements hold for $\sigma'$.} 

The Dirichlet boundary conditions mean that the surfaces  $S_{(1,\pm 1)}(\tau)$ for $\tau\in H_2(M_3, \Z_N)$ can be absorbed by fusing with the twist defect $V_{(0)}$ along the boundary $M_3$,
\begin{eqnarray}\label{eq:SSV}
S_{(1,-1)}(\tau)V_{(0)}(M_3,M_4)= V_{(0)}(M_3,M_4)~, \hspace{0.6cm} S_{(1,1)}(\tau')V_{(0)}(M_3,M_4)= V_{(0)}(M_3,M_4)~.
\end{eqnarray}
For the two conditions above to be compatible, it must be the case that the phase $\exp(4\pi i/N \braket{\tau,\tau'})$ coming from exchanging the order of $S_{(1,-1)}(\tau)$ and $S_{(1,1)}(\tau')$ is trivial for any $\tau,\tau'\in H_2(M_3,\Z_N)$.  This means that the intersection pairing must be trivial mod $N$.  To justify this, we note that $M_3$ is the boundary of $M_4$, and that one can regularize the intersection between $\tau$ and $\tau'$ within $M_3$ to be an intersection between $\tau$ in $M_3$ and $\tau'$ in $M_3'$, where $M_3'$ is obtained by parallel transporting $M_3$ along the direction orthogonal to $M_3$ into $M_4$. This ensures that $M_3$ and $M_3'$ do not share a common submanifold, and thus implies that $\braket{\tau,\tau'}=0$. Since the intersection pairing is between elements in homology, and since the regularization prescribed above amounts to choosing different representatives within the same homology class, we conclude that the pairing  always vanishes.  This justifies the simultaneous validity of the two equations in \eqref{eq:SSV}.

Using \eqref{eq:SSV}, we further find that there is only one type of twist defect. To see this, we fuse an arbitrary surface $S_{(e,m)}(\tau)$ with $V_{(0)}(M_3,M_4)$. Noting that we may decompose
\begin{eqnarray}\label{eq:Semdecompose}
S_{(e,m)}(\tau)= 
\begin{cases}
S_{(1,-1)}^{\frac{e-m}{2}}(\tau) S_{(1,1)}^{\frac{e+m}{2}}(\tau)~, & e\pm m\in 2\Z\\
S_{(1,-1)}^{\frac{e-m+N}{2}}(\tau) S_{(1,1)}^{\frac{e+m+N}{2}}(\tau)~, & e\pm m\in 2\Z+1\\
\end{cases}
\end{eqnarray}
where we have used $\braket{\tau,\tau}=0\mod N$ for $\tau\in H_2(M_3,\Z_N)$, we find that $V_{(0)}$ can absorb arbitrary $S_{(e,m)}(\tau)$ under fusion.

We define the orientation reversal of $V_{(0)}$ to be  
\begin{equation}\label{eq:Vbar}
\begin{split}
    \overline{V}_{(0)}(M_3,M_4) =& \chi[M_4^{\geq 0}, \Z_N\times \Z_N]^{-1} \frac{|H^0(M_4, M_3, \Z_N)|^2}{|H^1(M_4, M_3, \Z_N)|^2}\times 
    \\&\sum_{\sigma,\sigma'\in H_2(M_4,\Z_N)} \exp\left(-\frac{2\pi i}{N} (\braket{\sigma,\sigma}+ \braket{\sigma,\sigma'})\right) S_{(1,1)}(-\sigma') S_{(1,-1)}(-\sigma)
\end{split}
\end{equation}
where the right-hand side is, apart from the Euler counterterm, the Hermitian conjugate of $V_{(0)}$. The inclusion of the Euler counterterm is motivated by \eqref{eq:CNbarCN4d}. As shown in Appendix \ref{app:DEM}, including the same Euler counterterm in the orientation reversal of $D_{\text{EM}}$ ensures the expected behavior $D_{\text{EM}} \overline{D}_{\text{EM}}=1$.

Since the twist defect is defined to be the higher gauging of a two-form symmetry along half of the codimension-one submanifold $M_4$ of the spacetime with Dirichlet boundary conditions on $M_3$, it 
can be interpreted as a higher duality interface.

\subsubsection{Fusion rules of the twist defects} 

We now discuss the fusion rules involving twist defects for $N$ odd.

\paragraph{Fusion rule $S_{(e,m)}\times V_{(0)}$:}
As discussed in the previous subsection, 
any simple surface operator $S_{(e,m)}$  can be absorbed into the twist defect ${V}_{(0)}$, and hence we have the fusion rule 
\begin{eqnarray}\label{eq:4dSVV}
S_{(e,m)}(\tau) \times V_{(0)}(M_3,M_4) = V_{(0)}(M_3,M_4)~.
\end{eqnarray}

\paragraph{Fusion rule $V_{(0)}\times \overline{V}_{(0)}$:}
More interesting is the fusion rule between two twist defects. For simplicity, we first consider  $V_{(0)}\times \overline{V}_{(0)}$. It is useful to again specify the geometry near the boundary of $M_4$ as $M_3\times \mathbb{R}_+$, and to use the coordinate $x$ to parameterize the direction orthogonal to the boundary. We take $V_{(0)}$ to be on $M_4^{\geq 0}$ and $\overline{V}_{(0)}$ to be on $M_4^{\geq \eps}$.   From the definition of the twist defect \eqref{eq:tw4d} and its Hermitian conjugate \eqref{eq:Vbar}, we find that 
\begin{equation}
    \begin{split}
        &V_{(0)}(M_3,M_4)\times \overline{V}_{(0)}(M_3,M_4)= \chi[M_4^{\geq \eps}, \Z_N]^{-2} \frac{|H^0(M_4^{\geq 0}, M_3|_0, \Z_N)|^2}{|H^1(M_4^{\geq 0}, M_3|_0, \Z_N)|^2} \frac{|H^0(M_4^{\geq \eps}, M_3|_\eps, \Z_N)|^2}{|H^1(M_4^{\geq \eps}, M_3|_\eps, \Z_N)|^2}\times \\
    &
 \sum_{\substack{\sigma,\sigma',\in H_2(M_4^{\geq 0},\Z_N)\\ \tau,\tau',\in H_2(M_4^{\geq \eps},\Z_N)}} \exp\left(\frac{2\pi i}{N}(\braket{\sigma',\sigma'}+ \braket{\sigma,\sigma'}+ \braket{\sigma+\sigma',\tau'}+ \braket{\sigma-\sigma',\tau})\right)
 S_{(1,-1)}(\sigma) S_{(1,1)}(\sigma')~.
    \end{split}
\end{equation}
To evaluate this expression, we apply steps similar to those in Section \ref{sec:fusionruledualityinterfaces4d}, upon which the above expression simplifies to 
\begin{equation}
\frac{|H^0(M_4^{[0,\eps]},M_3|_0\cup M_3|_{\eps}, \Z_N)|^2}{|H^1(M_4^{[0,\eps]},M_3|_0\cup M_3|_{\eps}, \Z_N)|^2} \sum_{\sigma,\sigma'\in H_2(M_4^{[0,\eps]}, \Z_N)} \exp\left(\frac{2\pi i}{N}\braket{\sigma',\sigma'}\right)
    S_{(1,0)}(\sigma+\sigma') S_{(0,1)}(\sigma'-\sigma)~
\end{equation}
which is exactly the result one would obtain by placing the $D_{\text{EM}}$ defect on an interval $M_4^{[0,\eps]} \simeq M_3\times I_{[0,\eps]}$, with Dirichlet boundary conditions on both sides. Note that on the interval, the intersection pairing $\braket{\cdot,\cdot}$ is trivial, which also holds when shrinking the slab to $M_3$. After making a change of variables $\sigma+\sigma'\to \rho', \sigma'-\sigma\to \rho$ (which is possible for odd $N$), the fusion rule then simplifies to 
\begin{eqnarray}\label{eq:4dVVbar1}
\begin{split}
    V_{(0)}(M_3,M_4)\times \overline{V}_{(0)}(M_3,M_4) =\frac{1}{|H^0(M_3,\Z_N)|^2} \sum_{\rho,\rho'\in H_2(M_3,\Z_N)} S_{(1,0)}(\rho') S_{(0,1)}(\rho)~.
\end{split}
\end{eqnarray}

\paragraph{Fusion rule $V_{(0)}\times V_{(0)}$: }
We next consider the fusion between $V_{(0)}$ and itself, with both defined on $M_4^{\geq 0}$.  Taking the square of \eqref{eq:tw4d}, we find 
\begin{equation}\label{eq:VVfusion}
\begin{split}
    &V_{(0)}(M_3|_0,M_4^{\geq 0})\times V_{(0)}(M_3|_0,M_4^{\geq 0}) = \frac{|H^0(M_4^{\geq 0}, M_3|_0, \Z_N)|^4}{|H^1(M_4^{\geq 0}, M_3|_0, \Z_N)|^4} \left( \sum_{\tau\in H_2(M_4, \Z_N)} e^{2\pi i/N \braket{\tau,\tau}}
    \right)^2 \\&   \times\sum_{\sigma,\sigma'\in H_2(M_4, \Z_N)} \exp\left( \frac{2\pi i}{N} (-\braket{\sigma,\sigma}+ \braket{\sigma',\sigma'}+ 2\braket{\sigma,\sigma'})\right)S_{(1,-1)}(\sigma'-\sigma) S_{(1,1)}(\sigma'+\sigma) \\
    &= \chi[M_4^{\geq 0}, \Z_N] \CZ_{Y}[M_4^{\geq 0},\Z_N]^2 \frac{|H^0(M_4^{\geq 0}, M_3|_0, \Z_N)|^2}{|H^1(M_4^{\geq 0}, M_3|_0, \Z_N)|^2} \sum_{\sigma,\sigma'\in H_2(M_4, \Z_N)} \exp\left( \frac{2\pi i}{N} 2\braket{\sigma,\sigma'}\right)S_{(1,0)}(2\sigma') S_{(0,1)}(2\sigma) 
\end{split}
\end{equation}
where we define $\CZ_{Y}[M_4^{\geq 0},\Z_N]$ for an open $M_4^{\geq 0}$ to be 
\begin{eqnarray}
\CZ_{Y}[M_4^{\geq 0},\Z_N]= \chi[M_4^{\geq 0}, \Z_N]^{-\frac{1}{2}}   \frac{|H^0(M_4^{\geq 0}, M_3|_0, \Z_N)|}{|H^1(M_4^{\geq 0}, M_3|_0, \Z_N)|} \sum_{\tau\in H_2(M_4^{\geq 0},\Z_N)} e^{\frac{2\pi i}{N}\braket{\tau,\tau}}~.
\end{eqnarray}
Although $\CZ_Y$ is an invertible TQFT on a closed manifold, $\CZ_Y$ is no longer invertible on an open manifold.
Equation \eqref{eq:VVfusion} is precisely the charge conjugation defect \eqref{eq:Codd} on $M_4^{\geq 0}$, i.e. $C(M_4^{\geq 0})$.

\subsection{Duality defect in $(3+1)$d from twist defects in $(4+1)$d for odd $N$}
\label{sec:6.4}

Still restricting to $N$ odd, we now insert the extended defects into the $(4+1)$d slab and examine their fate  upon shrinking the slab. We will find the following correspondence,
\begin{eqnarray}
\begin{split}
    \text{Twist defect } V_{(0)} &\longleftrightarrow \text{Duality interface } \CN\\
    \text{Magnetic surface } S_{(0,1)} & \longleftrightarrow \text{$\Z_N^{(1)}$ symmetry defect } \eta\\
    \text{Electric line } S_{(1,0)} & \longleftrightarrow \text{$\Z_N^{(1)}$ Order parameter } L 
\end{split}
\end{eqnarray}

\subsubsection{$\Z_N^{(1)}$ symmetry defects and order parameters from bulk surface operators}

When inserting a surface operator $S_{(e,m)}$ into the $(4+1)$d slab, one can either place the surface parallel to the boundary $X_4|_0$, or with one of its directions orthogonal to the boundary. Because of the Dirichlet boundary conditions of $b$, i.e. $b|_{X_4|_0}=0$, the electric surface $S_{(1,0)}$ can either end on the boundary perpendicularly, or be absorbed into the boundary when parallel to it. Thus upon shrinking, the only way for it to survive as a non-trivial operator is to place it orthogonal to the boundary, with one of its boundaries terminating on the Dirichlet boundary and the other terminating on the non-topological boundary. As a consequence, $S_{(1,0)}$ becomes a non-topological line operator, which can act as the $\Z_N^{(1)}$ order parameter $L$.

On the other hand, the magnetic surface $S_{(0,1)}$ survives upon shrinking, and becomes the $\Z_N^{(1)}$ symmetry defect $\eta$ in the $(3+1)$d QFT $\CX$. The $\eta$ surface can link with $L$ on $X_4$ and produces a nontrivial phase measuring the charge of $L$.

\subsubsection{Duality interface from the twist defect}

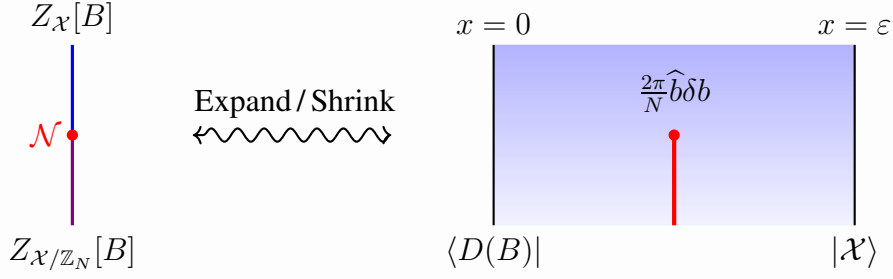
\begin{figure}[!tbp]
	\centering
	\begin{tikzpicture}[scale=0.8]
	
	\draw[very thick, violet] (-7,0) -- (-7,1.5);
	\draw[very thick, blue] (-7,1.5) -- (-7,3);
	\node[above] at (-7,3) {$Z_{\CX}[B]$};
	\node[below] at (-7,0) {$Z_{\CX/\ZZ_N}[B]$};
	\node at (-7,1.5) [circle,fill,red, inner sep=1.5pt]{};
	\node[left,red] at (-7,1.5) {$\cN$};

	\draw[thick, snake it, <->] (-1.7,1.5) -- (-5, 1.5);
	\node[above] at (-3.3,1.6) {Expand\,/\,Shrink};

	\shade[line width=2pt, top color=blue!30, bottom color=blue!5] 
	(0,0) to [out=90, in=-90]  (0,3)
	to [out=0,in=180] (6,3)
	to [out = -90, in =90] (6,0)
	to [out=180, in =0]  (0,0);
	
	\draw[thick] (0,0) -- (0,3);
	\draw[thick] (6,0) -- (6,3);
	\draw[ultra thick,red] (3,0) -- (3,1.5);
	\node[above] at (3,1.7) {$\frac{2\pi}{N} \widehat{b} \delta b$};
	\node at (3,1.5) [circle,fill,red, inner sep=1.5pt]{};
	\node[below] at (0,0) {$\langle D(B)| $};
	\node[below] at (6,0) {$|\cX\rangle $}; 
	
	\node[above] at (0,3) {$x=0$};
	\node[above] at (6,3) {$x=\varepsilon$};

	\end{tikzpicture}
	
	\caption{A $(3+1)$d QFT $\cX$ with $\Z_N^{(1)}$ one-form symmetry and another $(3+1)$d QFT $\cX/\Z_N$ with a quantum $\widehat{\Z}_N^{(1)}$ one-form symmetry are separated by a topological interface $\cN$. This setup can be expanded into a $(4+1)$d slab. The $(4+1)$d $\Z_N^{(1)}$ SymTFT has an insertion of a twist defect parallel to the Dirichlet boundary.  }
	\label{fig:twistdefshrink4d}
\end{figure}

We can also insert a twist defect $V_{(0)}$ into the $(4+1)$d slab.  Like in $(2+1)$d, colliding the $D_{\text{EM}}$ defect with the Dirichlet boundary yields Neumann boundary conditions. Hence colliding a twist defect with the Dirichlet boundary yields an interface between $\CX$ and $\CX/\Z_N$; see Figure \ref{fig:twistdefshrink4d}.

To reproduce the fusion rules of duality defects \eqref{eq:FR4d} from the twist defects, let us first consider the collision of the twist defect with the Dirichlet boundary condition.\footnote{We emphasize that there are two Dirichlet boundary conditions appearing in the current discussion. One is the Dirichlet boundary condition for $[\sigma]$ and $[\sigma']$ on the boundary $M_3$ of the twist defect, while the other is the Dirichlet boundary condition for the gauge field $b$ in the ambient SymTFT.} The Dirichlet boundary condition of the dynamical field $b$ of the SymTFT means that $b$ in $S_{(1,0)}(\tau)= e^{2\pi i/N \int_{\tau}b}$ is valued in relative cohomology $H^{2}(X_5, X_4, \Z_N)$. To integrate such $b$ on a two-cycle $\tau$, the two-cycle $\tau$ should be in relative homology $H_2(X_5, X_4, \Z_N)$. Moreover, when $\tau$ lives on the boundary $M_3$ of a twist defect, we require $\tau\in H_2(M_3,\Z_N)$. Hence upon moving the twist defect onto the Dirichlet boundary, $\tau$ is restricted to lie in
\begin{eqnarray}\label{eq:cohtorelcoh}
H_2(M_3,\Z_N) \cap H_2(X_5, X_4, \Z_N)~.
\end{eqnarray}
Since after colliding the twist defect with the Dirichlet boundary condition we have $M_3=\partial M_4 \subset  X_4=\partial X_5$, it follows that $\tau$ is relative to itself, i.e. $\tau\in H_2(M_3,M_3,\Z_N) = \varnothing$.
This fact will be important below.

Returning to the fusion, the fusion rule $\eta\times \CN= \CN$ clearly follows from \eqref{eq:4dSVV}. Concretely, bringing the operators to the Dirichlet boundary we have $S_{(e,m)} \to \eta$ and $V_{(0)}\to \CN$, and hence \eqref{eq:4dSVV} immediately leads to the desired fusion rule.

More interesting is the fusion rule of two duality interfaces $\CN\times \overline{\CN}$, which should descend from \eqref{eq:4dVVbar1}, which we reproduce here for convenience,
\begin{eqnarray}
V_{(0)}(M_3,M_4)\times \overline{V}_{(0)}(M_3,M_4) =\frac{1}{|H^0(M_3,\Z_N)|^2} \sum_{\rho,\rho'\in H_2(M_3,\Z_N)} S_{(1,0)}(\rho') S_{(0,1)}(\rho)~.
\end{eqnarray}
Upon colliding the twist defect with the Dirichlet boundary, $\rho'$ should be valued in the relative homology of $M_3$ with respect to itself as discussed above, meaning that it is trivial. We should also change one of the normalization factors to be in relative cohomology. The above fusion rule then simplifies to 
\begin{eqnarray}
\begin{split}
\label{eq:NNbcondensate}
     \CN\times \overline{\CN}&=\frac{1}{|H^0(M_3,\Z_N)||H^0(M_3,M_3,\Z_N)|} \sum_{\substack{\rho\in H_2(M_3,\Z_N)\\ \rho'\in H_2(M_3,M_3,\Z_N)}} S_{(1,0)}(\rho') S_{(0,1)}(\rho)\\
    &= \frac{1}{|H^0(M_3,\Z_N)|} \sum_{\substack{\rho\in H_2(M_3,\Z_N)}}  S_{(0,1)}(\rho)\\
    &= \frac{1}{|H^0(M_3,\Z_N)|} \sum_{\substack{\rho\in H_2(M_3,\Z_N)}}  \eta(\rho) ~.
\end{split}
\end{eqnarray}
This reproduces the results given in (\ref{eq:FR4d}).

\subsection{Twist defects and their fusion rules for $N=2$}
\label{sec:6.5}

We now proceed to a discussion of twist defects in $(4+1)$d $\Z_N^{(1)}$ gauge theory for $N=2$. Because the discussion is parallel to that in the previous subsection we will be brief, mainly just emphasizing the differences. The busy reader may wish to skip directly to Section \ref{sec:5dSymTFTTYZN}. 

\subsubsection{Twist defects}

We begin by defining the minimal twist defect by placing the $\Z_2^{\text{EM}}$ symmetry defect $D_{\text{EM}}$ defined in \eqref{eq:DEMN=2} on an open 4d manifold $M_4$, with Dirichlet boundary conditions on  $M_3=\partial M_4$,
\begin{eqnarray}
\label{eq:eqwithbeta}
V_{(0)}(M_3,M_4) = \frac{|H^0(M_4,M_3,\Z_N)|}{|H^1(M_4,M_3,\Z_N)|} \sum_{\sigma\in H_2(M_4,\Z_2)}S_{(1,1)}(\sigma) S_{(0,1)}([w_2^{TM}])~.
\end{eqnarray}
Above, $[w_2^{TM}]\in H^2(M_4,\Z_2)$ is the Lefschetz dual of the second Stiefel-whitney class $w_2^{TM}\in H^2(M_4,M_3,\Z_N)$. Note that $w_2^{TM}$ is trivialized on $M_3$ by a spin structure, and hence $V_{(0)}$ depends on the presence of a spin structure. 
Since the defect has Dirichlet boundary conditions on $M_3$, the surface $S_{(1,1)}(\tau)$ for $\tau\in H_2(M_3,\Z_2)$ can be absorbed by $V_{(0)}$, giving
\begin{eqnarray}
S_{(1,1)}(\tau) V_{(0)}(M_3,M_4) = V_{(0)}(M_3,M_4)~.
\end{eqnarray}
On the other hand, since $S_{(1,0)}(\tau)$ and $S_{(0,1)}(\tau)$ do not belong to the condensate, fusing them with $V_{(0)}$ on the boundary produces a \textit{new} twist defect,
\begin{eqnarray}\label{eq:VSV}
V_{(1)}(\tau, M_3,M_4):= S_{(1,0)}(\tau) V_{(0)}(M_3,M_4) = S_{(0,1)}(\tau) V_{(0)}(M_3,M_4)~.
\end{eqnarray}
Note that fusing $V_{(0)}$ with $S_{(1,0)}(\tau)$ and $S_{(0,1)}(\tau)$ yields the same defect, because they differ by a phase $(-1)^{\braket{\tau,\tau}}$ which vanishes for $\tau\in H_2(M_3, \Z_2)$; c.f. the discussion below \eqref{eq:SSV}. The presence of multiple twist defects here should be contrasted with the case of $N$ odd. Moreover, we note that $V_{(1)}$ depends explicitly on $\tau$, and hence it is best thought of as a surface operator living inside $M_3$, which is in turn the boundary of $M_4$. 

The orientation reversal of $V_{(0)}$ is given by
\begin{eqnarray}\label{eq:V0N=2orirev}
\overline{V}_{(0)}(M_3,M_4) =\chi[M_4,\Z_2]^{-1} \frac{|H^0(M_4,M_3,\Z_N)|}{|H^1(M_4,M_3,\Z_N)|} \sum_{\sigma\in H_2(M_4,\Z_2)}S_{(1,1)}(\sigma) S_{(0,1)}([w_2^{TM}])~
\end{eqnarray}
which is precisely the same as placing the orientation reversal of the $\Z_2^{\text{EM}}$ defect  \eqref{eq:DEMN=2orientrev} on half of the space.

\subsubsection{Fusion rules of the twist defects}

We now discuss fusion rules involving the twist defects for $N=2$.

\paragraph{Fusion rule: $S_{(e,m)}\times V_{(e')}$:}
The fusion rule between a surface operator and a twist defect follows almost by definition, 
\begin{eqnarray}
S_{(e,m)}(\tau) \times V_{(e')}(\tau,M_3,M_4) = V_{(e-m+e')}(\tau, M_3, M_4)~.
\end{eqnarray}
When $e-m+e'=0$ mod $N$, the right hand side is independent of the surface $\tau$.

\paragraph{Fusion rule: $V_{(0)}\times \overline{V}_{(0)}$:}
We further consider the fusion between two twist defects. Since all variants of twist defects can be obtained from the minimal one by fusing a surface operator, it suffices to only discuss the fusion rule for the minimal twist defect $V_{(0)}$. 
From \eqref{eq:eqwithbeta} and \eqref{eq:V0N=2orirev}, we compute the fusion to be
\begin{equation}
\begin{split}
    &V_{(0)}(M_3|_0,M_4^{\geq 0}) \times \overline{V}_{(0)}(M_3|_\eps, M_4^{\geq \eps})\\&= \chi[M_4^{\geq \eps}, \Z_2]^{-1} \frac{|H^0(M_4^{\geq 0},M_3|_0,\Z_2)|}{|H^1(M_4^{\geq 0},M_3|_0,\Z_2)|}\frac{|H^0(M_4^{\geq \eps},M_3|_\eps,\Z_2)|}{|H^1(M_4^{\geq \eps},M_3|_\eps,\Z_2)|}  \sum_{\substack{\sigma\in H_2(M_4^{\geq 0}, \Z_2)\\ \sigma'\in H_2(M_4^{\geq \eps}, \Z_2)}} e^{i\pi \braket{\sigma,\sigma'}} S_{(1,1)}(\sigma) ~.
\end{split}
\end{equation}
Here we have used the fact that $w_2^{TM}$ on $M_4^{[0,\eps]}= M_3\times I_{[0,\eps]}$ is always trivial.\footnote{This is because $w_2^{TM}$ is always trivial on any 3d oriented manifold, i.e. they are all spin manifolds. It follows that the product space $M_4^{[0,\eps]}$ is also spin.} Simplifying the above expression in the way done in previous subsections, we obtain exactly the $\Z_2^{\text{EM}}$ topological operator on an interval $M_4^{[0,\eps]}$ with two Dirichlet boundary conditions, which can be further simplified to 
\begin{eqnarray}\label{eq:VVN=2fusion}
\begin{split}
    &V_{(0)}\times \overline{V}_{(0)}= \frac{1}{|H^0(M_3,\Z_2)|} \sum_{\sigma\in H_2(M_3,\Z_2)} S_{(1,1)}(\sigma)~.
\end{split}
\end{eqnarray}
Using the relation $V_{(0)}= \chi[M_4,\Z_2]\cdot  \overline{V}_{(0)}$, the fusion rule for $V_{(0)}\times V_{(0)}$ can also be obtained.

\subsection{Duality defect in $(3+1)$d from twist defects in $(4+1)$d for $N=2$}
\label{sec:6.6}

Still restricting to $N=2$, we proceed to insert the extended defects into the $(4+1)$d slab and examine their fate upon shrinking the slab. We will find the following correspondence,
\begin{eqnarray}
\begin{split}
    \text{Twist defect } V_{(e)} &\longleftrightarrow \text{Duality interface } \CN\\
    \text{Magnetic surface } S_{(0,1)} & \longleftrightarrow \text{$\Z_2^{(1)}$ symmetry defect } \eta\\
    \text{Electric line } S_{(1,0)} & \longleftrightarrow \text{$\Z_2^{(1)}$ Order parameter } L 
\end{split}
\end{eqnarray}

\subsubsection{$\Z_2^{(1)}$ symmetry defects and order parameters from bulk surface operators}

We first consider inserting a surface operator $S_{(e,m)}$ into the $(4+1)$d slab. As before, by placing the surface operator along different directions, we obtain different operators upon shrinking. Summarizing the results, 
\begin{enumerate}
    \item The surface operator $S_{(1,0)}$ can terminate on the boundary orthogonally or be absorbed into the boundary parallelly. Hence upon shrinking, it survives as a line operator $L$, which is the order parameter of the $\Z_2^{(1)}$ one-form symmetry of the $(3+1)$d QFT $\CX$. 
    \item The surface operator $S_{(0,1)}$, when placed parallel to the boundary, survives upon shrinking. It becomes the symmetry defect $\eta$ for the $\Z_2^{(1)}$  one-form symmetry. The surface $\eta$ can link with $L$ with a nontrivial phase, which measures the $\Z_2^{(1)}$ charge of $L$. 
\end{enumerate}

\subsubsection{Duality interface from twist defects}

We can also insert a twist defect into the $(4+1)$d slab, giving the configuration shown in Figure \ref{fig:twistdefshrink4d}. We again find that the twist defect in the $(4+1)$d bulk becomes a duality interface upon shrinking. Moreover, thanks to the Dirichlet boundary conditions of the SymTFT, the resulting duality defect is independent of the type of the twist defect, 
\begin{eqnarray}\label{eq:NV}
\CN(M_3,M_4)= V_{(e)}(\tau, M_3,M_4)|_{x \rightarrow 0}~, \hspace{1cm} e=0,1~. 
\end{eqnarray}
In other words, there is only a single type of duality interface.

It is interesting to see how the fusion rules of the duality interfaces are reproduced from those of the twist defects. We start from \eqref{eq:VVN=2fusion} and collide the defects in the fusion rule with the Dirichlet boundary condition, which amounts to replacing $V_{(0)}\to \CN,$ $\overline{V}_{(0)}\rightarrow \overline{\cN}$, and $S_{(1,1)}(\sigma)\to \eta(\sigma)$.\footnote{One may wonder about the subtlety of absolute versus relative homology discussed around \eqref{eq:cohtorelcoh}. 
Note that $S_{(1,1)}(\sigma) = e^{i \pi \int_{\sigma} b+\widehat{b}}$. Because $b$ obeys Dirichlet boundary conditions while $\widehat{b}$ obeys Neumann boundary conditions, the sum $b+\widehat{b}$ is still Neumann, and hence $\sigma$ is still in absolute homology. The same comment also applies to the fusion rule of duality defects in $(1+1)$d in Section \ref{sec:symdef1+12+1}. } The fusion rule \eqref{eq:VVN=2fusion} then simplifies to 
\begin{eqnarray}
\CN \times \overline{\CN} = \frac{1}{|H^0(M_3,\Z_2)|} \sum_{\sigma\in H_2(M_3,\Z_2)}\eta(\sigma)~.
\end{eqnarray}
The right-hand side is exactly the condensation defect associated with higher gauging of $\Z_2^{(2)}$ one-form symmetry on a codimension one defect $M_3$. This agrees with the results of \eqref{eq:FR4d}.

\subsection{Twist defects and fusion rules for even $N$ and $N\geq 4$}
\label{sec:6.7}

We finally discuss the remaining case, namely twist defects in $(4+1)$d $\Z_N^{(1)}$ gauge theory with even $N$ and $N\geq 4$. This proceeds as before, and  the busy reader may wish to skip directly to Section \ref{sec:5dSymTFTTYZN}.

\subsubsection{Twist defects}

We start by defining the minimal twist defect by placing the $\Z_4^{\text{EM}}$ defect  \eqref{eq:DEMN>2} on an open 4d manifold with Dirichlet boundary conditions
\begin{equation}\label{eq:N>=4V0}
\begin{split}
    &V_{(0)}(M_3, M_4) = \,\frac{|H^0(M_4,M_3,(\Z_N\times \Z_N)/\Z_2)|}{|H^1(M_4,M_3,(\Z_N\times \Z_N)/\Z_2)|}\times \\& \sum_{(\sigma,\sigma')\in H_2(M_4, (\Z_N\times \Z_N)/\Z_2)} \exp\left( \frac{2\pi i}{N} (\braket{\sigma',\sigma'}+ \braket{\sigma,\sigma'})\right)   S_{(1,-1)}(\sigma) \,S_{(0,1)}\left(\frac{N}{2}[w_2^{TM}]\right) S_{(1,1)}(\sigma')~.
\end{split}
\end{equation}
The Dirichlet boundary conditions for $\sigma,\sigma'$ imply that fusing $S_{(1,\pm 1)}(\tau)$ with $(\tau, \tau')\in H_2(M_3, (\Z_N\times \Z_N)/\Z_2)$ does not change the twist defect, 
\begin{eqnarray}
S_{(1,-1)}(\tau) V_{(0)}(M_3, M_4) = V_{(0)}(M_3, M_4)~, \hspace{0.5cm} S_{(1,1)}(\tau') V_{(0)}(M_3, M_4)  = V_{(0)}(M_3, M_4)~. 
\end{eqnarray}
The two conditions can be simutaneously satisfied for the same reasons as given below \eqref{eq:SSV}. Moreover, by sequentially fusing with either $S_{(1,1)}$ or $S_{(1,-1)}$, a surface operator with arbitrary charge can be decomposed as
\begin{eqnarray}
S_{(e,m)}(\tau) = 
\begin{cases}
S_{(1,-1)}^{\frac{e-m}{2}}(\tau) S_{(1,1)}^{\frac{e+m}{2}}(\tau)~ & e+m \in 2\Z\\
S_{(1,0)}(\tau)S_{(1,-1)}^{\frac{e-m-1}{2}}(\tau) S_{(1,1)}^{\frac{e+m-1}{2}}(\tau)~ & e+m\in 2\Z+1
\end{cases}
\end{eqnarray}
Hence as in the $N=2$ case, there are \textit{two} types of the twist defects, $V_{(0)}$ and $V_{(1)}$, defined as
\begin{eqnarray}
V_{(1)} (\tau, M_3, M_4) = S_{(1,0)}(\tau) V_{(0)}(M_3,M_4)~.
\end{eqnarray}
When $e+m$ is even, fusing $S_{(e,m)}(\tau)$ with the minimal twist defect $V_{(0)}(M_3, M_4)$ does not give rise to a new operator, while when $e+m$ is odd fusing $S_{(e,m)}(\tau)$ with $V_{(0)}(M_3,M_4)$ yields $V_{(1)}(\tau, M_3, M_4)$.

We define the orientation reversal of $V_{(0)}$ as
\begin{equation}\label{eq:N>=4V0rev}
\begin{split}
    &\overline{V}_{(0)}(M_3,M_4)= \chi[M_4, (\Z_N\times \Z_N)/\Z_2]^{-1} \frac{|H^0(M_4,M_3,(\Z_N\times \Z_N)/\Z_2)|}{|H^1(M_4,M_3,(\Z_N\times \Z_N)/\Z_2)|}\times \\& \sum_{\substack{(\sigma,\sigma')\in\\ H_2(M_4, (\Z_N\times \Z_N)/\Z_2)}} \exp\left(- \frac{2\pi i}{N} (\braket{\sigma,\sigma}+ \braket{\sigma,\sigma'})\right) \times S_{(1,1)}(-\sigma')\, S_{(0,-1)}\left(\frac{N}{2}[w_2^{TM}]\right) S_{(1,-1)}(-\sigma)~,
\end{split}
\end{equation}
which is precisely the orientation reversal of the $\Z_4^{\text{EM}}$ defect \eqref{eq:orientationN>4} on an open 4d manifold.

\subsubsection{Fusion rules of twist defects}

 We now discuss fusion rules involving the twist defects for $N$ even and $N\geq 4$.

\paragraph{Fusion rule $S_{(e,m)}\times V_{(e')}$:}
The fusion rules between a surface operator and a twist defect are a straightforward generalization of those for $N=2$. When both of them depend on the same $\tau$, we have 
\begin{eqnarray}
S_{(e,m)}(\tau) \times V_{([e']_2)}(\tau,M_3,M_4) = V_{([e-m+e']_2)}(\tau,M_3,M_4)
\end{eqnarray}
where $[e]_2 \equiv e\mod 2$.

\paragraph{Fusion rule $V_{(0)}\times \overline{V}_{(0)}$:}
We next consider the fusion between two twist defects. Without loss of generality, we only discuss the fusion between minimal twist defects. From \eqref{eq:N>=4V0} and \eqref{eq:N>=4V0rev}, the fusion $V_{(0)}\otimes \overline{V}_{(0)}$ is exactly the $\Z_4^{\text{EM}}$ topological defect on an interval with Dirichlet boundary conditions on both sides, 
\begin{eqnarray}
\begin{split}
    &V_{(0)}(M_3|_0,M_4^{\geq 0})\times \overline{V}_{(0)}(M_3|_{\eps},M_4^{\geq \eps}) = \frac{|H^0(M_4^{[0,\eps]}, M_3|_0\cup M_3|_{\eps}, (\Z_N\times \Z_N)/\Z_2)|}{|H^1(M_4^{[0,\eps]}, M_3|_0\cup M_3|_{\eps}, (\Z_N\times \Z_N)/\Z_2)|}\times \\& \sum_{\substack{(\sigma,\sigma')\in \\ H_2(M_4^{[0,\eps]}, (\Z_N\times \Z_N)/\Z_2)}} \exp\bigg( \frac{2\pi i}{N}(\braket{\sigma',\sigma'}+ \braket{\sigma,\sigma'})\bigg) S_{(1,-1)}(\sigma) 
    S_{(1,1)}(\sigma')~,
\end{split}
\end{eqnarray}
where we have used $S_{(0,1)}(\frac{N}{2}[w_2^{TM}])=1$ on $M_3\times I$. The sum can further be reduced to one over $M_3$ by using $H_2(M_4^{[0,\eps]}, G)= H_2(M_3,G)$, which also allows us to drop the phase in the sum. This gives
\begin{equation}\label{eq:VVbarN>4}
\begin{split}
    V_{(0)}(M_3,M_4)\times \overline{V}_{(0)}(M_3,M_4)&= \frac{1}{|H^0(M_3,(\Z_N\times \Z_N)/\Z_2)|} \sum_{\substack{(\sigma,\sigma')\in \\ H_2(M_3, (\Z_N\times \Z_N)/\Z_2)}} 
     S_{(1,0)}(\sigma+\sigma') 
    S_{(0,1)}(\sigma'-\sigma)~.
\end{split}
\end{equation}

\subsection{Duality defect in $(3+1)$d from twist defects in $(4+1)$d for even $N$ and $N\geq 4$}
\label{sec:6.8}

We finally insert the extended defects into the $(4+1)$d slab and examine their fate upon shrinking the slab. Like in the previous cases, we  find the correspondence
\begin{eqnarray}
\begin{split}
    \text{Twist defects } V_{(e)} &\longleftrightarrow \text{Duality interface } \CN\\
    \text{Magnetic surface } S_{(0,1)} & \longleftrightarrow \text{$\Z_N^{(1)}$ symmetry defect } \eta\\
    \text{Electric line } S_{(1,0)} & \longleftrightarrow \text{$\Z_N^{(1)}$ order parameter } L 
\end{split}
\end{eqnarray}

\subsubsection{$\Z_N^{(1)}$ symmetry defects and order parameters from bulk surface operators}

For insertions of invertible surfaces, the discussion is identical to the previous cases.
In short, 
\begin{enumerate}
    \item The surface operator $S_{(1,0)}$ can terminate on the boundary orthogonally or be absorbed into the boundary parallelly. Upon shrinking, it survives as a line operator $L$, which is the order parameter of the $\Z_N^{(1)}$ one-form symmetry of the $(3+1)$d QFT $\CX$. 
    \item The surface operator $S_{(0,1)}$, when placed parallel to the boundary, survives upon shrinking. Hence it becomes the symmetry defect $\eta$ for the $\Z_N^{(1)}$ one-form symmetry on the boundary. The surface $\eta$ can have non-trivial linking with $L$, which measures the $\Z_N^{(1)}$ charge of $L$. 
\end{enumerate}

\subsubsection{Duality interface from twist defects}

We next insert a twist defect into the $(4+1)$d slab, giving the configuration shown in Figure \ref{fig:twistdefshrink4d}. From this we see that the twist defect in the $(4+1)$d bulk induces a duality interface upon shrinking. Moreover, thanks to the Dirichlet boundary condition of the SymTFT, the resulting duality interface is independent of the type of the twist defect, 
\begin{eqnarray}\label{eq:NVN>4}
\CN(M_3,M_4)= V_{(e)}(\tau, M_3,M_4)|_{x\rightarrow 0}~, \hspace{1cm} e=0,1~. 
\end{eqnarray}
In other words, there is only one type of duality interface.

It is interesting to see how the fusion rules of the duality interface are reproduced from those of the twist defects. We start from \eqref{eq:VVbarN>4}. It is useful to introduce the new variables, $\tau'=\sigma+\sigma'$ and $\tau=\sigma'-\sigma$. Then $\tau$ and $\tau'$ both belong to $H_2(M_3,\Z_N)$ subject to one constraint, namely that there exists a primitive $\lambda\in H_2(M_3,\Z_N)$ such that $\tau'- \tau=2\lambda$. Then summing over $(\sigma,\sigma')\in H_2(M_3,(\Z_N\times \Z_N)/\Z_2)$ is equivalent to summing over $\tau, \tau'\in H_2(M_3,\Z_N)$ with the constraint that $\tau'-\tau=2\lambda$ for certain $\lambda\in H_2(M_3,\Z_N)$. Because $\lambda$ and $\lambda+N/2\eta$ do not give rise to different $(\tau,\tau')$ for arbitrary $\eta$,  the sum over $H_2(M_3,(\Z_N\times \Z_N)/\Z_2)$ is further equivalent to summing over $\tau\in H_2(M_3,\Z_N)$ and $\lambda\in H_2(M_3,\Z_{N/2})$. In other words, we have
\begin{equation}
V_{(0)}(M_3,M_4)\times \overline{V}_{(0)}(M_3,M_4)= \frac{1}{|H^0(M_3,\Z_N)||H^0(M_3,\Z_{N/2})|} \sum_{\substack{\tau\in H_2(M_3, \Z_N)\\ \lambda\in H_2(M_3,\Z_{N/2})}} 
     S_{(1,0)}(2\lambda) 
    S_{(1,1)}(\tau)~.
\end{equation}
We further apply the discussion around \eqref{eq:cohtorelcoh}, demanding that  $\lambda\in H_2(M_3,M_3,\Z_N)$ when colliding with the Dirichlet boundary condition, and correspondingly also changing the normalization $|H^0(M_3,\Z_{N/2})| \to |H^0(M_3,M_3,\Z_{N/2})|=1$. Furthermore, $S_{(1,1)}(\tau) \to S_{(0,1)}(\tau)=\eta(\tau)$. Hence
\begin{equation}
\begin{split}
    \CN\times \overline{\CN}
    &= \frac{1}{|H^0(M_3,\Z_N)||H^0(M_3,M_3,\Z_{N/2})|} \sum_{\substack{\tau\in H_2(M_3, \Z_N)}} 
    S_{(0,1)}(\tau)\\
    &= \frac{1}{|H^0(M_3,\Z_N)|} \sum_{\substack{\tau\in H_2(M_3, \Z_N)}} 
    \eta(\tau)
\end{split}
\end{equation}
which is the condensation defect. This matches the fusion rule found in \eqref{eq:FR4d}.

\section{$(4+1)$d Symmetry TFT for duality defects}
\label{sec:5dSymTFTTYZN}

In Section \ref{sec:symTFT5d} we studied the SymTFT for a $(3+1)$d theory with $\ZZ_N^{(1)}$ global symmetry, and rederived the fusion rules for the duality interface implementing $\ZZ_N^{(1)}$ gauging. In the current section we demand that the $(3+1)$d theory be self-dual under gauging $\ZZ_N^{(1)}$, which means that the duality interface identified before becomes a duality \textit{defect} in a single theory. The full symmetry (higher-)category is now some analog of the Tambara-Yamagami category $TY(\ZZ_N)$ in $(1+1)$d. In order to obtain the SymTFT corresponding to this larger symmetry, we proceed as in $(2+1)$d by gauging the EM duality symmetry. As before, we do this in two separate ways: first, by tracing the fate of the various topological operators identified in Section \ref{sec:symTFT5d} under gauging; and second, by writing down an explicit expression for the action in terms of twisted cocycles. 

\subsection{Symmetry TFT for duality defects}
\label{sec:5dgaugingmethod1}
We begin by using the topological operators and fusion rules obtained in Section \ref{sec:symTFT5d} to identify the topological operators and fusion rules of the SymTFT after gauging the EM duality symmetry. In Section \ref{sec:5dcocycle} we give an independent cocycle rederivation of many of these results.

\subsubsection{Topological operators}
\label{sec:5dtopops}

Recall that in $(2+1)$d, after gauging the $\Z_2^{\text{EM}}$ symmetry the surface operator $D_{\text{EM}}$ became transparent and the twist defects $\Sigma_{(e)}$ became genuine line operators. As a result, the SymTFT involved only line operators.  In $(4+1)$d, the spectrum of topological operators after gauging the $\Z_4^{\text{EM}}$ symmetry is richer: the SymTFT will have line, surface, and three-manifold operators. Because of this diversity, it will be important to distinguish between local and global fusions of the operators. We begin in this subsection by listing the operators of various dimensions.

\paragraph{Three-manifold operators:} We begin with the three-manifold operators. These operators descend from the twist defects $V_{(e)}(\sigma, M_3, M_4)$ in the $\Z_N$ gauge theory, which were the boundaries of the four-manifold operator implementing EM duality transformations. Upon gauging EM duality this four-manifold becomes transparent, and $V_{(e)}(\sigma, M_3, M_4)$ becomes a genuine three-manifold operator $\widehat V_{(e)}(\sigma, M_3)$ (note that throughout this section, we denote the operators in the $\Z_4^{\text{EM}}$-gauged theory with a hat). Recall that for $N$ odd we have $e=0$, whereas for $N$ even we have $e=0,1$. When $e=0$, the dependence on the surface $\sigma$ drops out.

It is convenient to consider the operators defined on a local patch. Then there are two types of three-manifold operators, 
\begin{eqnarray}\label{eq:threemanifoldoperators5d}
\mathfrak{I}_3, \hspace{0.6cm} \widehat{V}_{(0)}~,
\end{eqnarray}
where $\mathfrak{I}_3$ is the trivial three-manifold operator. Note that this holds for both even and odd $N$; for $N$ even $\widehat V_{(0)}$ and $\widehat V_{(1)}$ are indistinguishable on a local three-dimensional patch, though they can be distinguished at the level of local surface operators, as described in the next paragraph.

\paragraph{Surface operators:} We now turn to surface operators. Before gauging $\Z_4^{\text{EM}}$ there were a total of $N^2$ surface operators $S_{(e,m)}$ defined in (\ref{eq:4dsigma}), which transformed under EM duality as in (\ref{eq:Z4EMoper}). In the current context it is useful to define an equivalent basis of surfaces $\widetilde S_{(e,m)}$ as 
\bea
\label{eq:firstdeftildeS}
\widetilde S_{(e,m)}(\sigma) \,\,\,:=\,\,\, e^{i {2 \pi \over N}\oint_\sigma (e \,b + m\, \widehat b)} \,\,\,=\,\,\, e^{{\pi i \over N} em \CP([\sigma])}\, S_{(e,m)} ~,
\eea
where $\CP([\sigma])$ is the Pontryagin square of the Poincar{\'e} dual of $\sigma$. 
Unlike for the lines encountered in $(2+1)$d, for surfaces in $(4+1)$d the self-pairing $\langle \sigma, \sigma \rangle$ (and hence $\CP([\sigma])$) is not in general trivial, and hence $\widetilde S_{(e,m)}(\sigma)$ is not identical to $S_{(e,m)}$. The surfaces $\widetilde S_{(e,m)}(\sigma)$ have the virtue of transforming more simply under EM duality,
\bea
\widetilde S_{(e,m)}(\sigma)\hspace{0.1 in} \rightarrow \hspace{0.1 in} \widetilde S_{(-m,e)}(\sigma)~. 
\eea
We may now search for EM invariant combinations of $\widetilde S_{(e,m)}(\sigma)$ which can survive the gauging. A single surface $\widetilde S_{(e,m)}(\sigma)$ is EM invariant in isolation when $(e,m) = (-m,e)$ mod $N$. For $N$ odd the only solution to this constraint is $(e,m) = (0,0)$, whereas for $N$ even there are two solutions $(e,m) = (0,0)$ and $(N/2,N/2)$. We will denote these surfaces as
\bea
\label{eq:TypeIsurfaces}
\widehat S_{(0,0)} := \widetilde S_{(0,0)}~, \hspace{0.5 in}\widehat S_{(N/2, N/2)} := \widetilde S_{(N/2, N/2)}~.
\eea
This notation can refer either to the operators defined locally on a patch or globally on a surface, though for the latter we will write the surface dependence explicitly, e.g. $\widehat{S}_{(0,0)}(\sigma)$.

We may next consider a sum of two surfaces $\widetilde S_{(e,m)} \oplus \widetilde S_{(e',m')}$. This is non-simple in the original theory, but after gauging can become simple. We will assume that neither $(e,m)$ nor $(e',m')$ is itself gauge invariant, lest we obtain a direct sum of the simple surfaces in (\ref{eq:TypeIsurfaces}). The condition for the sum to be EM invariant is then
\bea
(-m,e) = (e',m')~, \hspace{0.5 in} (-m', e') = (e,m)~. 
\eea
For $N$ odd there is only a trivial solution $(e,m)=(e',m')=(0,0)$ which reduces to the first entry in \eqref{eq:TypeIsurfaces}. For $N$ even, in addition to the trivial solution we may also choose $(e,m) = (N/2, 0)$ and $(e',m') = (0,N/2)$. We will denote the resulting surface by 
\bea
\widehat S_{\{N/2, 0\}} := \widetilde S_{(N/2, 0)} \oplus \widetilde S_{(0,N/2)}~. 
\eea
Likewise we may consider sums of four surfaces,
\bea
\widehat S_{[e,m]}:=  \widetilde S_{(e,m)}\oplus \widetilde S_{(-m,e)} \oplus \widetilde S_{(-e,-m)} \oplus \widetilde S_{(m,-e)}~.
\eea
For $N$ odd there are a total of ${1\over 4}(N^2 - 1)$ such surfaces, while for $N$ even there are a total of $\left({N\over 2} \right)^2 - 1$ of them (and in particular for $N=2$ there are no surfaces of this type).

Finally we can also consider the non-genuine surface operators that are constrained to live on three-manifolds. For odd $N$, there is a single type of non-trivial non-genuine surface operator \begin{eqnarray}
\mathfrak{I}_2(\widehat{V}_{(0)})~,
\end{eqnarray}
which is the identity surface living on $\widehat{V}_{(0)}$. For even $N$, there are two types of non-genuine surface operators, 
\begin{eqnarray}
\mathfrak{I}_2(\widehat{V}_{(0)})~, \hspace{1cm} \mathfrak{I}_2(\widehat{V}_{(1)})~.
\end{eqnarray}
The surface $\mathfrak{I}_2(\widehat{V}_{(1)})$  is obtained by fusing the identity surface living on the boundary $M_3$ of $V_{(0)}(M_3,M_4)$ with $S_{(1,0)}(\sigma)$ before gauging $\Z_4^{\text{EM}}$. Note that despite our notation here,  $\mathfrak{I}_2(\widehat{V}_{(1)})$ is not the identity surface on a three-manifold $\widehat{V}_{(1)}$, and indeed there is no three-manifold operator $\widehat{V}_{(1)}$. Instead, it is a non-identity surface on $\widehat{V}_{(0)}$.

In summary, the list of non-trivial surfaces is as follows,
\bea\label{eq:surfaceoperator4d}
N\,\,\mathrm{odd}&:& \hspace{0.3 in}\mathfrak{I}_2(\widehat V_{(0)})~, \,\, \widehat S_{(0,0)}~, \,\, \widehat S_{[e,m]}~,
\no\\
N\,\,\mathrm{even}&:& \hspace{0.3 in}\mathfrak{I}_2(\widehat V_{(0)})~, \,\, \mathfrak{I}_2(\widehat V_{(1)})~, \,\,\widehat S_{(0,0)}~, \,\,\widehat S_{({N \over 2}, {N \over 2})}~, \,\, \widehat S_{\{{N \over 2},0\}}~, \,\, \widehat S_{[e,m]}~.
\eea
The operators labelled with an $\mathfrak{I}_2$ are non-genuine surface operators, while those without are genunine surface operators.
When $N=2$, we drop the $\widehat S_{[e,m]}$ in the even $N$ case.

\paragraph{Line operators:} We finally proceed to the line operators. The spectrum of lines is even richer than for surfaces. Upon gauging the $\ZZ_4^\mathrm{EM}$ (or $\ZZ_2^\mathrm{EM}$ for $N=2$) duality symmetry, one obtains a quantum $\ZZ_4^{(3)}$ (or $\ZZ_2^{(3)}$) three-form symmetry. This symmetry is generated by a topological line in $(4+1)$d, which we denote by $K$ in analogy to $(2+1)$d. This is the only non-trivial genuine line operator in the SymTFT.

We should also consider the non-genuine line operators living on three-manifolds and surfaces. We can first consider the identity operators living on each three-manifold or surface, 
\begin{eqnarray}
\begin{split}
\label{eq:identitylines}
    N~\text{odd} :& \hspace{1cm} \mathfrak{I}^0_1(\widehat V_{(0)})~, \,\,  \mathfrak{I}^0_1(\widehat S_{(0,0)})~,  \,\,  \mathfrak{I}_1(\widehat S_{[e,m]})~,\\
    N~\text{even} : & \hspace{1cm} 
    \mathfrak{I}^0_1(\widehat V_{(0)})~, \,\, \mathfrak{I}^0_1(\widehat V_{(1)})~, \,\, \mathfrak{I}^0_1(\widehat S_{(0,0)})~,  \,\,  \mathfrak{I}^0_1(\widehat S_{({N \over 2}, {N \over 2})})~, \,\,  \mathfrak{I}^0_1(\widehat S_{\{{N \over 2},0\}})~, \,\, \mathfrak{I}_1(\widehat S_{[e,m]})~.
\end{split}
\end{eqnarray}
The meaning of superscripts $0$ will be explained below. Since $\widehat S_{(0,0)}$ is a trivial surface operator, the identity line $\mathfrak{I}^0_1(\widehat S_{(0,0)})$ living on it is also a trivial, genuine line. The other line operators are non-genuine.

One can now stack the non-trivial genuine line $K$ on top of the line operators in (\ref{eq:identitylines}). First, since the trivial line $\mathfrak{I}^0_1(\widehat S_{(0,0)})$ can be rewritten as $K^0$, it is useful to rewrite $K^q$ as $\mathfrak{I}^q_1(\widehat S_{(0,0)})$. Here $q$ runs from $0,\dots, 3$ for $N>2$, and through $0,1$ for $N=2$. We then consider stacking $K$ with the non-genuine line operators in (\ref{eq:identitylines}).  To determine whether such stacking will generate a new line operator, we apply a discussion similar to the one around Figure \ref{fig:LemabsorbK}. Since the pre-gauged counterparts of $\mathfrak{I}_1^0(\widehat{V}_{(0)})$,   $\mathfrak{I}_1^0(\widehat{V}_{(1)})$, and $\mathfrak{I}^0_1(\widehat S_{(N/2,N/2)})$ are $\Z_4^{\text{EM}}$ invariant, attaching a $K$ line to them generates new lines, which we denote by $\mathfrak{I}_1^q(\widehat{V}_{(0)})$,   $\mathfrak{I}_1^q(\widehat{V}_{(1)})$, and $\mathfrak{I}^q_1(\widehat S_{(N/2,N/2)})$, with $q\simeq q+4$.  On the other hand, since each of the pre-gauged counterparts of the constituents of $\mathfrak{I}_1^{0}(\widehat{S}_{[e,m]})$ is not $\Z_4^{\text{EM}}$ invariant, the $K$ line can be completely absorbed by $\mathfrak{I}_1^{0}(\widehat{S}_{[e,m]})$. In this case we simply drop the $0$ superscript. 

Somewhat more subtle is the fact that, although each of the pregauged counterparts of the constituents of $\mathfrak{I}_1^{0}(\widehat{S}_{\{N/2,0\}})$ is not $\Z_4^{\text{EM}}$ invariant, they are invariant under the $\Z_2^{\text{EM}}$ normal subgroup of $\Z_4^{\text{EM}}$. As discussed in Appendix \ref{app.absorbK}, this means that $\mathfrak{I}_1^{0}(\widehat{S}_{\{N/2,0\}})$ cannot absorb $K$, but can absorb $K^2$. Hence one can still stack $K$ and define $\mathfrak{I}_1^{p}(\widehat{S}_{\{N/2,0\}})$ with $p\simeq p+2$.

In summary, we have the following line operators, 
\bea\label{eq:lineoperator4d}
    N~\text{odd} :&& \hspace{1cm} \mathfrak{I}^q_1(\widehat V_{(0)})~, \,\,  \mathfrak{I}^q_1(\widehat S_{(0,0)})~,  \,\,  \mathfrak{I}_1(\widehat S_{[e,m]})~,\no\\
    N~\text{even} : && \hspace{1cm} 
    \mathfrak{I}^q_1(\widehat V_{(0)})~, \,\, \mathfrak{I}^q_1(\widehat V_{(1)})~, \,\, \mathfrak{I}^q_1(\widehat S_{(0,0)})~,  \,\,  \mathfrak{I}^q_1(\widehat S_{({N \over 2}, {N \over 2})})~, \,\,  \mathfrak{I}^p_1(\widehat S_{\{{N \over 2},0\}})~, \,\, \mathfrak{I}_1(\widehat S_{[e,m]})~,\no\\
    &&\hspace{0.5 in}q= 0,1,2,3, \hspace{1cm} p=0,1
\eea
For $N=2$, we drop $\mathfrak{I}_1(\widehat S_{[e,m]})$, take $q$ to be defined mod 2, i.e. $q=0,1$, and drop the label $p$ on $\mathfrak{I}^p_1(\widehat S_{\{{N \over 2},0\}})$.

The operators described in \eqref{eq:threemanifoldoperators5d}, \eqref{eq:surfaceoperator4d}, and \eqref{eq:lineoperator4d} together form the (non-condensate) objects, 1-morphisms, and 2-morphisms of the 3-category of the $(4+1)$d SymTFT.\footnote{
In general, a $(4+1)$d TFT can contain 0-, 1-, 2-, 3- and 4-dimensional operators forming a monoidal 4-category. The 3-category in the main text is the endo-category on the trivial 4-dimensional operator. By \cite[Theorem 4]{johnson2020classification}, when there is no nontrivial point operator in a $(4+1)$d TFT, all the 4-dimensional operators are condensation operators.
} We note that all line (resp. surface) operators are endomorphisms of surface (resp. volume) operators. For example, $\widehat{S}_{[e,m]}$ in \eqref{eq:surfaceoperator4d} is an endomorphism of $\mathfrak{I}_3$ in \eqref{eq:threemanifoldoperators5d}, and $\mathfrak{I}_1(\widehat{S}_{[e,m]})$ in \eqref{eq:lineoperator4d} is an endomorphism of $\widehat{S}_{[e,m]}$ in \eqref{eq:surfaceoperator4d}.

\subsubsection{Fusion rules for $N$ odd}

We now discuss the fusion rules of the topological operators discussed above, starting with the case of $N$ odd. We will be somewhat brief in our presentation here, postponing details (and derivation of the F-symbols) to future work. We will follow the method described in \cite{Bhardwaj:2022yxj}, where one first derives the local fusion rules, from which the global fusion rules can then be obtained.

\paragraph{Local fusion of three-manifolds:}
Beginning with local fusions of three-manifold operators, we have simply
\bea
\label{eq:3manifoldfusionaftergauging}
\widehat V_{(0)} \otimes \overline{\widehat V_{(0)}} = \mathfrak{I}_3~.
\eea

\paragraph{Local fusion of surface operators:}
The local fusion rules for the surface operators are
\bea
\label{eq:surfacefusionaftergauging}
\widehat S_{[e,m]} \otimes \widehat S_{[e',m']} &=& 
\widehat S_{[e + e, m+m']} \oplus \widehat S_{[e + m', m-e']} \oplus \widehat S_{[e - e', m-m']} \oplus \widehat S_{[e -m', m+e']}~,
\no\\
&& \hspace{4cm} (e\pm e', e\pm m', m\pm e', m\pm m'\neq 0)\no \\
\widehat S_{[e,m]} \otimes \widehat S_{[e,m]} &=& 
\widehat S_{[2e , 2m]} \oplus 2\widehat S_{[e + m, m-e]} \oplus 4\widehat S_{(0,0)}~,
\no \\
\mathfrak{I}_2(\widehat V_{(0)}) \otimes \overline{\mathfrak{I}_2(\widehat V_{(0)})} &=& \widehat S_{(0,0)} \oplus \bigoplus_{m=0}^{N-1 \over 2}\bigoplus_{n=1}^{N-1 \over 2} \widehat S_{[n,m]}~,
\no\\
\mathfrak{I}_2(\widehat V_{(0)})\otimes \widehat S_{[e,m]} &=& 4\, \mathfrak{I}_2(\widehat V_{(0)})~.
\eea
We make some remarks about these fusion rules:
\begin{enumerate}
    \item The fusion rules amongst $\widehat S_{[e,m]}$ follow straightforwardly from the fusion rules (\ref{eq:4dfusionrule}) amongst $S_{(e,m)}$ in the ungauged theory. Note that the result depends on the value of $(e,m)$ and $(e',m')$, and we  list only two representative cases. In the first case, the condition on $(e,m)$ and $(e',m')$ is such that none of the terms on the right-hand side reduce to $\widehat{S}_{(0,0)}$. When this condition is violated, such as in the fusion between two $\widehat{S}_{[e,m]}$ of the same charge (given in the second line), one must replace $\widehat{S}_{[0,0]}$ by $4\widehat{S}_{(0,0)}$. 
    
Let us add a cautionary remark that the first fusion rule in \eqref{eq:surfacefusionaftergauging} does not strictly hold since, before gauging $\Z_4^{\text{EM}}$,  the global fusion rule  $\widetilde{S}_{(e,m)}(\sigma)\times \widetilde{S}_{(e',m')}(\sigma)= \widetilde{S}_{(e+e',m+m')}(\sigma)$ does not hold. One should instead include a phase factor $e^{\frac{\pi i}{N}(me'-em')\CP([\sigma])}$ on the right hand side of the fusion rule, and we expect a similar phase in the local fusion. However, if we restrict these operators to a three dimensional submanifold $M_3$, the phase factor trivializes, and \eqref{eq:surfacefusionaftergauging} holds. For simplicity, we will assume this throughout this section.
    
    \item The fusion rule between $\mathfrak{I}_2(\widehat V_{(0)})$ and its orientation reversal follows from \eqref{eq:4dVVbar1} before gauging $\Z_4^{\text{EM}}$. Indeed, one can rewrite the global fusion rule \eqref{eq:4dVVbar1} as 
    \begin{eqnarray}
    V_{(0)}(M_3, M_4)\times \overline{V}_{(0)}(M_3, M_4) =  \frac{\mathfrak{I}_3(M_3)}{\Z_N^{(0)}\times \Z_N^{(0)}}~,
    \end{eqnarray}
    where the denominator on the right-hand side means gauging a $\Z_N^{(0)}\times \Z_N^{(0)}$ symmetry on $M_3$. Gauging the zero-form symmetry amounts to inserting a mesh of algebra objects, given by $\CA=\bigoplus_{m,n=0}^{N-1} \widetilde{S}_{(m,0)}\otimes  \widetilde{S}_{(0,n)}= \bigoplus_{m,n=0}^{N-1} \widetilde{S}_{(m,n)}$. After gauging $\Z_4^{\text{EM}}$, we need to recast the algebra object in terms of a linear combination of $\widehat{S}_{(0,0)}$ and $\widehat{S}_{[e,m]}$,
    \begin{eqnarray}
    \CA= \widehat S_{(0,0)} \oplus \bigoplus_{m=0}^{N-1 \over 2}\bigoplus_{n=1}^{N-1 \over 2} \widehat S_{[n,m]}~.
    \end{eqnarray}
    We thus conclude that the local fusion rule is $\mathfrak{I}_2(\widehat V_{(0)}) \otimes \overline{\mathfrak{I}_2(\widehat V_{(0)})} = \CA$, which is precisely the third fusion rule in \eqref{eq:surfacefusionaftergauging}. 
    
    \item The fusion rules between $\mathfrak{I}_2(\widehat V_{(0)})$ and $\widehat S_{[e,m]}$ follow from the fact that $S_{(e,m)}$ can be absorbed by the twist defect in the ungauged theory. 
    
    \item Apart from the fusions discussed above, one can also consider the fusion of two surfaces within the same three manifold (i.e. ``composition" of surface operators).  For simplicity, we will not consider those fusion rules in this work. 
\end{enumerate}

\paragraph{Local fusion of line operators:}
The local fusion rules between line operators are as follows, 
\bea
\mathfrak{I}^q_1(\widehat V_{(0)}) \otimes \overline{\mathfrak{I}^{q'}_1(\widehat V_{(0)})} &=&\mathfrak{I}_1^{q+q'}(\widehat S_{(0,0)}) \oplus \bigoplus_{m=0}^{N-1 \over 2} \bigoplus_{n=1}^{N-1 \over 2} \mathfrak{I}_1(\widehat S_{[n,m]})~,
\no\\
\mathfrak{I}^q_1(\widehat S_{(0,0)}) \otimes \mathfrak{I}^{q'}_1(\widehat S_{(0,0)}) &=& \mathfrak{I}^{q+q'}_1(\widehat S_{(0,0)})~,
\no\\
\mathfrak{I}^q_1(\widehat V_{(0)}) \otimes \mathfrak{I}^{q'}_1(\widehat S_{(0,0)}) &=& \mathfrak{I}^{q+q'}_1(\widehat V_{(0)})~,
\no\\
\mathfrak{I}_1(\widehat S_{[e,m]})\otimes \mathfrak{I}^q_1(\widehat S_{(0,0)})&=& \mathfrak{I}_1(\widehat S_{[e,m]})~,
\no\\
\mathfrak{I}_1(\widehat S_{[e,m]})\otimes \mathfrak{I}^q_1(\widehat V_{(0)})&=&\bigoplus_{q'=0}^3 \mathfrak{I}_1^{q'}(\widehat V_{(0)})~,
\no\\
\mathfrak{I}_1(\widehat S_{[e,m]})\otimes \mathfrak{I}_1(\widehat S_{[e,m]}) &=& \mathfrak{I}_1(\widehat S_{[2e,2m]}) \oplus 2\mathfrak{I}_1(\widehat S_{[e-m,e+m]})\oplus \bigoplus_{q=0}^3 \mathfrak{I}^q_1(\widehat S_{(0,0)})~,\no \\
\mathfrak{I}_1(\widehat S_{[e,m]}) \otimes \mathfrak{I}_1(\widehat S_{[e',m']}) &=& 
\mathfrak{I}_1(\widehat S_{[e + e, m+m']}) \oplus \mathfrak{I}_1(\widehat S_{[e + m', m-e']}) \oplus \mathfrak{I}_1(\widehat S_{[e - e', m-m']}) \oplus \mathfrak{I}_1(\widehat S_{[e -m', m+e']})~.
\no\\
&& \hspace{4cm} (e\pm e', e\pm m', m\pm e', m\pm m'\neq 0)\no \\
\eea
The general form of these fusion rules follows from the fusion between two surfaces given in \eqref{eq:surfacefusionaftergauging}. The only new feature for the line operators is that one is able to assign charges $q$ by stacking with the quantum line $K^q$. To ensure that the $q$ indices are assigned consistently on the two sides of the fusion rule, a useful trick is to stack a $K$ line on both sides and see whether it changes the fusion rule to another consistent one. For instance, consider the fifth fusion rule $\mathfrak{I}_1(\widehat S_{[e,m]})\otimes \mathfrak{I}^q_1(\widehat V_{(0)})=\bigoplus_{q'=0}^3 \mathfrak{I}_1^{q'}(\widehat V_{(0)})$. Stacking a $K$ line does not change the left-hand side because $\mathfrak{I}_1(\widehat S_{[e,m]})$ absorbs the $K$ line, as discussed above. Hence for consistency, the right-hand side should also be able to absorb $K$. The only way to achieve this is to sum over all $q'=0,1,2,3$. The same discussion also applies to the fusion between $\mathfrak{I}_1(\widehat{S}_{[e,m]})$ and itself (see the sixth fusion rule). Note that the outcome of the fusion $\mathfrak{I}_1(\widehat S_{[e,m]}) \otimes \mathfrak{I}_1(\widehat S_{[e',m']})$ depends on whether any of the following $e\pm e', e\pm m', m\pm e', m\pm m'$ are 0 mod $N$. For simplicity, we only listed two representative choices above, which are the last two fusion rules. Other cases can be similarly worked out.

\paragraph{Global fusion from local fusion:}

Given the local fusion rules, one can now construct the global fusion rules by combining local patches together. For instance, consider the global fusion rule between $\widehat{V}_{(0)}(M_3)$ and its orientation reversal on $M_3=S^1\times S^2$. Recall that the relevant local fusion rules are
\begin{eqnarray}
\begin{split}
    &\widehat{V}_{(0)}\otimes \overline{\widehat{V}_{(0)}}=\mathfrak{I}_3, \hspace{1cm} \mathfrak{I}_2(\widehat{V}_{(0)})\otimes \overline{\mathfrak{I}_2(\widehat{V}_{(0)})}=\CA\equiv \widehat S_{(0,0)} \oplus \bigoplus_{m=0}^{N-1 \over 2}\bigoplus_{n=1}^{N-1 \over 2} \widehat S_{[n,m]}~,\\
    & \mathfrak{I}^q_1(\widehat V_{(0)}) \otimes \overline{\mathfrak{I}^{q'}_1(\widehat V_{(0)})} =\mathfrak{I}_1^{q+q'}(\widehat S_{(0,0)}) \oplus \bigoplus_{m=0}^{N-1 \over 2} \bigoplus_{n=1}^{N-1 \over 2} \mathfrak{I}_1(\widehat S_{[n,m]})~,
\end{split}
\end{eqnarray}
which implies that the global fusion between $\widehat{V}_{(0)}(M_3)\times \overline{\widehat{V}_{(0)}}(M_3)$ is the sum of a mesh of the algebra object $\CA$ along every two-cycle of $M_3$. In particular, when $M_3= S^1\times S^2$, the only nontrivial two-cycle is $S^2$, and the only nontrivial one-cycle is $S^1$. We can define $\widehat{V}^q_{(0)}(S^2\times S^1)$ to be
\begin{eqnarray}
\widehat{V}^q_{(0)}(S^2\times S^1)= K^q(S^1)\times  \widehat{V}^0_{(0)}(S^2\times S^1)
\end{eqnarray}
similar to \eqref{eq:VSV}. 
Then the global fusion rule is simply
\begin{eqnarray}
\widehat{V}^q_{(0)}(S^2\times S^1)\times \overline{\widehat{V}^{q'}_{(0)}}(S^2\times S^1) = \frac{1}{N^2} K^{q+q'}(S^1)\times
\CA(S^2)~,
\end{eqnarray}
where the normalization $\frac{1}{N}$ is the standard normalization from gauging a zero form symmetry (i.e. condensing a surface) on a three manifold, $1/|H^0(S^2\times S^1, \Z_N)|^2 =1/N^2$.

From this single example of global fusion, we already encounter the general feature that operators of different dimensions enter the same fusion rule. In contrast, in local fusion rules all the operators entering  a given fusion rule are of the same dimension---objects fuse with objects, 1-morphisms fuse with 1-morphisms, etc. This distinction was not present in the $(2+1)$d case discussed in Section \ref{sec:3dDTYZN}, since there all operators were lines.

\subsubsection{Fusion rules for $N$ even}

We now move on to the case of $N$ even (in the case of $N=2$, one simply drops the fusion rules involving $\widehat S_{[e,m]}$ and restricts $q$ to be $\Z_2$-valued). We will again be brief, leaving details to future work.

\paragraph{Local fusion of three manifolds:}
Beginning with local fusions of three-manifold operators, we have 
\bea
\label{eq:3manifoldfusionaftergauging2}
\widehat V_{(0)} \otimes \overline{\widehat V}_{(0)} = \mathfrak{I}_3~.
\eea

\paragraph{Local fusion of surface operators:}
The local fusion rules between surface operators take the following form, 
\bea
\label{eq:surfacefusionaftergauging2}
&\vphantom{.}&\mathfrak{I}_2(\widehat V_{(e)}) \otimes \overline{\mathfrak{I}_2(\widehat V_{(e)})} =
\left\{\begin{matrix}
\widehat S_{(0,0)} \oplus \widehat S_{(N/2,N/2)} \oplus \substack{\bigoplus_{m=0}^{N/2}\bigoplus_{n=1}^{N/2-1}\\m + n \in 2 \ZZ} \widehat S_{[m,n]}&& N = 2\,\,\,\mathrm{mod}\,\,4
\\\\
\widehat S_{(0,0)} \oplus \widehat S_{(N/2,N/2)} \oplus \widehat S_{\{N/2,0\}} \oplus \substack{\bigoplus_{m=0}^{N/2}\bigoplus_{n=1}^{N/2-1}\\m + n \in 2 \ZZ}\widehat S_{[m,n]}&& N = 0\,\,\,\mathrm{mod}\,\,4
\end{matrix}\right. 
\no\\\no\\
&\vphantom{.}&\mathfrak{I}_2(\widehat V_{(1)}) \otimes \overline{\mathfrak{I}_2(\widehat V_{(0)})}=
\left\{\begin{matrix}
\widehat S_{\{N/2,0\}} \oplus \substack{\bigoplus_{m=0}^{N/2}\bigoplus_{n=1}^{N/2-1}\\m + n \in 2 \ZZ+1} \widehat S_{[m,n]}&& N = 2\,\,\,\mathrm{mod}\,\,4
\\\\
\substack{\bigoplus_{m=0}^{N/2}\bigoplus_{n=1}^{N/2-1}\\m + n \in 2 \ZZ+1}\widehat S_{[m,n]}&& N = 0\,\,\,\mathrm{mod}\,\,4
\end{matrix}\right. 
\no
\eea

\bea
&\vphantom{.}&
\mathfrak{I}_2(\widehat V_{(e)})\otimes \widehat S_{(N/2,N/2)} = \mathfrak{I}_2(\widehat V_{(e)})~,
\no\\
&\vphantom{.}&
\mathfrak{I}_2(\widehat V_{(e)})\otimes \widehat S_{\{N/2,0\}} = 2 \mathfrak{I}_2(\widehat V_{(e+N/2)})~,
\no\\
&\vphantom{.}&
\mathfrak{I}_2(\widehat V_{(e)})\otimes \widehat S_{[e',m']} = 4\, \mathfrak{I}_2(\widehat V_{(e+e'+m')})~,
\no\\
&\vphantom{.}&
\widehat S_{(N/2,N/2)}\otimes \widehat S_{(N/2,N/2)} = \widehat S_{(0,0)}~,
\no\\
&\vphantom{.}&\widehat S_{(N/2,N/2)}\otimes \widehat S_{\{N/2,0\}} = \widehat S_{\{N/2,0\}} ~,
\no\\
&\vphantom{.}& \widehat S_{\{N/2,0\}}\otimes  \widehat S_{\{N/2,0\}} = 2  \widehat S_{(0,0)}\oplus2  \widehat S_{(N/2,N/2)}~,
\no\\
&\vphantom{.}&\widehat S_{(N/2,N/2)} \otimes \widehat S_{[e,m]} = \widehat S_{[e+N/2,m+N/2]}~,
\no\\
&\vphantom{.}&\widehat S_{\{N/2,0\}} \otimes \widehat S_{[e,m]} = \widehat S_{[e+N/2,m]}\oplus \widehat S_{[e,m+N/2]}~,
\no\\
&\vphantom{.}&\widehat S_{[e,m]} \otimes  \widehat S_{[e',m']} = \widehat S_{[e + e', m+m']} \oplus \widehat S_{[e + m', m-e']} \oplus \widehat S_{[e - e', m-m']} \oplus \widehat S_{[e -m', m+e']}~. \\
&& \hspace{7cm} (e\pm e', e\pm m', m\pm e', m\pm m'\neq 0, N/2)\no
\eea
The derivation of these fusion rules is mostly the same as for the odd $N$ case. Let us comment only on some of the new features,
\begin{enumerate}
    \item There are two non-genuine surface operators coming from the twist defects, $\mathfrak{I}_2(\widehat V_{(0)})$ and $\mathfrak{I}_2(\widehat V_{(1)})$. Before gauging $\Z_4^{\text{EM}}$, we know from \eqref{eq:VVbarN>4} that the global fusion rule between $V_{(0)}$ and its orientation reversal is 
    \begin{eqnarray}
    V_{(0)}(M_3, M_4)\times \overline{V}_{(0)}(M_3, M_4) = \frac{\mathfrak{I}_3(M_3)}{(\Z_N^{(0)}\times \Z_N^{(0)})/\Z_2^{(0)}}~.
    \end{eqnarray}
    This implies that the right-hand side is a condensation of the algebra object of $(\Z_N^{(0)}\times \Z_N^{(0)})/\Z_2^{(0)}$, which is given by the surface 
    \begin{eqnarray}
    \CA= \bigoplus_{\substack{m,n=0\\ (m,n)\simeq (m+\frac{N}{2}, n+\frac{N}{2})}}^{N-1} \widetilde{S}_{(m+n, n-m)} = 
    \bigoplus_{\substack{m,n=0\\ m+n\in 2\Z}}^{N-1} \widetilde{S}_{(m,n)}~.
    \end{eqnarray}
    After gauging, one should express the right-hand side of the above in terms of the operators invariant under $\Z_4^{\text{EM}}$. It turns out that the result depends on whether $N/2$ is even or odd, and yields the expression in the first fusion rule of \eqref{eq:surfacefusionaftergauging2}. For the second fusion rule, the algebra object before gauging $\Z_4^{\text{EM}}$ should instead be $\CA'= \bigoplus_{\substack{m,n=0\\ m+n\in 2\Z+1}}^{N-1} \widetilde{S}_{(m,n)}$. Rewriting in terms of $\Z_4^{\text{EM}}$ invariant surfaces, one finds the right-hand side of the second fusion rule of \eqref{eq:surfacefusionaftergauging2}, which again depends on the parity of $N/2$.
    
    \item For even $N$, there are additional genuine surface operators $S_{(N/2, N/2)}$ and $S_{\{N/2,0\}}$. When fusing $S_{\{N/2,0\}}$ with $\mathfrak{I}_2(\widehat V_{(e)})$ one gets two copies of $\mathfrak{I}_2(\widehat V_{(e)})$ for even  $N/2$, and two copies of $\mathfrak{I}_2(\widehat V_{(e+1)})$ for odd $N/2$. This follows from the definition of $V_{(1)}$ before gauging. 
\end{enumerate}

\paragraph{Local fusion of line operators:}
The local fusion rules between lines are as follows, 
\begin{eqnarray}
&\vphantom{.}&\mathfrak{I}^q_1(\widehat V_{(e)}) \otimes \overline{\mathfrak{I}^{q'}_1(\widehat V_{(e)})}=
\no\\
&\vphantom{.}&\hspace{0.1 in}\,\,\,
\left\{\begin{matrix}
\mathfrak{I}_1^{q+q'}(\widehat S_{(0,0)}) \oplus \mathfrak{I}_1^{q+q'{+2e}}(\widehat  S_{(N/2,N/2)}) \oplus \substack{\bigoplus_{m=0}^{N/2}\bigoplus_{n=1}^{N/2-1}\\m+n \in 2 \ZZ}\mathfrak{I}_1(\widehat S_{[m,n]})&& N = 2\,\,\,\mathrm{mod}\,\,4
\\\\
\mathfrak{I}_1^{q+q'}(\widehat  S_{(0,0)}) \oplus \mathfrak{I}_1^{q+q'+2{e}}(\widehat S_{(N/2,N/2)}) \oplus \mathfrak{I}_1^{[q+q']_2}(\widehat S_{\{N/2,0\}}) \oplus \substack{\bigoplus_{m=0}^{N/2}\bigoplus_{n=1}^{N/2-1}\\m+n \in 2 \ZZ}\mathfrak{I}_1(\widehat S_{[m,n]})&& N = 0\,\,\,\mathrm{mod}\,\,4
\end{matrix}\right.
\no\\\no\\
&\vphantom{.}&\mathfrak{I}^q_1(\widehat V_{(1)}) \otimes \overline{\mathfrak{I}^{q'}_1(\widehat V_{(0)})}=\left\{\begin{matrix}
\mathfrak{I}_1^{[q+q']_2}(\widehat S_{\{N/2,0\}}) \oplus \substack{\bigoplus_{m=0}^{N/2}\bigoplus_{n=1}^{N/2-1}\\m+n \in 2 \ZZ+1}\mathfrak{I}_1(\widehat S_{[m,n]})&& N = 2\,\,\,\mathrm{mod}\,\,4
\\\\
\substack{\bigoplus_{m=0}^{N/2}\bigoplus_{n=1}^{N/2-1}\\m+n \in 2 \ZZ+1}\mathfrak{I}_1(\widehat S_{[m,n]})&& N = 0\,\,\,\mathrm{mod}\,\,4
\end{matrix}\right.
\no\\
&\vphantom{.}&\mathfrak{I}^q_1(\widehat V_{(e)}) \otimes \mathfrak{I}^{q'}_1(\widehat S_{(0,0)}) = \mathfrak{I}^{q+q'}_1(\widehat V_{(e)})~,
\no\\
&\vphantom{.}&\mathfrak{I}^q_1(\widehat V_{(e)}) \otimes \mathfrak{I}^{q'}_1(\widehat S_{(N/2,N/2)}) = \mathfrak{I}^{q+q'+2e}_1(\widehat V_{(e)})~,
\no\\
&\vphantom{.}&\mathfrak{I}^q_1(\widehat V_{(e)}) \otimes \mathfrak{I}^{[q']_2}_1(\widehat S_{\{N/2,0\}}) = \mathfrak{I}^{q+q'}_1(\widehat V_{(e+N/2)})\oplus\mathfrak{I}^{q+q'+2}_1(\widehat V_{(e+N/2)}) ~,
\no\\
&\vphantom{.}&\mathfrak{I}^q_1(\widehat V_{(e)})\otimes \mathfrak{I}_1(\widehat S_{[e',m']})= \bigoplus_{q'=0}^3\mathfrak{I}_1^{q'}(\widehat V_{e+e'+m'})~, \hspace{1.5 in}~,
\no\\
&\vphantom{.}&\mathfrak{I}^q_1(\widehat S_{(0,0)}) \otimes \mathfrak{I}^{q'}_1(\widehat S_{(0,0)}) = \mathfrak{I}^{q+q'}_1(\widehat S_{(0,0)})~,
\no\\
&\vphantom{.}&\mathfrak{I}^q_1(\widehat S_{(N/2,N/2)}) \otimes \mathfrak{I}^{q'}_1(\widehat S_{(N/2,N/2)}) = \mathfrak{I}^{q+q'}_1(\widehat S_{(0,0)})~,
\no\\
&\vphantom{.}&\mathfrak{I}^q_1(\widehat S_{(0,0)}) \otimes \mathfrak{I}^{q'}_1(\widehat S_{(N/2,N/2)}) = \mathfrak{I}^{q+q'}_1(\widehat S_{(N/2,N/2)})~,
\no\\
&\vphantom{.}&\mathfrak{I}^q_1(\widehat S_{(N/2,N/2)})\otimes \mathfrak{I}^{[q']_2}_1(\widehat S_{\{N/2,0\}}) =  \mathfrak{I}^{[q+q']_2}_1(\widehat S_{\{N/2,0\}})~,
\no\\
&\vphantom{.}&\mathfrak{I}^{[q]_2}_1(\widehat S_{\{N/2,0\}}) \otimes \mathfrak{I}^{[q']_2}_1(\widehat S_{\{N/2,0\}}) = 2 \mathfrak{I}^{q+q'}_1(\widehat S_{(0,0)}) + 2\mathfrak{I}^{q+q'+2}_1(\widehat S_{(0,0)})  ~,
\no
\\
&\vphantom{.}&\mathfrak{I}^{[q]_2}_1(\widehat S_{\{N/2,0\}}) \otimes \mathfrak{I}^{q'}_1(\widehat S_{(0,0)}) = \mathfrak{I}^{[q+q']_2}_1(\widehat S_{\{N/2,0\}})~,
\no\\
&\vphantom{.}&\mathfrak{I}^q_1(\widehat S_{(0,0)})\otimes \mathfrak{I}_1(\widehat S_{[e,m]})= \mathfrak{I}_1(\widehat S_{[e,m]})~,
\no\\
&\vphantom{.}&\mathfrak{I}^q_1(\widehat S_{(N/2,N/2)})\otimes \mathfrak{I}_1(\widehat S_{[e,m]}) = \mathfrak{I}_1(\widehat S_{[e+N/2, m+N/2]}~,
\no\\
&\vphantom{.}&\mathfrak{I}^{[q]_2}_1(\widehat S_{\{N/2,0\}})\otimes \mathfrak{I}_1(\widehat S_{[e,m]}) = \mathfrak{I}_1(\widehat S_{[e+N/2,m]}) \oplus\mathfrak{I}_1(\widehat S_{[e,m+N/2]})~,\no
\\
&\vphantom{.}&\mathfrak{I}_1(\widehat S_{[e,m]})\otimes \mathfrak{I}_1(\widehat S_{[e,m]}) = \mathfrak{I}_1(\widehat S_{[2e,2m]}) \oplus 2\mathfrak{I}_1(\widehat S_{[e-m,e+m]})\oplus \bigoplus_{q=0}^3 \mathfrak{I}^q_1(\widehat S_{(0,0)}),\no\\
&&\hspace{10cm} (2e, 2m, e\pm m\neq 0, N/2)
\end{eqnarray}
Many of these fusion rules, including the first two, follow from those for surfaces in (\ref{eq:surfacefusionaftergauging2}). Let us make some remarks about the features that were not present in the cases already discussed:
\begin{enumerate}
    \item Note that the non-genuine line $\mathfrak{I}_1(\widehat S_{\{N/2,0\}})$ can be assigned a $\Z_2$ charge $[q]_2$. Here, we take $q\in\Z_4$, and denote its mod 2 reduction as $[q]_2$, i.e. $[q]_2= q\mod 2$. This is described in Appendix \ref{app.absorbK}.
    \item It is useful to check that both sides of the fusion rule $ \mathfrak{I}^q_1(\widehat V_{(e)}) \otimes \mathfrak{I}^{[q']_2}_1(\widehat S_{\{N/2,0\}}) = \mathfrak{I}^{q+q'}_1(\widehat V_{(e+N/2)})\oplus\mathfrak{I}^{q+q'+2}_1(\widehat V_{(e+N/2)})$ depend on $q'$ only via its mod 2 reduction. In other words, one can check that the result does not change under shifting $q'\to q'+2$. 
    
    \item We should make a cautionary remark that the $+2e$ shift of the $q$ charge assignment on the right-hand side of the fusion rule ${\mathfrak{I}^q_1(\widehat V_{(e)}) \otimes \mathfrak{I}^{q'}_1(\widehat S_{(N/2,N/2)}) = \mathfrak{I}^{q+q'+2e}_1(\widehat V_{(e)})}$ is at present conjectural, and is motivated by the lower-dimensional calculation in Section \ref{sec:ZTYZN}. To actually prove the presence of this shift, one must measure the charge of the junction between the three lines by wrapping a 4d condensation defect around it. This is analogous to the computation performed in Appendix \ref{app:junctioncharge} for $(2+1)$d. We will leave this calculation to future work. 
\end{enumerate}

\subsection{Twisted cocycle description}
\label{sec:5dcocycle}
We close our analysis of the $(4+1)$d SymTFT by presenting an explicit twisted cocycle description of the theory. As in $(2+1)$d, our starting point is the BF theory (\ref{eq:5dBFtheory}). We will find it more useful to write this in K-matrix form,
\begin{eqnarray}\label{5dZN}
S = \frac{2\pi}{2N}\int \mathbf{b}^T \cup K\, \delta \mathbf{b}
\end{eqnarray}
where $\mathbf{b}= (b,\widehat b)$ is a two component cochain valued in $\Z$ and $K=i\sigma^y$.\footnote{Once again, we hope that the reader will not confuse the $K$ matrix here with the generator $K$ of the quantum symmetry.} Note that $K$ is no longer symmetric as it was in $(2+1)$d, due to the anticommutativity properties of form fields in $(4+1)$d.

As before, our goal is to gauge $\Z_4^{\mathrm{EM}}$ (or $\Z_2^{\mathrm{EM}}$ for $N=2$). We begin by coupling to a background gauge field $C$, which as in $(2+1)$d amounts to promoting $\mathbf{b}$ to a twisted cocycle. This modifies the gauge transformation of $\mathbf{b}$ to 
\begin{eqnarray}\label{5dGT}
\mathbf{b} _{ijk}\to \mathbf{b} _{ijk} + K^{C_{ij}} \mathbf{h}_{jk} - \mathbf{h}_{ik}+ \mathbf{h}_{ij}~.
\end{eqnarray}
The action then becomes 
\begin{eqnarray}
S[C]= \frac{2\pi}{2N} \int \mathbf{b}^T \cup_C K \delta_C \mathbf{b}~,
\end{eqnarray}
where the integrand in components reads
\begin{eqnarray}
\left(\mathbf{b}^T \cup_C K \delta_C \mathbf{b}\right)_{ijklpq}= \mathbf{b}_{ijk}^T K^{C_{ik}+1} \left(K^{C_{kl}} \mathbf{b}_{lpq} - \mathbf{b}_{kpq} + \mathbf{b}_{klq} - \mathbf{b}_{klp}\right)~.
\end{eqnarray}
The action is invariant under the dynamical gauge transformation \eqref{5dGT} as well as the background gauge transformations 
\begin{eqnarray}
C_{ij} \to C_{ij} + \gamma_j - \gamma_i~,\hspace{0.5 in} \mathbf{b} _{ijk}\to K^{-\gamma_i } \mathbf{b}_{ijk}~,\hspace{0.5 in} \mathbf{h}_{ij}\to K^{-\gamma_i} \mathbf{h}_{ij}~.
\end{eqnarray}
Gauging the $\Z_4^{\mathrm{EM}}$ symmetry amounts to promoting $C$ to a dynamical field $c$, giving
\begin{eqnarray}
\label{eq:5dSymTFTaction}
S_{\mathrm{SymTFT}}=\frac{2\pi}{N} \int \mathbf{b}^T \cup_c K \delta_c \mathbf{b} + {\pi\over 2} \int x \delta c~,
\end{eqnarray}
where $x$ is a $\Z_4$-valued 3-cochain. See Appendix B of \cite{Benini:2018reh} for more details on twisted cocycles. 

As in $(2+1)$d, the complicated dependence on $c$ in the kinetic term means that $c$ is now flat only on-shell, and hence the action is no longer invariant under the gauge transformations in (\ref{5dGT}). Instead, we must restrict to gauge transformations satisfying the analog of (\ref{eq:constraintong}), namely
\bea
\label{eq:hgtconst}
(K^{\delta c_{ijk}} - \mathds{1}) \mathbf{h}_{k \ell} = 0~.
\eea
This constraint will allow surfaces to end on sources of flux for $c$, which will give rise to  junctions. Note however that unlike in the case of $(2+1)$d for which the constraint on $\mathbf{g}$ in the presence of non-trivial $c$ flux was $(1,-1)\mathbf{g}=0$, the above constraint on $\mathbf{h}$ instead requires $\mathbf{h}=(0,0)^T$ for $N$ odd, and $\mathbf{h} = (0,0)^T$ or $(N/2, N/2)^T$ for $N$ even. This will give rise to more potential junctions than in lower dimensions. 

\subsubsection{Operators in Symmetry TFT}

We now give a description of the gauge-invariant operators of the theory, beginning with surface operators. 
The full gauge transformations for the theory are 
\bea
\label{eq:5dGT}
&\vphantom{.}&\mathbf{b}_{ijk}\to K^{-\gamma_i}\left(\mathbf{b}_{ijk} + (\delta_c \mathbf{h})_{i j k} \right)~,\hspace{0.4 in} c_{ij} \rightarrow c_{ij} + (\delta \gamma)_{ij}~,
\no\\
&\vphantom{.}& x^{(3)}_{ijk\ell} \rightarrow x^{(3)}_{ijk\ell} + (\delta \eta^{(2)})_{i j k \ell}~,
\eea
with $\mathbf{h}$ constrained to satisfy (\ref{eq:hgtconst}). 

\paragraph{Surface operators} We may first consider the invertible surface operators, given by
\begin{eqnarray}\label{5dLn}
\widetilde{S}_{\mathbf{n}}(\sigma)= e^{i\frac{2\pi}{N} \oint_\sigma \mathbf{n}^T \mathbf{b}}~, \hspace{0.5 in} \mathbf{n}\in \Z_N\times \Z_N~.
\end{eqnarray}
This operator is identical to the one defined in (\ref{eq:firstdeftildeS}). We recall that, due to the non-commutativity of $b$ and $\widehat b$, the surface $\widetilde S_{\mathbf{n}}$ differs from the surface $S_{\mathbf{n}}$ defined in (\ref{eq:4dsigma}) by a phase, i.e. $\widetilde{S}_{\mathbf{n}}(\sigma)= {S}_{\mathbf{n}}(\sigma) e^{\frac{\pi i}{N} n_1 n_2 \CP([\sigma])}$, where $\CP([\sigma])$ is the Pontryagin square of the Poincar{\'e} dual of $\sigma$ and $n_{1,2}$ are the two components of $\mathbf{n}$. We prefer to use $\widetilde{S}_{\mathbf{n}}(\sigma)$ because it transforms simply as $\widetilde{S}_{\mathbf{n}}\to \widetilde{S}_{K\mathbf{n}}$ under $\Z_4^{\text{EM}}$, and hence it is $\Z_4^{\text{EM}}$ invariant if and only if $\mathbf{n}^T K = \mathbf{n}^T$. A straightforward analysis paralleling that in Section \ref{sec:5dtopops} then reproduces the spectrum of gauge-invariant surfaces $\widehat S_{[e,m]}$.

\paragraph{Line and 3-manifold operators} 

In addition to the surface operators studied above, there is also a topological line operator 
\bea
K(\gamma) := e^{{\pi i \over 2} \oint_\gamma c}
\eea
generating the $\widehat \ZZ_4$ symmetry quantum dual to $\ZZ_4^{\mathrm{EM}}$. There is also a three-manifold operator 
\bea
\chi(M_3):=e^{{\pi i \over 2} \oint_{M_3} x}~,
\eea
but as in $(2+1)$d this is generically non-topological since $x$ is not closed in general. However, with appropriate restrictions on $b$ and $\widehat b$ the field $x$ can become closed. For odd $N$, we require $b = \widehat b=0$ at the locus $M_3$.\footnote{This is chosen such that $K$ acts as the identity on $\mathbf{b}=(b , \widehat b)$, and hence such that $c$ drops out of the kinetic term of (\ref{eq:5dSymTFTaction}) and can act as a proper Lagrange multiplier field for $x$.} We may then construct three-manifold operators $\widehat V_{(0)}(M_3)$ via
\bea
\widehat V_{(0)} (M_3) := \frac{|H^1(M_3, \Z_N)|}{|H^0(M_3,\Z_N)|}\chi(M_3)\, \delta_{\Z_N}(\,b\,)\big|_{M_3}\, \delta_{\Z_N}(\,\widehat b\,)\big|_{M_3}~,
\eea
which are the images of the twist defects upon gauging $\ZZ_4^{\mathrm{EM}}$. The normalization is chosen to match the fusion rules derived in Section \ref{sec:5dgaugingmethod1}. The delta functions may alternatively be rewritten as 
\bea
\label{eq:deltafuncts5d}
\delta_{\Z_N}(\,b\,)\big|_{M_3}\, \delta_{\Z_N}(\,\widehat b\,)\big|_{M_3} &=& {1\over |H^1(M_3, \ZZ_N)|^2}\sum_{[\sigma]\in H^1(M_3, \ZZ_N)} e^{{2 \pi i \over N} \int [\sigma] \,\cup\, b} \times \sum_{[\sigma']\in H^1(M_3, \ZZ_N)} e^{{2 \pi i \over N} \int [\sigma'] \,\cup\, \widehat b}
\no\\
&=& {1\over |H^1(M_3, \ZZ_N)|^2} \sum_{\sigma, \sigma' \in H_2(M_3, \ZZ_N)} S_{(1,0)}(\sigma) S_{(0,1)}(\sigma')~.
\eea
For even $N$, we instead require that $(b,\widehat{b})$ be trivial as an element in $H^2(M_3, (\Z_N\times \Z_N)/\Z_2)$ on the defect locus. We may then construct the three-manifold operator $\widehat{V}_{(0)}(M_3)$ via 
\begin{eqnarray}
\widehat V_{(0)} (M_3) := \frac{|H^1(M_3, (\Z_N\times \Z_N)/\Z_2)|^{\frac{1}{2}}}{|H^0(M_3,(\Z_N\times \Z_N)/\Z_2)|^{\frac{1}{2}}}\chi(M_3)\, \delta_{(\Z_N\times \Z_N)/\Z_2}(\,(b,\widehat{b})\,)\big|_{M_3}~,
\end{eqnarray}
where the delta function can be rewritten as 
\begin{equation}
\delta_{(\Z_N\times \Z_N)/\Z_2}(\,(b,\widehat{b})\,)\big|_{M_3} = {1\over |H^1(M_3, (\Z_N\times \Z_N)/\Z_2)|} \sum_{(\sigma, \sigma') \in H_2(M_3, (\Z_N\times \Z_N)/\Z_2)} S_{(1,0)}(\sigma+\sigma') S_{(0,1)}(\sigma'-\sigma)~.
\end{equation}
The remaining twist defects $\widehat V_{(1)} (\sigma, M_3)$ can then be obtained by stacking with $S_{(1,0)}$ as before. We can also consider the 1-endomorphisms and 2-endomorphisms on the three manifolds and surfaces, which give rise to non-genuine surfaces  and lines. The discussion is similar to that in Section \ref{sec:5dtopops}, and hence we will not repeat it here. 

\subsubsection{Junctions and local fusion rules}

We now ask about gauge-invariant junctions between the operators above. The analysis here is done following that in Section \ref{sec:3djunc}. We begin by considering $\widetilde S_{(e,m)}$ on a 2-chain $\sigma$ with boundary. Such a configuration is not gauge-invariant, but instead transforms as 
\bea\label{eq:Semtransform}
\widetilde S_{(e,m)}(\sigma) \rightarrow \widetilde S_{(e,m)}(\sigma) e^{{2 \pi i \over N} (e,m) \mathbf{h} |_{\partial \sigma} }~.
\eea
Since the gauge-invariant operators $\widehat S_{[e,m]}$ are built out of these surfaces, in order for them to be well-defined in the presence of a boundary, one of the following must be satisfied: 
\begin{itemize}
\item The surface ends on a line together with other surfaces $\widehat S_{[e,m]}$ such that the total charge cancels.
\item The surface ends on a locus with non-zero $c$ flux. 
\end{itemize}
The first of these allows for gauge-invariant junctions between three surface defects,
\bea
\widetilde S_{(e+e',m+m')}\subset \widetilde S_{(e,m)} \otimes \widetilde S_{(e',m')} ~.
\eea
This then gives rise to a number of junctions between gauge-invariant surface operators. When all are accounted for, we obtain precisely the same fusion rules as in (\ref{eq:surfacefusionaftergauging}) and (\ref{eq:surfacefusionaftergauging2}).

On the other hand, we may also consider the surfaces ending on a locus with non-zero $c$ flux. In this case equation (\ref{eq:hgtconst}) enforces $\mathbf{h}= (0,0)^T$ or $(N/2, N/2)^T$, with the latter only possible when $N$ is even. Assuming first that $N$ is odd, we have $\mathbf{h}= (0,0)^T$ and hence \textit{any} surface is allowed to end on $\widehat V_{(0)}$. However, we cannot really discuss trivalent junctions of $\widehat S_{[e,m]}$ and $\widehat V_{(0)}$ directly, since the former is supported on a surface while the latter is supported on a three-manifold. Instead, we can summarize the above observation by saying that the identity surface in $\widehat V_{(0)}$, which we previously denoted by $\mathfrak{I}_2(\widehat V_{{(0)}})$, admits a trivalent junction with $\widehat S_{[e,m]}$ of the form 
\bea
\mathfrak{I}_2(\widehat V_{{(0)}})\subset \widehat S_{[e,m]} \otimes  \mathfrak{I}_2(\widehat V_{{(0)}}) ~, \hspace{1cm}  N\in 2\Z+1~.
\eea
This is consistent with the last local fusion rule in \eqref{eq:surfacefusionaftergauging}.

For $N$ even, we instead allow both $\mathbf{h}= (0,0)^T$ and $(N/2, N/2)^T$ at loci of non-zero $c$ flux. Under the latter, we have from \eqref{eq:Semtransform} that $\widetilde S_{(e,m)}(\sigma)$ transforms as 
\bea
\label{eq:SemgthN2}
\widetilde S_{(e,m)}(\sigma) \rightarrow \widetilde S_{(e,m)}(\sigma) e^{{\pi i} (e+m)}~,
\eea
which means that only operators $\widehat S_{(e,m)}$ with $e+m$ even can end on $\widehat V_{(0)}$, giving trivalent junctions of the form $\mathfrak{I}_2(\widehat V_{{(0)}})\subset \widehat S_{[e,m]} \otimes  \mathfrak{I}_2(\widehat V_{{(0)}}) $. Contrarily, when $e+m$ is odd $\widehat S_{[e,m]}$ cannot end alone, but can form a trivalent junction 
\bea
\mathfrak{I}_2(\widehat V_{{(1)}})\subset \widehat S_{[e,m]} \otimes \mathfrak{I}_2(\widehat V_{{(0)}})~, \hspace{1cm} N\in 2\Z~, \hspace{0.5cm} e+m\in 2\Z+1
\eea
since the gauge non-invariance of $\mathfrak{I}_2(\widehat V_{{(1)}})$ at the boundary (coming from the gauge non-invariance of $S_{(1,0)}$) precisely cancels that in (\ref{eq:SemgthN2}). These reproduce the local fusion rule of the surfaces in \eqref{eq:surfacefusionaftergauging2}, and we see that a trivalent junction is allowed (by gauge invariance) whenever the three surface operators appear in the same fusion rule in \eqref{eq:surfacefusionaftergauging} and \eqref{eq:surfacefusionaftergauging2}. Similar techniques could be used to study trivalent junctions between lines, but we leave this to future work. 

Let us close by mentioning that one might aim to use the Hopf linking of the operators above, which are readily computed, to define some higher analog of the S-matrix, along the lines of \cite{Johnson-Freyd:2021chu,reuttertalk}. We do not comment on this further here.

\section{Application: Intrinsic vs. non-intrinsic non-invertible duality defects}
\label{sec:SymTFTspinnonspin}

In this final section we briefly discuss one application of SymTFTs, namely to the determination of whether a given duality defect is intrinsically non-invertible or not. We will only discuss the case of $(1+1)$d bosonic theories, with the results for spin theories or theories in $(3+1)$d left for future work.

\subsection{Topological manipulations}

Given any $(1+1)$d QFT $\CX$ with a non-anomalous $\Z_N^{(0)}$ zero-form symmetry, one can define a non-trivial topological manipulation $\sigma_{r}$ for any $r$ dividing $N$, defined by gauging the $\Z_r$ normal subgroup of $\Z_N$. In terms of partition functions, we have
\begin{eqnarray}
Z_{\sigma_r\CX} [X_2, (A,A')]= \frac{1}{|H^0(X_2, \Z_r)|} \sum_{a \in H^1(X_2, \Z_r)} Z_{\CX}\left[X_2, \frac{N}{r} a + \tilde{A}'\right]\, e^{\frac{2\pi i}{r} \int_{X_2} a A}
\end{eqnarray}
where $A$ is a $\Z_r$ background field, $A'$ is a $\Z_{N/r}$ background field, and $\tilde{A}'$ is its $\Z_N$ lift. The combination $\frac{N}{r}a + \tilde{A}'$ is a $\Z_N$ gauge field.  After gauging $\Z_{r}$, the non-anomalous $\Z_N$ symmetry becomes a product symmetry $\Z_r\times \Z_{N/r}$ with a nontrivial mixed anomaly between them \cite{Tachikawa:2017gyf}. The anomaly is given by $e^{\frac{2\pi i}{r}\int_{X_3} A \beta(A') }$
where $\beta(A')$ is defined by the symmetry extension $1\to \Z_r\to \Z_N \to \Z_{N/r}\to 1$ specified by $\delta a = \beta(A)$.

Starting from $\sigma_r \CX$, there are two topological manipulations that one can perform. The first, which we denote by $\tau$ is stacking with a $\Z_r\times \Z_{N/r}$ SPT, 
\begin{eqnarray}
Z_{\tau \sigma_r \CX}[X_2, (A,A')] = Z_{\sigma_r \CX}[X_2, (A,A')] e^{\frac{2\pi i}{\gcd(r,N/r)} \int_{X_2} AA'}~.
\end{eqnarray}
The second is gauging of a $\Z_{r'}$ subgroup of $\Z_{r}\times \Z_{N/r}$. The resulting theory has the global symmetry $\Z_{r'}\times \frac{\Z_{r}\times \Z_{N/r}}{\Z_r'}$ with an appropriate mixed anomaly. One can then perform suitable $\sigma$ or $\tau$ transformations on the resulting theory.

Suppose that the theory $\CX$ is invariant under gauging $\Z_N$ (up to an Euler counterterm), i.e. 
\begin{eqnarray}
\CX = \CX/\Z_N \equiv \sigma_{N} \CX~.
\end{eqnarray}
If this is the case, then as we have seen above $\CX$ admits a Tambara-Yamagami extension of its $\Z_N$ symmetry. We now ask if, starting from $\CX$, we can perform a sequence of topological manipulations $\phi$ such that the $TY(\Z_N)$ category of $\CX$ is replaced by a group-like category $\phi(TY(\Z_N))$ in a theory $\phi(\CX)$. 
The non-invertible symmetry is referred to as non-intrinsic or intrinsic depending on if such a $\phi$ exists or does not exist,
\begin{eqnarray}
\begin{split}
    \phi \text{ exists }:&\hspace{0.2 in} TY(\Z_N)~\text{is non-intrinsically non-invertible}~,\\
    \phi \text{ doesn't exist }:&\hspace{0.2 in} TY(\Z_N)~\text{is intrinsically non-invertible}~.
\end{split}
\end{eqnarray}
Because the possible set of topological manipulations is complicated when $N$ is large, it is difficult to enumerate all possible chains of $\sigma_r$ and $\tau$ operations and demonstrate that such a topological manipulation $\phi$ does or does not exist.

On the other hand, the SymTFT  allows us to answer this question rather straightforwardly. This is due to two basic facts, both of which have been used throughout this paper: 
\begin{enumerate}
    \item The SymTFT is an invariant under topological manipulations,
    \item The SymTFT for group-like symmetries is a Dijkgraaf-Witten theory.
\end{enumerate}
What this means is that the duality defect in $TY(G)$ is non-intrinsically non-invertible \textit{if and only if the SymTFT is a bosonic Dijkgraaf-Witten theory}.
Using the explicit form of the SymTFT obtained in Section \ref{sec:3dDTYZN}, we may now propose a sufficient criteria for when the duality defect of $TY(\Z_N)$ is non-intrinsically non-invertible.

\subsection{A sufficient condition for non-intrinsically non-invertible symmetry}

Our goal is to see whether the SymTFT of $TY(\Z_N)$ obtained in Section \ref{sec:3dDTYZN} is a DW theory. Recall that a $(2+1)$d DW theory is a finite group gauge theory with gauge group $G$ (which can be non-abelian), and which is specified by an element in the group cohomology $H^3(BG, U(1))$. A line in a DW theory is specified by the data \cite{Hu:2012wx,Dijkgraaf:1989pz,deWildPropitius:1995cf}
\begin{eqnarray}
([g], \rho_{[g]})~,
\end{eqnarray}
where $[g]$ is an element of a conjugacy class of $G$, and $\rho_{g}$ is an irreducible projective representation of $G$ satisfying
\begin{eqnarray}
\rho(h) \rho(k) =\beta_{g}(h,k) \rho(hk)~, \hspace{1cm} \beta_g(h,k)= \frac{\omega_3(g,h,k) \omega_3(h,k,k^{-1} h^{-1} g h k)}{\omega_3(h,h^{-1}gh, k)}~,
\end{eqnarray}
with $\omega_3(g,h,k)\in H^3(BG, U(1))$. The irreducible projective representation depends only on the conjugacy class $[g]$, and not on the choice of specific element $g$ inside a given conjugacy class. The quantum dimension of the line is given by the dimension of the irreducible projective representation, 
\begin{eqnarray}
d_{([g], \rho_{[g]})} = \dim(\rho_{[g]})~.
\end{eqnarray}
As a consequence, the quantum dimension of any line in a DW theory is an integer. For example, the DW theory with gauge group $\Z_2\times \Z_2\times \Z_2$ and group cohomology class $\omega_{3}(g,h,k)=e^{i \pi g_1 h_2 k_3}$ has 8 lines of quantum dimension 1 and 7 lines of quantum dimension 2 \cite{deWildPropitius:1995cf}. The field theory and its line operators are discussed in \cite{He:2016xpi}.

In contrast,  the SymTFT for $TY(\Z_N)$ contains non-invertible lines with quantum dimension $\sqrt{N}$, which is not an integer unless $N=n^2$ is a perfect square. We therefore conclude that when $N$ is not a perfect square, a QFT with $TY(\Z_N)$ symmetry category \textit{cannot} be mapped to another QFT with only invertible symmetry via topological manipulations, which is consistent with the result in \cite{gelaki2009centers}. In other words, the duality defect for $TY(\Z_N)$ with $N$ not a perfect square is always intrinsically non-invertible.

\section*{Acknowledgements}

We would like to thank Andrea Antinucci, Christian Copetti, Ryohei Kobayashi, Ho Tat Lam, Linhao Li, Emily Nardoni, Sakura Schafer-Nameki, Sahand Seifnashri, Shu-Heng Shao, Zhengdi Sun, Yuji Tachikawa, Matthew Yu, and  Gabi Zafrir for helpful discussions. J.K. would like to thank Kavli IPMU for their generous hospitality and support. 
KO is supported in part by JSPS KAKENHI Grant-in-Aid, No.22K13969 and the Simons Collaboration on Global Categorical Symmetries.
Y.Z. is partially supported by WPI Initiative, MEXT, Japan at IPMU, the University of Tokyo.

\newpage

\appendix

\section{Correlation functions of $k$-dimensional operators in $(2k+1)$d}
\label{app:linecorr}

In this appendix, we provide derivations for equations (\ref{eq:linking}), \eqref{eq:corrLL}, and \eqref{eq:4dcomm}; namely we derive the linking of $k$-dimensional operators in $(2k+1)$-dimensional $\Z_N$ gauge theory, as well as their commutation relations in $2k$-dimensions. We begin with the action for $\Z_N$ gauge theory in $(2k+1)$ dimensions,
\begin{eqnarray}
S= \frac{2\pi}{N}\int_{X_{2k+1}} b^{(k)} \cup \delta  a^{(k)}~.
\end{eqnarray}
This theory has $N^2$ $k$-dimensional operators given by 
\begin{eqnarray}
L_{(e,m)}(M_k) = \exp\left( \frac{2\pi i}{N} \oint_{M_k} e a^{(k)} \right) \exp\left( \frac{2\pi i}{N} \oint_{M_k} m b^{(k)}\right), \hspace{1cm} (e,m)\in \Z_N\times \Z_N
\end{eqnarray}
with $L_{(1,0)}$ and $L_{(0,1)}$ together generating a $\Z_N^{(k)}\times \Z_N^{(k)}$ $k$-form symmetry. When $k=1$, we obtain a standard $\Z_N$ gauge theory in $(2+1)$d, as discussed in Section \ref{sec:symTFT3dZN}. When $k=2$, we obtain a $\Z_N^{(1)}$ gauge theory in $(4+1)$d, as discussed in Section \ref{sec:symTFT5d}.\footnote{In Section \ref{sec:symTFT5d}, since the operators are surfaces, we use the notation $S_{(e,m)}$ instead of $L_{(e,m)}$. }

We consider the mutual braiding between two operators labeled by $(e,m)$ and $(e',m')$. To do so, 
we evaluate the correlation functions of these two operators on manifolds $M_k$ and $M_k'$ which form a Hopf link. In practice, this means that we have insertions in the action of the form
\bea
\label{eq:actionwithloopsapp}
S &=&{2 \pi \over N} \int_{X_{2k+1}} b^{(k)}  \delta a^{(k)} + {2 \pi \over N}  \int_{M_k} (e a^{(k)} + m b^{(k)} ) + {2 \pi \over N}  \int_{M_k'} (e' a^{(k)} + m' b^{(k)})
\\
&=&{2 \pi \over N} \int_{X_{2k+1}} b^{(k)}  \delta a^{(k)} + {2 \pi \over N}  \int_{X_{2k+1}} ( e \omega_{M_k}+ e' \omega_{M_k'})  a^{(k)} + {2 \pi \over N}  \int_{X_{2k+1}}  ( m \omega_{M_k}+ m' \omega_{M_k'})  b^{(k)}~\no,
\eea
where $\omega_{M_k}$ is the Poincar{\'e} dual of the $k$-cycle $M_k$ with respect to $X_{2k+1}$, and hence is a cocycle of degree $k+1$. Integrating out the field $b^{(k)}$ enforces 
\bea\label{eq:deltaak}
\delta a^{(k)} =  -(m \omega_{M_k}+ m' \omega_{M_k'})~.
\eea 
Defining $V_{k+1}$ such that $\partial V_{k+1} = m M_k + m' M_k'$, we have $a^{(k)} =- \mathrm{PD}(V_{k+1})$. Plugging this back into the action then gives contribution
\bea 
 -{2 \pi \over N}  \int_{X_{2k+1}}  ( e \omega_{M_k}+ e' \omega_{M_k'}) \cup  \mathrm{PD}(V_{k+1})  = - {2 \pi \over N}  \int_{X_{2k+1}}  \mathrm{PD}( (e M_k+ e' M_k') \cap  V_{k+1}) ~. 
\eea
In general, the intersection pairing $M_k \cap V_{d-k}$ between $M_k \in H_k(X_{d}, \ZZ)$ and $V_{d-k} \in H_{d-k}(X_d, \ZZ)$ satisfies 
\bea
\label{eq:intersectionpairing}
M_k\cap V_{d-k} = (-1)^{k(d-k)} V_{d-k} \cap M_k~, 
\eea
and thus in the current case we may write 
\bea\label{eq:braidingZNk}
-{2 \pi \over N}  \int_{X_{2k+1}}  \left[  e \,\mathrm{PD}(  M_k\cap  V_{k+1}) + e'\, \mathrm{PD}( V_{k+1} \cap M_k' )\right]~.
\eea
It is convenient to formally rewrite $V_{k+1}=m \partial^{-1}M_k + m' \partial^{-1}M_k'$ by moving the boundary operator $\partial$ to the right hand side in the definition below \eqref{eq:deltaak}. Then $\mathrm{PD}(  M_k\cap  V_{k+1})= \mathrm{PD}(  M_k\cap  \partial^{-1} M_k') \equiv \link(M_k, M_k')$. Moreover, by using integration by parts, we have $\mathrm{PD}( V_{k+1} \cap M_k' ) = \mathrm{PD}( \partial^{-1}M_{k} \cap M_k' ) = (-1)^{k+1}\mathrm{PD}(  M_k\cap  \partial^{-1} M_k') \equiv (-1)^{k+1}\link(M_k, M_k')$. Substituting these into \eqref{eq:braidingZNk} gives
\bea
\label{eq:Hopflinkgen}
-{2 \pi \over N}  (e m' + (-1)^{k+1}m e')\, \mathrm{link}(M_k, M_k')~.
\eea
We thus conclude that the braiding between two $k$-dimensional operators is 
\begin{eqnarray}\label{eq: ZNKbraiding}
\langle L_{(e,m)}(M_k) L_{(e',m')}(M_k')...\rangle = \exp\left( - \frac{2\pi i}{N} (em' + (-1)^{k+1} me') \link(M_k, M_k')\right) \langle ...\rangle
\end{eqnarray}

We next derive the equal time correlation function between generic dyonic lines $(e,m)$ and $(e',m')$, given for $k=1$ and $k=2$ in \eqref{eq:corrLL} and \eqref{eq:5dcomm}. We start with the commutation relation 
\begin{eqnarray}\label{eq:A1}
L_{(1,0)}(M_k) L_{(0,1)}(M_k') = \exp\left(-{2 \pi i \over N} \langle M_k, M_k'\rangle \right) L_{(0,1)}(M_k')  L_{(1,0)}(M_k)
\end{eqnarray}
which can be obtained, for example,  by canonical quantization.\footnote{To carry out the canonical quantization, we switch back to differential form notation. Let us assume $k=1$ for simplicity; higher $k$ can be similarly derived, but with more indices. Canonical quantization in the usual way yields $[b_x(p), a_y(p')]=iN/2\pi \delta(p-p')$, and then using the Baker?Campbell?Hausdorff formula, we obtain $e^{2\pi i/N \oint_{x} a_x} e^{2\pi i/N \oint_{y} b_y}= e^{(2\pi i/N)^2 i (N/2\pi)}e^{2\pi i/N \oint_{x} b_y} e^{2\pi i/N \oint_{x} a_x}=e^{-2\pi i/N }e^{2\pi i/N \oint_{y} b_y} e^{2\pi i/N \oint_{x} a_x}$. } The pairing $\langle M_k, M_k'\rangle$ is the intersection pairing between two $k$-manifolds in $2k$-dimensions.
Moving the phase to the left side and relabeling $M_k\leftrightarrow M_k'$, we have 
\begin{eqnarray}\label{eq:A2}
\begin{split}
    L_{(0,1)}(M_k) L_{(1,0)}(M_k') &= \exp\left({2 \pi i \over N} \langle M_k', M_k\rangle \right) L_{(1,0)}(M_k')  L_{(0,1)}(M_k)\\
    &=\exp\left({2 \pi i \over N}(-1)^k \langle M_k, M_k'\rangle \right) L_{(1,0)}(M_k')  L_{(0,1)}(M_k)\\
\end{split}
\end{eqnarray}
where in the second line we used $\braket{M_k,M_k'}= (-1)^k \braket{M_k',M_k}$.\footnote{Note that the symmetry properties of the linking number and the intersection pairing are opposite. We have $\braket{M_k,M_k'}= (-1)^k \braket{M_k',M_k}$, but $\link(M_k, M_k')= (-1)^{k+1} \link(M_k',M_k)$. Hence when the intersection pairing is even (odd), the linking number is odd (even). }
Using \eqref{eq:A1} and \eqref{eq:A2} and the definition of the dyonic line \eqref{eq:beforegaugingLs}, we may then determine the equal time commutation relation between dyonic lines, 
\begin{equation}\label{eq:intersectionZNk}
\begin{split}
    &L_{(e,m)}(M_k) L_{(e',m')}(M_k') \,=\, \left(L_{(1,0)}(M_k)\right)^{\otimes e} \left(L_{(0,1)}(M_k)\right)^{\otimes m} \left(L_{(1,0)}(M_k')\right)^{\otimes e'} \left(L_{(0,1)}(M_k')\right)^{\otimes m'}\\
    &\,\,\,\,=e^{-\frac{2\pi i}{N}(e m' +(-1)^{k+1} m e')\braket{M_k, M_k'}} \left(L_{(1,0)}(M_k')\right)^{\otimes e'} \left(L_{(0,1)}(M_k')\right)^{\otimes m'}\left(L_{(1,0)}(M_k)\right)^{\otimes e} \left(L_{(0,1)}(M_k)\right)^{\otimes m}\\
    &\,\,\,\,= e^{-\frac{2\pi i}{N}(e m' +(-1)^{k+1} m e')\braket{M_k, M_k'}} L_{(e',m')}(M_k') L_{(e,m)}(M_k)
\end{split}
\end{equation}
which reproduces \eqref{eq:corrLL} and \eqref{eq:5dcomm}. Note that the phase from the equal time commutation relation in \eqref{eq:intersectionZNk} coincides with the phase from the linking in \eqref{eq: ZNKbraiding}, with the intersection pairing number being replaced by the linking number.

\section{Measuring EM charges of defect junctions}
\label{app:junctioncharge}

\begin{figure}[tbp]
\begin{center}

\begin{tikzpicture}[baseline=-10]
  \shade[ball color = white!60!red] (0,0) circle (0.6cm);
 \draw[red] (0,0) circle (0.6cm);
 \draw[red] (-0.6,0) arc (180:360:0.6 and 0.2);
\draw[red,dashed] (0.6,0) arc (0:180:0.6 and 0.2);

\draw [dgreen, thick] (0,-1.5) -- (0,1.5);
\draw [blue, thick, decoration={markings, mark=at position 0.8 with {\arrow{<}}}, postaction={decorate}] (0,0) to[out=45,in=200 ]  (1.5,0);
\node[isosceles triangle,scale=0.4,
    isosceles triangle apex angle=60,
    draw,fill=violet!60,
    rotate=90,
    minimum size =0.01cm] at (0,0){};
\node[blue, above] at (1.7,0) {$L_{(N/2,N/2)}$};
\node[dgreen, below] at (0,-1.5) {$\Sigma_{(e)}$};
\node[red, below] at (-0.5,-0.45) {$D_{\mathrm{EM}}$};
\end{tikzpicture}
$\,\,\,\,\,=\,\,\,\,\,$
\begin{tikzpicture}[baseline=-10]
\draw[red,thick] (-0.6,-0.0) arc (180:120:0.6 and 0.65);
\draw[red,thick] (-0.3,-0.565) arc (240:180:0.6 and 0.65);
\draw[red,thick] (0.575,0.33) arc (30:66:0.6 and 0.65);
\draw[red,thick] (0.58,-0.26) arc (-23:-66:0.6 and 0.65);

\shade[ball color = white!60!red] (0,0.6) ellipse (0.3cm and 0.08cm);
\draw[thick, red] (0,0.6) ellipse (0.3cm and 0.08cm);

\shade[ball color = white!60!red] (0,-0.6) ellipse (0.3cm and 0.08cm);
\draw[thick,red] (0,-0.6) ellipse (0.3cm and 0.08cm);

\shade[ball color = white!60!red] (0.6,0.05) ellipse (0.08cm and 0.3cm );
\draw[thick,red] (0.6,0.05) ellipse (0.08cm and 0.3cm );

\draw [dgreen, thick] (0,-1.5) -- (0,1.5);
\draw [blue, thick, decoration={markings, mark=at position 0.8 with {\arrow{<}}}, postaction={decorate}] (0,0) to[out=45,in=200 ]  (1.5,0);
\node[isosceles triangle,scale=0.4,
    isosceles triangle apex angle=60,
    draw,fill=violet!60,
    rotate=90,
    minimum size =0.01cm] at (0,0){};
\node[blue, above] at (1.7,0) {$L_{(N/2,N/2)}$};
\node[dgreen, below] at (0,-1.5) {$\Sigma_{(e)}$};
\node[red] at (-0.9,0) {$\cA$};
\end{tikzpicture}
$\,\,\,\,\,=\,\,\,\,\,$
\begin{tikzpicture}[baseline=-10]
\draw [dgreen, thick] (0,-1.5) -- (0,1.5);
\draw [blue, thick, decoration={markings, mark=at position 0.8 with {\arrow{<}}}, postaction={decorate}] (0,0) to[out=45,in=200 ]  (1.5,0);
\node[isosceles triangle,scale=0.4,
    isosceles triangle apex angle=60,
    draw,fill=violet!60,
    rotate=90,
    minimum size =0.01cm] at (0,0){};
\draw[red, thick] (0,0) circle (0.6 cm);
\node[blue, above] at (1.7,0) {$L_{(N/2,N/2)}$};
\node[dgreen, below] at (0,-1.5) {$\Sigma_{(e)}$};
\node[red, below] at (-0.5,-0.45) {$\mathcal{A}$};
\filldraw[violet!60] (0.58,0.12) circle (2pt);
\filldraw[orange] (0,0.6) circle (2pt);
\filldraw[orange] (0,-0.6) circle (2pt);
\end{tikzpicture}

\caption{The $\ZZ_2^{\mathrm{EM}}$ charge of the trivalent junction denoted by a triangle can be measured by enclosing the junction with the surface $D_{\mathrm{EM}}$. This surface is a condensate of the algebra anyon $\cA$, and may be decomposed as in the middle image. The remaining discs of $D_{\mathrm{EM}}$ can be shrunk to give orange and purple junctions, which will be studied in the text. }
\label{fig:psisigsigcharge}
\end{center}
\end{figure}
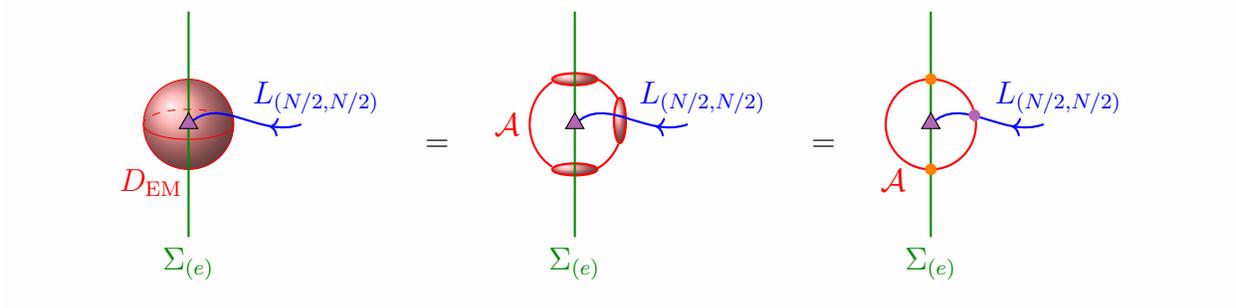

The goal of this appendix is to compute the $\ZZ_2^{\mathrm{EM}}$ charge of the junctions involving $L_{(N/2,N/2)}$ and $\Sigma_{(e)}$. Recall that there are various such junctions depending on the angle between the line and the surface anchored on $\Sigma_{(e)}$, c.f. Figure \ref{fig:squarevstrianglejunc}, and we will focus here on the charge of the triangle junction. The $\ZZ_2^{\mathrm{EM}}$ charge of this junction can be measured by enclosing the junction with the surface $D_{\mathrm{EM}}$, as shown in the left of Figure \ref{fig:psisigsigcharge}. This surface is a condensate of the algebra anyon $\cA$, and we may simplify the configuration to the one in the middle of Figure \ref{fig:psisigsigcharge}. This leaves us with discs of $D_{\mathrm{EM}}$ surrounding each of the outgoing lines, which may in general involve complicated intersections of $\cA$ with the external lines. Instead of studying the details of these discs, we may instead shrink them to point-like junctions as shown in the right of Figure \ref{fig:psisigsigcharge}. Our goal will now be to understand the junctions appearing here, which will allow us to evaluate the configuration and obtain the charge. Throughout we will neglect real number normalization factors, since these will in any case cancel out to give the final $\ZZ_2$-valued charge.

\begin{figure}[htbp]
\begin{center}
 {\begin{tikzpicture}[baseline=0]
\draw [red, thick] (0,-1) -- (0,1);
\draw [red, thick] (0,0) to[out=0,in=90 ]  (1,-1);
\filldraw[red] (0,0) circle (1.5pt);
\node[red,left] at (0,0) {$\mu$};

\node[red,below] at (0,-1) {$\mathcal{A}$};
\node[red,above] at (0,1) {$\mathcal{A}$};
\node[red,below] at (1,-1) {$\mathcal{A}$};
\end{tikzpicture}}\hspace{0.2 in}$=\,\,\,\,\,\sum_{p,q=0}^{N-1}$\hspace{0.1 in}
 {\begin{tikzpicture}[baseline=0]
\draw [blue, thick,decoration={markings, mark=at position 0.5 with {\arrow{>}}},
        postaction={decorate}] (0,-1) -- (0,0);
\draw [blue, thick,decoration={markings, mark=at position 0.5 with {\arrow{>}}},
        postaction={decorate}] (0,0) -- (0,1);
\draw [blue, thick, decoration={markings, mark=at position 0.5 with {\arrow{>}}},
        postaction={decorate}]  (1,-1) to[out=90,in=0 ]  (0,0);

\node[below,blue] at (-0.3,-1) {$L_{(p,-p)}$};
\node[above,blue] at (0,1)  {$L_{(p+q,-p-q)}$};
\node[below,blue] at (1.2,-1){$L_{(q,-q)}$};
\filldraw[blue] (0,0) circle (1.5pt);
\end{tikzpicture}}
\caption{Definition of the trivalent junction $\mu$ between three $\cA$ lines. }
\label{eq:mujuncdef}
\end{center}
\end{figure}
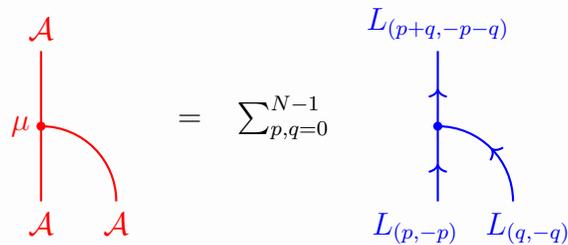

To begin, we consider the 4-valent junction between $L_{(e,m)}$ and the algebra object $\cA$. In this appendix, we will always draw invertible $L_{(e,m)}$ lines in blue, $\cA$ lines in red, and the junction between the two as a purple dot. The junction between $L_{(e,m)}$ and $\cA$ is the one encountered when e.g. $L_{(e,m)}$ pierces the $\ZZ_2^{\mathrm{EM}}$ surface as in Figure \ref{fig:psiaajunc}. We note that, in Figure \ref{fig:psiaajunc}, the configuration should not depend on precisely where along the surface $L_{(e,m)}$ intersects. As such, we obtain the consistency condition on the junction shown in Figure \ref{fig:psiAAconsistency}. Note that we have introduced an associative (co)multiplication junction $\mu$, defined in Figure \ref{eq:mujuncdef}, which every algebra object $\cA$ is automatically equipped with.

\begin{figure}[tbp]
\begin{center}
    {\begin{tikzpicture}[baseline=35]
 \shade[top color=red!40, bottom color=red!10,rotate=90]  (0,-0.7) -- (2,-0.7) -- (2.6,-0.2) -- (0.6,-0.2)-- (0,-0.7);
 \draw[thick,rotate=90] (0,-0.7) -- (2,-0.7);
\draw[thick,rotate=90] (0,-0.7) -- (0.6,-0.2);
\draw[thick,rotate=90]  (0.6,-0.2)--(2.6,-0.2);
\draw[thick,rotate=90]  (2.6,-0.2)-- (2,-0.7);
\draw[thick,blue,postaction={on each segment={mid arrow=blue}}] (-0.9,1.4)--(0.4,1.4);
\draw[thick,blue,dotted] (0.4,1.4)--(0.68,1.4);
\draw[thick,blue,postaction={on each segment={mid arrow=blue}}] (0.72,1.4)--(1.7,1.4);
\node[blue,right] at (-2.2,1.4) {$L_{(e,m)}$};
\node[blue,right] at (1.7,1.4) {$L_{(m,e)}$};
\node[red,below] at (0.4, -0) {$D_{\text{EM}}$};

\filldraw[violet!60] (0.4,1.4) circle (2pt);

\end{tikzpicture}}
\qquad$=$\qquad
   {\begin{tikzpicture}[baseline=35]
\draw[thick,rotate=90] (0,-0.7) -- (2,-0.7);
\draw[thick,rotate=90] (0,-0.7) -- (0.6,-0.2);
\draw[thick,rotate=90]  (0.6,-0.2)--(2.6,-0.2);
\draw[thick,rotate=90]  (2.6,-0.2)-- (2,-0.7);
\draw[thick,blue,postaction={on each segment={mid arrow=blue}}] (-0.9,1.4)--(0.4,1.4);
\draw[thick,blue] (0.4,1.4)--(0.68,1.4);
\draw[thick,blue,postaction={on each segment={mid arrow=blue}}] (0.72,1.4)--(1.7,1.4);
\node[blue,right] at (-2.2,1.4) {$L_{(e,m)}$};
\node[blue,right] at (1.7,1.4) {$L_{(m,e)}$};
\node[red,below] at (0.4, -0) {$D_{\text{EM}}$};

\draw[thick,red](0.4,1.4)--(0.4,1.8);
\draw[thick,red](0.4,1.8)--(0.2,2);
\draw[thick,red](0.2,2)--(0.5,2.2);
\draw[thick,red](0.4,1.8)--(0.7,2);
\draw[thick,red](0.2,0.9)--(0.4,0.7);
\draw[thick,red] (0.4,0.7)--(0.55,0.2);
\draw[thick,red] (0.4,0.7)--(0.6,0.9);
\draw[thick,red] (0.6,0.9)--(0.4,1.4);
\draw[thick,red] (0.6,0.9)--(0.7,0.8);

\filldraw[violet!60] (0.4,1.4) circle (2pt);

\end{tikzpicture}}
\caption{The intersection of $L_{(m,e)}$ and $D_{\mathrm{EM}}$, with the latter resolved into a mesh of $\mathcal{A}$.}
\label{fig:psiaajunc}
\end{center}
\end{figure}
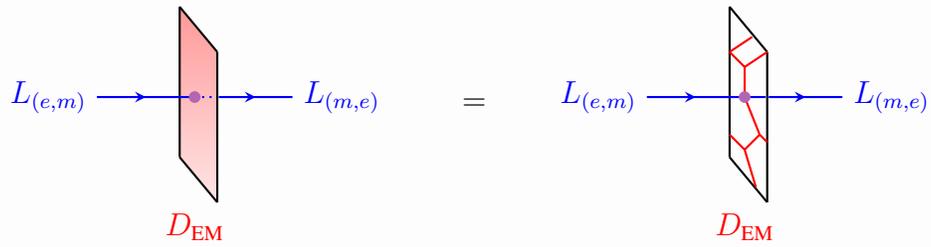

\begin{figure}[htbp]
\begin{center}
{\begin{tikzpicture}[baseline=0]

\draw [red, thick] (0,0) to[out=0,in=90 ]  (1,-1);
\filldraw[red] (0,0) circle (1.5pt);
\node[red,left] at (0,0) {$\mu$};

\node[blue, left] at (-1,-1) {$L_{(e,m)}$};
\node[red,below] at (0,-1) {$\mathcal{A}$};
\node[red,above] at (0,1) {$\mathcal{A}$};
\node[red,below] at (1,-1) {$\mathcal{A}$};
\node[blue, left] at (2.6,0.1) {$L_{(m,e)}$};

\draw[white, line width=5pt] (-1,-1) to [out=20,in = 200] (1.3,0);
\draw [red, thick] (0,-1) -- (0,1);
\draw[blue, thick,decoration={markings, mark=at position 0.2 with {\arrow{>}},mark=at position 0.9 with {\arrow{>}}},
        postaction={decorate}] (-1,-1) to [out=20,in = 200] (1.3,0);
\filldraw[violet!60] (0,-0.575) circle (2pt);

\end{tikzpicture}}
\hspace{0.4 in}$=$\hspace{0.4 in}
 {\begin{tikzpicture}[baseline=0]

\draw [red, thick] (0,0) to[out=0,in=90 ]  (1,-1);
\filldraw[red] (0,0) circle (1.5pt);
\node[red,left] at (0,0) {$\mu$};

\node[blue, left] at (-1,0) {$L_{(e,m)}$};
\node[blue, left] at (2.6,1) {$L_{(m,e)}$};
\node[red,below] at (0,-1) {$\mathcal{A}$};
\node[red,above] at (0,1) {$\mathcal{A}$};
\node[red,below] at (1,-1) {$\mathcal{A}$};

\draw [red, thick] (0,-1) -- (0,1);
\draw[blue, thick,decoration={markings, mark=at position 0.2 with {\arrow{>}},mark=at position 0.8 with {\arrow{>}}},
        postaction={decorate}] (-1,0) to [out=20,in = 200] (1.3,1);
\filldraw[violet!60] (0,0.42) circle (2pt);

\end{tikzpicture}}

\caption{Consistency condition involving the purple junction between $L_{(e,m)}$ and $\cA$, and the trivalent junction $\mu$ between three $\cA$ lines. }
\label{fig:psiAAconsistency}
\end{center}
\end{figure}
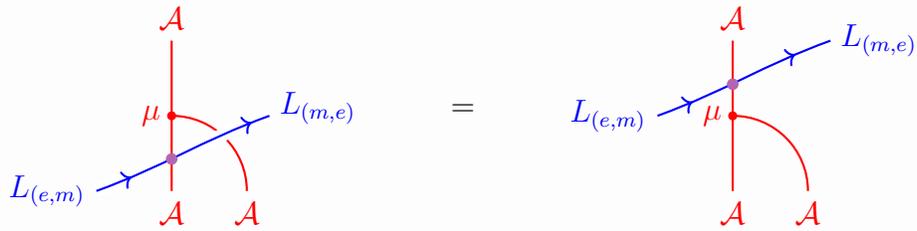

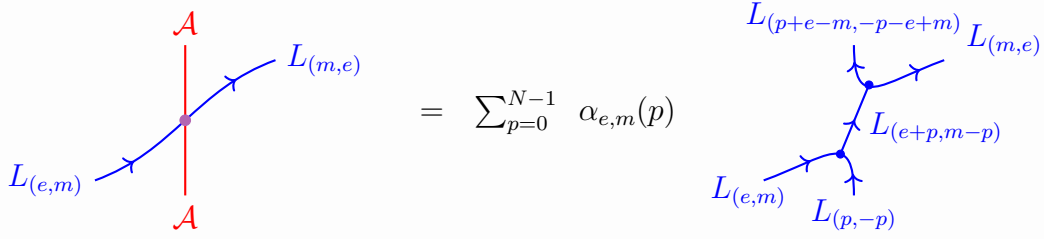
\begin{figure}[tbp]
\begin{center}
 {\begin{tikzpicture}[baseline=0]
\draw [red, thick] (0,-1) -- (0,1);
\draw[blue, thick,decoration={markings, mark=at position 0.2 with {\arrow{>}},mark=at position 0.8 with {\arrow{>}}},
        postaction={decorate}] (-1.2,-0.8) to [out=20,in = 200] (1.2,0.8);
\filldraw[violet!60] (0,0) circle (2pt);
\node[below]at (0,-1) {$\color{red}\cA$};
\node[above]at (0,1) {$\color{red}\cA$};
\node[left] at (-1.2,-0.8) {$\color{blue}L_{(e,m)}$};
\node[right] at (1.2,0.8) {$\color{blue}L_{(m,e)}$};
\end{tikzpicture}}
\hspace{0.2 in}$=\,\,\,\,\sum_{p=0}^{N-1}\,\,\, \alpha_{e,m}(p)\,\,\,$
 {\begin{tikzpicture}[baseline=0]
\draw [blue, thick, decoration={markings, mark=at position 0.2 with {\arrow{<}},mark=at position 0.8 with {\arrow{>}}},
        postaction={decorate},distance = 0.9cm] (0,1)  to [out=270,in = 200] (1.2,0.8);
\draw [blue, thick, decoration={markings, mark=at position 0.4 with {\arrow{>}},mark=at position 0.9 with {\arrow{<}}},
        postaction={decorate}, distance = 0.9cm]  (-1.2,-0.8)  to [out=20,in = 90] (0,-1);
\draw [blue, thick, decoration={markings, mark=at position 0.5 with {\arrow{>}}},
        postaction={decorate}] (-0.18,-0.45)--(0.2,0.47);
        \filldraw[blue] (-0.18,-0.45) circle (1.5pt);
        \filldraw[blue] (0.2,0.47) circle (1.5pt);
  \node[below] at (-1.4,-0.6){$\color{blue}L_{(e,m)}$};
   \node[below] at (0,-0.9){$\color{blue}L_{(p,-p)}$};
    \node[above] at (2,0.7){$\color{blue}L_{(m,e)}$};
     \node[above] at (0,1){$\color{blue}L_{(p+e-m,-p-e+m)}$};
      \node[right] at (0.05,-0.1){$\color{blue}L_{(e+p,m-p)}$};
\end{tikzpicture}}
\caption{The general form of the junction between $L_{(e,m)}$ and $\cA$, with $\alpha_{e,m}(p)$ a series of to-be-determined constants.}
\label{fig:genformofpsiAA}
\end{center}
\end{figure}

On general grounds, the junction between $L_{(e,m)}$ and $\cA$ must be of the form shown in Figure \ref{fig:genformofpsiAA}, with $\alpha_{e,m}(p)$ a series of undetermined constants. In fact, imposing the consistency condition in Figure \ref{fig:psiAAconsistency} is sufficient to fix these constants. This may be shown by expanding both sides of Figure \ref{fig:psiAAconsistency} and making use of Figures \ref{eq:mujuncdef} and \ref{fig:genformofpsiAA}. Once expanded, the right-hand side is given as in Figure \ref{fig:new1}, whereas the left-hand side is given in the first line of Figure \ref{fig:new2}. One may rearrange the configuration in Figure \ref{fig:new2} via a series of F-moves to get the second line in Figure \ref{fig:new2}, and upon using the half-braid and more F-moves one may put it in the form shown in the last line of Figure \ref{fig:new2}. This configuration may now be compared to the one in Figure \ref{fig:new1}. Equating the two gives 
\bea
\alpha_{e,m}(p+q) = e^{{2 \pi i \over N}m q}\alpha_{e,m}(p)~, 
\eea
and choosing $\alpha_{e,m}(0)=1$ (which is simply a choice of convention) we derive that 
\bea
\alpha_{e,m}(p) =e^{{2 \pi i \over N}m p}~. 
\eea
This completely specifies the junction in Figure \ref{fig:psiaajunc}.

\begin{figure}[tbp]
\begin{center}
 {\begin{tikzpicture}[baseline=0]

\draw [red, thick] (0,0) to[out=0,in=90 ]  (1,-1);
\filldraw[red] (0,0) circle (1.5pt);
\node[red,left] at (0,0) {$\mu$};

\node[blue, left] at (-1,0) {$L_{(e,m)}$};
\node[blue, left] at (2.6,1) {$L_{(m,e)}$};
\node[red,below] at (0,-1) {$\mathcal{A}$};
\node[red,above] at (0,1) {$\mathcal{A}$};
\node[red,below] at (1,-1) {$\mathcal{A}$};

\draw [red, thick] (0,-1) -- (0,1);
\draw[blue, thick,decoration={markings, mark=at position 0.2 with {\arrow{>}},mark=at position 0.8 with {\arrow{>}}},
        postaction={decorate}] (-1,0) to [out=20,in = 200] (1.3,1);
\filldraw[violet!60] (0,0.42) circle (2pt);

\end{tikzpicture}}
\hspace{0.1 in}$= \,\,\sum_{p,q=0}^{N-1}\,\,\,\alpha_{e,m}(p+q)$\hspace{0.1 in}
 {\begin{tikzpicture}[baseline=0]

\draw [blue, thick,,decoration={markings, mark=at position 0.5 with {\arrow{<}}},
        postaction={decorate}] (0,0) to[out=0,in=90 ]  (1,-1);
\filldraw[blue] (0,0) circle (1.5pt);

\node[blue, left] at (-1,0) {$L_{(e,m)}$};
\node[blue, left] at (2.6,1) {$L_{(m,e)}$};
\node[blue,below] at (-0.2,-1) {$L_{(p,-p)}$};
\node[blue,above] at (0,1.5) {$L_{(p+q+e-m,m-e-p-q)}$};
\node[blue,below] at (1.2,-1) {$L_{(q,-q)}$};

\draw [blue, thick,decoration={markings, mark=at position 0.2 with {\arrow{>}},mark=at position 0.9 with {\arrow{>}}},
        postaction={decorate}] (0,-1) -- (0,1.5);
\draw[blue, thick,decoration={markings, mark=at position 0.5 with {\arrow{>}}},postaction={decorate}] (-1,0) to [out=20,in = 200] (0,0.42);
\draw[blue, thick,decoration={markings, mark=at position 0.5 with {\arrow{>}}},postaction={decorate}] (0,0.7) to [out=20,in = 200] (1.3,1);
\filldraw[blue] (0,0.42) circle (1.5pt);
\filldraw[blue] (0,0.7) circle (1.5pt);

\end{tikzpicture}}

\caption{The right-hand side of Figure \ref{fig:psiAAconsistency}, expanded out using Figure \ref{fig:genformofpsiAA}. }
\label{fig:new1}
\end{center}
\end{figure}
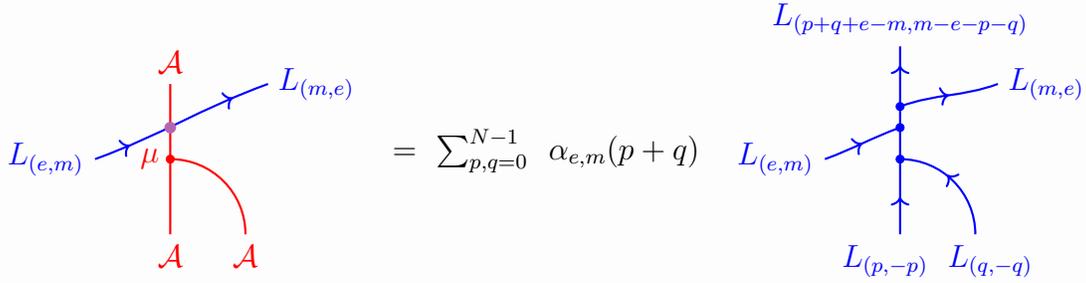

\begin{figure}[htbp]
\begin{center}
{\begin{tikzpicture}[baseline=0]

\draw [red, thick] (0,0) to[out=0,in=90 ]  (1,-1);
\filldraw[red] (0,0) circle (1.5pt);
\node[red,left] at (0,0) {$\mu$};

\node[blue, left] at (-1,-1) {$L_{(e,m)}$};
\node[red,below] at (0,-1) {$\mathcal{A}$};
\node[red,above] at (0,1) {$\mathcal{A}$};
\node[red,below] at (1,-1) {$\mathcal{A}$};
\node[blue, left] at (2.6,0.1) {$L_{(m,e)}$};

\draw[white, line width=5pt] (-1,-1) to [out=20,in = 200] (1.3,0);
\draw [red, thick] (0,-1) -- (0,1);
\draw[blue, thick,decoration={markings, mark=at position 0.2 with {\arrow{>}},mark=at position 0.9 with {\arrow{>}}},
        postaction={decorate}] (-1,-1) to [out=20,in = 200] (1.3,0);
\filldraw[violet!60] (0,-0.575) circle (2pt);

\end{tikzpicture}}
\hspace{0.1 in}$= \,\,\sum_{p,q=0}^{N-1}\,\,\,\alpha_{e,m}(p)$\hspace{0.0 in}
 {\begin{tikzpicture}[baseline=0]
\draw [blue, thick,decoration={markings, mark=at position 0.8 with {\arrow{<}}},postaction={decorate}] (0,0) to[out=0,in=90 ]  (1,-1.2);
\filldraw[blue] (0,0) circle (1.5pt);

\node[blue, left] at (-1,-1.2) {$L_{(e,m)}$};
\node[blue,below] at (-0.2,-1.2) {$L_{(p,-p)}$};
\node[blue,above] at (0,1) {$L_{(p+q+e-m,m-e-p-q)}$};
\node[blue,below] at (1.2,-1.2) {$L_{(q,-q)}$};
\node[blue, left] at (2.6,0.1) {$L_{(m,e)}$};

\draw[white, line width=5pt] (-1,-1) to [out=20,in = 200] (1.3,0);
\draw [blue, thick,decoration={markings, mark=at position 0.8 with {\arrow{>}},mark=at position 0.08 with {\arrow{>}}},postaction={decorate}] (0,-1.2) -- (0,1);
\draw[blue, thick,decoration={markings, mark=at position 0.8 with {\arrow{>}}},postaction={decorate}] (0,-0.575) to [out=20,in = 200] (1.3,0);
\draw[blue, thick,decoration={markings, mark=at position 0.5 with {\arrow{>}}},postaction={decorate}] (-1,-1.2) to [out=20,in = 200] (0,-0.9);
\filldraw[blue] (0,-0.9) circle (1.5pt);
\filldraw[blue] (0,-0.575) circle (1.5pt);

\end{tikzpicture}}
\\
\hspace{0.1 in}$= \,\,\sum_{p,q=0}^{N-1}\,\,\,\alpha_{e,m}(p)$\hspace{0.0 in}
 {\begin{tikzpicture}[baseline=0]
\draw [blue, thick,decoration={markings, mark=at position 0.5 with {\arrow{>}}},postaction={decorate}] (0,-1)--(0,0);
\draw [blue, thick] (0,0)--(-0.6,0.6);
\draw [blue, thick,decoration={markings, mark=at position 0.5 with {\arrow{>}}},postaction={decorate}] (-1.2,0)--(-0.6,0.6);
\draw [blue, thick,decoration={markings, mark=at position 0.5 with {\arrow{>}}},postaction={decorate}] (-0.6,0.6)--(-0.8,1.3);
\draw [blue, thick] (0,0)--(0.6,0);
\filldraw[blue] (0,0) circle (1.5pt);
\filldraw[blue] (0.6,0) circle (1.5pt);
\filldraw[blue] (-0.6,0.6) circle (1.5pt);

\draw [blue,thick,decoration={markings, mark=at position 0.6 with {\arrow{<}}},postaction={decorate}] (0.6,0) arc [radius=.3, start angle=240-90, end angle=130-90];
\draw [blue,thick] (1.2-0.05,-0.1+0.05)--(1.2+0.2,-0.15-0.2) ;
\draw [blue,thick,decoration={markings, mark=at position 0.6 with {\arrow{>}}},postaction={decorate}] (0.6,0) arc [radius=.3, start angle=-60-90, end angle=42-90];
\draw [blue,thick] (1.05,-0.08)--(1.05+0.4,-0.08+0.4) ;

\node[blue, left] at (-1.2,0) {$L_{(e,m)}$};
\node[blue,below] at (-0.2,-0.9) {$L_{(p,-p)}$};
\node[blue,above] at (-0.8,1.3) {$L_{(p+q+e-m,m-e-p-q)}$};
\node[blue,right] at (1.2+0.2,-0.15-0.2) {$L_{(q,-q)}$};
\node[blue, right] at (1.05+0.4,-0.08+0.4) {$L_{(m,e)}$};

\end{tikzpicture}}
\\
\hspace{0.1 in}$= \,\,\sum_{p,q=0}^{N-1}\,\,\,\alpha_{e,m}(p)\,e^{{2 \pi i \over N} m q}$\hspace{0.0 in}
 {\begin{tikzpicture}[baseline=0]

\draw [blue, thick,,decoration={markings, mark=at position 0.5 with {\arrow{<}}},
        postaction={decorate}] (0,0) to[out=0,in=90 ]  (1,-1);
\filldraw[blue] (0,0) circle (1.5pt);

\node[blue, left] at (-1,0) {$L_{(e,m)}$};
\node[blue, left] at (2.6,1) {$L_{(m,e)}$};
\node[blue,below] at (-0.2,-1) {$L_{(p,-p)}$};
\node[blue,above] at (0,1.5) {$L_{(p+q+e-m,m-e-p-q)}$};
\node[blue,below] at (1.2,-1) {$L_{(q,-q)}$};

\draw [blue, thick,decoration={markings, mark=at position 0.2 with {\arrow{>}},mark=at position 0.9 with {\arrow{>}}},
        postaction={decorate}] (0,-1) -- (0,1.5);
\draw[blue, thick,decoration={markings, mark=at position 0.5 with {\arrow{>}}},postaction={decorate}] (-1,0) to [out=20,in = 200] (0,0.42);
\draw[blue, thick,decoration={markings, mark=at position 0.5 with {\arrow{>}}},postaction={decorate}] (0,0.7) to [out=20,in = 200] (1.3,1);
\filldraw[blue] (0,0.42) circle (1.5pt);
\filldraw[blue] (0,0.7) circle (1.5pt);

\end{tikzpicture}}

\caption{The left-hand side of Figure \ref{fig:psiAAconsistency}, expanded out using Figure \ref{fig:genformofpsiAA}. A series of F-moves and a half-braid give the configuration in the last line. Equating this with Figure \ref{fig:new2} fixes the constants $\alpha_{e,m}(p)$. }
\label{fig:new2}
\end{center}
\end{figure}

Having understood the junction between $L_{(e,m)}$ and $\cA$, we next study the junctions between $L_{(e,m)}$ and $\Sigma_{(e')}$. As mentioned in the main text, this junction depends on the angle between $L_{(e,m)}$ and the EM duality defect $D_{\mathrm{EM}}$ which is anchored on $\Sigma_{(e')}$. This is illustrated in Figure \ref{fig:squarevstrianglejunc}. There we have highlighted two particular cases: first, the case in which $\theta = 0^+$, which we denote by a square; and second, the case in which $\theta = \pi$, which we denote by a triangle. These junctions are subject to consistency conditions illustrated in Figure \ref{fig:squaretriangconsconds}, which physically correspond to the statement that the junction should not depend on where along $\Sigma_{(e')}$ the line $L_{(e,m)}$ is anchored. We will now solve these consistency conditions, beginning with the triangle junction.

  \begin{figure}[tbp]
\begin{center}
 {\begin{tikzpicture}[baseline=0]
\draw [dgreen, thick] (0,-1) -- (0,1);
\draw [blue, thick,decoration={markings, mark=at position 0.5 with {\arrow{<}}},
        postaction={decorate}] (0,0) to[out=0,in=90 ]  (1,-1);
\node[isosceles triangle,scale=0.4,
    isosceles triangle apex angle=60,
    draw,fill=violet!60,
    rotate=90,
    minimum size =0.01cm] at (0,0){};

\node[dgreen,below] at (0,-1) {${\Sigma_{(e')}}$};
\node[dgreen,above] at (0,1) {${\Sigma_{(e'+e+m)}}$};
\node[blue,below] at (1,-1) {${L_{(e,m)}}$};
\end{tikzpicture}}\hspace{0.2 in}$=\,\,\,\,\,\sum_{p=0}^{N-1}\,\,\beta_{e,m,e'}(p)\,\,$
 {\begin{tikzpicture}[baseline=0]
\draw [blue, thick,decoration={markings, mark=at position 0.5 with {\arrow{>}}},
        postaction={decorate}] (0,-1) -- (0,0);
\draw [blue, thick,decoration={markings, mark=at position 0.5 with {\arrow{>}}},
        postaction={decorate}] (0,0) -- (0,1);
\draw [blue, thick, decoration={markings, mark=at position 0.5 with {\arrow{>}}},
        postaction={decorate}]  (1,-1) to[out=90,in=0 ]  (0,0);

\node[blue,below] at (-0.3,-1) {$L_{(e'+p,-p)}$};
\node[blue,above] at (0,1)  {$L_{(e+e'+p,m-p)}$};
\node[blue,below] at (1.2,-1){$L_{(e,m)}$};
\filldraw[blue] (0,0) circle (1.5pt);
\end{tikzpicture}}
\caption{The definition of the triangle junction, given in terms of a series of undetermined coefficients $\beta_{e,m,e'}(p)$. These coefficients may be fixed by imposing the consistency condition in Figure \ref{fig:squaretriangconsconds}.}
\label{fig:new3}
\end{center}
\end{figure}

\begin{figure}[!tbp]
\begin{center}
{\begin{tikzpicture}[baseline=0]

\draw[dgreen,thick](0,-1) -- (0,1); 
\draw [blue, thick,decoration={markings, mark=at position 0.7 with {\arrow{<}}},
        postaction={decorate}] (0,0) to[out=0,in=90 ]  (1,-1);
\draw [red, thick] (0,-0.5) to[out=180,in=270 ]  (-0.5,0);
\draw [red, thick] (-0.5,0) to[out=90,in=180 ]  (0,0.5);

\node[red, left] at (-0.5,0) {\footnotesize{$\mathcal{A}$}};
\node[dgreen, below] at (0,-1) {\footnotesize{$\Sigma_{(e')}$}};
\node[blue, below] at (1.5,-1) {\footnotesize{$L_{(e,m)}$}};
\node[red, right] at (0,0.5) {\footnotesize{$\mu_L$}};
\node[red, right] at (0,-0.5) {\footnotesize{$\mu_L^\vee$}};

\filldraw[red] (0,0.5) circle (1.5pt);
\filldraw[red] (0,-0.5) circle (1.5pt);

\node[isosceles triangle,scale=0.4,
    isosceles triangle apex angle=60,
    draw,fill=violet!60,
    rotate=90,
    minimum size =0.01cm] at (0,0){};
    
\end{tikzpicture}}\hspace{0.2 in}$=\,\,\,\,\,\sum_{p,q=0}^{N-1}\,\,\beta_{e,m,e'}(p+q)\,\,$
 {\begin{tikzpicture}[baseline=0]
\draw [blue, thick,decoration={markings, mark=at position 0.3 with {\arrow{>}}},
        postaction={decorate}] (0,-1) -- (0,0);
\draw [blue, thick,decoration={markings, mark=at position 0.8 with {\arrow{>}}},
        postaction={decorate}] (0,0) -- (0,1);
\draw [blue, thick, decoration={markings, mark=at position 0.5 with {\arrow{>}}},
        postaction={decorate}]  (1,-1) to[out=90,in=0 ]  (0,0);

\filldraw[blue] (0,0.5) circle (1.5pt);
\filldraw[blue] (0,-0.5) circle (1.5pt);

\node[blue,below] at (-0.3,-1) {$L_{(e'+p,-p)}$};
\node[blue,above] at (0,1)  {$L_{(e+e'+p,m-p)}$};
\node[blue,below] at (1.2,-1){$L_{(e,m)}$};
\node[blue,left] at (-0.6,0){$L_{(q.-q)}$};
\filldraw[blue] (0,0) circle (1.5pt);

\draw [blue, thick] (-0.5,0) to[out=90,in=180 ]  (0,0.5);
\draw [blue, thick, decoration={markings, mark=at position 1 with {\arrow{<}}},
        postaction={decorate}] (0,-0.5) to[out=180,in=-90 ] (-0.5,0) ;

\end{tikzpicture}}
$=\,\,\,\,\,\sum_{p=0}^{N-1}\,\,\beta_{e,m,e'}(p+q)\,\,$
 {\begin{tikzpicture}[baseline=0]
\draw [blue, thick,decoration={markings, mark=at position 0.5 with {\arrow{>}}},
        postaction={decorate}] (0,-1) -- (0,0);
\draw [blue, thick,decoration={markings, mark=at position 0.5 with {\arrow{>}}},
        postaction={decorate}] (0,0) -- (0,1);
\draw [blue, thick, decoration={markings, mark=at position 0.5 with {\arrow{>}}},
        postaction={decorate}]  (1,-1) to[out=90,in=0 ]  (0,0);

\node[blue,below] at (-0.3,-1) {$L_{(e'+p,-p)}$};
\node[blue,above] at (0,1)  {$L_{(e+e'+p,m-p)}$};
\node[blue,below] at (1.2,-1){$L_{(e,m)}$};
\filldraw[blue] (0,0) circle (1.5pt);
\end{tikzpicture}}
\caption{The left-hand side of the consistency condition in Figure \ref{fig:squaretriangconsconds}. Comparing this to Figure \ref{fig:new3} fixes the undetermined coefficients $\beta_{e,m,e'}(p)$. }
\label{fig:new4}
\end{center}
\end{figure}
The general form of the triangle junction is given in Figure \ref{fig:new3}, which depends on a series of undetermined coefficients $\beta_{e,m,e'}(p)$. We may solve for these coefficients by imposing the constraint in Figure \ref{fig:squaretriangconsconds}. The right-hand side of the equation for the triangle junction is precisely the configuration in Figure \ref{fig:new3}, whereas the left-hand side is as shown in the first line of Figure \ref{fig:new4} (we use the fact that the junctions $\mu_L$ and $\mu^\vee_L$ do not involve any phases in their expansions). A series of F-moves and shrinking of a loop gives the second line in Figure \ref{fig:new4}. From this we obtain the constraint
\bea
\beta_{e,m,e'}(p+q) = \beta_{e,m,e'}(p)~, 
\eea
and choosing the convention $\beta_{e,m,e'}(0)=1$, we see that all phases can be taken to be trivial,  
\begin{eqnarray}
\beta_{e,m,e'}(p)=1~.
\end{eqnarray}
This completely fixes the triangle junction.

  \begin{figure}[tbp]
\begin{center}
{\begin{tikzpicture}[baseline=0,square/.style={regular polygon,regular polygon sides=4}]

\draw[dgreen,thick](0,-1) -- (0,1); 
\draw [white, line width=5pt] (0,0) to[out=180,in=90 ]  (-1,-1);
\draw [blue, thick, decoration={markings, mark=at position 0.5 with {\arrow{<}}},
        postaction={decorate}] (0,0) to[out=180,in=90 ]  (-1,-1);

\node[dgreen, below] at (0,-1) {\footnotesize{$\Sigma_{(e')}$}};
\node[blue, below] at (-1.5,-1) {\footnotesize{$L_{(e,m)}$}};

\node at (0,0) [square,draw,fill=violet!60,scale=0.5] {}; 
   
\end{tikzpicture}}\hspace{0.2 in}$=\,\,\,\,\,\sum_{p=0}^{N-1}\,\,\gamma_{e,m,e'}(p)\,\,$
 {\begin{tikzpicture}[baseline=0]
\draw [blue, thick,decoration={markings, mark=at position 0.5 with {\arrow{>}}},
        postaction={decorate}] (0,-1) -- (0,0);
\draw [blue, thick,decoration={markings, mark=at position 0.5 with {\arrow{>}}},
        postaction={decorate}] (0,0) -- (0,1);

\draw [blue, thick, decoration={markings, mark=at position 0.5 with {\arrow{<}}}, postaction={decorate}] (0,0) to[out=180,in=90 ]  (-1,-1);

\node[blue,below] at (0.2,-1) {$L_{(e'+p,-p)}$};
\node[blue,above] at (0,1)  {$L_{(e+e'+p,m-p)}$};
\node[blue,below] at (-1.2,-1){$L_{(e,m)}$};
\filldraw[blue] (0,0) circle (1.5pt);
\end{tikzpicture}}
\caption{The definition of the square junction, given in terms of a series of undetermined coefficients $\gamma_{e,m,e'}(p)$. 
}
\label{fig:new5}
\end{center}
\end{figure}
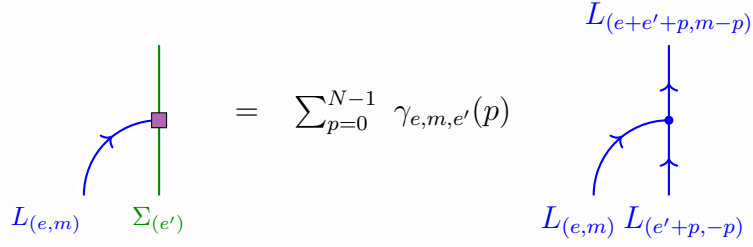

\begin{figure}[tbp]
\begin{center}
{\begin{tikzpicture}[baseline=0,square/.style={regular polygon,regular polygon sides=4}]

\draw[dgreen,thick](0,-1) -- (0,1); 
\draw [red, thick] (0,-0.5) to[out=180,in=270 ]  (-0.5,0);
\draw [red, thick] (-0.5,0) to[out=90,in=180 ]  (0,0.5);
\draw [white, line width=5pt] (0,0) to[out=180,in=90 ]  (-1,-1);

\node[red, left] at (-0.5,0) {\footnotesize{$\mathcal{A}$}};
\node[dgreen, below] at (0,-1) {\footnotesize{$\Sigma_{(e')}$}};
\node[blue, below] at (-1.5,-1) {\footnotesize{$L_{(e,m)}$}};
\node[red, right] at (0,0.5) {\footnotesize{$\mu_L$}};
\node[red, right] at (0,-0.5) {\footnotesize{$\mu_L^\vee$}};

\filldraw[red] (0,0.5) circle (1.5pt);
\filldraw[red] (0,-0.5) circle (1.5pt);

 \draw [blue, thick, decoration={markings, mark=at position 0.6 with {\arrow{<}}},
        postaction={decorate}] (0,0) to[out=180,in=90 ]  (-1,-1);
   \node at (0,0) [square,draw,fill=violet!60,scale=0.5] {};
\end{tikzpicture}}\hspace{0.2 in}$=\,\,\,\,\,\sum_{p,q=0}^{N-1}\,\,\gamma_{e,m,e'}(p+q)\,\,$
 {\begin{tikzpicture}[baseline=0]
\draw [blue, thick,decoration={markings, mark=at position 0.3 with {\arrow{>}}},
        postaction={decorate}] (0,-1) -- (0,0);
\draw [blue, thick,decoration={markings, mark=at position 0.8 with {\arrow{>}}},
        postaction={decorate}] (0,0) -- (0,1);
\draw [blue, thick, decoration={markings, mark=at position 0.5 with {\arrow{<}}}, postaction={decorate}] (0,0) to[out=180,in=90 ]  (-1,-1);

\node[blue,below] at (0.2,-1) {$L_{(e'+p,-p)}$};
\node[blue,above] at (0,1)  {$L_{(e+e'+p,m-p)}$};
\node[blue,below] at (-1.2,-1){$L_{(e,m)}$};
\node[blue,left] at (-0.6,0.2){$L_{(q,-q)}$};

\draw [blue, thick, decoration={markings, mark=at position 0.5 with {\arrow{<}}},
        postaction={decorate}] (-0.5,0) to[out=90,in=180 ]  (0,0.5);
\draw [blue, thick] (0,-0.5) to[out=180,in=-90 ] (-0.5,0) ;

\draw [white, line width=0.25cm] (0,0) to[out=180,in=90 ]  (-1,-1);
\draw [blue, thick, decoration={markings, mark=at position 0.6 with {\arrow{<}}},
        postaction={decorate}] (0,0) to[out=180,in=90 ]  (-1,-1);

\filldraw[blue] (0,0) circle (1.5pt);
\filldraw[blue] (0,0.5) circle (1.5pt);
\filldraw[blue] (0,-0.5) circle (1.5pt);

\end{tikzpicture}}\\
\hspace{-0.5 in}
$=\,\,\sum_{p,q=0}^{N-1}\,\,\gamma_{e,m,e'}(p+q)$
 {\begin{tikzpicture}[baseline=0]
\draw [blue, thick,decoration={markings, mark=at position 0.3 with {\arrow{>}}},
        postaction={decorate}] (0,-1) -- (0,0);
\draw [blue, thick,decoration={markings, mark=at position 0.8 with {\arrow{>}}},
        postaction={decorate}] (0,0) -- (0,1);
\draw [blue, thick, decoration={markings, mark=at position 0.5 with {\arrow{<}}}, postaction={decorate}] (-0.5,0) to[out=180,in=90 ]  (-1.5,-1);

\draw [blue, thick, decoration={markings, mark=at position 0.5 with {\arrow{>}}}, postaction={decorate}] (0,0.5) to[out=180,in=120 ]  (-1.1,-0.1);

\draw [blue, thick, distance = 0.4cm]  (-1,-0.25) to[out=300,in=260 ]  (-0.5,0);

\draw[blue,thick] (0,0)--(-0.5,0);

\filldraw[blue] (0,0.5) circle (1.5pt);
\filldraw[blue] (0,-0.5) circle (1.5pt);

\node[blue,below] at (0.2,-1) {$L_{(e'+p,-p)}$};
\node[blue,above] at (0,1)  {$L_{(e+e'+p,m-p)}$};
\node[blue,below] at (-1.7,-1){$L_{(e,m)}$};
\node[blue,left] at (-1.1,0.5){$L_{(q,-q)}$};
\filldraw[blue] (0,0) circle (1.5pt);
\filldraw[blue] (-0.5,0) circle (1.5pt);

\end{tikzpicture}}
\hspace{-0.1 in}$=\,\,\sum_{p,q=0}^{N-1}\,\,\gamma_{e,m,e'}(p+q)\,e^{-{2\pi i \over N} e q}$\hspace{-0.4in}
 {\begin{tikzpicture}[baseline=0]
\draw [blue, thick,decoration={markings, mark=at position 0.5 with {\arrow{>}}},
        postaction={decorate}] (0,-1) -- (0,0);
\draw [blue, thick,decoration={markings, mark=at position 0.5 with {\arrow{>}}},
        postaction={decorate}] (0,0) -- (0,1);

\draw [blue, thick, decoration={markings, mark=at position 0.5 with {\arrow{<}}}, postaction={decorate}] (0,0) to[out=180,in=90 ]  (-1,-1);

\node[blue,below] at (0.2,-1) {$L_{(e'+p,-p)}$};
\node[blue,above] at (0,1)  {$L_{(e+e'+p,m-p)}$};
\node[blue,below] at (-1.2,-1){$L_{(e,m)}$};
\filldraw[blue] (0,0) circle (1.5pt);
\end{tikzpicture}}
\caption{The left-hand side of the consistency condition in Figure \ref{fig:squaretriangconsconds}. Comparing this to Figure \ref{fig:new3} fixes the undetermined coefficients $\beta_{e,m,e'}(p)$. }
\label{fig:new6}
\end{center}
\end{figure}
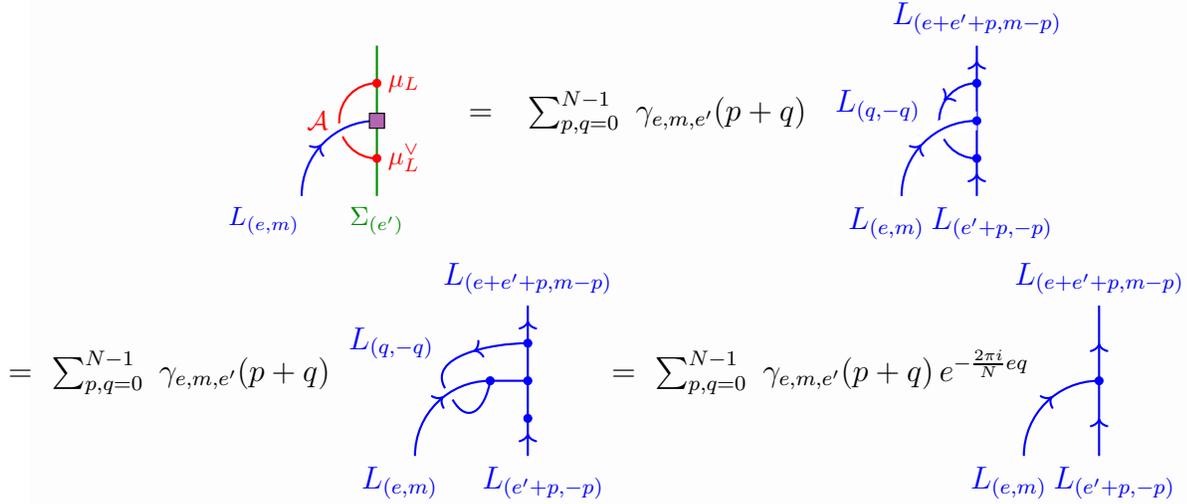

Having fixed the triangle junction, we may now fix the square junction in an analogous way. The most general form of the expansion is given in Figure \ref{fig:new5}, with the factors $\gamma_{e,m,e'}(p)$ to be determined. On the other hand, the left-hand side of the consistency condition in Figure \ref{fig:squaretriangconsconds} is as shown in Figure \ref{fig:new6}. A series of F-moves and braids gives the final line of Figure \ref{fig:new6}, from which we read off a constraint on  $\gamma_{e,m,e'}(p)$. Choosing conventions such that $\gamma_{e,m,e'}(0)=1$, the solution is 
\bea
\gamma_{e,m,e'}(p) = e^{ {2 \pi i \over N} e p}~. 
\eea
This completely fixes the square junction.  We note that, by the braiding properties of the lines $L_{(e,m)}$ and $L_{(e'+n,-n)}$, the square junction obtained here is equivalent to a half-braided triangle junction, as shown in Figure \ref{fig:tritosquare}. This was used in Section \ref{sec:3dFsymbols} of the main text, but will not be needed here. 

We have now successfully understood the purple circle and triangle junctions appearing in Figure \ref{fig:psisigsigcharge}. We have also understood the purple square junction, which does not appear in Figure \ref{fig:psisigsigcharge} explicitly but which will be important for its computation. All that remains is to evaluate the orange circle junctions, i.e. the four-fold intersection of $\cA$ and $\Sigma_{(e)}$.

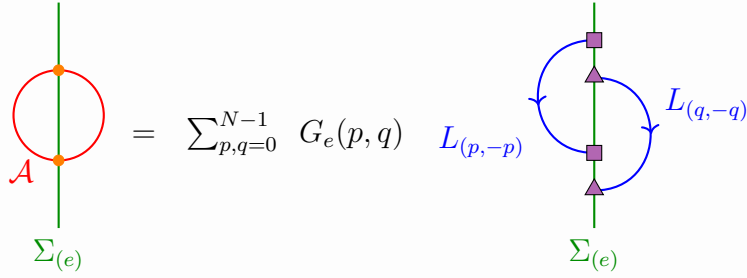
\begin{figure}[tbp]
\begin{center}
\begin{tikzpicture}[baseline=-10]
\draw [dgreen, thick] (0,-1.5) -- (0,1.5);
\draw[red, thick] (0,0) circle (0.6 cm);
\node[dgreen, below] at (0,-1.5) {$\Sigma_{(e)}$};
\node[red, below] at (-0.5,-0.45) {$\mathcal{A}$};
\filldraw[orange] (0,0.6) circle (2pt);
\filldraw[orange] (0,-0.6) circle (2pt);
\end{tikzpicture}
\hspace{0.1 in}$=\,\,\,\,\sum_{p,q=0}^{N-1} \,\,\,G_e(p,q)\,\,\,$
\begin{tikzpicture}[baseline=-10,square/.style={regular polygon,regular polygon sides=4}]
\draw [dgreen, thick] (0,-1.5) -- (0,1.5);

\draw[blue,thick,distance=1cm, decoration={markings, mark=at position 0.55 with {\arrow{>}}}, postaction={decorate}] (0,1) to[out=180,in=180] (0,-0.5); 

\draw[blue,thick,distance=1cm, decoration={markings, mark=at position 0.55 with {\arrow{<}}}, postaction={decorate}] (0,-1) to[out=0,in=0] (0,0.5); 

\node at (0,1) [square,draw,fill=violet!60,scale=0.5] {};
\node at (0,-0.5) [square,draw,fill=violet!60,scale=0.5] {};
\node[isosceles triangle,scale=0.4,
    isosceles triangle apex angle=60,
    draw,fill=violet!60,
    rotate=90,
    minimum size =0.01cm] at (0,-1){};
\node[isosceles triangle,scale=0.4,
    isosceles triangle apex angle=60,
    draw,fill=violet!60,
    rotate=90,
    minimum size =0.01cm] at (0,0.5){};

\node[dgreen, below] at (0,-1.5) {$\Sigma_{(e)}$};
\node[blue, below] at (-1.5,0) {$L_{(p,-p)}$};
\node[blue, below] at (1.5,0.5) {$L_{(q,-q)}$};

\end{tikzpicture}

\caption{Definition of the orange junctions between $\cA$ and $\Sigma_{(e)}$ in terms of the square and triangle junctions. For simplicity we only define the orange junctions in pairs, as shown. The coefficients $G_e(p,q)$ may be determined by imposing the consistency condition in Figure \ref{fig:new8}.}
\label{fig:new7}
\end{center}
\end{figure}

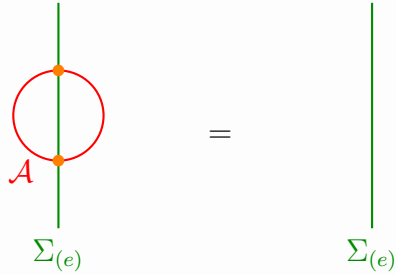
\begin{figure}[tbp]
\begin{center}
\begin{tikzpicture}[baseline=-10]
\draw [dgreen, thick] (0,-1.5) -- (0,1.5);
\draw[red, thick] (0,0) circle (0.6 cm);
\node[dgreen, below] at (0,-1.5) {$\Sigma_{(e)}$};
\node[red, below] at (-0.5,-0.45) {$\mathcal{A}$};
\filldraw[orange] (0,0.6) circle (2pt);
\filldraw[orange] (0,-0.6) circle (2pt);
\end{tikzpicture}
\hspace{0.5 in}$=$\hspace{0.5 in}
\begin{tikzpicture}[baseline=-10,square/.style={regular polygon,regular polygon sides=4}]
\draw [dgreen, thick] (0,-1.5) -- (0,1.5);
\node[dgreen, below] at (0,-1.5) {$\Sigma_{(e)}$};
\end{tikzpicture}

\caption{A consistency condition on the orange junctions. This follows from the fact that the $\Z_2^{\text{EM}}$ charge of the trivial junction between $\Sigma_{(e)}$, $\Sigma_{(e)}$, and $L_{(0,0)}$ is trivial. }
\label{fig:new8}
\end{center}
\end{figure}

In fact, instead of computing the four-fold junction itself, for our purposes it suffices to compute a \textit{pair} of orange junction, as shown in the left-hand side of Figure \ref{fig:new7}. This in general admits an expansion in terms of square and triangle junctions as in the right-hand side of Figure \ref{fig:new7}. Here $G_e(p,q)$ are a series of undertermined constants, which are subject to the consistency condition shown in Figure \ref{fig:new8}. Physically, what this condition says is that any point on $\Sigma_{(e)}$ is $\ZZ_2^{\mathrm{EM}}$ invariant. 

By using the definitions of the square and triangle junctions obtained above, it is straightforward to derive an expression for $G_e(p,q)$ from the consistency condition in Figure \ref{fig:new8}. We will only give the result, 
\bea
G_e(p,q) = e^{-{2\pi i \over N} (pq-p^2)}\, \delta_{p+q+e \text{ mod } 2}~,
\eea
where the delta function is included to enforce that the line in the interior of the bubble is the EM dual of the incoming line (since it passes $D_{\mathrm{EM}}$ in between). 
This is the final piece of data needed to compute the charge of the junction in Figure \ref{fig:psisigsigcharge}.

We finally return to the original calculation of interest, namely the computation of the charge of the junction between $L_{(N/2,N/2)}$ and $\Sigma_{(e)}$. The computation is illustrated in Figure \ref{fig:finalcomputation}. In words, we begin by using the definition of a pair of orange circle junctions given in Figure \ref{fig:new7}, together with the definition of the purple circle junction given in Figure \ref{fig:genformofpsiAA}. This gives a factor of 
\bea
G_e(p,q) \,\alpha_{{N\over 2},{N\over 2}}(-q) = e^{-{2 \pi i \over N}(pq-p^2)} e^{-\pi i q} \delta_{p+q+e \text{ mod } 2}~.
\eea
We may then use the definition of the square and triangle junctions given in Figures \ref{fig:new3} and \ref{fig:new5}, which gives a factor of
\bea
\gamma_{p,-p,e}(r+q)\times \gamma_{-p,p,e}\left(r+p+{N\over 2}\right) = e^{{2\pi i \over N}p(r+q)}e^{-{2 \pi i \over N} p (r + p + N/2)}~.  
\eea

\begin{figure}[!tbp]
\begin{center}
\begin{tikzpicture}[baseline=-10]
\draw [dgreen, thick] (0,-1.5) -- (0,1.5);
\draw [blue, thick, decoration={markings, mark=at position 0.8 with {\arrow{<}}}, postaction={decorate}] (0,0) to[out=45,in=200 ]  (1.5,0);
\node[isosceles triangle,scale=0.4,
    isosceles triangle apex angle=60,
    draw,fill=violet!60,
    rotate=90,
    minimum size =0.01cm] at (0,0){};
\draw[red, thick] (0,0) circle (0.6 cm);
\node[blue, above] at (1.7,0) {$L_{(N/2,N/2)}$};
\node[dgreen, below] at (0,-1.5) {$\Sigma_{(e)}$};
\node[red, below] at (-0.5,-0.45) {$\mathcal{A}$};
\filldraw[violet!60] (0.58,0.12) circle (2pt);
\filldraw[orange] (0,0.6) circle (2pt);
\filldraw[orange] (0,-0.6) circle (2pt);
\end{tikzpicture}
\hspace{0.0 in}$=\,\,\,\sum_{\substack{p,q=0\\ p+q+e=0\mod 2}}^{N-1} \,\,\,e^{-{2 \pi i \over N}(pq-p^2)} e^{-\pi i q} \,\,\,$
\begin{tikzpicture}[baseline=-10,square/.style={regular polygon,regular polygon sides=4}]
\draw [dgreen, thick] (0,-1.5) -- (0,1.5);

\draw[blue,thick,distance=1cm, decoration={markings, mark=at position 0.55 with {\arrow{>}}}, postaction={decorate}] (0,1) to[out=180,in=180] (0,-0.5); 

\draw[blue,thick,distance=1cm, decoration={markings, mark=at position 0.55 with {\arrow{<}}}, postaction={decorate}] (0,-1) to[out=0,in=0] (0,0); 

\draw [blue, thick, decoration={markings, mark=at position 0.8 with {\arrow{<}}}, postaction={decorate}] (0,0.5) to[out=45,in=200 ]  (1.5,0);

\node at (0,1) [square,draw,fill=violet!60,scale=0.5] {};
\node at (0,-0.5) [square,draw,fill=violet!60,scale=0.5] {};
\node[isosceles triangle,scale=0.4,
    isosceles triangle apex angle=60,
    draw,fill=violet!60,
    rotate=90,
    minimum size =0.01cm] at (0,-1){};
\node[isosceles triangle,scale=0.4,
    isosceles triangle apex angle=60,
    draw,fill=violet!60,
    rotate=90,
    minimum size =0.01cm] at (0,0.5){};

\node[isosceles triangle,scale=0.4,
    isosceles triangle apex angle=60,
    draw,fill=violet!60,
    rotate=90,
    minimum size =0.01cm] at (0,0){};

\draw[blue,thick] (0.5,-0.1)--(0.5+0.32,-0.1+0.32);

\filldraw[blue] (0.5,-0.1) circle (1.5pt);
\filldraw[blue] (0.5+0.32,-0.1+0.32) circle (1.5pt);

\node[dgreen, below] at (0,-1.5) {$\Sigma_{(e)}$};
\node[blue, below] at (-1.4,0) {$L_{(p,-p)}$};
\node[blue, below] at (1.5,-0.4) {$L_{(q,-q)}$};
\node[blue, below] at (2,0.7) {$L_{(N/2,N/2)}$};
\end{tikzpicture}
\\
$=\,\,\sum_{\substack{p,q,r=0\\ q+p+e+2r=0\,\,\mathrm{mod}\,\,N}}^{N-1} e^{-{2 \pi i \over N}(pq-p^2)} e^{-\pi i (q+p)} e^{{2\pi i \over N}p(q-p)}$
\begin{tikzpicture}[baseline=-10,square/.style={regular polygon,regular polygon sides=4}]
\draw [blue, thick] (0,-1.5) -- (0,1.5);

\draw[blue,thick,distance=1cm, decoration={markings, mark=at position 0.55 with {\arrow{>}}}, postaction={decorate}] (0,1) to[out=180,in=180] (0,-0.5); 

\draw[blue,thick,distance=1cm, decoration={markings, mark=at position 0.55 with {\arrow{<}}}, postaction={decorate}] (0,-1) to[out=0,in=0] (0,0);

\draw [blue, thick, decoration={markings, mark=at position 0.8 with {\arrow{<}}}, postaction={decorate}] (0,0.5) to[out=45,in=200 ]  (1.5,0);

\draw[blue,thick] (0.5,-0.1)--(0.5+0.32,-0.1+0.32);

\filldraw[blue] (0,-1) circle (1.5pt);
\filldraw[blue] (0,-0.5) circle (1.5pt);
\filldraw[blue] (0,0.5) circle (1.5pt);
\filldraw[blue] (0,1) circle (1.5pt);
\filldraw[blue] (0,0) circle (1.5pt);
\filldraw[blue] (0.5,-0.1) circle (1.5pt);
\filldraw[blue] (0.5+0.32,-0.1+0.32) circle (1.5pt);

\node[blue, below] at (0,-1.5) {$L_{(e+r,-r)}$};
\node[blue, below] at (-1.4,0) {$L_{(p,-p)}$};
\node[blue, below] at (1.5,-0.4) {$L_{(q,-q)}$};
\node[blue, below] at (2,0.7) {$L_{(N/2,N/2)}$};
\end{tikzpicture}\\
$\,\,\,=\,\,\,(-1)^e\,\,\sum_{r=0}^{N-1}$ \begin{tikzpicture}[baseline=-10]
\draw [blue, thick, decoration={markings, mark=at position 0.8 with {\arrow{>}},mark=at position 0.2 with {\arrow{>}}}, postaction={decorate}] (0,-1.5) -- (0,1.5);
\draw [blue, thick, decoration={markings, mark=at position 0.8 with {\arrow{<}}}, postaction={decorate}] (0,0) to[out=45,in=200 ]  (1.5,0);
\filldraw[blue] (0,0) circle (1.5pt);

\node[blue, above] at (1.7,0) {$L_{(N/2,N/2)}$};
\node[blue, below] at (0,-1.5) {$L_{(e+r,-r)}$};

\end{tikzpicture}
$\,\,\,=\,\,\,(-1)^e$\,\,\, \begin{tikzpicture}[baseline=-10]
\draw [dgreen, thick] (0,-1.5) -- (0,1.5);
\draw [blue, thick, decoration={markings, mark=at position 0.8 with {\arrow{<}}}, postaction={decorate}] (0,0) to[out=45,in=200 ]  (1.5,0);
\node[isosceles triangle,scale=0.4,
    isosceles triangle apex angle=60,
    draw,fill=violet!60,
    rotate=90,
    minimum size =0.01cm] at (0,0){};

\node[blue, above] at (1.7,0) {$L_{(N/2,N/2)}$};
\node[dgreen, below] at (0,-1.5) {$\Sigma_{(e)}$};

\end{tikzpicture}

\caption{The final computation of the junction charge. Details are given in the text. }
\label{fig:finalcomputation}
\end{center}
\end{figure}

Noting further that the line $L_{(e+r+q+p,-r-q-p)}$ in the interior of the bubble must be the EM dual of the incoming line $L_{(e+r,-r)}$ (since it passes $D_{\mathrm{EM}}$ in between) forces us to restrict to triplets $r,p,q$ satisfying
\bea
e+r+q+p = -r\,\,\,\mathrm{mod}\,\,N~,
\eea
or in other words $p+q = -e-2r$ mod $N$. Thus in total we produce a phase of 
\bea
e^{-{2 \pi i \over N}(pq-p^2)} e^{-\pi i q}\times e^{{2\pi i \over N}p(r+q)}e^{-{2 \pi i \over N} p (r + p + N/2)} = e^{-\pi i (p+q)} = (-1)^e~,
\eea
which is the final result for the charge of the junction.  In summary, for even $e$ the junction in Figure \ref{fig:psisigsigcharge} is $\Z_2^{\text{EM}}$ even.  On the other hand, for odd $e$ the junction is $\Z_2^{\text{EM}}$ odd, and hence there are an odd number of $K$ lines  terminating on the junction.

\section{Gauge invariance of twisted $\Z_N$ gauge theory}
\label{app:2dactioninvariance}

In this appendix we provide a bit more detail on the proof of gauge invariance of the action \eqref{3dZNC}. We first note that  $\delta_C \mathbf{a}$ is gauge invariant under \eqref{3dGT}. This means that the gauge variation of \eqref{3dgaugeinvariant} is 
\begin{eqnarray}\label{A1}
\begin{split}
(\delta_C \mathbf{g}^T)_{ij}  K^{C_{ij}+1}  (\delta_C \mathbf{a})_{jkl} &= \left(\mathbf{g}^T_{j} K^{C_{ij}}- \mathbf{g}^T_{i}\right) K^{C_{ij}+1}  (\delta_C \mathbf{a})_{jkl} \\
&= \left(\mathbf{g}^T_{j} - \mathbf{g}^T_{i} K^{C_{ij}} \right) K (\delta_C \mathbf{a})_{jkl}~.
\end{split}
\end{eqnarray}
We further use 
\begin{eqnarray}
(\delta_C \delta_C \mathbf{a})_{ijkl} = K^{C_{ij}}(\delta_C \mathbf{a})_{jkl} - (\delta_C \mathbf{a})_{ikl} + (\delta_C \mathbf{a})_{ijl} - (\delta_C \mathbf{a})_{ijk}=0
\end{eqnarray}
to replace $K^{C_{ij}} (\delta_C \mathbf{a})_{jkl}$ by $ (\delta_C \mathbf{a})_{ikl} - (\delta_C \mathbf{a})_{ijl} + (\delta_C \mathbf{a})_{ijk}$ in \eqref{A1}. The second line in \eqref{A1} becomes 
\begin{equation}
\begin{split}
\mathbf{g}^T_{j} K (\delta_C \mathbf{a})_{jkl} - \mathbf{g}^T_{i} K (\delta_C \mathbf{a})_{ikl} + \mathbf{g}^T_{i} K  (\delta_C \mathbf{a})_{ijl} - \mathbf{g}^T_{i} K  (\delta_C \mathbf{a})_{ijk} \equiv \left(\delta (\mathbf{g}^T K \delta_C \mathbf{a})\right)_{ijkl}
\end{split}
\end{equation} 
which is a total derivative. This shows that \eqref{3dZNC} is gauge invariant.

The invariance of \eqref{3dZNC} under background gauge transformation \eqref{3dbgdTransformation} is also straightforward. Under \eqref{3dbgdTransformation}, the term \eqref{3dgaugeinvariant} becomes 
\begin{eqnarray}
(\mathbf{a}_{ij})^T K^{\gamma_i} K^{C_{ij}+ \gamma_j - \gamma_i +1} \left(K^{C_{jk}+ \gamma_k-\gamma_j} K^{-\gamma_k}\mathbf{a}_{kl}- K^{-\gamma_j}\mathbf{a}_{jl} + K^{-\gamma_j} \mathbf{a}_{jk}\right)
\end{eqnarray}
and all of the $\gamma_i$ manifestly cancel.

\section{Fusion rules of the $(2+1)$d SymTFT of $TY(\Z_N)$ from modular $S$ matrices}
\label{app:fusionDCTY}

In this appendix, we collect the data of the Drinfeld center $\CZ(TY(\Z_N))$ of $TY(\Z_N)$ obtained in the mathematics literature  \cite{izumi2001structure,gelaki2009centers}. In particular, we record the explicit form of the modular $S$ matrices, which can be used to rederive the fusion rules of Section \ref{sec:ZTYZN} via the Verlinde formula.

\subsection{Objects and modular $S$ matrices}
\label{app.Smatrix}

In the notation of \cite{izumi2001structure}, the objects in the category $\CZ(TY(\Z_N))$ include 
\begin{enumerate}
    \item $2N$ invertible lines: $X_{g,i}, g\in \Z_N, i\in \Z_2$.
    \item $N(N-1)/2$ non-invertible lines of quantum dimension $2$: $Y_{[g,h]}, g,h\in \Z_N, g<h$.
    \item $2N$ non-invertible lines of quantum dimension $\sqrt{N}$: $Z_{g,i}, g\in \Z_N, i\in \Z_2$. 
\end{enumerate}
They are in one-to-one correspondence with $L_{(e)}^{q}, L_{[e,m]}, \Sigma_{(e)}^{q}$ in the main text. 
In this notation, the modular $S$ matrices are given by\footnote{For simplicity, we work only with the case in which the Frobenius-Schur indicator is trivial. }
\begin{eqnarray}
\label{eq:mathSmatrices}
\begin{split}
    &S_{X_{g,i}, X_{h,j}}=\frac{1}{2N}e^{- \frac{4\pi i}{N} gh}~,\\
    &S_{X_{g,i}, Z_{h,j}}= S_{Z_{h,j}, X_{g,i}} = \frac{(-1)^i}{2 \sqrt{N}} e^{- \frac{2\pi i}{N} gh}~,\\
    &S_{X_{g,i}, Y_{[h,k]}}= S_{ Y_{[h,k]}, X_{g,i}}= \frac{1}{N} e^{-\frac{2\pi i}{N} g (h+k)}~,\\
    &S_{Z_{g,i}, Y_{[h,k]}}= S_{ Y_{[h,k]}, Z_{g,i}} = 0~,\\
    &S_{Z_{g,i},Z_{h,j}}= \frac{(-1)^{i+j}}{2N} \omega_g \omega_h \sum_{k=0}^{N-1} e^{\frac{2\pi i}{N}(k-g-h)k}~,\\
    &S_{Y_{[g,h]},Y_{[g',h']}}=\frac{1}{N} \left( e^{-\frac{2\pi i}{N}(gh'+hg')}+ e^{-\frac{2\pi i}{N}(gg' + hh')}\right)~,
\end{split}
\end{eqnarray}
where 
\begin{eqnarray}
\omega_g= \left( \frac{1}{\sqrt{N}} \sum_{h=0}^{N-1} e^{-\frac{2\pi i}{N}gh}\gamma(h)\right)^{1/2}, \hspace{1cm} \gamma(h)
=\begin{cases}
(-1)^h e^{-\frac{\pi i}{N} h^2}, & N\in 2\Z+1\\
e^{-\frac{\pi i}{N} h^2}, & N\in 2\Z
\end{cases}
\end{eqnarray}
For completeness, we also record the spins of the lines here, 
\bea
\theta_{X_{g,i}} = e^{-{2 \pi i \over N} g^2}~, \hspace{0.5 in} \theta_{Y_{[g,h]}}=e^{-{2 \pi i \over N} gh}~, \hspace{0.5 in} \theta_{Z_{g,i}} = (-1)^i\, \omega^*_g~. 
\eea

The Verlinde formula \cite{Verlinde:1988sn} enables us to derive the fusion rules between $W_1, W_2, W_3\in \{X_{g,i}, Y_{[g,h]}, Z_{g,i}\}$ from the modular $S$ matrices,
\begin{eqnarray}
\label{eq:Verlindeformula}
N_{W_1, W_2}^{W_3} = \sum_{W} \frac{S_{W_1, W} S_{W_2, W} S^*_{W_3, W}}{S_{X_{0,0}, W}}~.
\end{eqnarray}
As a consistency check, we will now apply these formulae to small values of $N$ in order to reproduce the fusion rules found in the main text.

\subsection{Fusion rules for small $N$}

We emphasize that the operators $X_{g,i}, Y_{[g,h]}, Z_{g,i}$ in Appendix \ref{app.Smatrix} are not trivially identified with the operators $\widehat{L}_{(g)}^{(-1)^i}, \widehat{L}_{[g,h]}, \widehat{\Sigma}_{(g)}^{(-1)^i}$ in Section \ref{sec:3dDTYZN}. Rather, there is a non-trivial map depending on $N$.  Below, we work out this map for $N=2,3,4$ by comparing the fusion rules. It would be useful to determine this map for the general $N$, but we will not do so here.

\paragraph{$N=2$:}
The fusion rule between invertible lines is $X_{g,i} \otimes X_{h,j} = X_{g+h, i+j}$, and we therefore reproduce the results in the text upon identifying $X_{g,i} \leftrightarrow \widehat{L}_{(g)}^{(-1)^i}$. The fusion rules involving non-invertible lines are more interesting. The S-matrix elements and Verlinde formula give rise to,
\begin{eqnarray}
\begin{split}
    &X_{1,i} \otimes Z_{0,j} = Z_{0, i+j+1}~, \hspace{1cm} X_{1,i} \otimes Z_{1,j} = Z_{1, i+j}~,\\
    &Z_{g,i}\otimes Z_{g,j} = X_{0,i+j} \oplus X_{1,i+j+g+1}~.
\end{split}
\end{eqnarray}
Comparing with the fusion rule \eqref{eq:LSfusion}, we find the identification $Z_{1,j}\leftrightarrow \widehat{\Sigma}_{(0)}^{(-1)^j}, Z_{0,j}\leftrightarrow \widehat{\Sigma}_{(1)}^{(-1)^j}$. 
As there is only one line with quantum dimension 2, we also have the identification $Y_{[0,1]}\leftrightarrow \widehat{L}_{[0,1]}$. Indeed, the following fusion rules
\begin{eqnarray}
X_{g,i} \otimes Y_{[0,1]}= Y_{[0,1]}~, \hspace{1cm} Z_{0,i}\otimes Z_{1,j} = Y_{[0,1]}~,
\end{eqnarray}
match with those observed in Section \ref{sec:TYZ2}.

\paragraph{$N=3$:}
The fusion rules between invertible lines are again $X_{g,i} \otimes X_{h,j} = X_{g+h, i+j}$, and we therefore match the results in the text upon identify $X_{g,i} \leftrightarrow \widehat{L}_{(g)}^{(-1)^i}$. The fusion rules involving non-invertible lines are more interesting. The ones involving quantum dimension $\sqrt{3}$ lines are
\begin{eqnarray}\label{eq:Z1Z2}
\begin{split}
    &X_{0,i}\otimes Z_{0,j}= Z_{0,i+j}~, \hspace{1cm} X_{1,i}\otimes Z_{0,j}= Z_{2,i+j+1}~, \hspace{1cm} X_{2,i}\otimes Z_{0,j}= Z_{1,i+j+1}~.
\end{split}
\end{eqnarray}
Comparing with the fusion rule \eqref{eq:LSfusionodd}, we find the identification $Z_{0,j}\leftrightarrow \widehat{\Sigma}_{0}^{(-1)^j}$,  $Z_{1,j}\leftrightarrow \widehat{\Sigma}_{(1)}^{-(-1)^j}$, and $ Z_{2,j}\leftrightarrow \widehat{\Sigma}_{(2)}^{-(-1)^j}$. 
Other fusion rules between $X$ and $Z$ can be derived by starting with \eqref{eq:Z1Z2} and fusing additional invertible lines on both sides. The fusion rule between $X$ and $Y$ is 
\begin{eqnarray}
X_{g,i}\otimes Y_{[g',h']}= Y_{[g+g', g+h']}
\end{eqnarray}
which leads to the idenfication $Y_{[g,h]}\leftrightarrow \widehat{L}_{[g,h]}$. Other fusion rules can be similarly worked out, and we find that they match with those from Section \ref{sec:ZTYZN} upon using the above identifications.

\paragraph{$N=4$:}
The fusion rules between invertible lines are again $X_{g,i} \otimes X_{h,j} = X_{g+h, i+j}$, and we therefore identify $X_{g,i} \leftrightarrow \widehat{L}_{(g)}^{(-1)^i}$ once again.
The fusion between $X$ and $Z$ lines are
\begin{eqnarray}
\begin{split}
    X_{1,i}\otimes Z_{0,j}= Z_{2,i+j}~, \hspace{1cm} X_{1,i}\otimes Z_{1,j}= Z_{3,i+j+1}~, \hspace{1cm} X_{2,i}\otimes Z_{g,j}= Z_{g,i+j+g}~.
\end{split}
\end{eqnarray}
Note that upon fusing $X_{h,j}$, the $Z_{g,i}$'s for even $g$ form a closed orbit, while those for odd $g$ form another closed orbit. We therefore identify $Z_{0,j}\leftrightarrow \widehat{\Sigma}_{(0)}^{(-1)^{j}}$, $Z_{1,j}\leftrightarrow \widehat{\Sigma}_{(1)}^{(-1)^{j}}$, $Z_{2,j}\leftrightarrow \widehat{\Sigma}_{(2)}^{(-1)^{j}}$, and  $Z_{3,j}\leftrightarrow \widehat{\Sigma}_{(3)}^{-(-1)^{j}}$. The fusion rule between $X$ and $Y$ is 
\begin{eqnarray}
X_{g,i}\otimes Y_{[g',h']}= Y_{[g+g', g+h']}
\end{eqnarray}
which leads to the idenfication $Y_{[g,h]}\leftrightarrow \widehat{L}_{[g,h]}$. Other fusion rules can be similarly worked out, and we find that they match with those from Section \ref{sec:ZTYZN} upon using the above identifications.

\section{Condensation defects for the EM exchange symmetry in $(4+1)$d}
\label{app:DEM}

In this appendix, we discuss the four-dimensional defect $D_{\text{EM}}$ generating the $\Z_4^{\text{EM}}$ (or $\Z_2^{\text{EM}}$ for $N=2$) symmetry in $(4+1)$d
$\Z_N^{(1)}$ gauge theory. Since the defect is different depending on the value of $N$, we must discuss three separate cases.

\subsection{Odd $N$}

For $N$ odd the operator $D_{\text{EM}}$ is given by \eqref{eq:Z4EMcond}, which we reproduce here for convenience,
\begin{equation}\label{eq:DEModdN}
D_{\text{EM}}(M_4) =
\frac{|H^0(M_4, \Z_N)|^2}{|H^1(M_4, \Z_N)|^2} \sum_{\sigma, \sigma'\in H_2(M_4, \Z_N)} \exp\left(\frac{2\pi i}{N} (\braket{\sigma', \sigma'} + \braket{\sigma, \sigma'})\right) S_{(1,-1)}(\sigma) S_{(1,1)}(\sigma')~.
\end{equation}
From \eqref{eq:Z4EMoper}, the expected fusion rule would be $S_{(e,m)}(\tau) D_{\text{EM}}(M_4) = D_{\text{EM}}(M_4) S_{(-m,e)}(\tau)\exp(-\frac{2\pi i em}{N}\braket{\tau,\tau})$. We will check this now, together with the invertibility of the defect.

\subsubsection{Fusion rule $S_{(e,m)} D_{\mathrm{EM}}= D_{\mathrm{EM}} S_{(-m,e)} e^{-2\pi i em/N \braket{\tau,\tau}}$}

We begin by verifying the fusion rule above. We have
\begin{equation}
\begin{split}
    &S_{(e,m)}(\tau) D_{\text{EM}}(M_4)\\
    &= \frac{|H^0(M_4, \Z_N)|^2}{|H^1(M_4, \Z_N)|^2} \sum_{\sigma, \sigma'\in H_2(M_4, \Z_N)} \exp\left(\frac{2\pi i}{N} (\braket{\sigma', \sigma'} + \braket{\sigma, \sigma'})\right) S_{(e,m)}(\tau) S_{(1,-1)}(\sigma) S_{(1,1)}(\sigma')\\
    &=\frac{|H^0(M_4, \Z_N)|^2}{|H^1(M_4, \Z_N)|^2} \sum_{\sigma, \sigma'\in H_2(M_4, \Z_N)} \exp\left(\frac{2\pi i}{N} (\braket{\sigma', \sigma'} + \braket{\sigma, \sigma'}+ (e+m)\braket{\tau, \sigma}+ (m-e)\braket{\tau,\sigma'})\right)\\&\hspace{4cm} \times S_{(1,-1)}(\sigma) S_{(1,1)}(\sigma')S_{(e,m)}(\tau)~.
\end{split}
\end{equation}
Now we use the fusion rule \eqref{eq:4dfusionrule} to rewrite $S_{(e,m)}(\tau)$ as $e^{2\pi i m^2/N \braket{\tau,\tau}}S_{(e,-e)}(\tau) S_{(m,m)}(\tau) S_{(-m,e)}(\tau)$. The above expression then becomes 
\begin{equation}
\begin{split}
    &S_{(e,m)}(\tau) D_{\text{EM}}(M_4)\\
    &=\frac{|H^0(M_4, \Z_N)|^2}{|H^1(M_4, \Z_N)|^2} \sum_{\sigma, \sigma'\in H_2(M_4, \Z_N)} \exp\bigg(\frac{2\pi i}{N} (\braket{\sigma', \sigma'} + \braket{\sigma, \sigma'}+ (e+m)\braket{\tau, \sigma}+ (m-e)\braket{\tau,\sigma'}\\& \hspace{1cm}+ m^2 \braket{\tau,\tau})\bigg)  S_{(1,-1)}(\sigma) S_{(1,1)}(\sigma')S_{(e,-e)}(\tau) S_{(m,m)}(\tau) S_{(-m,e)}(\tau)~.
\end{split}
\end{equation}
Note that $(e,-e)=e(1,-1), (m,m)=m(1,1)$, and hence one can switch the order of the surface operators and combine the terms using the quantum torus algebra \eqref{eq:4dtorusquantumalgebra}. The result is \begin{equation}
\begin{split}
    &S_{(e,m)}(\tau) D_{\text{EM}}(M_4)\\
    &=\frac{|H^0(M_4, \Z_N)|^2}{|H^1(M_4, \Z_N)|^2} \sum_{\sigma, \sigma'\in H_2(M_4, \Z_N)} \exp\bigg(\frac{2\pi i}{N} (\braket{\sigma', \sigma'} + \braket{\sigma, \sigma'}+ m\braket{\tau, \sigma}+ (e+2m)\braket{\tau,\sigma'}\\&\hspace{1cm}+ m^2 \braket{\tau,\tau})\bigg) S_{(1,-1)}(\sigma+e \tau) S_{(1,1)}(\sigma'+m\tau) S_{(-m,e)}(\tau)~. 
\end{split}
\end{equation}
We finally make a change of variables $\sigma\to \sigma-e\tau$, $\sigma'\to \sigma'-m\tau$, upon which the above expression then becomes 
\begin{eqnarray}
\begin{split}
    &S_{(e,m)}(\tau) D_{\text{EM}}(M_4)\\
    &=\frac{|H^0(M_4, \Z_N)|^2}{|H^1(M_4, \Z_N)|^2} \sum_{\sigma, \sigma'\in H_2(M_4, \Z_N)} \exp\bigg(\frac{2\pi i}{N} (\braket{\sigma', \sigma'} + \braket{\sigma, \sigma'} -em \braket{\tau,\tau} )\bigg) \\&\hspace{4cm}\times  S_{(1,-1)}(\sigma) S_{(1,1)}(\sigma') S_{(-m,e)}(\tau) \\
    &= D_{\text{EM}}(M_4)S_{(-m,e)}(\tau)  \exp\left(-\frac{2\pi i em}{N}\braket{\tau,\tau}\right)~. 
\end{split}
\end{eqnarray}
Using the commutation relation \eqref{eq:5dcomm}, the surface operator in the last step can finally be re-expressed as 
\begin{eqnarray}
\begin{split}
    S_{(-m,e)}(\tau)  \exp\left(-\frac{2\pi i em}{N}\braket{\tau,\tau}\right)&= S_{(-m,0)}(\tau) S_{(0,e)}(\tau)  \exp\left(-\frac{2\pi i em}{N}\braket{\tau,\tau}\right)\\
    &= S_{(0,e)}(\tau) S_{(-m,0)}(\tau)~.
\end{split}
\end{eqnarray}
Thus $D_{\text{EM}}$ maps $S_{(e,0)}$ to $S_{(0,e)}$ and $S_{(0,m)}$ to $S_{(-m,0)}$, exactly as expected in \eqref{eq:Z4EMoper}.

\subsubsection{Charge conjugation operator $C=D_{\text{EM}}^2$}

Before checking $D_{\text{EM}}^4=1$, we first compute $C=D_{\text{EM}}^2$, which should be the charge conjugation operator.\footnote{Note that $C$ is the 4d charge conjugation defect, which should not be confused with the 3d condensation defect $\CC$ defined in \eqref{eq:NNbcondensate}. } To this effect, we have
\begin{equation}
\begin{split}
    D_{\text{EM}}(M_4)^2 &= \frac{|H^0(M_4, \Z_N)|^4}{|H^1(M_4, \Z_N)|^4} \sum_{\sigma, \sigma',\tau,\tau'\in H_2(M_4, \Z_N)} \exp\left(\frac{2\pi i}{N} (\braket{\sigma', \sigma'} + \braket{\sigma, \sigma'}+ \braket{\tau', \tau'} + \braket{\tau, \tau'})\right)\\&\hspace{4cm}\times  S_{(1,-1)}(\sigma) S_{(1,1)}(\sigma')S_{(1,-1)}(\tau) S_{(1,1)}(\tau')~.
\end{split}
\end{equation}
To simplify the above expression, we switch the order of surface operators, combine the terms with the same charge using the quantum torus algebra, and make a change of variables. The final expression is 
\begin{equation}
\begin{split}
    C(M_4)=D_{\text{EM}}(M_4)^2 &= \frac{|H^0(M_4, \Z_N)|^4}{|H^1(M_4, \Z_N)|^4} \sum_{\sigma, \sigma',\tau,\tau'\in H_2(M_4, \Z_N)} \exp\bigg(\frac{2\pi i}{N} (\braket{\sigma', \sigma'} + \braket{\sigma, \sigma'}+ \braket{\tau', \tau'}\\& \hspace{1cm} + \braket{\tau, \tau} - \braket{\sigma',\tau'}- \braket{\sigma,\tau'}+ \braket{\sigma',\tau}-\braket{\sigma,\tau} )\bigg) S_{(1,-1)}(\sigma) S_{(1,1)}(\sigma')~.
\end{split}
\end{equation}
Since $\tau, \tau'$ only appear in the exponent and not in the surface operators, we can perform the sum over them to simplify the expression. We first sum over $\tau$. The relevant part is
\begin{eqnarray}\label{eq:tautau}
\sum_{\tau\in H_2(M_4, \Z_N)} \exp\left( \frac{2\pi i}{N} (\braket{\tau, \tau}+ \braket{\sigma'-\sigma, \tau})\right) ~.
\end{eqnarray}
To complete the square, it is useful to introduce $\rho$ satisfying $2\rho=\sigma'-\sigma$. Note that $\rho$ exists since the equality is defined modulo $N\lambda$ where $\lambda$ is the generator of $H_2(M_4, \Z_N)$ and $N$ is odd. Hence \eqref{eq:tautau} can be simplified to 
\begin{eqnarray}
\sqrt{|H_2(M_4, \Z_N)|} \CZ_{Y}[M_4, \Z_N] \exp\left(- \frac{2\pi i}{N} \braket{\rho, \rho}\right)~,
\end{eqnarray}
where 
\begin{eqnarray}\label{eq:Yfunc}
\CZ_{Y}[M_4, \Z_N]:= \frac{1}{\sqrt{|H_2(M_4, \Z_N)|}}\sum_{\tau\in H_2(M_4, \Z_N)} \exp\left( \frac{2\pi i}{N} \braket{\tau, \tau}\right)
\end{eqnarray}
defines an invertible TQFT, which trivializes on a spin manifold. Similarly, we can sum over $\tau'$ by introducing $2\omega=\sigma+\sigma'$, which gives
\begin{eqnarray}
\sqrt{|H_2(M_4, \Z_N)|} \CZ_{Y}[M_4, \Z_N] \exp\left(- \frac{2\pi i}{N} \braket{\omega, \omega}\right)~.
\end{eqnarray}
The charge conjugation defect thus becomes 
\begin{eqnarray}
\begin{split}
    C(M_4) &= \frac{|H^0(M_4, \Z_N)|^4 |H_2(M_4, \Z_N)|}{|H^1(M_4, \Z_N)|^4} \CZ_{Y}[M_4, \Z_N]^2 \sum_{\sigma, \sigma'\in H_2(M_4, \Z_N)} \exp\bigg(\frac{2\pi i}{N} (\braket{\sigma', \sigma'} \\& \hspace{1cm} +\braket{\sigma,\sigma'}-\braket{\rho,\rho}-\braket{\omega,\omega} )\bigg) S_{(1,-1)}(\sigma) S_{(1,1)}(\sigma')~.
\end{split}
\end{eqnarray}
To further simplify the expression, it is useful to replace $\sigma,\sigma'$ in terms of $\rho,\omega$ via 
\begin{eqnarray}
\sigma=\omega-\rho, \hspace{1cm} \sigma'= \omega+ \rho~.
\end{eqnarray}
Summing over $\sigma,\sigma'$ is equivalent to summing over $\rho,\omega$. Hence the expression for the charge conjugation defect  simplifies to 
\begin{equation}\label{eq:Codd}
\begin{split}
    &C(M_4)\\&= \frac{|H^0(M_4, \Z_N)|^4 |H_2(M_4, \Z_N)|}{|H^1(M_4, \Z_N)|^4} \CZ_{Y}[M_4, \Z_N]^2 \sum_{\rho, \omega\in H_2(M_4, \Z_N)} \exp\bigg(\frac{2\pi i}{N} (-\braket{\rho,\rho}+\braket{\omega,\omega}+ 2 \braket{\rho,\omega})\bigg) \\& \hspace{8cm}\times S_{(1,-1)}(\omega-\rho) S_{(1,1)}(\omega+\rho)\\ 
    &= \frac{|H^0(M_4, \Z_N)|^4 |H_2(M_4, \Z_N)|}{|H^1(M_4, \Z_N)|^4} \CZ_{Y}[M_4, \Z_N]^2 \sum_{\rho, \omega\in H_2(M_4, \Z_N)} \exp\bigg(-\frac{4\pi i}{N} \braket{\rho, \omega}\bigg) S_{(0,2)}(\rho) S_{(2,0)}(\omega)~.
\end{split}
\end{equation}

\subsubsection{Invertibility of $D_{\text{EM}}$}

We finally check the invertibility of $D_{\text{EM}}$ by computing $D_{\text{EM}}^4$, and conforming it is the identity, up to a local counterterm.  Since we have already computed $C=D_{\text{EM}}^2$, we only need to compute $C^2$ as follows. 
\begin{equation}
\begin{split}
  &C(M_4)^2\\&= \frac{|H^0(M_4, \Z_N)|^8 |H_2(M_4, \Z_N)|^2}{|H^1(M_4, \Z_N)|^8} \CZ_{Y}[M_4, \Z_N]^4 \sum_{\sigma,\sigma',\tau,\tau'\in H_2(M_4, \Z_N)} \exp\bigg(-\frac{4\pi i}{N} (\braket{\sigma, \sigma'}+ \braket{\tau,\tau')}\bigg)\\& \hspace{6cm} \times S_{(0,2)}(\sigma) S_{(2,0)}(\sigma')S_{(0,2)}(\tau) S_{(2,0)}(\tau')\\
  &=\frac{|H^0(M_4, \Z_N)|^8 |H_2(M_4, \Z_N)|^2}{|H^1(M_4, \Z_N)|^8} \CZ_{Y}[M_4, \Z_N]^4 \sum_{\sigma,\sigma',\tau,\tau'\in H_2(M_4, \Z_N)}\exp\bigg(\frac{2\pi i}{N} (-2\braket{\sigma, \sigma'}-2 \braket{\tau,\tau')}\\& \hspace{1cm}-4\braket{\sigma',\tau}\bigg)  S_{(0,2)}(\sigma+\tau) S_{(2,0)}(\sigma'+\tau')\\
  &= \frac{|H^0(M_4, \Z_N)|^8 |H_2(M_4, \Z_N)|^2}{|H^1(M_4, \Z_N)|^8} \CZ_{Y}[M_4, \Z_N]^4 \sum_{\sigma,\sigma',\tau,\tau'\in H_2(M_4, \Z_N)}\exp\bigg(\frac{2\pi i}{N} (-2\braket{\sigma, \sigma'}+2 \braket{\sigma,\tau')}\\& \hspace{1cm}-2\braket{\sigma',\tau}\bigg)  S_{(0,2)}(\sigma) S_{(2,0)}(\sigma')\\
  &= \frac{|H^0(M_4, \Z_N)|^8 |H_2(M_4, \Z_N)|^4}{|H^1(M_4, \Z_N)|^8} \CZ_{Y}[M_4, \Z_N]^4 = \chi[M_4,\Z_N]^4 \CZ_{Y}[M_4, \Z_N]^4~.
\end{split}
\end{equation}
From the second to the last line, we have combined the surface operators with the same charge, made a change of variables $\sigma\to \sigma-\tau, \sigma'\to \sigma'-\tau'$,  integrated out $\tau$ and $\tau'$ which enforced $\sigma,\sigma'$ to be trivial 2-cycles in $H_2(M_4, \Z_N)$, and finally used the definition of the 4d Euler counterterm in \eqref{eq:Euler4d}. Note that $\chi[M_4,\Z_N]$ is a local counterterm and $\CZ_{Y}[M_4, \Z_N]$ is a phase, which proves that $C(M_4)^2= D_{\text{EM}}(M_4)^4$ is an invertible operator.

\subsubsection{Summary of algebra of co-dimension one}

We now summarize the algebra involving the $D_{\text{EM}}$ defect. We may define the orientation reversal of $D_{\text{EM}}$ as  $\overline{D}_{\text{EM}}$ via 
\begin{equation}
\begin{split}
    \overline{D}_{\text{EM}}(M_4) = \chi[M_4,\Z_N]^{-2}
\frac{|H^0(M_4, \Z_N)|^2}{|H^1(M_4, \Z_N)|^2} \sum_{\sigma, \sigma'\in H_2(M_4, \Z_N)} \exp\left(-\frac{2\pi i}{N} (\braket{\sigma, \sigma} + \braket{\sigma, \sigma'})\right) S_{(1,1)}(\sigma') S_{(1,-1)}(\sigma)
\end{split}
\end{equation}
where motivated by \eqref{eq:CNbarCN4d} the Euler counterterm is included. This also renders $\overline{D}_{\text{EM}}\times D_{\text{EM}}=1$. The fusion rules are then
\begin{eqnarray}
\begin{split}
    &C(M_4)= D_{\text{EM}}(M_4)^2~,\\ 
    &D_{\text{EM}}(M_4)^4=\chi[M_4, \Z_N]^4 \CZ_{Y}[M_4, \Z_N]^4~,\\
    &\overline{D}_{\text{EM}}(M_4)\times D_{\text{EM}}(M_4)=1~,\\
    & \overline{D}_{\text{EM}}(M_4)= \chi[M_4,\Z_N]^{-4} \CZ_{Y}[M_4, \Z_N]^{-4} C(M_4) D_{\text{EM}}(M_4)^2~.
\end{split}
\end{eqnarray}

\subsection{$N=2$}

When $N=2$, there is only one type of condensate, and the EM symmetry is $\Z_2^{\text{EM}}$. The topological defect for $\Z_2^{\text{EM}}$ is 
\begin{eqnarray}
\begin{split}
    D_{\text{EM}}(M_4) = \frac{|H^0(M_4,\Z_2)|}{|H^1(M_4,\Z_2)|} \sum_{\sigma\in H_2(M_4,\Z_2)} S_{(1,0)}(\sigma) S_{(0,1)}(\sigma+[w_2^{TM}])~,
\end{split}
\end{eqnarray}
where $[w_2^{TM}]$ is the Poincar{\'e} dual of the second Stiefel-Whitney class of the tangent bundle of the spacetime manifold $w_2^{TM}$.\footnote{We thank Ryohei Kobayashi for discussions and a note which lead us to this construction. } To confirm that this is the correct result, we check the commutation relations with the surface operators $S_{(e,m)}(\tau)$, as well as that $D_{\text{EM}}(M_4)^2=1$  up to Euler counterterm.

\subsubsection{Fusion with $S_{(e,m)}$}

We begin by computing 
\begin{equation}
    \begin{split}
        &S_{(e,m)}(\tau) D_{\text{EM}}(M_4)\\
        &= \frac{|H^0(M_4,\Z_2)|}{|H^1(M_4,\Z_2)|} \sum_{\sigma\in H_2(M_4,\Z_2)} S_{(e,m)}(\tau) S_{(1,0)}(\sigma) S_{(0,1)}(\sigma+[w_2^{TM}])\\
        &= \frac{|H^0(M_4,\Z_2)|}{|H^1(M_4,\Z_2)|} \sum_{\sigma\in H_2(M_4,\Z_2)} \exp\left(i \pi (m \braket{\tau,\sigma}+ e \braket{\tau, \sigma+[w_2^{TM}]})\right) S_{(1,0)}(\sigma) S_{(0,1)}(\sigma+[w_2^{TM}])S_{(e,m)}(\tau)~.\\
    \end{split}
\end{equation}
Using $S_{(e,m)}(\tau)= S_{(1,0)}((e-m)\tau)S_{(0,1)}((m-e)\tau) S_{(m,e)}(\tau)e^{i\pi m(m-e)\braket{\tau,\tau}}$ and combining the surface operators with the same charge, the above expression simplifies to 
\begin{equation}
\begin{split}
    &S_{(e,m)}(\tau) D_{\text{EM}}(M_4)\\
    &= \frac{|H^0(M_4,\Z_2)|}{|H^1(M_4,\Z_2)|} \sum_{\sigma\in H_2(M_4,\Z_2)} \exp\Big(i \pi (m \braket{\tau,\sigma}+ e \braket{\tau, \sigma+[w_2^{TM}]}+m(m-e)\braket{\tau,\tau}\\&\hspace{0.5cm}+(e-m)\braket{\tau,\sigma+[w_2^{TM}]})\Big) S_{(1,0)}(\sigma+(e-m)\tau) S_{(0,1)}(\sigma+[w_2^{TM}]+(e-m)\tau)S_{(m,e)}(\tau)~.\\
\end{split}
\end{equation}
We further make a change of variable $\sigma\to \sigma-(e-m)\tau$, and use $\braket{\tau,\tau}=\braket{\tau,w_2^{TM}}\,\,\mathrm{mod}\,\, 2$, upon which the expression simplifies to
\begin{equation}
\begin{split}
    &S_{(e,m)}(\tau) D_{\text{EM}}(M_4)\\
    &= \frac{|H^0(M_4,\Z_2)|}{|H^1(M_4,\Z_2)|} \sum_{\sigma\in H_2(M_4,\Z_2)} \exp\left(i \pi em \braket{\tau,\tau}\right) S_{(1,0)}(\sigma) S_{(0,1)}(\sigma+[w_2^{TM}])S_{(m,e)}(\tau)\\
    &=D_{\text{EM}}(M_4) S_{(m,e)}(\tau) \exp\left(i \pi em \braket{\tau,\tau}\right)
\end{split}
\end{equation}
as expected.\footnote{Note that this surface operator exchanges $e\leftrightarrow m$. It can be shown that this operator is a composition of $D_{m\psi} D_{e\psi} D_{m\psi}$, where $D_{m\psi}$ and $D_{e\psi}$ are operators exchanging $m \leftrightarrow \psi$ and $e\leftrightarrow \psi$, respectively. One has
$D_{m\psi}= \frac{|H^0(M_4,\Z_2)|}{|H^1(M_4,\Z_2)|} \sum_{\sigma \in H_2(M_4,\Z_2)} \exp\left( \frac{i \pi}{2} \int \cP(\text{PD}(\sigma)) \right) S_{(1,0)}(\sigma)$, and $D_{e\psi}= \frac{|H^0(M_4,\Z_2)|}{|H^1(M_4,\Z_2)|} \sum_{\sigma \in H_2(M_4,\Z_2)} \exp\left( \frac{i \pi}{2} \int \cP(\text{PD}(\sigma)) \right) S_{(0,1)}(\sigma)$, where $\text{PD}(\sigma)$ is the Poincar{\'e} dual of $\sigma$. 
}

\subsubsection{Invertibility of $D_{\mathrm{EM}}(M_4)$}

We next compute
\begin{equation}
    \begin{split}
        D_{\text{EM}}(M_4)^2 &= \frac{|H^0(M_4,\Z_2)|^2}{|H^1(M_4,\Z_2)|^2} \sum_{\sigma,\tau} S_{(1,0)}(\sigma) S_{(0,1)}(\sigma+[w_2^{TM}]) S_{(1,0)}(\tau) S_{(0,1)}(\tau+[w_2^{TM}]) \\
        &= \frac{|H^0(M_4,\Z_2)|^2}{|H^1(M_4,\Z_2)|^2} \sum_{\sigma,\tau} S_{(1,0)}(\sigma+\tau) S_{(0,1)}(\sigma+\tau) \exp\left( i\pi \braket{\sigma+[w_2^{TM}], \tau}\right)\\
        &=\frac{|H^0(M_4,\Z_2)|^2}{|H^1(M_4,\Z_2)|^2} \sum_{\sigma,\tau} S_{(1,1)}(\sigma)  \exp\left( i\pi \braket{\sigma+[w_2^{TM}]-\tau, \tau}\right)~.\\
    \end{split}
\end{equation}
Using $\braket{\tau,\tau}= \braket{[w_2^{TM}],\tau}\,\,\mathrm{mod}\,\,2$, this expression can be simplified to 
\begin{equation}\label{eq:N=2fusionDD}
    \begin{split}
        D_{\text{EM}}(M_4)^2 &=\frac{|H^0(M_4,\Z_2)|^2}{|H^1(M_4,\Z_2)|^2} \sum_{\sigma,\tau} S_{(1,1)}(\sigma)  \exp\left( i\pi \braket{\sigma, \tau}\right)
        \\&= \frac{|H^0(M_4,\Z_2)|^2 |H_2(M_4,\Z_2)|}{|H^1(M_4,\Z_2)|^2} = \chi[M_4,\Z_2]~.
    \end{split}
\end{equation}
We can also define the orientation reversal as 
\begin{eqnarray}\label{eq:DEMN=2orientrev}
\overline{D}_{\text{EM}}= \chi[M_4, \Z_2]^{-1} D_{\text{EM}}
\end{eqnarray}
upon which \eqref{eq:N=2fusionDD} can be rewritten as 
\begin{eqnarray}
D_{\text{EM}}(M_4) \times \overline{D}_{\text{EM}}(M_4) = 1
\end{eqnarray}
Hence $D_{\text{EM}}(M_4)$ is invertible, and generates a $\Z_2$ symmetry.

\subsection{Even $N$ and $N\geq 4$:}
\label{app:DEMN>4}

The symmetry defect in this case is almost identical to that for odd $N$, but with some minor modifications. In particular, we now have
\begin{equation}\label{eq:DEMevenN}
\begin{split}
    D_{\text{EM}}(M_4)& =
\frac{|H^0(M_4, (\Z_N\times \Z_N)/\Z_2)|}{|H^1(M_4, (\Z_N\times \Z_N)/\Z_2))|} \sum_{(\sigma, \sigma')\in H_2(M_4, (\Z_N\times \Z_N)/\Z_2)} \exp\left(\frac{2\pi i}{N} (\braket{\sigma', \sigma'} + \braket{\sigma, \sigma'})\right) \\& \hspace{4cm} \times S_{(1,-1)}(\sigma)S_{(0,1)}\left(\frac{N}{2}[w_2^{TM}]\right) S_{(1,1)}(\sigma')~.
\end{split}
\end{equation}
As in the case of $N=2$, for generic even $N$, we are allowed to turn on the background field $w_2^{TM}$.

\subsubsection{Gauge invariance}

As explained in Section \ref{sec:Z4EM4d}, summing over $H_2(M_4, (\Z_N\times \Z_N)/\Z_2)$ requires  that the defect should be invariant under the gauge transformation $\sigma\to \sigma+\frac{N}{2}\lambda, \sigma'\to \sigma'+\frac{N}{2}\lambda$, which follows from the identification $S_{(N/2,N/2)}(\sigma)= S_{(N/2,-N/2)}(\sigma)$. Indeed, it is straightforward to check that \eqref{eq:DEMevenN} is invariant under this gauge transformation, due to the proper coupling to the background field $[w_2^{TM}]$. This means that summing over the elements $(\sigma,\sigma')$ in $H_2(M_4, (\Z_N\times \Z_N)/\Z_2)$ amounts to summing over $\sigma$ and $\sigma'$ in $H_2(M_4,\Z_N)$ separately, but only over half of the total domain.\footnote{As an analogy, summing over $(a,b)\in (\Z_N\times \Z_N)/\Z_2$ amounts to choosing one element in the pair $(a,b) \sim (a+\frac{N}{2}, b+\frac{N}{2})$ and summing over the $N^2/2$ choices of pairs. The sum is independent of the choice thanks to gauge invariance. }

The $\Z_4^{\text{EM}}$ defect \eqref{eq:DEMevenN} differs from that for the odd $N$ case by coupling to the background field $[w_2^{TM}]$. If we do not couple to the background field in \eqref{eq:DEMevenN} and use \eqref{eq:DEModdN} instead, then under the gauge transformation $\sigma\to \sigma+\frac{N}{2}\lambda, \sigma'\to \sigma'+\frac{N}{2}\lambda$ we would find that $D_{\text{EM}}$ transforms as $D_{\text{EM}}\to (-1)^{\frac{N}{2}\braket{\lambda,\lambda}} D_{\text{EM}} = (-1)^{\frac{N}{2}\braket{\lambda,[w_2^{TM}]}}D_{\text{EM}}$, which is not invariant and renders the sum over $H_2(M_4, (\Z_N\times \Z_N)/\Z_2)$ ill-defined.  The coupling to $[w_2^{TM}]$ in \eqref{eq:DEMevenN} precisely compensates this non-invariance.

\subsubsection{Fusing with $S_{(e,m)}(\tau)$}

To justify that \eqref{eq:DEMevenN} is the correct condensation defect of $\Z_4^{\text{EM}}$, we may check that the commutation relation with the surface operators is of the desired form \eqref{eq:Z4EMoper},
\begin{eqnarray}
S_{(e,m)}(\tau) D_{\text{EM}}(M_4) = D_{\text{EM}}(M_4) S_{(-m,e)}(\tau) \exp\left( -\frac{2\pi i e m }{N} \braket{\tau,\tau}\right)~.
\end{eqnarray}
The proof is identical to the case for odd $N$, and will not be reproduced here.

\subsubsection{Two useful summation formulas}

Before discussing the charge conjugation operator and verifying the invertibility of $D_{\text{EM}}$, it is useful to discuss a summation formula which will be used later on. We start with the discrete Fourier transformation,
\begin{eqnarray}\label{eq:DFT}
\frac{1}{\sqrt{N}} \sum_{x\in \Z_N} \exp\left( \frac{2\pi i}{2N} x^2 + \frac{2\pi i}{N} x p \right) = \exp\left( \frac{\pi i}{4}- \frac{\pi i p^2}{N}\right)
\end{eqnarray}
and 
\begin{eqnarray}\label{eq:C27}
\begin{split}
    &\frac{1}{N} \sum_{x,y\in \Z_N, x+y\in 2\Z} \exp\left(\frac{2\pi i}{2N} (x^2+y^2)\right) = 
    \begin{cases}
    i, & N\in 4\Z\\
    0, & N\in 4\Z+2
    \end{cases}\\
    &\frac{1}{N} \sum_{x,y\in \Z_N, x+y\in 2\Z+1} \exp\left(\frac{2\pi i}{2N} (x^2+y^2)\right) = \begin{cases}
    0, & N\in 4\Z\\
    i, & N\in 4\Z+2
    \end{cases}
\end{split}
\end{eqnarray}

\paragraph{First Formula:}
We now prove the first of two summation formulas of interest. To motivate the first, we begin by considering the function
\begin{eqnarray}
f(a,b)= \frac{1}{N}\sum_{\substack{x\in \Z_N, y\in \Z_N\\ x+y\in 2\Z}} \exp\left( \frac{2\pi i}{N} \bigg( \frac{x^2+y^2}{2}- ax - by\bigg)\right)
\end{eqnarray}
where $N$ is an even number, and $a,b\in \Z_N$. 
Note that $f(a,b)$ is invariant under shifting $a,b$ by $N/2$ simultaneously, i.e.
\begin{eqnarray}\label{eq:ff}
f(a,b)= f\left(a+\frac{N}{2}, b+\frac{N}{2}\right)
\end{eqnarray}
which is guaranteed by $x+y\in 2\Z$ in the range of sum. To perform the sum, we complete the square and make a change of variable. Then $f(a,b)$ simplifies to 
\begin{eqnarray}\label{eq:fff}
f(a,b)= \exp\left(- \frac{2\pi i}{2N} (a^2+b^2)\right) \frac{1}{N}\sum_{\substack{x\in \Z_N, y\in \Z_N\\ x+y+a+b\in 2\Z}} \exp\left( \frac{2\pi i}{2N} (x^2+y^2)\right)~.
\end{eqnarray}
The sum is invariant under the shift $(a,b)\to (a+\frac{N}{2}, b+\frac{N}{2})$, but the phase in  front is not. The phase transforms as 
\begin{eqnarray}
\exp\left(- \frac{2\pi i}{2N} (a^2+b^2)\right) \to e^{- i \pi (a+b) - i \pi N/2}\exp\left(- \frac{2\pi i}{2N} (a^2+b^2)\right)~.
\end{eqnarray}
On the other hand, $f(a,b)$ satisfies \eqref{eq:ff}, and hence $f(a,b)=0$ whenever the phase $e^{- i \pi (a+b) - i \pi N/2}\neq 1$. In particular, this implies that $f(a,b)=0$ when $(N,a+b)\in (4\Z, 2\Z)$ and $(N,a+b)\in (4\Z+2, 2\Z+1)$. Further evaluating the summation \eqref{eq:fff} via \eqref{eq:C27}, we obtain 
\begin{eqnarray}\label{eq:DFT2}
f(a,b)= \exp\left(-\frac{2\pi i}{2N} (a^2+b^2) + \frac{\pi i}{2}\right) \delta_{a+b+N/2\,\,\mathrm{mod}\,\,2}~.
\end{eqnarray}
Comparing this to \eqref{eq:DFT}, this is the generalized discrete Fourier transformation on a space with constraints.

We would like to generalize \eqref{eq:DFT2} so that the variables are the elements in $H_2(M_4,\Z_N)$, and the product is replaced by the intersection pairing or the Poincar{\'e} dual of the Pontryagin square. Concretely, we are interested in evaluating the generalized discrete Fourier transformation 
\begin{eqnarray}\label{eq:useful1}
\begin{split}
    \frac{1}{|H_2(M_4,\Z_N)|}\sum_{\substack{\tau,\tau'\in H_2(M_4,\Z_N),\\ \tau+\tau'= 2\eta\in H_2(M_4,\Z_N)}} \exp\left(\frac{2\pi i}{2N} (\CP([\tau]) + \CP([\tau'])) -\frac{2\pi i}{N} (\braket{a,\sigma} + \braket{b,\tau})\right)~,
\end{split}
\end{eqnarray}
where $a,b\in H_2(M_4,\Z_N)$, and $[\tau]$ is the Poincar{\'e} dual of $\tau$. The condition $\tau+\tau'= 2\eta\in H_2(M_4,\Z_N)$ simply means that $\tau+\tau'$ is an even element in $H_2(M_4,\Z_N)$, i.e. $\tau+\tau'$ is trivial when regarded as an element in $H_2(M_4,\Z_2)$. The sum \eqref{eq:useful1} is invariant under the gauge transformation 
\begin{eqnarray}
a\to a+ \frac{N}{2}\lambda, \hspace{1cm} b\to b+ \frac{N}{2}\lambda~,
\end{eqnarray}
and hence $(a,b)$ can be regarded as an element in $H_2(M_4,(\Z_N\times \Z_N)/\Z_2)$. Following the same discussion which lead to \eqref{eq:DFT2}, we can complete the square in \eqref{eq:useful1}, and find
\begin{equation}\label{eq:C36}
\begin{split}
    \exp\left(-\frac{2\pi i}{2N}(\CP([a])+ \CP([b]))\right) \frac{1}{|H_2(M_4,\Z_N)|}\sum_{\substack{\tau,\tau'\in H_2(M_4,\Z_N),\\ \tau+\tau'+a+b= 2\eta\in H_2(M_4,\Z_N)}} \exp\left(\frac{2\pi i}{2N} (\CP([\tau]) + \CP([\tau'])) \right)~.
\end{split}
\end{equation}
The sum is gauge invariant, and defines an invertible TQFT $\CZ_Y(M_4)$ as follows,
\begin{equation}\label{eq:Euler4dNeven}
\begin{split}
     \frac{1}{|H_2(M_4,\Z_N)|}\sum_{\substack{\tau,\tau'\in H_2(M_4,\Z_N),\\ \tau+\tau'= 2\eta\in H_2(M_4,\Z_N)}} \exp\left(\frac{2\pi i}{2N} (\CP([\tau]) + \CP([\tau'])) \right)=
     \begin{cases}
     \CZ_{Y}[M_4, \frac{\Z_N\times \Z_N}{\Z_2}]~, & N\in 4\Z\\
     0, & N\in 4\Z+2
     \end{cases}
     \\
     \frac{1}{|H_2(M_4,\Z_N)|}\sum_{\substack{\tau,\tau'\in H_2(M_4,\Z_N),\\ \tau+\tau'+\xi= 2\eta\in H_2(M_4,\Z_N)}} \exp\left(\frac{2\pi i}{2N} (\CP([\tau]) + \CP([\tau'])) \right)=
     \begin{cases}
     0, & N\in 4\Z\\
     \CZ_{Y}[M_4, \frac{\Z_N\times \Z_N}{\Z_2}]~, & N\in 4\Z+2
     \end{cases}
     \\
\end{split}
\end{equation}
where $\xi$ is an generator of $H_2(M_4,\Z_N)$, and the sum is independent of the choice of $\xi$. In particular, we can choose $\xi=[w_2^{TM}]\mod 2$. 
The above formula is a generalization of \eqref{eq:C27} and is similar to $\CZ_Y$ for the odd $N$ case, c.f. \eqref{eq:Yfunc}. But the phase in \eqref{eq:C36}  is not gauge invariant. This enforces that \eqref{eq:useful1} vanishes for certain $(N,a,b)$. To find the vanishing condition, we perform the gauge transformation 
\begin{equation}
\exp\left(-\frac{2\pi i}{2N}(\CP([a])+ \CP([b]))\right) \to e^{-\pi i \braket{a+b,\lambda} - \pi i \frac{N}{2} \braket{w_2^{TM}, \lambda}} \exp\left(-\frac{2\pi i}{2N}(\CP([a])+ \CP([b]))\right)~.
\end{equation}
Because $\lambda$ is arbitrary, we require $a+b+\frac{N}{2}w_2^{TM}$ to be a trivial element in $H_2(M_4,\Z_2)$ for the sum \eqref{eq:useful1} to be nontrivial. In summary, the sum \eqref{eq:useful1} equals
\begin{equation}\label{eq:usefulresult1}
    \begin{split}
         \exp\left(-\frac{2\pi i}{2N}(\CP([a])+ \CP([b]))\right) \CZ_{Y}\left[M_4, \frac{\Z_N\times \Z_N}{\Z_2}\right] \delta_{a+b+\frac{N}{2}w_2^{TM}\,\,\mathrm{mod}\,\,2}~.
    \end{split}
\end{equation}

\paragraph{Second Formula:}
We further consider the summation
\begin{eqnarray}
\frac{1}{N^2/4} \sum_{\substack{(x,y)\in (\Z_N\times \Z_N)/\Z_2,\\ x+y=0\mod 2}} \exp\left( \frac{2\pi i}{N} (-x a + y b)\right) \delta_{a+b\mod 2} = \delta_{a} \delta_{b}~.
\end{eqnarray}
Upgrading the summation variables to elements in homology, we find 
\begin{equation}\label{eq:useful2}
\frac{|H_2(M_4,\Z_2)|^2}{|H_2(M_4,\Z_N)|^2} \sum_{\substack{(\tau,\tau')\in H_2(M_4,(\Z_N\times \Z_N)/\Z_2)\\ \tau+\tau'=0\mod 2}} \exp\left( \frac{2\pi i}{N} (-\braket{\tau', \sigma}+ \braket{\tau, \sigma'})\right) \delta_{\sigma+\sigma'\mod 2} = \delta_{\sigma}\delta_{\sigma'} ~.
\end{equation}

\subsubsection{Charge conjugation $C= D_{\text{EM}}^2$}

We now return to the evaluation of powers of $D_{\text{EM}}$. Taking the square of \eqref{eq:DEMevenN}, we obtain 
\begin{equation}
\begin{split}
    &C(M_4)= D_{\text{EM}}(M_4)^2 \\
    &=  \frac{|H^0(M_4, (\Z_N\times \Z_N)/\Z_2)|^2}{|H^1(M_4, (\Z_N\times \Z_N)/\Z_2))|^2} \sum_{\substack{(\sigma, \sigma')\in \\ H_2(M_4, (\Z_N\times \Z_N)/\Z_2)}} \exp\left(\frac{2\pi i}{N} (\braket{\sigma', \sigma'} + \braket{\sigma, \sigma'}+ \frac{N}{2}\braket{[w_2^{TM}], \sigma'})\right) \\& \hspace{1cm} \times S_{(1,-1)}(\sigma) S_{(1,1)}(\sigma')  \sum_{\substack{(\tau, \tau')\in \\ H_2(M_4, (\Z_N\times \Z_N)/\Z_2)}} \exp\bigg( \frac{2\pi i}{N} (\braket{\tau',\tau'}+ \braket{\tau,\tau}- \braket{\sigma+\sigma'+\frac{N}{2}[w_2^{TM}], \tau'} \\&\hspace{1cm}- \braket{\sigma-\sigma'+\frac{N}{2}[w_2^{TM}], \tau})\bigg)~.
\end{split}
\end{equation}
Summing over $(\tau,\tau')\in H_2(M_4, (\Z_N\times \Z_N)/\Z_2))$ can be reorganized as summing over $\widetilde{\tau}:=\tau'+\tau$ and $\widetilde{\tau}':=\tau'-\tau$ in $H_2(M_4, \Z_N)$ respectively, but with the constraint $\widetilde{\tau}\pm \widetilde{\tau}'$ being an even element in $H_2(M_4,\Z_N)$, or equivalently a trivial element in $H_2(M_4,\Z_2)$. This gives 
\begin{equation}
\begin{split}
    &\sum_{\substack{(\tau, \tau')\in \\ H_2(M_4, (\Z_N\times \Z_N)/\Z_2)}} \exp\Bigg( \frac{2\pi i}{N}\bigg( \braket{\tau',\tau'}+ \braket{\tau,\tau}- \braket{\sigma+\sigma'+\frac{N}{2}[w_2^{TM}], \tau'} - \braket{\sigma-\sigma'+\frac{N}{2}[w_2^{TM}], \tau}\bigg)\Bigg)\\
    &=\sum_{\substack{\widetilde{\tau}, \widetilde{\tau}'\in H_2(M_4, \Z_N)\\ \widetilde{\tau}+\widetilde{\tau}'=2\eta\in H_2(M_4,\Z_N)}}\exp\Bigg(\frac{2\pi i}{2N} (\CP([\widetilde{\tau}])+ \CP([\widetilde{\tau}']))  -\frac{2\pi i}{N}(\braket{\sigma+\frac{N}{2}[w_2^{TM}], \widetilde{\tau}}+\braket{\sigma', \widetilde{\tau}'})\Bigg)~.  \\
\end{split}
\end{equation}
In particular,  the gauge invariance under $(\sigma,\sigma')\to (\sigma+\frac{N}{2}\lambda, \sigma'+ \frac{N}{2}\lambda)$ is manifest for the first line, and can be seen in the second line from the constraint in the summation domain $\widetilde{\tau}+\widetilde{\tau}'=2\eta\in H_2(M_4,\Z_N)$. 
The last expression is of precisely the form \eqref{eq:useful1}, where $(a,b)= (\sigma+\frac{N}{2}[w_2^{TM}], \sigma')$, and hence we can apply the result \eqref{eq:usefulresult1} here. The sum simplifies to 
\begin{eqnarray}
|H_2(M_4,\Z_N)| \CZ_{Y}[M_4, \frac{\Z_N\times \Z_N}{\Z_2}] \exp\left( - \frac{2\pi i}{2N} (\CP([\sigma]+\frac{N}{2}w_2^{TM})+ \CP([\sigma']))\right) \delta_{\sigma+\sigma'\,\,\mathrm{mod}\,\,2}~.\no\\
\end{eqnarray}
Substituting the above calculations into the expression of $C(M_4)$, we get
\begin{equation}\label{eq:Ceven}
    \begin{split}
        &C(M_4)=  \frac{|H^0(M_4, (\Z_N\times \Z_N)/\Z_2)|^2 |H_2(M_4,\Z_N)|}{|H^1(M_4, (\Z_N\times \Z_N)/\Z_2))|^2} \CZ_{Y}[M_4, \frac{\Z_N\times \Z_N}{\Z_2}]  \\& \hspace{2cm} \times \sum_{\substack{(\sigma, \sigma')\in  H_2(M_4, (\Z_N\times \Z_N)/\Z_2),\\ \sigma+\sigma' =0 \in H_2(M_4,\Z_2)}} \exp\bigg(\frac{2\pi i}{N} (\braket{\sigma', \sigma'} + \braket{\sigma, \sigma'}+ \frac{N}{2}\braket{[w_2^{TM}], \sigma'} \\&\hspace{2cm}-\frac{1}{2} \CP([\sigma]+ \frac{N}{2}w_2^{TM})-\frac{1}{2} \CP([\sigma']))\bigg)  S_{(1,-1)}(\sigma)S_{(1,1)}(\sigma') ~.
    \end{split}
\end{equation}

\subsubsection{Invertibility of $D_{\text{EM}}$}
We finally show that $D_{\text{EM}}^4 = C^2$ is the identity up to an Euler counterterm and an invertible phase $\CZ_{Y}$. Taking the square of \eqref{eq:Ceven}, we find 
\begin{equation}
    \begin{split}
        &D_{\text{EM}}(M_4)^4 = C(M_4)^2\\
        &= \frac{|H^0(M_4, (\Z_N\times \Z_N)/\Z_2)|^4 |H_2(M_4,\Z_N)|^2}{|H^1(M_4, (\Z_N\times \Z_N)/\Z_2))|^4} \CZ_{Y}[M_4, \frac{\Z_N\times \Z_N}{\Z_2}]^2 \\&\times \sum_{\substack{(\sigma, \sigma'),(\tau,\tau')\in \\ H_2(M_4, (\Z_N\times \Z_N)/\Z_2),\\ \sigma+\sigma', \tau+\tau'=0 \in\\ H_2(M_4,\Z_2)}} \exp\bigg(\frac{2\pi i}{N} (\frac{1}{2}\CP([\sigma'])- \frac{1}{2}\CP([\sigma])+ \braket{\sigma,\sigma'}- \frac{N}{2}\braket{\sigma-\sigma',[w_2^{TM}]}\\&- \frac{N^2}{4}\braket{[w_2^{TM}],[w_2^{TM}]}-\braket{\tau',\sigma}+ \braket{\tau,\sigma'})\bigg)    S_{(1,-1)}(\sigma)S_{(1,1)}(\sigma') ~.
    \end{split}
\end{equation}
We further sum over $\tau,\tau'$ by applying the formula \eqref{eq:useful2}, which constrains $\sigma,\sigma'$ to be trivial provided $\sigma+\sigma'=0\mod 2$. Hence the above expression simplifies to 
\begin{equation}
    \begin{split}
        &D_{\text{EM}}(M_4)^4 = C(M_4)^2\\
        &= \frac{|H^0(M_4, (\Z_N\times \Z_N)/\Z_2)|^4 |H_2(M_4,\Z_N)|^2}{|H^1(M_4, (\Z_N\times \Z_N)/\Z_2))|^4} \CZ_{Y}[M_4, \frac{\Z_N\times \Z_N}{\Z_2}]^2 \frac{|H_2(M_4,\Z_N)|^2}{|H_2(M_4,\Z_2)|^2}\\
        &=\frac{|H^0(M_4, (\Z_N\times \Z_N)/\Z_2)|^4 |H^2(M_4,(\Z_N\times \Z_N)/\Z_2)|^2}{|H^0(M_4, (\Z_N\times \Z_N)/\Z_2)|^4}\CZ_{Y}[M_4, \frac{\Z_N\times \Z_N}{\Z_2}]^2 \\
        &= \chi\left[M_4, \frac{\Z_N\times \Z_N}{\Z_2}\right]^2 \CZ_{Y}\left[M_4, \frac{\Z_N\times \Z_N}{\Z_2}\right]^2~.
    \end{split}
\end{equation}

\subsubsection{Summary of algebra of co-dimension one}

We close by summarizing the algebra involving the $D_{\text{EM}}$ defect. Noting that the orientation reversal of $D_{\text{EM}}$ is
\begin{equation}\label{eq:orientationN>4}
\begin{split}
    \overline{D}_{\text{EM}}(M_4)& = \chi[M_4,(\Z_N\times \Z_N)/\Z_2]^{-1}
\frac{|H^0(M_4, (\Z_N\times \Z_N)/\Z_2)|}{|H^1(M_4, (\Z_N\times \Z_N)/\Z_2))|}\times \\&\sum_{(\sigma, \sigma')\in H_2(M_4, (\Z_N\times \Z_N)/\Z_2)} \exp\left(-\frac{2\pi i}{N}  \braket{\sigma, \sigma'}\right) S_{(1,1)}(\sigma') S_{(0,-1)}(\sigma+\frac{N}{2}[w_2^{TM}])  S_{(1,0)}(\sigma)~,
\end{split}
\end{equation}
we have 
\begin{eqnarray}
\begin{split}
    &C(M_4)= D_{\text{EM}}(M_4)^2~,\\
    & D_{\text{EM}}(M_4)^4= \chi[M_4, (\Z_N\times \Z_N)/\Z_2]^2 \CZ_{Y}[M_4, (\Z_N\times \Z_N)/\Z_2]^2~,\\
    & \overline{D}_{\text{EM}}(M_4)\times D_{\text{EM}}(M_4)=1~,\\
    & \overline{D}_{\text{EM}}(M_4)=\chi[M_4, (\Z_N\times \Z_N)/\Z_2]^{-2} \CZ_{Y}[M_4, (\Z_N\times \Z_N)/\Z_2]^{-2} C(M_4) D_{\text{EM}}(M_4)^2~.
\end{split}
\end{eqnarray}

\section{More on $\widehat S_{\{N/2,0\}}$}
\label{app.absorbK}

Given an operator $\mathcal{O}$ invariant under a symmetry $G$, upon gauging $G$ we may get a series of operators $\widehat{\mathcal{O}}$ transforming in representations of the quantum dual $\widehat G$.
In this final appendix, we discuss how to assign representations of subgroups of the quantum symmetry $\widehat G$ to operators $\widehat{\mathcal{O}}$ descending from $\mathcal{O}$ that are invariant under only a subgroup $H \subset G$. We will focus only on the case of interest to us in this paper, namely $G= \ZZ_4^{\mathrm{EM}}$, $H = \ZZ_2^{\mathrm{EM}}$, and the operator $\widehat O = \widehat S_{\{N/2,0\}}$. 

Recall from the discussion in Section \ref{sec:5dtopops} that, unlike the surface $\widehat S_{(N/2,N/2)}$ which is invariant under $\ZZ_4^{\mathrm{EM}}$, or the surface $\widehat S_{[e,m]}$ whose constituents are not invariant under any subgroup of $\ZZ_4^{\mathrm{EM}}$, the constituents of $\widehat S_{\{N/2,0\}}$ are invariant under a $\ZZ_2^{\mathrm{EM}}$ subgroup of $\ZZ_4^{\mathrm{EM}}$. That is, under $(e,m) \rightarrow (-e,-m)$ we see that both $S_{(N/2,0)}$ and $S_{(0,N/2)}$ are left unchanged. This means that $\widehat S_{[e,m]}$ can be assigned a representation of the quantum $\widehat \ZZ_2^{\mathrm{EM}}\subset \widehat \ZZ_4^{\mathrm{EM}}$, and hence that the identity line in $\widehat S_{[e,m]}$ carries a two-fold index, i.e. $\mathfrak{I}_1^p(\widehat S_{[e,m]})$ for $p = 0,1$. 

Physically, the statement is that the bare identity line $\mathfrak{I}_1^0(\widehat S_{\{N/2,0\}})$ cannot absorb a single copy of $K$, and hence when stacked with $K$ gives a distinct line $\mathfrak{I}_1^1(\widehat S_{\{N/2,0\}})$. However, it \textit{can} absorb $K^2$, and hence there are no other distinct lines generated in this way. To see how these statements arise, let us focus on the global fusion. We first consider coincident loops of $\mathfrak{I}_1^0(\widehat S_{\{N/2,0\}})$ and $K$, and ask whether this configuration can be distinguished from a loop of $\mathfrak{I}_1^0(\widehat S_{\{N/2,0\}})$ in isolation (c.f. Figure \ref{fig:LemabsorbK} and the discussion surrounding it). A loop of $K$ gives a non-trivial contribution to correlation functions if and only if it links with $n$ units of EM flux, with $n$ non-zero modulo 4. If $n=1,3$, then $K$ gives a non-trivial result, but the loop of $\mathfrak{I}_1^0(\widehat S_{\{N/2,0\}})$ evaluates to zero since the constituent lines $\mathfrak{I}_1^0(S_{(N/2,0)})$ and $\mathfrak{I}_1^0(S_{(0,N/2)})$ (which are not lines in the EM-gauged theory, but which do exist in the pregauged theory) are not invariant under the primitive generator of $\ZZ_4^{\mathrm{EM}}$. Hence in the presence of $n=1,3$ units of EM flux, one cannot distinguish the configurations with $\mathfrak{I}_1^0(S_{\{N/2,0\}})$ stacked with $K$ from the configuration with $\mathfrak{I}_1^0(S_{\{N/2,0\}})$ in isolation. However, if we consider the case of $n=2$ units of EM flux, then the loop of $K$ is again non-trivial, but now $\mathfrak{I}_1^0(S_{(0,N/2)})$ does not evaluate to zero since the constituent lines $\mathfrak{I}_1^0(S_{(N/2,0)})$ and $\mathfrak{I}_1^0(S_{(0,N/2)})$ are invariant under $\ZZ_2^{\mathrm{EM}} \subset \ZZ_4^{\mathrm{EM}}$. Thus by considering the configuration with $n=2$ units of EM flux, one \textit{can} distinguish between the configurations with $\mathfrak{I}_1^0(S_{\{N/2,0\}})$ stacked with $K$ and with $\mathfrak{I}_1^0(S_{\{N/2,0\}})$ in isolation. In other words, $\mathfrak{I}_1^0(S_{\{N/2,0\}})$ cannot absorb $K$.

We may now ask whether $\mathfrak{I}_1^0(S_{\{N/2,0\}})$ can absorb $K^2$. As before, we consider coincident loops of $\mathfrak{I}_1^0(S_{\{N/2,0\}})$ and $K^2$, and ask whether this configuration can be distinguished from a loop of $\mathfrak{I}_1^0(S_{\{N/2,0\}})$ in isolation. The main difference from before is that the loop of $K^2$ gives a non-trivial result only in the presence of an \textit{odd} number of units of EM flux. However, as we have argued before, in the presence of an odd number of units of EM flux, the loop of $\mathfrak{I}_1^0(S_{\{N/2,0\}})$ evaluates to zero. Hence $\mathfrak{I}_1^0(S_{\{N/2,0\}})$ can absorb $K^2$.

\newpage

\bibliographystyle{ytphys}
\baselineskip=.95\baselineskip
\bibliography{bib}

\end{document}